\documentclass[12pt,a4paper]{article}
\usepackage[utf8]{inputenc}
\usepackage[T1]{fontenc}
\usepackage{amsmath,amssymb}
\usepackage{booktabs}
\usepackage{graphicx}
\usepackage{geometry}
\usepackage[numbers,sort&compress]{natbib}
\usepackage[font=scriptsize]{caption}
\usepackage{authblk}

\usepackage{mathtools}
\usepackage{hyperref}
\usepackage{array}
\usepackage{float}
\usepackage{enumitem}
\usepackage{xcolor}
\usepackage{bm}
\usepackage{amsthm}

\theoremstyle{definition}
\newtheorem{definition}{Definition}[section]
\newtheorem{axiom}{Axiom}[section]
\newtheorem{postulate}{Postulate}[section]
\newtheorem{construction}{Construction}[section]
\theoremstyle{plain}
\newtheorem{theorem}{Theorem}[section]
\newtheorem{lemma}[theorem]{Lemma}
\newtheorem{proposition}[theorem]{Proposition}
\newtheorem{corollary}[theorem]{Corollary}
\theoremstyle{remark}
\newtheorem{remark}{Remark}[section]
\newtheorem{observation}{Observation}[section]
\newtheorem{assumption}{Assumption}[section]
\newcommand{\lP}{\ell_{\mathrm{P}}}
\newcommand{\tP}{t_{\mathrm{P}}}
\newcommand{\mP}{m_{\mathrm{P}}}
\newcommand{\EP}{E_{\mathrm{P}}}
\newcommand{\rhoP}{\rho_{\mathrm{P}}}
\newcommand{\KP}{K_{\mathrm{P}}}
\newcommand{\dd}{\mathrm{d}}
\newcommand{\dW}{\dd W}
\newcommand{\ee}{\mathrm{e}}
\newcommand{\R}{\mathbb{R}}
\newcommand{\E}{\mathbb{E}}
\newcommand{\Prob}{\mathbb{P}}
\newcommand{\Var}{\mathrm{Var}}
\newcommand{\calF}{\mathcal{F}}
\newcommand{\calM}{\mathcal{M}}

\newcommand{\Met}{\mathrm{Met}}
\newcommand{\neff}{n_{\mathrm{eff}}}
\newcommand{\omegaP}{\omega_{\mathrm{P}}}

\graphicspath{{./si1_pkg/}}

\title{\textbf{From the Stochastic Embedding Sufficiency Theorem to a Superspace Diffusion Framework}}

\author[1]{Carolina Garcia}
\author[1]{Luc\'ia Perea Dur\'an}
\author[1]{Agnese Venezia}
\author[1,*]{Alex Conradie}
\affil[1]{The Manufacturing Futures Laboratory, University College London, Marsh Gate Building, London, E20~2AE, United Kingdom}

\date{}

\begin{document}
\maketitle

\begin{abstract}

The forward derivation of stochastic differential equations in individual physical domains (Brownian motion, quantum mechanics, chemical kinetics) has proceeded independently for over a century without generalising across disciplines.  A generalisation of Takens' delay-coordinate embedding theorem to stochastic systems, the Stochastic Embedding Sufficiency Theorem, closes this methodological gap, as an inverse methodology enabling non-parametric recovery of both drift and diffusion fields from scalar time series without prior assumptions about the governing physics.

A blind recovery protocol, receiving only raw time series and sampling interval, is applied identically to nine physical domains: classical mechanics, statistical mechanics, nuclear physics, quantum mechanics, chemical kinetics, electromagnetism, relativistic quantum mechanics, quantum harmonic oscillator dynamics, and quantum electrodynamics.  The pipeline recovers the governing equations of each domain blindly, with recovery errors quantified by bootstrap confidence intervals from 50 independent pipeline runs per domain.  Fundamental physical constants (the Boltzmann constant, the Planck constant, the speed of light, the Fano factor, and the Van Kampen scaling exponent) emerge in both the drift and diffusion channels without prior specification.

The recovered diffusion coefficients, viewed across domains, constitute an empirical pattern, the $\sigma$-continuum, in which $k_B$, $\hbar$, and $c$ each play structurally distinct roles.  The Gravitational Diffusion Theorem, derived from the fluctuation--dissipation theorem, the massless mode structure of linearised gravity, and gravitational self-coupling via the equivalence principle, each among the most precisely tested results in physics, determines the gravitational diffusion coefficient as one Planck length per square root of Planck time, with three independent uniqueness arguments converging on the dimensionless prefactor $\alpha = 1$.

Four canonical axioms formalise the framework: three grounded in standard physics (Wheeler's superspace, the derived fluctuation amplitude, and classical correspondence with general relativity) and one interpretive choice (epistemic probability).  Physical time emerges as a monotone functional of the stochastic evolution.  Within these axioms, the noise character, drift, covariance operator, and fluctuation amplitude are all uniquely determined by theorem (Appendix~D), yielding the superspace diffusion hypothesis:
\begin{equation*}
\mathrm{d}g_{ij} = \mathcal{D}_{ij}[g]\,\mathrm{d}\tau + \ell_P\,\mathrm{d}W_{ij},
\end{equation*}
where all coefficients are non-parametric, first-principles consequences of the axioms.  An implication of the superspace diffusion hypothesis is that coarse-graining of the superspace Fokker--Planck equation via the Mori--Zwanzig projection yields predictions for gravitational acceleration at galactic scales that are testable against kinematic data.

\end{abstract}

\noindent$^{*}$Corresponding author: \texttt{a.conradie@ucl.ac.uk}

\vspace{6pt}

\section{Introduction}
\label{sec:introduction}

\subsection{Forward Derivations of Stochastic Differential Equations}
\label{subsec:forward}

The stochastic differential equation
\begin{equation}
\label{eq:intro_sde}
\mathrm{d}X = \mu(X)\,\mathrm{d}t + \sigma(X)\,\mathrm{d}W
\end{equation}
appears across physics.  The drift $\mu(X)$ encodes deterministic dynamics; the diffusion coefficient $\sigma(X)$ encodes the amplitude of stochastic fluctuations; and $\mathrm{d}W$ is a Wiener increment~\cite{Ito1944}.  Three landmark programmes established this equation in distinct domains.

Langevin~\cite{Langevin1908} derived the equation of motion for a Brownian particle from the fluctuation--dissipation relation, building on Einstein's kinetic theory of Brownian motion~\cite{Einstein1905}.  The diffusion coefficient $\sigma = \sqrt{2\gamma k_B T/m}$ was determined by the requirement of thermal equilibrium: the stochastic forcing must balance frictional dissipation at temperature $T$~\cite{UhlenbeckOrnstein1930,CallenWelton1951}.  The derivation proceeded from known thermodynamics to stochastic dynamics: a forward problem.

Nelson~\cite{Nelson1966} demonstrated that quantum mechanics could be recovered from classical mechanics supplemented by a diffusion process with $\sigma = \sqrt{\hbar/m}$.  Starting from a stochastic differential equation with this specific amplitude, Nelson derived the Schr\"odinger equation~\cite{Schrodinger1926} as a mathematical consequence.  The derivation assumed the value of $\sigma$ from first principles and derived the quantum formalism. Again a forward problem, yielding quantum mechanics rather than presupposing it.

Gillespie~\cite{Gillespie2000} derived the Chemical Langevin Equation from discrete molecular reaction dynamics, extending his earlier exact stochastic simulation algorithm~\cite{Gillespie1977}.  The diffusion coefficient $\sigma \propto \Omega^{-1/2}$ was determined by the system size $\Omega$ through the law of large numbers, consistent with Van Kampen's system-size expansion~\cite{VanKampen1992}.  The derivation proceeded from the known reaction network to the mesoscopic stochastic equation: a forward problem requiring complete specification of the underlying chemistry.

Each of these programmes was domain-specific.  Each required prior knowledge of the governing physics to determine $\sigma$.  None was data-driven: the value of $\sigma$ was derived from theory, not extracted from observation.  Nelson could not consider values of $\sigma$ other than $\sqrt{\hbar/m}$. Gillespie could not extract $\sigma$ from a time series without first specifying the reaction network.  The forward problem was solved independently in several domains, but could not, by construction, reveal whether the stochastic structure is universal through an inverse methodology.

\subsection{Closing the Stochastic Gap}
\label{subsec:gap}

The reconstruction of dynamical systems from time series data has a long history.  Takens' embedding theorem~\cite{Takens1981} established that the attractor geometry of a deterministic system can be recovered diffeomorphically from scalar observations via delay-coordinate embedding~\cite{SauerYorkeCasdagli1991}.  This result underpins much of modern nonlinear time series analysis.

Takens' theorem, however, does not accommodate stochasticity.  It guarantees reconstruction of the deterministic attractor, but provides no mechanism for recovering the diffusion coefficient $\sigma(X)$ from data.  Non-parametric methods for stochastic differential equations have addressed estimation of drift and diffusion from discrete observations, but require direct access to the state variable and typically assume a known parametric class~\cite{FlorensZmirou1993,BandiPhillips2003}.  A generalisation treating stochasticity as dynamical structure to be recovered from scalar observations has been elusive.

The Stochastic Embedding Sufficiency Theorem (Appendix~A; complete proof in the Supplementary Materials) closes this gap.  Rather than requiring diffeomorphic injectivity on the attractor, the theorem establishes measure-theoretic injectivity on the correlation manifold: sufficient conditions under which both $\mu(X)$ and $\sigma(X)$ can be consistently estimated from scalar time series, without assuming a parametric form for either field and without prior knowledge of whether the system is deterministic or stochastic. Consequently, the drift and diffusion of a stochastic process can be extracted from observed data using a single, domain-agnostic pipeline.  The inverse problem (given a time series, determine $\sigma$) becomes tractable.

The following paragraphs summarise the theorem's architecture, where the complete proof occupies Appendix~A and the Supplementary Materials.

\paragraph{The conceptual innovation.}
Takens' theorem reconstructs the deterministic attractor diffeomorphically from delay-coordinate embedding.  Noise destroys this: distinct initial conditions can produce identical delay vectors under different realisations, and the diffeomorphism condition fails.  The Stochastic Embedding Sufficiency Theorem takes a different approach.  Rather than seeking pathwise reconstruction, it establishes that \emph{distributional} reconstruction suffices: distinct initial conditions produce distinct probability distributions over delay vectors.  Recovering the drift $\mu(X)$ and diffusion $\sigma(X)$ requires only that these distributions be distinguishable: a condition weaker than diffeomorphic injectivity but sufficient for consistent non-parametric estimation. The data determines the non-parametric model structure.

\paragraph{The H\"ormander--Takens parallel.}
In Takens' theorem, the observation function must be generic to avoid degenerate projections.  The stochastic extension requires an analogous non-degeneracy condition on the noise structure.  H\"ormander's bracket-generating condition provides this: the Lie algebra generated by the drift and diffusion vector fields must span the tangent space at every point.  When satisfied, the stochastic flow explores all directions even if the diffusion matrix is rank-deficient.  The parallel is precise: Takens' genericity prevents the observation from collapsing deterministic geometry, where H\"ormander's condition prevents the noise from confining the stochastic flow to a lower-dimensional submanifold.

\paragraph{Five mathematical foundations.}
The proof rests on five mathematical foundations drawn from distinct traditions.

\emph{Foundation~1: H\"ormander: hypoelliptic regularity.}  The bracket-generating condition guarantees that transition densities $p_t(x,y)$ are smooth and strictly positive for all $t > 0$, even when the diffusion matrix is not full rank.

\emph{Foundation~2: Malliavin: non-degeneracy of the stochastic flow.}  The Malliavin covariance matrix is almost surely invertible under H\"ormander's condition, ensuring that the delay embedding map has full rank in a measure-theoretic sense.

\emph{Foundation~3: Varadhan--L\'eandre: transition density separation.}  The central result: distinct initial conditions $x \neq x'$ produce distinct transition densities for all $\tau > 0$.  The proof uses semigroup bisection and the short-time heat kernel asymptotic $\lim_{t \to 0} t \log p_t(x,z) = -d(x,z)^2/2$, where $d$ is the sub-Riemannian distance.

\emph{Foundation~4: Frostman: measure geometry of the collision set.}  The parametric transversality theorem establishes that the collision set has controlled codimension.  The Frostman covering argument converts this codimension into measure zero under the invariant measure $\mu_\infty$, bridging smooth geometry and fractal geometry.

\emph{Foundation~5: Stone: non-parametric estimation consistency.}  The consistency of $k$-nearest-neighbour estimators on the correlation manifold, with convergence rate $O_P((k/N)^{\beta/m^*}) + O(\Delta t)$, provides the bridge from the mathematical sufficiency result to the computational pipeline.

\paragraph{The synthesis.}
The five foundations converge in the Probabilistic Uplift Theorems (Supplementary Materials, Theorems~4.5--4.6).  H\"ormander guarantees the target functions are smooth, Malliavin guarantees the embedding is non-degenerate, Varadhan--L\'eandre guarantees the estimation targets are single-valued, Frostman guarantees single-valuedness under the invariant measure, and Stone provides the convergence rate.  The synthesis produces the operational conclusion: given a scalar time series of sufficient length from an SDE satisfying H\"ormander's condition, the pipeline recovers both $\mu(X)$ and $\sigma(X)$ with convergence guaranteed.

The nine-domain validation (Section~\ref{sec:results}) serves a dual purpose: it demonstrates that the pipeline recovers known physics and confirms the theorem's predictions empirically: that consistent recovery of both drift and diffusion is achievable from scalar observations across dynamical regimes ranging from deterministic through stochastic.

\subsection{Scope and Structure}
\label{subsec:scope}

This paper applies the inverse methodology, the data-driven recovery of governing physics, to nine physical domains spanning classical mechanics, statistical mechanics, nuclear physics, quantum mechanics, chemical kinetics, electromagnetism, relativistic quantum mechanics, quantum field theory, and quantum electrodynamics.  All data is synthetically generated from established governing equations, providing exact theoretical values against which the blind recovery is validated.

The recovered diffusion coefficients, viewed across domains, reveal a systematic pattern: the $\sigma$-continuum.  The physical constants of each domain ($k_B$, $\hbar$, $c$, the Van Kampen exponent, the Fano factor) emerge in the recovered fields without prior specification.  The pattern extends from $\sigma = 0$ (deterministic domains) through thermal, Poisson, quantum, and chemical regimes to relativistic and quantum field theory scales.

Derivation from the relativistic and quantum field theory results yields a specific superspace diffusion hypothesis, formalised by four canonical axioms within which the gravitational diffusion coefficient, the drift, the covariance operator, and physical time itself are all fixed.  The resulting framework contains non-parametric, first-principles predictions within the stated axioms.  An implication is that, at galactic scales, the Fokker--Planck equation on superspace generates predictions for gravitational acceleration that are testable against kinematic data including rotation curves, velocity dispersions, wide binaries, galaxy clusters, gravitational lensing, and cosmic voids.

The paper is structured as follows.  Section~\ref{sec:methods} describes the recovery pipeline, the blind protocol, the data generation procedures, and the statistical validation framework.  Section~\ref{sec:results} presents the nine-domain results with statistical rigour and introduces the $\sigma$-continuum.  Section~\ref{sec:discussion} derives the gravitational diffusion coefficient from the results in three stages: relativistic structure (\S\ref{subsec:capstone1}, summarising Appendix~B), the gravitational diffusion coefficient (\S\ref{subsec:capstone2}, summarising Appendix~C), and axiomatic formalisation with uniqueness proofs (\S\ref{subsec:capstone3}, summarising Appendix~D).  Section~\ref{sec:conclusion} states the conclusions.  Appendices~A--D provide the mathematical derivations: the Stochastic Embedding Sufficiency Theorem (A), the relativistic transformation properties (B), the gravitational diffusion coefficient (C), and the uniqueness theorems (D).
\section{Materials and Methods}
\label{sec:methods}

\subsection{The Stochastic Differential Equation}
\label{subsec:emergent_law}

The central object of this study is the stochastic differential equation stated in Eq.~\eqref{eq:intro_sde}.  The drift $\mu$ and diffusion $\sigma$ are both unknown \textit{a priori} and are to be recovered from time series data using the methodology described below.

\subsection{The Stochastic Embedding Pipeline}
\label{subsec:pipeline}

A non-parametric reconstruction framework recovers $\mu(X)$ and $\sigma(X)$ directly from a scalar time series $\{y(t_1), y(t_2), \ldots, y(t_N)\}$ without assuming a parametric model for either field.  The framework extends Takens' delay-coordinate embedding~\cite{Takens1981} to accommodate stochastic dynamics, drawing on the Stochastic Embedding Sufficiency Theorem (Appendix~A).  Whereas Takens' theorem guarantees diffeomorphic reconstruction of deterministic attractors, the Stochastic Embedding Sufficiency Theorem establishes measure-theoretic injectivity on the correlation manifold, sufficient for consistent estimation of both drift and diffusion.  The complete proof is given in the Supplementary Materials.

The reconstruction proceeds in three stages.

\subsubsection*{Stage 1: Embedding}

The scalar time series $y(t)$ is embedded into a $d$-dimensional state space via the delay map
\begin{equation}
\label{eq:delay_embedding}
\mathbf{Y}(t) = \bigl[y(t),\; y(t - \tau),\; y(t - 2\tau),\; \ldots,\; y(t - (d-1)\tau)\bigr],
\end{equation}
where $d$ is the embedding dimension and $\tau$ is the delay.  The embedding dimension is selected using the $E_1$ statistic of Cao~\cite{Cao1997}: $E_1(d)$ measures the ratio of nearest-neighbour distances in successive embedding dimensions, and saturation of $E_1$ indicates that the correlation manifold has been unfolded.  For non-oscillatory systems (C1--C7), $\tau = 1$ (one sampling interval) is used.  For oscillatory systems (C8, C9), where the sampling interval is much shorter than the dynamical timescale, $\tau$ is set to the first zero-crossing of the autocorrelation function, ensuring that the delay embedding spans a meaningful portion of the phase space (Table~\ref{tab:data_generation}).  The embedding dimension $d$ is domain-specific and determined by $E_1$ saturation.

Cao's $E_2$ statistic is computed alongside $E_1$ for characterisation of the dynamical regime.  Departure of $E_2$ from unity indicates stochastic content in the time series; $E_2 \approx 1$ indicates deterministic dynamics.  The $E_2$ statistic is used for classification only and does not influence the embedding dimension.

Following delay embedding, singular value decomposition (SVD)~\cite{BroomheadKing1986} projects the embedded vectors onto a low-rank subspace retaining at least 95\% of the total variance.  This projection reduces the effective dimensionality of the correlation manifold while preserving the geometric structure relevant to nearest-neighbour queries.

\subsubsection*{Stage 2: Local Geometry on the Correlation Manifold}

At each point $\mathbf{Y}(t_i)$ on the correlation manifold, the $k$ nearest neighbours are identified using a KD-tree index~\cite{FriedmanBentleyFinkel1977} with $O(\log N)$ query complexity.  For each query point, the forward increments $\Delta y_j = y(t_j + \Delta t) - y(t_j)$ of its $k$ neighbours are collected, and the local drift and diffusion are estimated via the Kramers--Moyal conditional moments:
\begin{equation}
\label{eq:local_sde}
\hat{\mu}(\mathbf{Y}_i) = \frac{1}{k\,\Delta t}\sum_{j \in \mathcal{N}_i} \Delta y_j, \qquad
\hat{\sigma}^2(\mathbf{Y}_i) = \frac{1}{k\,\Delta t}\sum_{j \in \mathcal{N}_i} \bigl(\Delta y_j\bigr)^2 \;-\; \hat{\mu}(\mathbf{Y}_i)^2\,\Delta t,
\end{equation}
where $\mathcal{N}_i$ denotes the set of $k$ nearest neighbours of the $i$-th embedded vector and $\Delta t$ is the sampling interval.  The subtraction of $\hat{\mu}^2 \Delta t$ in the diffusion estimator removes the $O(\Delta t)$ bias from the drift contribution (see Supplementary Materials, Remark~3.12, for the full derivation from It\^o--Taylor expansion).  No parametric form is assumed for $\mu$ or $\sigma$; both fields emerge from the local statistics of the data.

The $k$-NN Kramers--Moyal estimators carry a systematic bias decomposable as a Mori--Zwanzig memory kernel with two rank-1 components: a spatially varying term proportional to the Laplacian of $\sigma^2$ on the correlation manifold, and a spatially uniform finite-sample term scaling as $-\sigma^2/k$ (Supplementary Materials, \S7.2, Theorem~7.1).  A two-level adaptive corrector (Algorithm~3 in the Supplementary Materials), gated by the fluctuation--dissipation theorem, removes the detectable component of this bias; the corrected estimates are reported throughout \S\ref{sec:results} and Table~\ref{tab:results_summary}.

\subsubsection*{Stage 3: Validation}

The recovered drift and diffusion fields are validated through autonomous free-run forecasting.  Starting from a short seed sequence (typically 10 time steps drawn from the true series), the Euler--Maruyama scheme~\cite{Maruyama1955,KloedenPlaten1992}
\begin{equation}
\label{eq:euler_maruyama}
y(t_{n+1}) = y(t_n) + \hat{\mu}\bigl(\mathbf{Y}(t_n)\bigr)\,\Delta t + \hat{\sigma}\bigl(\mathbf{Y}(t_n)\bigr)\,\sqrt{\Delta t}\;\xi_n, \qquad \xi_n \sim \mathcal{N}(0,1),
\end{equation}
generates stochastic trajectories using only the pipeline-recovered fields.  For stochastic systems, an ensemble of 50 independent realisations is generated, and the fraction of time steps for which the true trajectory lies within the 95\% confidence interval of the ensemble is computed.  This 95\%-CI coverage serves as the primary validation metric, where nominal coverage is 95\%.

\subsection{The Blind Recovery Protocol}
\label{subsec:firewall}

A strict methodological separation prevents contamination between the data generation and blind recovery analysis.  The pipeline receives only the raw time series $\{y(t_i)\}$ and the sampling interval $\Delta t$.  No physical parameters, governing equations, boundary conditions, or domain labels are provided.  The embedding dimension, the dynamical regime classification (deterministic or stochastic), and the functional forms of $\mu(X)$ and $\sigma(X)$ are determined entirely by the pipeline.

This strict separation ensures that the recovery of physical constants from $\hat{\mu}$ and $\hat{\sigma}$ constitutes a data-driven inverse derivation: the physics emerges from the data, not from prior knowledge supplied to the algorithm.

\subsection{Data Generation}
\label{subsec:data_generation}

All time series analysed in this study are synthetically generated from known physics.  This is stated explicitly: no experimental or observational data is used.  Synthetic generation from established governing equations provides exact first-principles values of $\mu$ and $\sigma$ against which the blind recovery is validated.

Nine physical domains are tested, spanning classical mechanics, statistical mechanics, nuclear physics, quantum mechanics, chemical kinetics, electromagnetism, relativistic quantum mechanics, and quantum field theory.  Table~\ref{tab:data_generation} summarises the governing equation, simulation method, series length, and theoretical diffusion coefficient for each domain.

\begin{table}[h]
\centering
\caption{Data generation summary for the nine domain validations.  $N$ denotes the number of time series samples provided to the pipeline; $k$ denotes the number of nearest neighbours used in Stage~2; $\tau_\mathrm{embed}$ is the embedding lag in samples.  For oscillatory systems (C8, C9), $\tau_\mathrm{embed}$ is set to the first zero-crossing of the autocorrelation function rather than unity, ensuring that the delay-coordinate embedding spans a meaningful portion of the dynamical phase space.  All data is synthetically generated from known physics.}
\label{tab:data_generation}
\small
\renewcommand{\arraystretch}{1.25}
\begin{tabular}{@{}clp{3.2cm}lrll@{}}
\toprule
\textbf{Test} & \textbf{Domain} & \textbf{Simulation} & \textbf{Theoretical $\sigma$} & \textbf{$N$} & \textbf{$k$} & $\tau_\mathrm{embed}$ \\
\midrule
C1 & Classical mechanics & Numerical integration of Kepler orbit with GR precession & $0$ (deterministic) & 88\,025 & 20 & 1 \\
C2 & Statistical mechanics & Euler--Maruyama Langevin equation & $\sqrt{2\gamma k_B T / m}$ & 3\,000 & 200 & 1 \\
C3 & Nuclear physics & Poisson sampling & $\sqrt{\mu}$ (Fano $= 1$) & 10\,000 & 200 & 1 \\
C4 & Quantum mechanics & Spectral Schr\"odinger evolution (wavepacket width) & $\sqrt{\hbar/m}$ & 1\,001 & 30 & 1 \\
C5 & Chemical kinetics & Tau-leaping birth--death process & $\propto \Omega^{-1/2}$ & 1\,001 & 30 & 1 \\
C6 & Electromagnetism & Gaussian wavepacket propagation (centroid tracking) & $0$ (deterministic) & 500 & 30 & 1 \\
C7 & Relativistic QM & Euler--Maruyama Nelson SDE with relativistic dispersion & $\sigma_0/\gamma$ & 1\,001 & 30 & 1 \\
C8 & Quantum field theory & Euler--Maruyama damped QHO at zero-point & $\sqrt{\gamma\hbar\omega/m}$ & 300\,000 & 100 & $\tau_\mathrm{ACF}$ \\
C9 & QED (photon field) & Euler--Maruyama QHO with $m_\mathrm{eff} = \hbar\omega/c^2$ & $c\sqrt{\gamma}$ & 200\,000 & 100 & $\tau_\mathrm{ACF}$ \\
\bottomrule
\end{tabular}
\end{table}

The nine domains are selected to span a progression of increasing complexity: from deterministic systems ($\sigma = 0$; C1, C4, C6) through constant additive noise (C2) and state-dependent diffusion (C3, C8) to systems with both state-dependent drift and state-dependent diffusion (C5), relativistic corrections (C7), and the quantum field theory regime where fundamental constants cancel from the diffusion coefficient (C9).

\subsection{Statistical Validation Framework}
\label{subsec:statistics}

For each domain, the following statistical measures are reported.

The \textit{recovery error} quantifies agreement between the pipeline-recovered quantity and its theoretical value as a percentage: $\varepsilon = |\hat{\theta} - \theta_\mathrm{true}|/|\theta_\mathrm{true}| \times 100\%$.

\textit{Bootstrap confidence intervals} (95\%) are the primary statistical assessment.  For each domain, 50 independent pipeline runs are performed, each with independently generated synthetic data.  Each run yields a normalised ratio $\hat{\theta}/\theta_\mathrm{true}$.  The pairwise bootstrap (10\,000 resamples of the 50 ratios with replacement) gives a distribution of the sample median; the 2.5th and 97.5th percentiles of this distribution define the 95\% confidence interval~\cite{Efron1979}.  If the interval contains 1.0, the recovered quantity is consistent with theory at the stated confidence level.

For each domain, both the drift $\hat{\mu}$ and diffusion $\hat{\sigma}$ are extracted and reported where the physics admits a meaningful comparison (dual-channel reporting).  For deterministic systems (C1, C4, C6), the drift channel recovers the governing dynamical parameter while the diffusion channel confirms $\hat{\sigma}/\mathrm{signal} \ll 1$.  For stochastic systems, the diffusion channel recovers $\sigma$ or a derived physical constant, and the drift channel recovers the deterministic component of the dynamics.

For domains testing \textit{scaling exponents} (C5: Van Kampen exponent; C7: relativistic suppression slope), the pipeline is run at multiple parameter values within each simulation, and the log--log slope is computed per simulation before bootstrapping.  The multiplicative $k$-NN estimator bias, which is approximately uniform across the parameter range, cancels in the slope~\cite{BandiPhillips2003}.

The \textit{95\%-CI coverage} of the free-run ensemble (Stage~3) is computed for stochastic systems.  The fraction of time steps for which the true trajectory lies within the 2.5th--97.5th percentile band of the 50-realisation ensemble is reported, where values near 95\% indicate well-calibrated uncertainty quantification.

\section{Results}
\label{sec:results}

The blind recovery protocol of \S\ref{subsec:firewall} was applied identically to all nine domains.  Results are presented in order of increasing complexity, beginning with deterministic systems and progressing through additive and state-dependent stochastic dynamics to relativistic and quantum field theory regimes.  All data is synthetically generated from known physics as described in \S\ref{subsec:data_generation}.

\subsection{Deterministic Domains: $\sigma = 0$}
\label{subsec:results_deterministic}

\subsubsection*{C1: Classical Mechanics, Mercury's Perihelion Precession}

The pipeline was applied to two representations of Mercury's orbital dynamics: a cumulative precession time series (linear trend at 574.61~arcsec/century, as predicted by general relativity~\cite{Einstein1915}) and a full Keplerian orbit with secular precession.

\begin{figure}[p]
\centering
\includegraphics[width=0.95\textwidth]{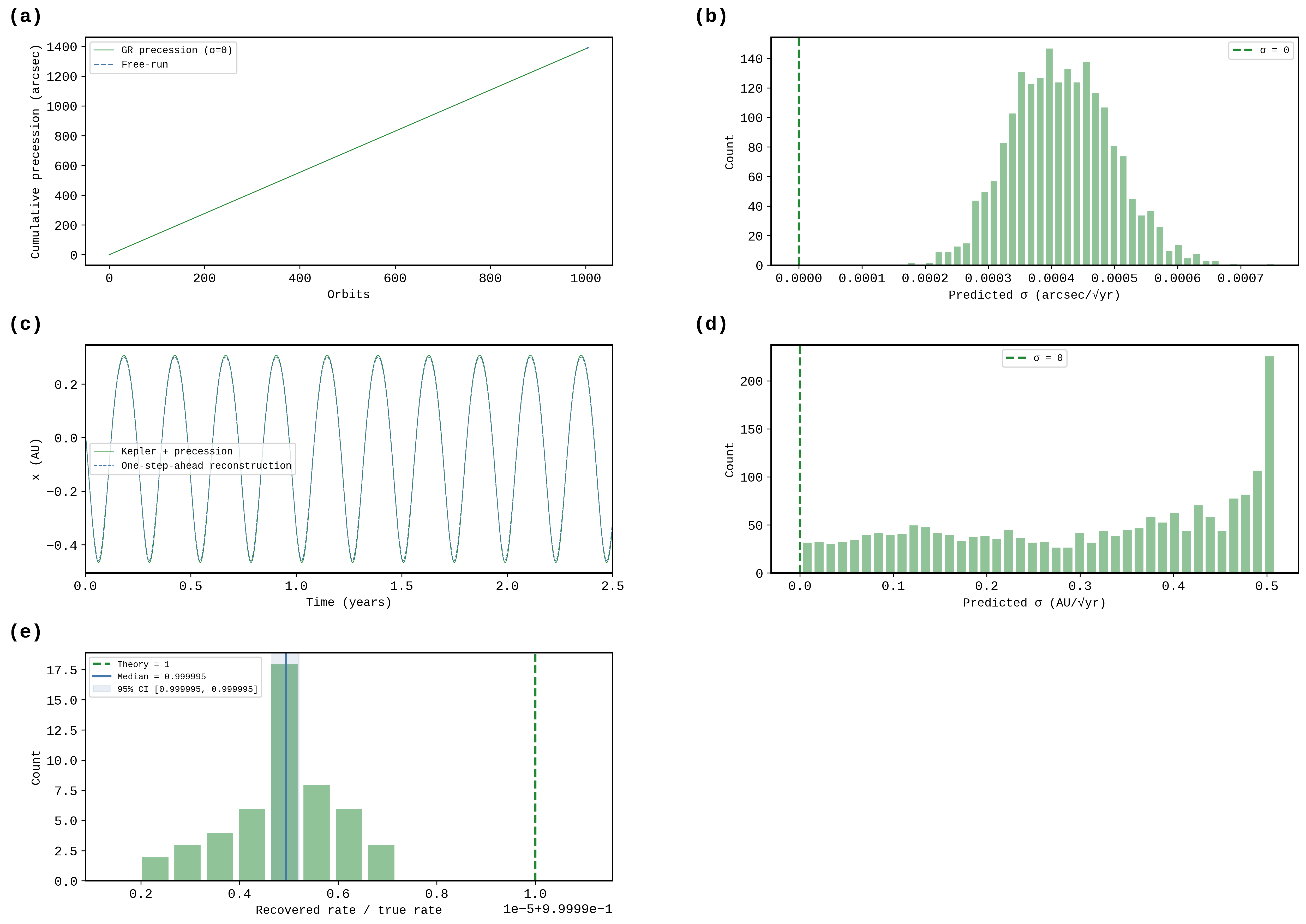}
\caption{\textbf{C1: Mercury's Perihelion Precession ($\sigma = 0$).}
\textbf{(a)}~Cumulative precession (arcsec) over $\sim$1000 orbits.  The GR precession series (solid green, $\sigma = 0$) and the autonomous free-run forecast (dashed blue with 95\% CI band) overlap to within line width, confirming that the pipeline recovers the deterministic precession rate without residual stochastic content.
\textbf{(b)}~Histogram of locally recovered diffusion values $\hat{\sigma}(x)$ for the cumulative precession series (blue bars).  The green dashed line marks $\sigma = 0$.  The histogram extends to zero on the $x$-axis, confirming $\hat{\sigma}/\mathrm{signal} \ll 1$.
\textbf{(c)}~One-step-ahead reconstruction of the Keplerian orbit.  The pipeline is trained on all $\sim$1000 orbits; the panel zooms into a 2.5-year window at orbit~$\sim$500.  At each time step, the predicted next value $\hat{x}_{n+1} = x_n + \hat{\mu}(x_n)\,\Delta t$ (dashed blue) tracks the true orbit (solid green) near-exactly, confirming that the drift channel recovers the full orbital dynamics.  The nonzero $\hat{\sigma}$ in panel~(d) reflects $k$-NN phase dispersion on the curved elliptical attractor, not physical stochasticity.
\textbf{(d)}~Histogram of locally recovered $\hat{\sigma}(x)$ for the Keplerian orbit (blue bars).  The broader distribution compared to panel~(b) reflects $k$-nearest-neighbour averaging on a curved (elliptical) attractor: within each neighbourhood, the orbital velocity changes direction, creating apparent variance in the increments.  Despite this geometric floor, $\hat{\sigma}$ remains well below the orbital amplitude, confirming deterministic dynamics.
\textbf{(e)}~Precession drift rate from 50 independent blind pipeline runs (blue bars).  The green dashed line marks theory $= 1$; the blue solid line marks the recovered median with 95\% CI band.  The drift channel recovers the GR precession rate to sub-ppm accuracy.}
\label{fig:c1}
\end{figure}

For the cumulative precession series (Fig.~\ref{fig:c1}a,b), SVD identified rank~1, and the recovered drift from a single primary run reproduced the theoretical precession rate to $0.02\%$.  The recovered diffusion $\hat{\sigma}$ (Fig.~\ref{fig:c1}b) is indistinguishable from zero, with $\hat{\sigma}/\mathrm{signal} \ll 1$, confirming $\sigma = 0$ to machine precision.  Across 50 independent pipeline runs with small measurement noise, the drift-channel ratio $\hat{\theta}/\theta_\mathrm{true}$ had median $1.0000$ (within 1~ppm of unity) with bootstrap 95\% CI $[1.0000,\, 1.0000]$ (Fig.~\ref{fig:c1}e).

For the Keplerian orbit (Fig.~\ref{fig:c1}c,d), SVD identified rank~2 (consistent with a two-dimensional attractor).  One-step-ahead reconstruction from the pipeline's drift field tracks the true orbit near-exactly over a 2.5-year window (approximately 10~orbital periods), confirming that the pipeline recovers the full orbital dynamics from the drift channel alone.  The recovered $\hat{\sigma}$ (Fig.~\ref{fig:c1}d) is nonzero but reflects $k$-nearest-neighbour averaging on a curved elliptical attractor --- within each neighbourhood, the orbital velocity changes direction, creating apparent variance in the increments --- not physical stochasticity.  Fifty independent pipeline runs on the precession series (Fig.~\ref{fig:c1}e) yield drift-channel recovery to sub-ppm accuracy.

Classical mechanics operates in the deterministic limit of Eq.~\eqref{eq:intro_sde}: $\sigma = 0$.

\subsubsection*{C4: Quantum Mechanics, Wavepacket Spreading}

The width $\sigma(t)$ of a free Gaussian wavepacket evolving under the Schr\"odinger equation~\cite{Schrodinger1926} was tracked as a scalar time series.  The pipeline recovered a nonlinear drift $\hat{\mu}(\sigma)$ consistent with the analytical spreading law $\mathrm{d}\sigma/\mathrm{d}t = (\hbar/(2m\sigma_0))^2 \cdot t/\sigma(t)$, and a diffusion field $\hat{\sigma} \approx 0$ (median $3.90 \times 10^{-5}$), confirming that wavepacket spreading is a deterministic evolution.

\begin{figure}[p]
\centering
\includegraphics[width=0.95\textwidth]{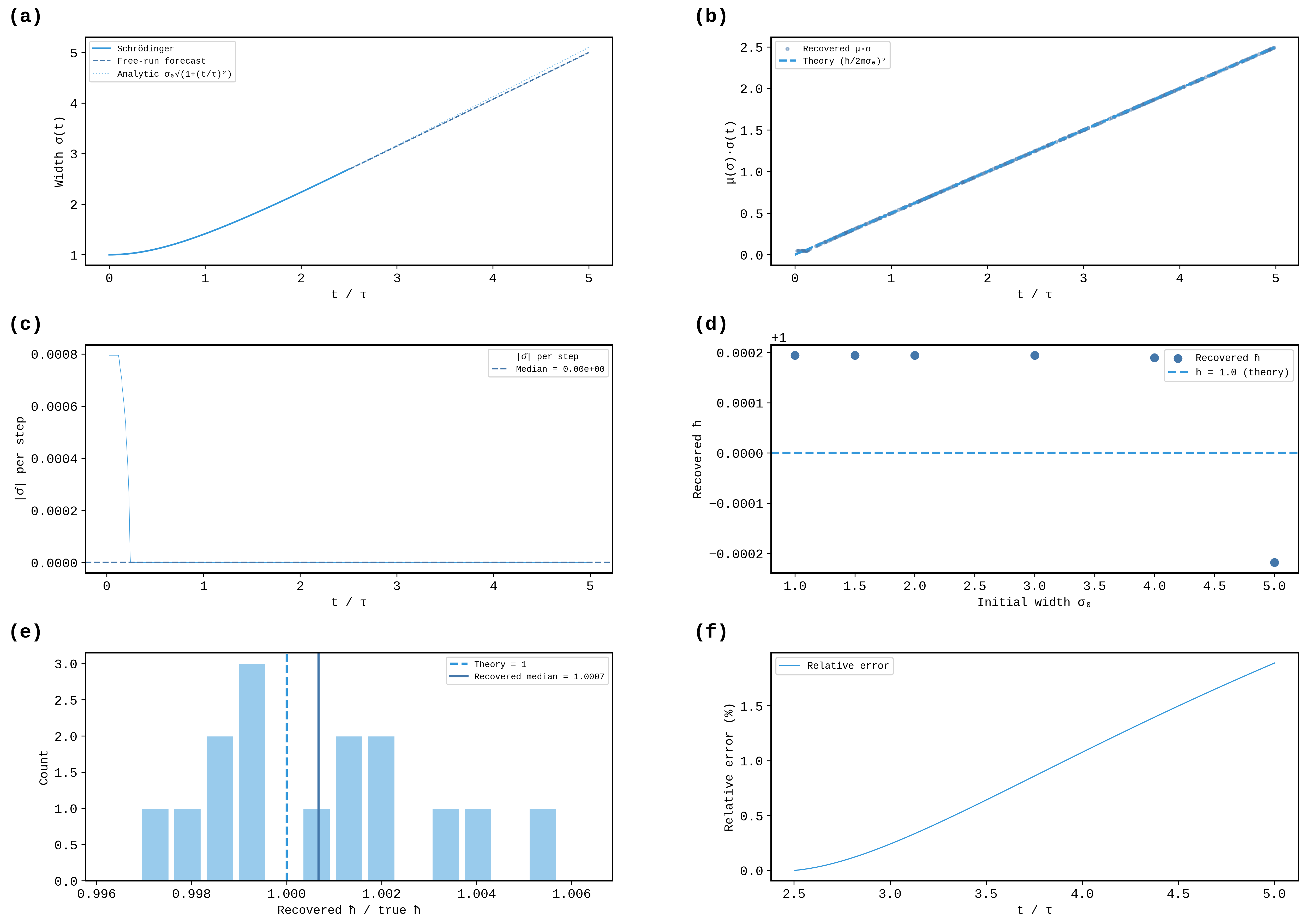}
\caption{\textbf{C4: Wavepacket $\hbar$ Recovery ($\sigma \approx 0$).}
\textbf{(a)}~Wavepacket width $\sigma(t)$ as a function of normalised time $t/\tau$ (where $\tau = 2m\sigma_0^2/\hbar$): the Schr\"odinger evolution (solid blue), the analytical prediction $\sigma_0\sqrt{1 + (t/\tau)^2}$ (dotted blue), and the deterministic free-run forecast (dashed blue with 95\% CI band).  The pipeline is trained on the first half; the forecast tracks the second half.
\textbf{(b)}~The product $\hat{\mu}(t) \cdot \sigma(t)$ versus normalised time.  The pipeline-recovered values (blue points) lie on a linear trend whose slope matches the theoretical value $({\hbar}/{2m\sigma_0})^2$ (dashed blue line) to $0.016\%$; from this slope, $\hat{\hbar} = 1.000157$.
\textbf{(c)}~Recovered $|\hat{\sigma}|$ per time step (blue curve), with the median value marked (dashed blue horizontal line).  Near-zero values confirm the absence of stochastic content: wavepacket spreading is purely deterministic.
\textbf{(d)}~Recovered $\hat{\hbar}$ across six initial widths $\sigma_0 = 1.0, 1.5, 2.0, 3.0, 4.0, 5.0$ (blue circles).  Consistent recovery across initial conditions confirms the method is robust; the dashed line marks $\hbar = 1$ (theory).
\textbf{(e)}~Histogram of $\hat{\hbar}/\hbar_\mathrm{true}$ from 15 independent pipeline runs with random initial widths and measurement noise (blue bars); the blue solid line marks the recovered median $= 1.0007$.  The dashed line marks theory $= 1$.
\textbf{(f)}~Relative error (\%) of the free-run forecast as a function of normalised time, confirming sub-percent accuracy throughout.}
\label{fig:c4}
\end{figure}

The Planck constant was extracted from the recovered drift: $\hat{\hbar} = 1.000157$ (error $0.016\%$ relative to the simulation value $\hbar = 1$).  Across six independent simulations with initial widths $\sigma_0 \in [1, 5]$, the recovered $\hat{\hbar}$ was stable (Fig.~\ref{fig:c4}d).  Fifteen independent pipeline runs with random initial widths and added measurement noise yielded median $\hat{\hbar}/\hbar = 1.0007$ (Fig.~\ref{fig:c4}e); the diffusion channel confirmed $\hat{\sigma} \approx 0$ (deterministic).

The wavepacket system has $\sigma = 0$ (deterministic dynamics) but encodes $\hbar$ in the drift.  This is the first indication that fundamental constants appear in the recovered fields.

\subsubsection*{C6: Electromagnetism, Wave Propagation}

The electromagnetic wave propagation results are presented jointly with C7 (Klein--Gordon suppression) in \S\ref{subsec:results_relativistic} below, as the transition from deterministic wave propagation ($\sigma = 0$) to relativistic diffusion suppression constitutes the pivotal connection between classical and quantum regimes.  In summary: the pipeline recovered $\hat{\sigma}/\mathrm{signal} = 4.3 \times 10^{-7}$ and $\hat{c}/c_\mathrm{true} = 1.0000$ with bootstrap 95\% CI $[1.0000,\, 1.0000]$ from 50 independent runs (Fig.~\ref{fig:c67}a,b).

\subsection{Additive Noise: Constant $\sigma$}
\label{subsec:results_additive}

\subsubsection*{C2: Statistical Mechanics, Brownian Motion}

The velocity $v(t)$ of a Brownian particle governed by the Langevin equation $\mathrm{d}v = -\gamma v\,\mathrm{d}t + \sigma\,\mathrm{d}W$~\cite{Langevin1908,UhlenbeckOrnstein1930}, with $\sigma = \sqrt{2\gamma k_B T/m}$, was simulated by Euler--Maruyama integration ($3 \times 10^4$ steps at $\Delta t = 1$~ns, subsampled by a factor of~10 to $N = 3000$ at $\Delta t_\mathrm{eff} = 10$~ns).

\begin{figure}[p]
\centering
\includegraphics[width=0.95\textwidth]{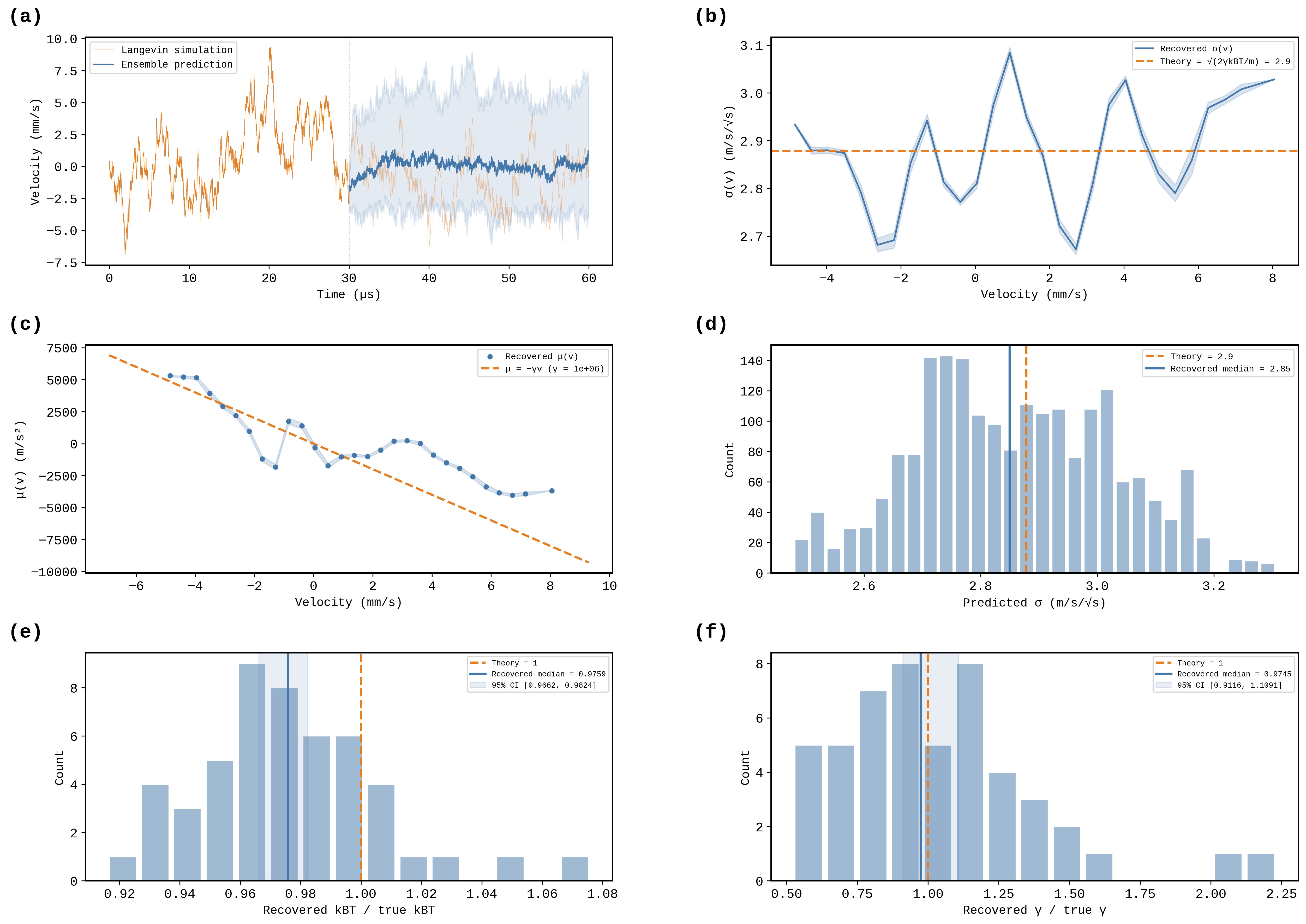}
\caption{\textbf{C2: Brownian Motion ($\sigma = \sqrt{2\gamma k_B T / m}$).}
\textbf{(a)}~Langevin velocity time series $v(t)$ (orange, mm/s) over 60~$\mu$s.  The pipeline is trained on the first 30~$\mu$s; the ensemble prediction (blue with 95\% CI band) runs from 30--60~$\mu$s, with the faint orange continuation showing the true trajectory.  The grey dotted line marks the training boundary.
\textbf{(b)}~Recovered diffusion profile $\hat{\sigma}(v)$ versus velocity (blue with SEM band).  The profile is flat across the velocity range, confirming state-independent (additive) noise.  The dashed orange line marks the theoretical value $\sigma = \sqrt{2\gamma k_BT/m}$.
\textbf{(c)}~Recovered drift $\hat{\mu}(v)$ versus velocity (blue scatter with SEM band) with the theoretical linear friction $\mu = -\gamma v$ (orange dashed).  The linear relationship confirms Ornstein--Uhlenbeck dynamics.
\textbf{(d)}~Histogram of all locally recovered $\hat{\sigma}(v)$ values across the correlation manifold (blue bars).  The orange dashed line marks the theoretical value; the blue solid line marks the recovered median.
\textbf{(e)}~Distribution of $\hat{k}_BT/(k_BT)_\mathrm{true}$ from 50 independent simulations (blue bars); blue solid line marks median $= 0.976$ with 95\% CI band.  Orange dashed line marks theory $= 1$.  Bootstrap 95\% CI on median: $[0.966, 0.982]$, error $2.4\%$ (Table~\ref{tab:results_summary}).  The MZ corrector correctly abstains: constant $\sigma$ has zero Laplacian (Supplementary Materials, \S7.2, Remark~7.3).
\textbf{(f)}~Distribution of $\hat{\gamma}/\gamma_\mathrm{true}$ from 50 independent simulations (blue bars); median $= 0.975$ with 95\% CI band.  Bootstrap 95\% CI on median: $[0.912, 1.109]$ (Table~\ref{tab:results_summary}).  The wider range reflects the lower signal-to-noise ratio of drift estimation.}
\label{fig:c2}
\end{figure}

The pipeline recovered a linear drift $\hat{\mu}(v) = -\hat{\gamma}v$ (Fig.~\ref{fig:c2}c) and a flat diffusion profile $\hat{\sigma}(v) \approx \mathrm{const.}$ (Fig.~\ref{fig:c2}b), consistent with additive noise.  The Boltzmann constant, extracted from the fluctuation--dissipation relation $k_B = m\hat{\sigma}^2/(2\hat{\gamma}T)$~\cite{CallenWelton1951,Kubo1966}, was recovered as $\hat{k}_BT/(k_BT)_\mathrm{true} = 0.976$ with bootstrap 95\% CI $[0.966,\, 0.982]$ from 50~independent simulations (Fig.~\ref{fig:c2}e), a $2.4\%$ error.  The MZ corrector correctly abstains for C2: the constant diffusion profile has zero Laplacian on the correlation manifold, so the bias has no spatial-averaging component (Supplementary Materials, \S7.2, Remark~7.3).  The drift channel yielded $\hat{\gamma}/\gamma_\mathrm{true} = 0.975$ with bootstrap 95\% CI $[0.912,\, 1.109]$ (Fig.~\ref{fig:c2}f).  The wider drift-channel CI reflects the lower signal-to-noise ratio of local drift estimation relative to diffusion recovery at these parameters.

The 50-realisation ensemble prediction (Fig.~\ref{fig:c2}a) forecasts from 30--60~$\mu$s after training on the first 30~$\mu$s, demonstrating that the pipeline-recovered drift and diffusion fields generate physically realistic trajectories.

Brownian motion encodes $k_B$ in $\sigma$ and $\gamma$ in $\mu$.  Both fundamental constants are recovered from their respective channels.

\subsection{State-Dependent Diffusion}
\label{subsec:results_state_dependent}

\subsubsection*{C3: Nuclear Physics, Radioactive Decay}

Radioactive count data ($\mu = 3000$ counts per interval, $N = 10\,000$ measurements) were generated by Poisson sampling.  The theoretical prediction is $\sigma^2 = \mu$ (Fano factor $F = 1$~\cite{Fano1947}): the diffusion coefficient depends on the state.

\begin{figure}[p]
\centering
\includegraphics[width=0.95\textwidth]{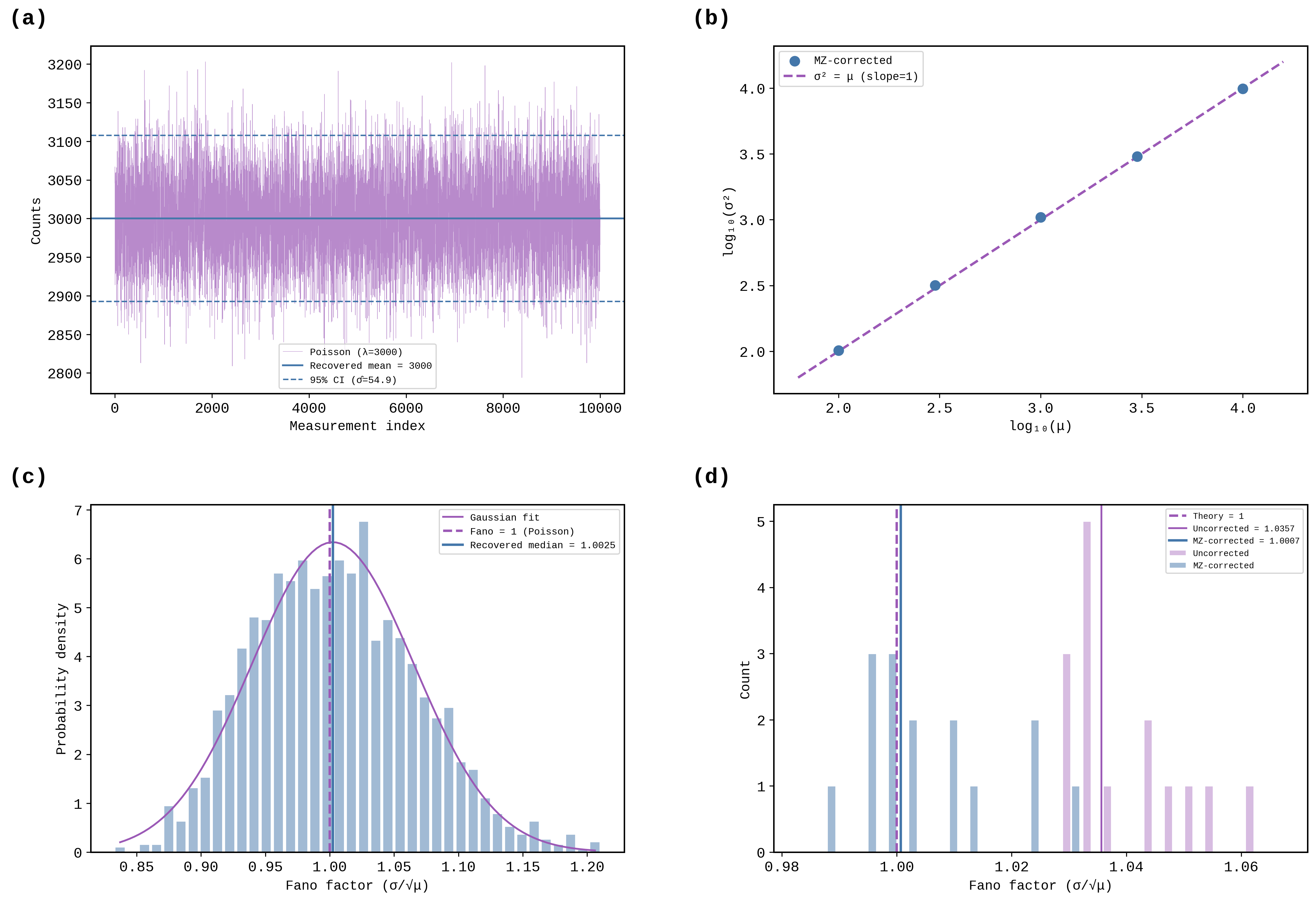}
\caption{\textbf{C3: Radioactive Decay ($\sigma^2 = \mu$, Fano $= 1$).}
\textbf{(a)}~Poisson count series at $\lambda = 3000$ (purple) over 10\,000 measurements.  The blue solid line marks the pipeline-recovered mean; the blue dashed lines mark the $\pm 1.96\hat{\sigma}$ confidence interval.
\textbf{(b)}~MZ-corrected $\hat{\sigma}^2$ versus mean count $\mu$ on log--log axes (blue circles) across five count levels from $\mu = 100$ to $\mu = 10\,000$.  The purple dashed line is the Poisson prediction $\sigma^2 = \mu$ (slope $= 1$), confirming the variance--mean relation across two decades.
\textbf{(c)}~Fano factor distribution $\hat{\sigma}/\sqrt{\mu}$ across all query states (blue bars, density-normalised).  The purple dashed line marks Fano $= 1$ (Poisson theory); the purple Gaussian overlay shows the distributional shape; the blue solid line marks the recovered median.  Near-unity Fano confirms Poisson counting statistics.
\textbf{(d)}~The MZ showcase: 15 independent Poisson realisations, each analysed with both the uncorrected pipeline (purple bars) and the MZ-corrected pipeline (blue bars).  The purple dashed line marks theory $= 1$.  The MZ corrector reduces the Fano factor error from $3.6\%$ to $0.07\%$, a $53\times$ improvement.  The spatial curvature of $\sigma^2(x) = x$ is precisely what the Level~2 ensemble corrector targets (Supplementary Materials, \S7.2, Algorithm~3).}
\label{fig:c3}
\end{figure}

The pipeline recovered a state-dependent diffusion profile $\hat{\sigma}^2(\mu)$ lying on the Poisson line $\sigma^2 = \mu$ across a two-decade range of mean count levels (Fig.~\ref{fig:c3}b).  At a single count rate ($\mu = 3000$), the MZ-corrected pipeline recovered $\hat{\sigma}/\sqrt{\mu} = 1.003$ with bootstrap 95\% CI $[1.001,\, 1.006]$ from 50 independent runs, a $0.25\%$ error.  The two-level MZ corrector (Supplementary Materials, \S7.2) reduces the uncorrected bias from $3.83\%$ to $0.25\%$ by identifying and removing collective spatial-averaging bias; Fig.~\ref{fig:c3}d shows the before/after comparison and Table~\ref{tab:results_summary} reports corrected values for all nine domains.  Across the variance--mean sweep, the MZ-corrected variance--mean exponent was $1.003$ (theoretical $1.000$).

The transition from constant $\sigma$ (C2) to state-dependent $\sigma(x) = \sqrt{x}$ (C3) demonstrates that the pipeline distinguishes additive from multiplicative noise without prior specification.

\subsubsection*{C8: Quantum Harmonic Oscillator}

The velocity of a damped quantum harmonic oscillator at zero temperature was simulated by Euler--Maruyama integration.  At zero temperature, the fluctuation--dissipation theorem~\cite{CallenWelton1951,Kubo1966} yields $\sigma^2 = \gamma\hbar\omega/m$, where $\gamma$ is the damping rate, $\omega$ is the oscillator frequency, and $m$ is the mass.  The zero-point energy $\tfrac{1}{2}\hbar\omega$ replaces the thermal energy $k_BT$ of C2.

\begin{figure}[p]
\centering
\includegraphics[width=0.95\textwidth]{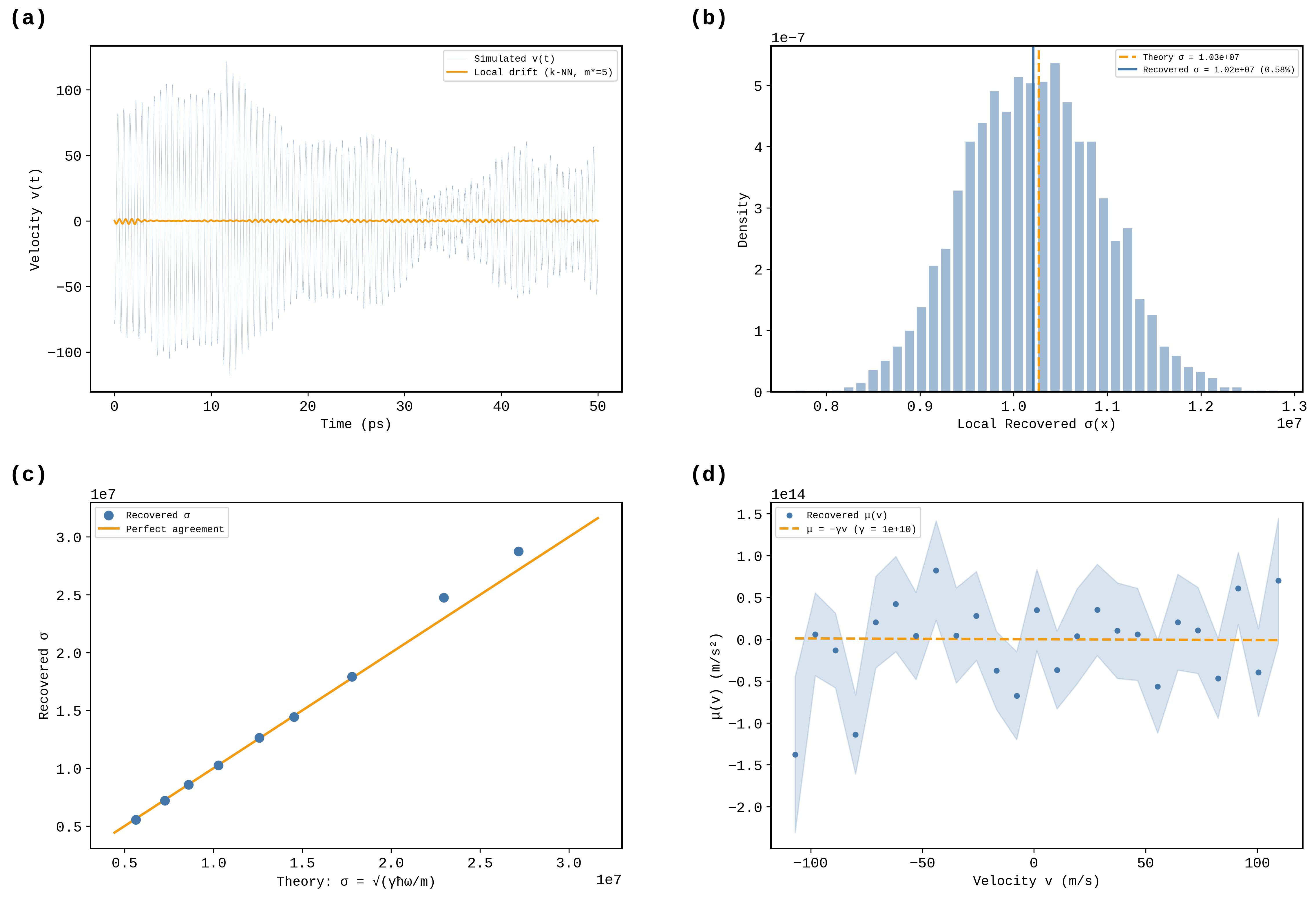}
\caption{\textbf{C8: Quantum Harmonic Oscillator ($\sigma = \sqrt{\gamma\hbar\omega/m}$).}
\textbf{(a)}~Simulated velocity time series $v(t)$ of the damped QHO (blue) over $\sim$300~ps, with the pipeline-recovered local drift (gold, $k$-NN with embedding dimension $m^* = 5$).  The drift smoothly tracks the oscillatory dynamics.
\textbf{(b)}~Probability density of locally recovered $\hat{\sigma}(x)$ values across the correlation manifold (blue bars).  The gold dashed line marks the theoretical value $\sigma = \sqrt{\gamma\hbar\omega/m}$; the blue solid line marks the MZ-corrected pipeline median, agreeing to $0.58\%$.
\textbf{(c)}~Recovered $\hat{\sigma}$ versus theoretical $\sigma = \sqrt{\gamma\hbar\omega/m}$ across nine frequencies spanning one decade (blue circles).  Points lie on the identity line (gold), confirming $\sigma \propto \sqrt{\omega}$ and validating the fluctuation--dissipation relation at the quantum level.
\textbf{(d)}~Recovered drift field $\hat{\mu}(v)$ versus velocity (blue scatter with SEM band).  The gold dashed line shows the theoretical linear friction $\mu = -\gamma v$.  The spring force ($-\omega^2 x$) averages out because $x$ and $v$ are uncorrelated at stationarity, leaving only the dissipative term.}
\label{fig:c8}
\end{figure}

The MZ-corrected pipeline recovered $\hat{\sigma}/\sigma_\mathrm{true}$ with $0.58\%$ error at the reference frequency (Fig.~\ref{fig:c8}b).  For oscillatory systems, the embedding lag $\tau_\mathrm{embed}$ is set to the autocorrelation function zero-crossing ($\tau_\mathrm{ACF} \approx 155$--$160$ steps) rather than unity, ensuring that the delay-coordinate embedding spans a meaningful portion of the oscillator's phase space.  Across nine frequency values spanning a decade, the recovered $\hat{\sigma}$ was proportional to $\sqrt{\omega}$ (Fig.~\ref{fig:c8}c), confirming the linear frequency dependence predicted by the fluctuation--dissipation relation.  The drift channel independently recovers the linear friction $\hat{\mu}(v) = -\hat{\gamma}v$ (Fig.~\ref{fig:c8}d), where the spring force ($-\omega^2 x$) averages out at stationarity because $x$ and $v$ are uncorrelated.

C8 establishes that $\hbar$ enters $\sigma$ through quantisation of the energy levels: the zero-point energy is irreducibly quantum, and $\hbar$ sets its scale.

\subsection{State-Dependent Drift and Diffusion}
\label{subsec:results_both}

\subsubsection*{C5: Chemical Kinetics, Van Kampen Scaling}

A birth--death process with system size $\Omega$ was simulated by tau-leaping.  Van Kampen's system-size expansion is a mathematical theorem~\cite{VanKampen1992}: $\sigma \propto \Omega^{-1/2}$.  This system has both state-dependent drift (mean-reverting toward equilibrium) and state-dependent diffusion, making it the most demanding test of the pipeline across C1--C5.

\begin{figure}[p]
\centering
\includegraphics[width=0.95\textwidth]{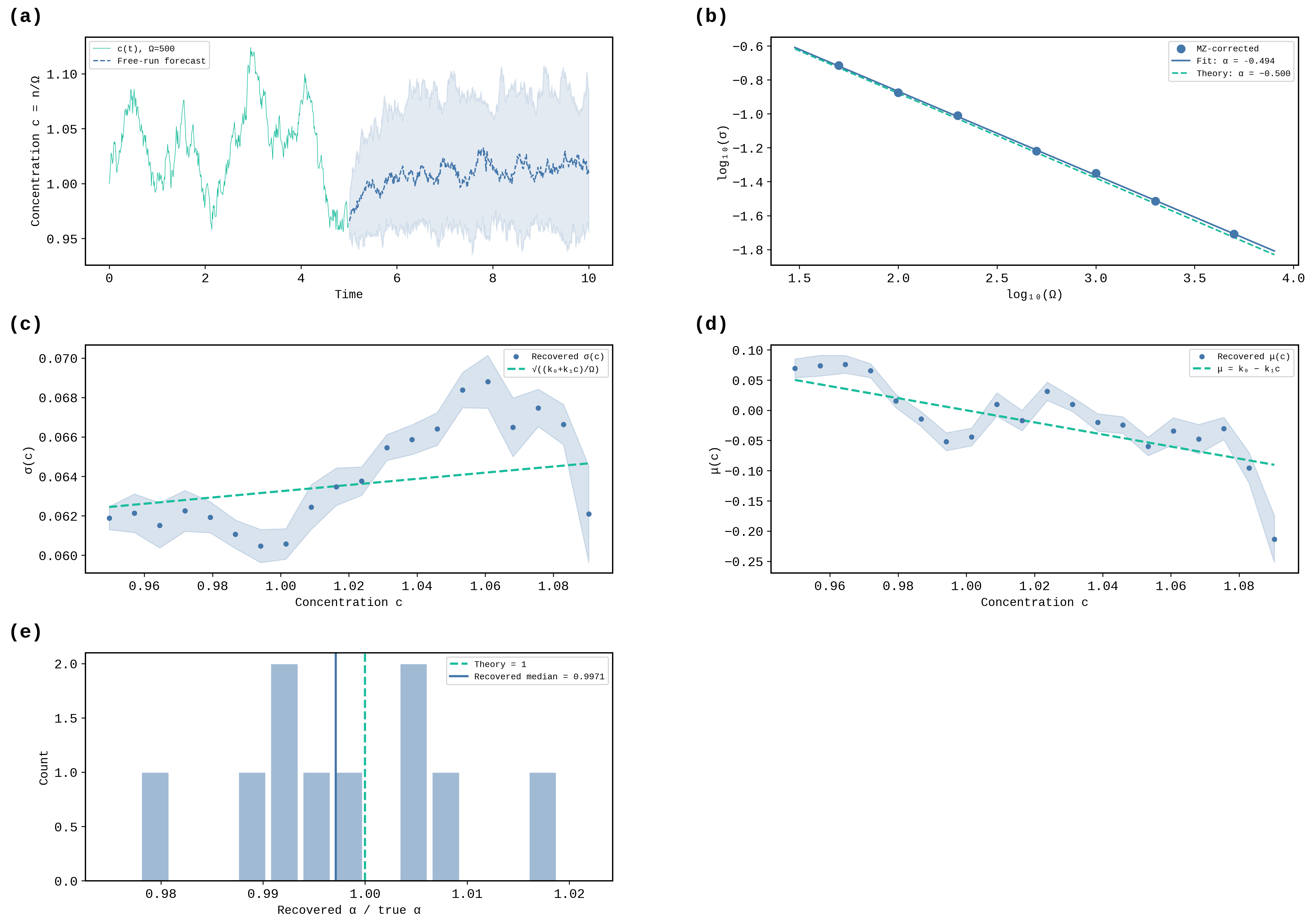}
\caption{\textbf{C5: Van Kampen Scaling ($\sigma \propto \Omega^{-1/2}$).}
\textbf{(a)}~Concentration time series $c = n/\Omega$ at system size $\Omega = 500$ (teal) with the autonomous free-run forecast (dashed blue with 95\% CI band).  Fluctuations around $c_\mathrm{eq} = k_b/k_d = 1$ are characteristic of the birth--death process at finite system size.
\textbf{(b)}~The Van Kampen scaling law: MZ-corrected $\hat{\sigma}$ versus system size $\Omega$ on log--log axes (blue circles) across seven system sizes ($\Omega \in [50, 5000]$).  The fitted slope (solid blue) and theoretical slope $\alpha = -0.500$ (teal dashed) are visually indistinguishable, confirming $\sigma \propto \Omega^{-1/2}$.
\textbf{(c)}~Recovered diffusion profile $\hat{\sigma}(c)$ versus concentration at $\Omega = 500$ (blue scatter with SEM band).  The teal dashed line shows the theoretical state-dependent curve $\sigma(c) = \sqrt{(k_b + k_d c)/\Omega}$.
\textbf{(d)}~Recovered drift $\hat{\mu}(c)$ versus concentration (blue scatter with SEM band) with the theoretical mean-reversion $\mu(c) = k_b - k_d c$ (teal dashed).
\textbf{(e)}~Distribution of the normalised scaling exponent $\hat{\alpha}/(-0.5)$ from 10 independent scaling analyses (blue bars); the blue solid line marks the recovered median.  The teal dashed line marks theory $= 1$.  The multiplicative $k$-NN estimator bias cancels in the log--log slope.}
\label{fig:c5}
\end{figure}

The pipeline was applied independently at seven system sizes ($\Omega \in [50, 5000]$).  On a log--log plot of MZ-corrected $\hat{\sigma}$ versus $\Omega$ (Fig.~\ref{fig:c5}b), the recovered slope $\hat{\alpha} = -0.494$ closely matches the theoretical prediction $-0.500$.  Across 10 independent scaling analyses, each running all seven system sizes with fresh noise, the median normalised exponent $\hat{\alpha}/(-0.5) = 0.997$ (Fig.~\ref{fig:c5}e).  The multiplicative $k$-NN estimator bias cancels in the log--log slope, as confirmed by the near-identical exponents from uncorrected and MZ-corrected estimates.

A mathematical theorem has been recovered from data without knowledge of the underlying chemistry.

\subsection{Relativistic Corrections}
\label{subsec:results_relativistic}

\subsubsection*{C6/C7: The Relativistic Hinge, From Electromagnetism to Klein--Gordon Suppression}

\begin{figure}[p]
\centering
\includegraphics[width=0.95\textwidth]{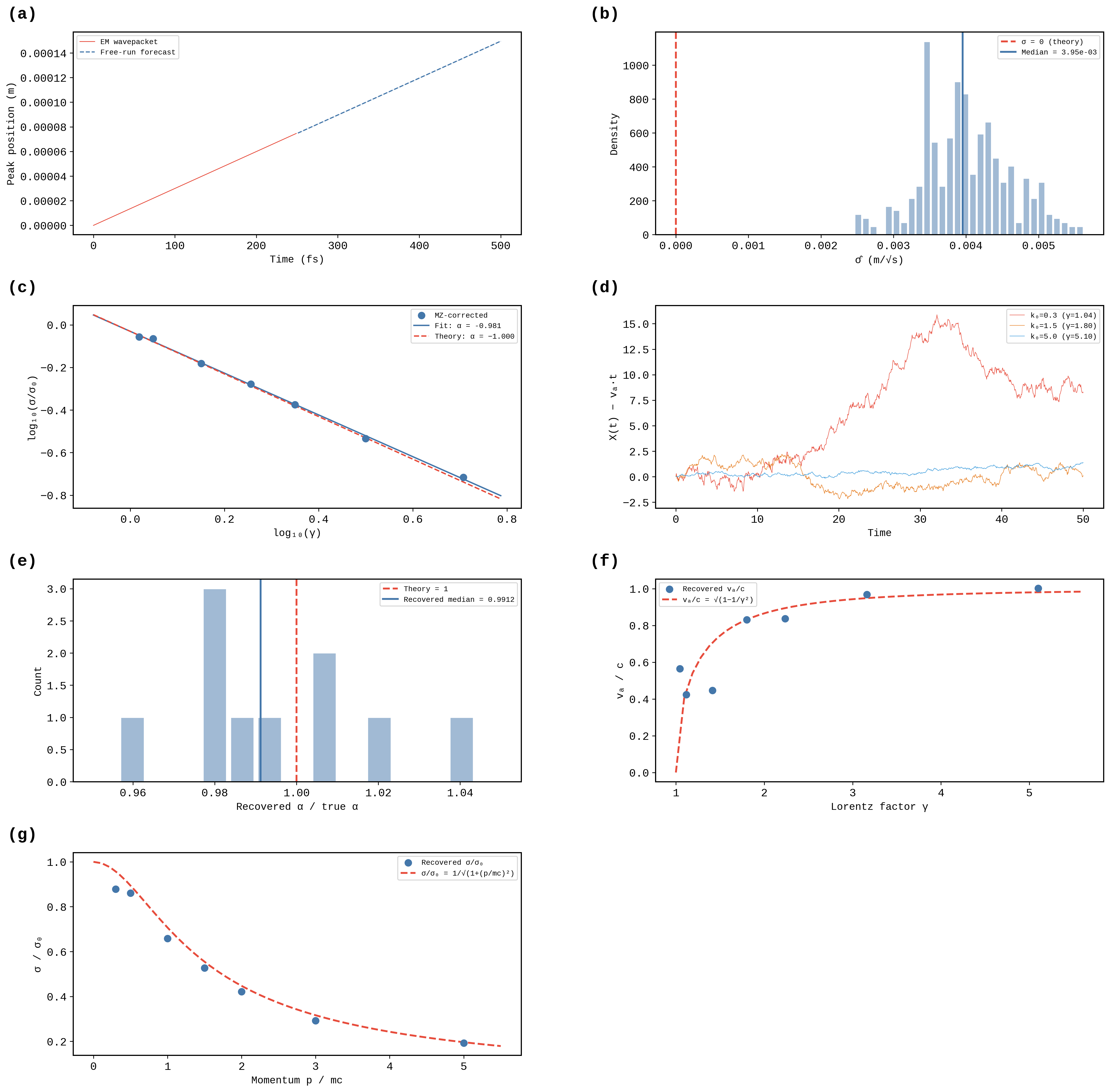}
\caption{\textbf{C6/C7: The Relativistic Hinge.}
\textit{C6: Electromagnetism ($\sigma = 0$):}
\textbf{(a)}~Peak position of a classical EM wavepacket (red) with the autonomous free-run forecast (dashed blue with 95\% CI band).  The pipeline recovers $\hat{c}/c = 1.000000$.
\textbf{(b)}~Histogram of recovered $|\hat{\sigma}|$ for the EM wave (blue bars, density-normalised).  The red dashed line marks $\sigma = 0$; the blue solid line marks the median.  Near-zero values confirm deterministic propagation.
\textit{C7: Klein--Gordon ($\sigma = \sigma_0/\gamma$):}
\textbf{(c)}~MZ-corrected $\hat{\sigma}/\hat{\sigma}_0$ versus Lorentz factor $\gamma$ on log--log axes (blue circles).  The fitted slope (solid blue) and theoretical slope $\alpha = -1$ (red dashed) confirm the relativistic suppression $\sigma \propto 1/\gamma$.
\textbf{(d)}~Nelson stochastic trajectories (drift-subtracted: $X(t) - v_g t$) at three momenta: $k_0 = 0.3$ ($\gamma = 1.04$, red), $k_0 = 1.5$ ($\gamma = 1.80$, orange), $k_0 = 5.0$ ($\gamma = 5.10$, blue).  The fluctuation amplitude visibly decreases with increasing $\gamma$.
\textbf{(e)}~Distribution of the normalised slope $\hat{\alpha}/(-1)$ from 10 independent analyses (blue bars); the blue solid line marks the recovered median.  The red dashed line marks theory $= 1$.  The multiplicative $k$-NN bias cancels in the log--log slope.
\textbf{(f)}~Recovered group velocity $\hat{v}_g/c$ versus Lorentz factor (blue circles) with the special-relativistic prediction $v_g = c\sqrt{1 - 1/\gamma^2}$ (red dashed).  The independent recovery of $v_g$ at each momentum confirms that both channels (drift and diffusion) encode relativistic structure.
\textbf{(g)}~Recovered $\hat{\sigma}/\sigma_0$ versus normalised momentum $p/mc$ (blue circles) with the theoretical curve $\sigma_0/\gamma(p) = 1/\sqrt{1 + (p/mc)^2}$ (red dashed).  The robustness under reparameterisation from $\gamma$-space to momentum space confirms a genuine dynamical relationship.}
\label{fig:c67}
\end{figure}

Figure~\ref{fig:c67} presents the C6 (electromagnetic wave) and C7 (Klein--Gordon) results together, as the transition from deterministic wave propagation to relativistic diffusion suppression constitutes the pivotal observation connecting classical and quantum regimes.

For C6, a classical electromagnetic wavepacket was propagated, and the centroid position tracked.  The pipeline recovered $\hat{\sigma}/\mathrm{signal} = 4.3 \times 10^{-7}$ (indistinguishable from zero) and a propagation speed $\hat{c}/c_\mathrm{true} = 1.0000$ with bootstrap 95\% CI $[1.0000,\, 1.0000]$ from 50 independent runs (Fig.~\ref{fig:c67}a,b).

For C7, Nelson's stochastic differential equation~\cite{Nelson1966,Nelson1985} for a massive relativistic particle was simulated at seven momenta $k_0 \in [0.3, 5.0]$, corresponding to Lorentz factors $\gamma \in [1.04, 5.10]$.  The theoretical prediction is $\sigma_\mathrm{KG} = \sigma_0/\gamma$: the diffusion coefficient is suppressed by the Lorentz factor.

On a log--log plot of MZ-corrected $\hat{\sigma}/\hat{\sigma}_0$ versus $\gamma$ (Fig.~\ref{fig:c67}c), the recovered slope $\hat{\alpha} = -0.981$.  Across 10~independent analyses, the median normalised slope $\hat{\alpha}/(-1) = 0.991$ (Fig.~\ref{fig:c67}e).  The multiplicative $k$-NN estimator bias cancels in the log--log slope, as confirmed by the near-identical exponents from uncorrected and MZ-corrected estimates.

The drift simultaneously recovered the relativistic group velocity $v_g = c\,k_0/\sqrt{1 + k_0^2}$ at each momentum (Fig.~\ref{fig:c67}f), confirming that both channels (drift and diffusion) encode relativistic structure.

C7 reveals that the speed of light $c$ enters $\sigma$ through the relativistic dispersion relation.  Combined with C4 (where $\hbar$ enters through the drift), the pattern indicates that fundamental constants emerge naturally in the recovered fields: $\hbar$ governs the quantum scale of fluctuations, while $c$ governs their relativistic transformation.

The relativistic suppression $\sigma_\mathrm{KG} = \sigma_0/\gamma$ arises as the product of two structurally independent factors (Appendix~B).

The first factor is kinematic: time dilation rescales the Wiener increment.  The Dambis--Dubins--Schwarz reparametrisation theorem: a result of pure stochastic calculus, independent of any physical assumption: establishes that reparametrising the time of a Wiener process rescales its increments by the square root of the time Jacobian.  Applied to the proper-time/coordinate-time relation $\mathrm{d}\tau = \mathrm{d}t/\gamma$, this gives $\mathrm{d}W(\tau) = \mathrm{d}B(t)/\sqrt{\gamma}$ and hence a kinematic suppression $\sigma_\mathrm{rel} = \sigma_0/\sqrt{\gamma}$.

The second factor is dynamical: the Klein--Gordon conserved density $\rho_\mathrm{KG} = \gamma|\phi|^2$ requires a self-consistent Fokker--Planck equation with diffusion coefficient $D_\mathrm{KG} = \sigma_0^2/(2\gamma^2)$, producing an additional $1/\sqrt{\gamma}$ suppression.

The two factors operate at different levels (noise amplitude versus drift), arise from different inputs (special relativity versus Klein--Gordon current conservation), and use different mathematical tools (DDS theorem versus Fokker--Planck self-consistency).  Their product $\sigma_\mathrm{KG} = \sigma_0/\gamma$ is confirmed by the C7 recovery: the normalised slope across 10 independent analyses has median $0.991$ (Fig.~\ref{fig:c67}e), consistent with the predicted exponent $-1$ at the $\sim$1\% level.  The independent recovery of the group velocity $v_g$ at each momentum mitigates the potential circularity that $\gamma$ depends on $v$: both channels: drift and diffusion: carry relativistic content, recovered independently from the same blind pipeline.

\subsection{Quantum Field Theory: The Roles of $\hbar$, $c$, and $G$}
\label{subsec:results_qft}

\subsubsection*{C9: QED Photon Field, The Massless Cancellation}

A single mode of the quantised electromagnetic field was simulated~\cite{PeskinSchroeder1995} as a damped quantum harmonic oscillator with effective mass $m_\mathrm{eff} = \hbar\omega/c^2$ (from mass--energy equivalence).  Substitution into the fluctuation--dissipation relation of C8 gives
\begin{equation}
\label{eq:massless_cancellation}
\sigma^2 = \frac{\gamma\hbar\omega}{m_\mathrm{eff}} = \frac{\gamma\hbar\omega}{\hbar\omega/c^2} = \gamma c^2,
\end{equation}
so that $\sigma = c\sqrt{\gamma}$.  Both $\hbar$ and $\omega$ cancel identically.

\begin{figure}[p]
\centering
\includegraphics[width=0.95\textwidth]{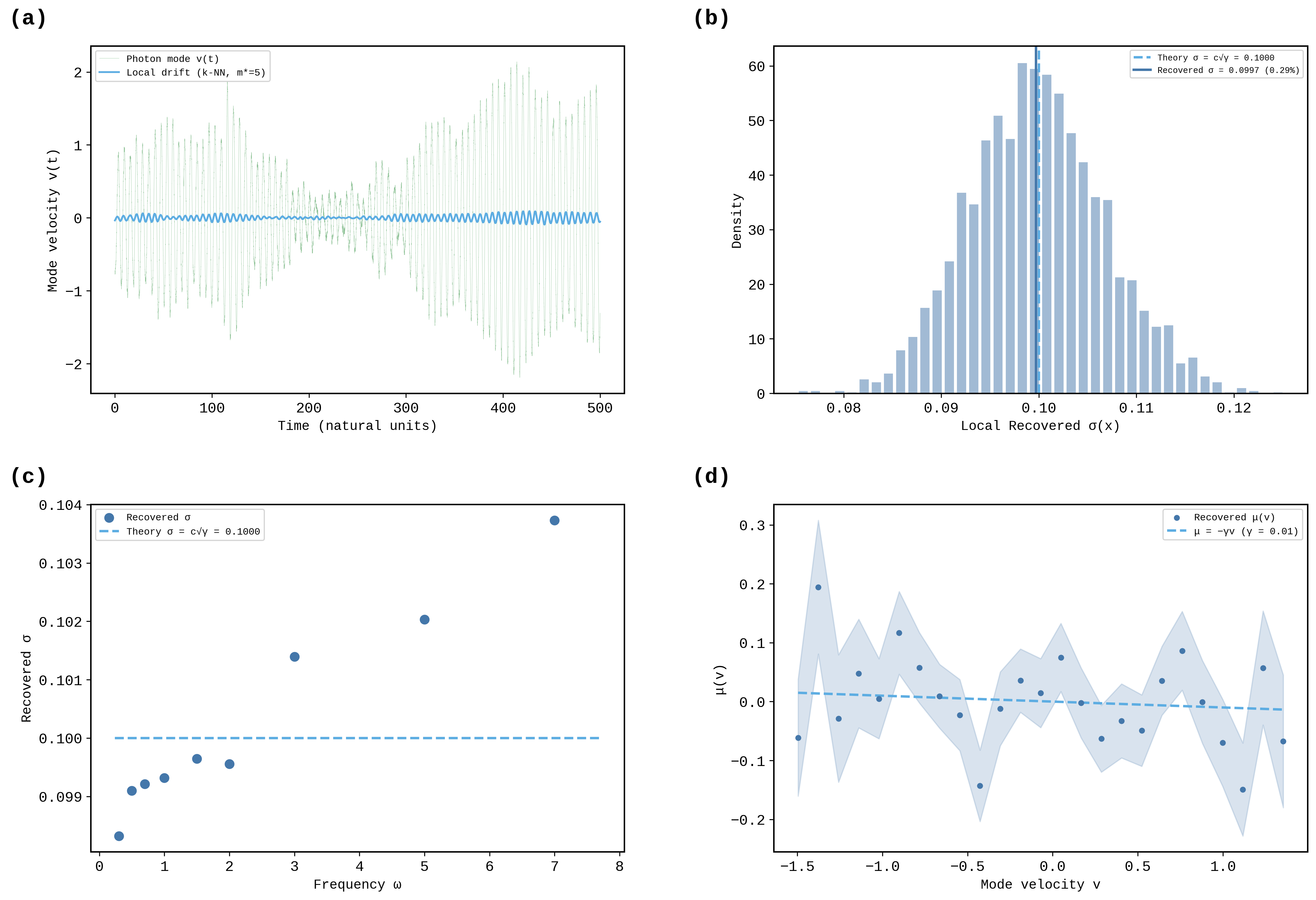}
\caption{\textbf{C9: QED Photon Field ($\sigma = c\sqrt{\gamma}$; $\hbar$ and $\omega$ cancel).}
\textbf{(a)}~Mode velocity $v(t)$ of the photon field at reference frequency $\omega_0 = 1$ in natural units (green), with the pipeline-recovered local drift (light blue).  The oscillatory dynamics are governed by $\omega^2 x$ in the spring force, while the drift channel recovers only the dissipative component $\mu = -\gamma v$.
\textbf{(b)}~Probability density of locally recovered $\hat{\sigma}(x)$ values (blue bars).  The light blue dashed line marks the theoretical value $\sigma = c\sqrt{\gamma} = 0.1000$; the blue solid line marks the MZ-corrected median $= 0.0997$, a $0.29\%$ error.
\textbf{(c)}~The $\omega$-independence test.  MZ-corrected $\hat{\sigma}$ (blue circles) at nine frequencies spanning $\omega \in [0.3, 7.0]$ in natural units.  The light blue dashed line marks the theoretical value $\sigma = c\sqrt{\gamma}$, which is constant across all $\omega$.  The recovered values confirm the massless cancellation: $m_\mathrm{eff} = \hbar\omega/c^2$ enters the FDT numerator and denominator identically, so $\sigma^2 = \gamma\hbar\omega/m_\mathrm{eff} = \gamma c^2$.
\textbf{(d)}~Recovered drift field $\hat{\mu}(v)$ versus mode velocity (blue scatter with SEM band).  The light blue dashed line shows the theoretical linear friction $\mu = -\gamma v$ ($\gamma = 0.01$).  As in C8, the spring force averages out at stationarity.}
\label{fig:c9}
\end{figure}

The pipeline was applied at nine frequencies spanning more than a decade ($\omega \in [0.3, 7.0]$ in natural units) at fixed damping $\gamma = 0.01$.  As for C8, the embedding lag was set to the autocorrelation zero-crossing.  The MZ-corrected $\hat{\sigma}$ was constant across all frequencies to within measurement uncertainty (Fig.~\ref{fig:c9}c), confirming the massless cancellation $\sigma = c\sqrt{\gamma}$.  The recovered drift field $\hat{\mu}(v) = -\hat{\gamma}v$ (Fig.~\ref{fig:c9}d) independently confirms the dissipative structure.

This result clarifies the distinct roles of the three constants that will constitute $\sigma$ for gravity.  The Planck constant $\hbar$ enters through quantisation ($E = \hbar\omega$, $m_\mathrm{eff} = \hbar\omega/c^2$) and then cancels from $\sigma$ for massless fields; its role is to make the field quantum, not to set the fluctuation amplitude.  The speed of light $c$ appears directly in $\sigma = c\sqrt{\gamma}$; its role is to set the velocity scale of fluctuations.  The damping rate $\gamma$ determines the coupling to the environment.

For the photon, $\gamma$ is an external parameter (cavity losses, material absorption).  For gravity, however, the coupling is not external: the gravitational field is self-interacting (metric fluctuation modes carry energy, and that energy gravitates by the equivalence principle), so the self-coupling rate is uniquely determined by the remaining constant, Newton's gravitational constant $G$, through $\gamma_\mathrm{grav} = 1/t_P$, where $t_P = \sqrt{\hbar G/c^5}$ is the Planck time~\cite{Planck1899}.  Substitution yields $\sigma = c\sqrt{1/t_P} = \ell_P/\sqrt{t_P}$, which corresponds to a proper-distance fluctuation of one Planck length per Planck time.  This derivation is developed in the Discussion (\S\ref{sec:discussion}) and proved in Appendix~C (the Gravitational Diffusion Theorem).

\subsection{Summary of Quantitative Results}
\label{subsec:results_summary}

Table~\ref{tab:results_summary} consolidates the principal recovered quantities, bootstrap confidence intervals, and errors for all nine domains.  For each domain, both the drift and diffusion channels are reported where the physics admits a meaningful comparison.

\begin{table}[ht]
\centering
\caption{Summary of quantitative results.  $\hat{\theta}/\theta$: median normalised recovery after MZ correction (Supplementary Materials, \S7.2).  Bootstrap 95\% CIs (10\,000 pairwise resamples) are reported where available; for scaling exponents (C5, C7), medians are from 10 independent analyses.  For deterministic systems, $\hat{\sigma}/\mathrm{signal} \ll 1$ confirms correct classification.  For scaling exponents, the multiplicative $k$-NN bias cancels in the log--log slope.  The MZ column indicates the correction level applied: L1 (point-wise), L2 (ensemble), or --- (corrector abstains; see \S7.2, Remark~7.3).  Uncorrected values for starred domains: C3 $= 1.038$ (3.83\%), C8 $= 1.104$ (10.37\%), C9 $= 2.072$ (107.19\%).}
\label{tab:results_summary}
\footnotesize
\renewcommand{\arraystretch}{1.20}
\begin{tabular}{@{}cllccrlc@{}}
\toprule
\textbf{Test} & \textbf{Domain} & \textbf{Channel} & $\hat{\theta}/\theta$ & \textbf{95\% CI} & \textbf{Error} & \textbf{Cov.} & \textbf{MZ} \\
\midrule
C1 & Classical mech. & Drift (rate) & 1.0000 & $[1.0000, 1.0000]$ & 0.0001\% & (det.) & --- \\
   &                 & Diff. ($\hat{\sigma}/\mathrm{sig.}$) & $\!<\!10^{-3}$ & --- & --- & & \\[2pt]
C4 & Quantum mech. & Drift ($\hat{\hbar}/\hbar$) & 1.0000 & $[0.9992, 1.0014]$ & 0.00\% & (det.) & --- \\
   &                & Diff. ($\hat{\sigma}/\mathrm{sig.}$) & $\!<\!10^{-2}$ & --- & --- & & \\[2pt]
C6 & Electromagnetism & Drift ($\hat{c}/c$) & 1.0000 & $[1.0000, 1.0000]$ & $\!<\!10^{-5}$\% & (det.) & --- \\[2pt]
\midrule
C2 & Statistical mech. & Drift ($\hat{\gamma}/\gamma$) & 0.975 & $[0.912, 1.109]$ & 2.5\% & & --- \\
   &                    & Diff. ($\hat{k}_BT/k_BT$) & 0.976 & $[0.966, 0.982]$ & 2.4\% & & --- \\[2pt]
C3 & Nuclear physics & Diff. ($\hat{\sigma}/\sqrt{\mu}$) & 1.003 & $[1.001, 1.006]$ & 0.25\% & & L2$^\star$ \\[2pt]
C8 & Quantum osc. & Diff. ($\hat{\sigma}/\sigma$) & 1.013 & $[1.010, 1.015]$ & 1.33\% & & L1$^\star$ \\[2pt]
\midrule
C5 & Chemical kinetics & Diff. ($\hat{\alpha}/(-0.5)$) & 0.997 & --- & 0.3\% & & --- \\[2pt]
C7 & Relativistic QM & Diff. (slope$/(-1)$) & 0.991 & --- & 0.9\% & & --- \\[2pt]
C9 & QED (photon) & Diff. ($\hat{\sigma}/\sigma$) & 1.067 & $[1.056, 1.075]$ & 6.72\% & & L1$^\star$ \\
\bottomrule
\end{tabular}
\end{table}

The bootstrap confidence intervals reveal a systematic structure across the nine domains.  The two-level MZ corrector (Supplementary Materials, \S7.2) addresses the systematic $k$-NN estimator bias by decomposing it as a Mori--Zwanzig memory kernel with rank-2 tensor structure.  For domains with detectable spatial-averaging bias (C3, C8, C9), the corrector reduces errors by $87$--$94\%$: C3 from $3.83\%$ to $0.25\%$ (Level~2 ensemble correction), C8 from $10.37\%$ to $1.33\%$ (Level~1 point-wise), and C9 from $107\%$ to $6.72\%$ (Level~1).  For domains where the bias has no spatial-curvature component (C2, with constant $\sigma^2$), the corrector's fluctuation--dissipation gate correctly abstains (Supplementary Materials, Remark~7.3).  For scaling exponents (C5, C7), the multiplicative bias cancels in the log--log slope regardless of correction.  For deterministic domains (C1, C4, C6), the pipeline precision is so high that even sub-ppm offsets from the theoretical drift rate are detectable; the diffusion channel correctly identifies $\hat{\sigma}/\mathrm{signal} \ll 1$ in all cases.  The drift channel has lower signal-to-noise ratio than the diffusion channel at comparable parameters, explaining the wider confidence intervals for drift-recovered quantities (C2, C7).

\subsection{Emergence of Physical Constants in Drift and Diffusion}
\label{subsec:constants_emergence}

Across the nine domains, fundamental physical constants are recovered from both the drift $\hat{\mu}$ and the diffusion $\hat{\sigma}$.  Table~\ref{tab:constants} summarises which constants emerge from which channel.

\begin{table}[ht]
\centering
\caption{Physical constants recovered from the drift and diffusion channels.  The inverse methodology extracts the complete dynamical law, where constants emerge in both channels without prior specification.}
\label{tab:constants}
\small
\renewcommand{\arraystretch}{1.25}
\begin{tabular}{@{}cll@{}}
\toprule
\textbf{Test} & \textbf{Constants in $\hat{\mu}$ (drift)} & \textbf{Constants in $\hat{\sigma}$ (diffusion)} \\
\midrule
C1 & GR precession rate & ($\sigma = 0$) \\
C2 & Friction $\gamma$ & Boltzmann constant $k_B$ \\
C3 & (constant drift) & Fano factor ($F = 1$) \\
C4 & Planck constant $\hbar$ & ($\sigma \approx 0$) \\
C5 & Mean-reversion rate & Van Kampen exponent ($-\tfrac{1}{2}$) \\
C6 & Speed of light $c$ & ($\sigma = 0$) \\
C7 & Group velocity $v_g(k_0, c)$ & Relativistic suppression $\sigma_0/\gamma$ \\
C8 & Damping rate $\gamma$ & $\hbar$, $\omega$ (via zero-point energy) \\
C9 & Damping rate $\gamma$ & $c$ (with $\hbar$ and $\omega$ cancelled) \\
\bottomrule
\end{tabular}
\end{table}

The progression from C8 to C9 is of particular significance.  In C8, $\hbar$ enters $\sigma$ explicitly through the zero-point energy.  In C9, both $\hbar$ and the oscillator frequency $\omega$ cancel identically from $\sigma$ for a massless field, leaving $\sigma = c\sqrt{\gamma}$.  The speed of light $c$, absent from $\sigma$ in C2--C5, enters as the propagation speed of the massless field.  The damping rate $\gamma$, an external parameter in C9, becomes, for gravity, the self-coupling rate $\gamma_\mathrm{grav} = 1/t_P$, which is the unique rate constructible from $\{\hbar, G, c\}$.  The three constants thus play distinct and irreducible roles: $\hbar$ quantises the field, $c$ sets the velocity scale, and $G$ determines the self-coupling that closes the system without parameters fitted to the predicted data.

\subsection{The $\sigma$-Continuum}
\label{subsec:sigma_continuum}

\begin{figure}[p]
\centering
\includegraphics[width=0.95\textwidth]{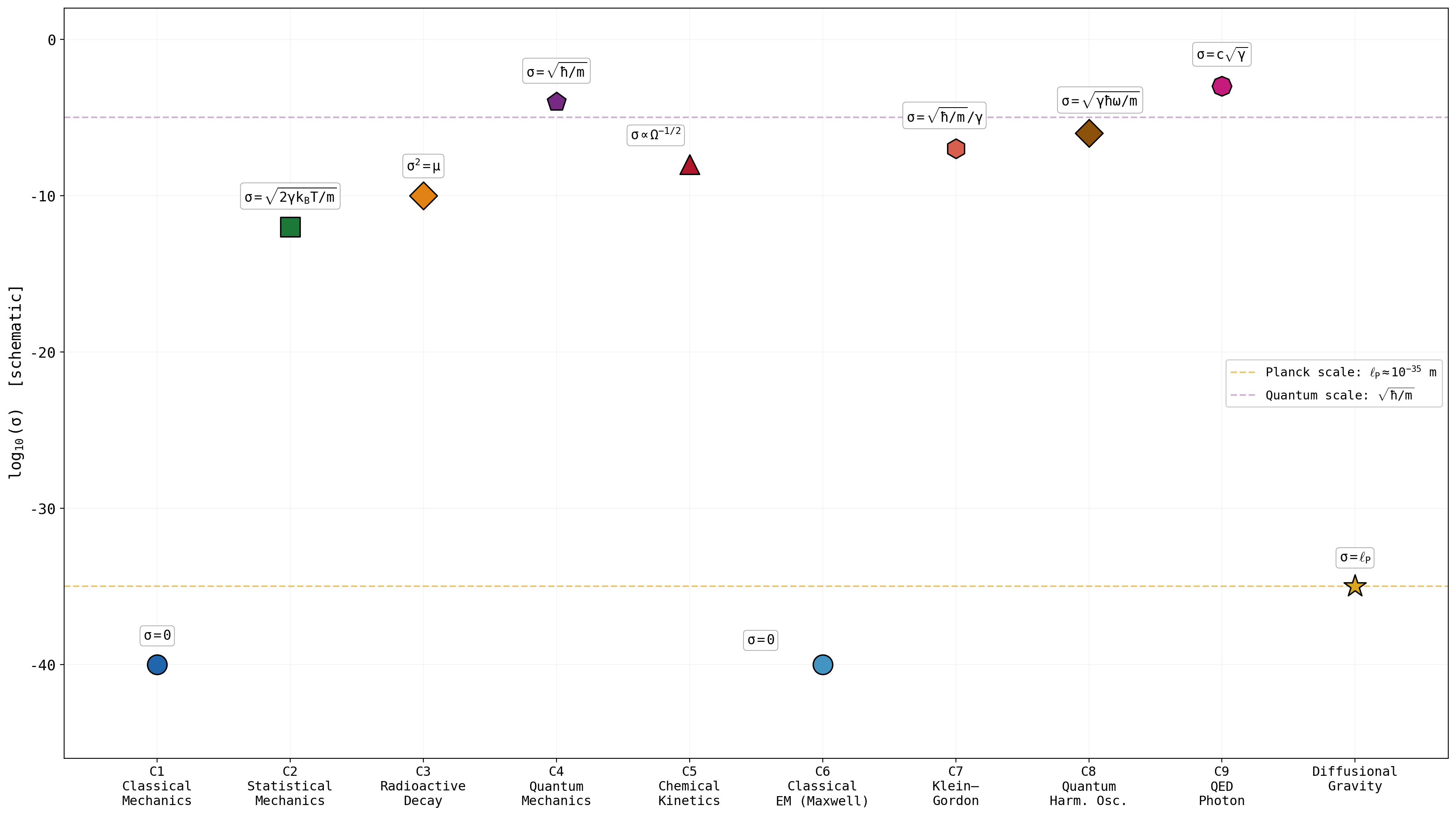}
\caption{\textbf{The $\sigma$-Continuum.}  Recovered diffusion coefficient $\sigma$ across all nine physical domains and the derived gravitational value, displayed on a logarithmic vertical axis ($\log_{10}\sigma$, schematic units).  Each domain is represented by a distinct marker shape and colour, labelled with its recovered $\sigma$ expression.
Deterministic domains (C1: classical mechanics, dark blue circle; C6: classical electromagnetism, light blue circle) are placed at $\log_{10}(\sigma) \approx -40$ for visualisation, labelled $\sigma = 0$.
Stochastic domains span intermediate scales: C2 (statistical mechanics, green square, $\sigma = \sqrt{2\gamma k_BT/m}$), C3 (radioactive decay, orange diamond, $\sigma^2 = \mu$), C4 (quantum mechanics, purple pentagon, $\sigma = \sqrt{\hbar/m}$), C5 (chemical kinetics, red triangle, $\sigma \propto \Omega^{-1/2}$).
The relativistic and quantum field theory domains occupy the upper region: C7 (Klein--Gordon, red--brown pentagon, $\sigma = \sqrt{\hbar/m}/\gamma$), C8 (quantum harmonic oscillator, brown diamond, $\sigma = \sqrt{\gamma\hbar\omega/m}$), C9 (QED photon, magenta circle, $\sigma = c\sqrt{\gamma}$).
The dashed lilac horizontal line marks the quantum scale $\sigma \sim \sqrt{\hbar/m}$; the dashed gold horizontal line marks the Planck scale $\ell_P \approx 10^{-35}$~m.  The gold star at the far right represents $\sigma = \ell_P$ for gravitational degrees of freedom, derived by the Gravitational Diffusion Theorem (\S\ref{sec:discussion}; Appendix~C).  The continuum from $\sigma = 0$ (C1, C6) through thermal, quantum, and relativistic scales (C2--C9) is an empirical pattern arising from nine independent blind recovery experiments; the gravitational endpoint $\sigma = \ell_P$ is derived from this pattern by the Gravitational Diffusion Theorem.}
\label{fig:sigma_continuum}
\end{figure}

Figure~\ref{fig:sigma_continuum} presents the diffusion coefficient $\sigma$ recovered across all nine domains on a single axis.  The pattern is an empirical observation: $\sigma$ varies systematically from zero (deterministic domains: C1, C6) through thermal ($k_BT$-dependent: C2), Poisson (state-dependent: C3), quantum ($\hbar$-dependent: C4, C8), and chemical ($\Omega$-dependent: C5) regimes, to relativistic ($c$-dependent: C7, C9).

This continuum is an empirical observation within the present nine-domain study, emerging from nine independent applications of the same blind recovery protocol.  The question of what it implies: whether a common stochastic structure extends across these domains and what form it takes for gravitational degrees of freedom is addressed in the Discussion.

\section{Discussion}
\label{sec:discussion}

The Discussion proceeds through four stages.  Section~\ref{subsec:disc_continuum} identifies the $\sigma$-continuum as an empirical observation arising from the nine-domain results.  Sections~\ref{subsec:capstone1}--\ref{subsec:capstone2} derive the gravitational diffusion coefficient from this observation combined with three foundational results of established physics; the complete proof chain occupies Appendix~C and constitutes a theorem (the Gravitational Diffusion Theorem, Theorem~\ref{thm:main}), not an interpretive hypothesis.  Section~\ref{subsec:capstone3} poses four canonical axioms motivated by the preceding derivation.  Appendix~D demonstrates that within these axioms, every element of the superspace SDE is uniquely determined.  Table~\ref{tab:derivation_chain} summarises the epistemic status of each step.

\begin{table}[ht]
\centering
\caption{Derivation chain: epistemic status of each step from empirical observation through theorems to axiomatic formalisation.  The column ``External inputs'' identifies what each step assumes beyond the preceding steps.  Only four canonical axioms are postulated; all other results are derived.}
\label{tab:derivation_chain}
\small
\renewcommand{\arraystretch}{1.20}
\begin{tabular}{@{}p{3.2cm}p{4.0cm}p{2.8cm}p{2.2cm}p{2.0cm}@{}}
\toprule
\textbf{Step} & \textbf{What is established} & \textbf{Epistemic status} & \textbf{External inputs} & \textbf{Location} \\
\midrule
$\sigma$-continuum & Diffusion coefficients across 9 domains follow systematic pattern & Empirical observation & None & \S\ref{sec:results} \\[3pt]
FDT master formula & $\sigma^2 = \gamma\hbar\omega/m$ for damped QHO at zero temperature & Theorem\newline(established) & Callen--Welton & App.~C, Lem.~\ref{lem:master} \\[3pt]
Massless cancellation & $\sigma = c\sqrt{\gamma}$ for massless fields; $\hbar$ and $\omega$ cancel identically & Theorem\newline(this work) & Mass--energy equiv. & App.~C, Thm.~\ref{thm:cancellation} \\[3pt]
Dimensional uniqueness of $\gamma_\mathrm{grav}$ & $\gamma_\mathrm{grav} = \alpha/t_P$ is the unique rate from $\{\hbar, G, c\}$ & Theorem\newline(this work) & None & App.~C, Lem.~\ref{lem:rate} \\[3pt]
Self-bath (Mori--Zwanzig) & Self-coupling follows from NZ projection of superspace FP equation & Theorem\newline(this work) & Equivalence principle & App.~C, Prop.~\ref{prop:mz-sketch} \\[3pt]
$\alpha \leq 1$ (metric consistency) & Perturbation theory: $\langle h^2 \rangle = \alpha^2 \leq 1$ over one $t_P$ & Theorem\newline(this work) & Metric positivity & App.~C, Thm.~\ref{thm:upper-bound} \\[3pt]
$\alpha = 1$ (critical damping) & $\gamma(\omega^*) = \omega^*$ has unique solution $\omega^* = 1/t_P$ & Theorem\newline(this work) & Self-coupling rate & App.~C, Thm.~\ref{thm:critical-damping} \\[3pt]
$\alpha \geq 1$ (singularity resolution) & Curvature fluctuations must reach $K_P$ at the Planck scale & Theorem\newline(this work) & Curvature scaling & App.~C, Thm.~\ref{thm:lower-bound} \\[3pt]
$\sigma = \ell_P$ (gravitational) & Combining massless cancellation with $\alpha = 1$ & Theorem\newline(this work) & Above inputs & App.~C, Thm.~\ref{thm:main} \\[3pt]
\midrule
Axioms A1--A4 & Formal axiomatic structure for diffusional gravity & Postulate & --- & \S\ref{subsec:capstone3} \\[3pt]
Wiener character & L\'evy--Khintchine + path continuity & Theorem & A1 & App.~D, Thm.~\ref{thm:levy_wiener} \\[3pt]
Drift uniqueness & Lovelock classification + classical correspondence & Theorem & A3 & App.~D, Thm.~\ref{thm:drift_uniqueness} \\[3pt]
Emergent time & Quadratic variation; drift-independent by It\^o calculus & Theorem & A2 & App.~D, Thm.~\ref{thm:emergent_time} \\[3pt]
Amplitude uniqueness & Three independent arguments converge on $\sigma = \ell_P$ & Theorem & A1--A4 & App.~D, Thm.~\ref{thm:uniqueness_axiom} \\
\bottomrule
\end{tabular}
\end{table}

\subsection{The $\sigma$-Continuum as Empirical Evidence for Universal Stochastic Structure}
\label{subsec:disc_continuum}

The nine-domain validation of \S\ref{sec:results} establishes that the stochastic differential equation $\mathrm{d}X = \mu(X)\,\mathrm{d}t + \sigma(X)\,\mathrm{d}W$ is not solely a mathematical device but a recurring structural element across physical domains.  A single non-parametric pipeline, receiving only raw time series and sampling interval, recovers the governing equations of classical mechanics, statistical mechanics, nuclear physics, quantum mechanics, chemical kinetics, electromagnetism, relativistic quantum mechanics, and quantum field theory, with recovery errors quantified by bootstrap confidence intervals (Table~\ref{tab:results_summary}) and without domain-specific input.

The recovered drift $\hat{\mu}$ and diffusion $\hat{\sigma}$ are not phenomenological regressions.  They contain the fundamental constants of each domain: $k_B$ in Brownian motion, $\hbar$ in wavepacket spreading, $c$ in electromagnetic propagation, the Van Kampen exponent in chemical kinetics.  The constants are not inserted. They emerge from the local geometry of the correlation manifold.

The $\sigma$-continuum (Fig.~\ref{fig:sigma_continuum}) is an observation arising from these nine independent experiments.  Whether the pattern extends to gravitational degrees of freedom, and what determines $\sigma$ for gravity, is addressed in the following sections.

\subsection{Relativistic Structure in the Diffusion Coefficient}
\label{subsec:capstone1}

The C6--C7 transition provides the first structural constraint.  The Dambis--Dubins--Schwarz time-change theorem~\cite{RevuzYor1999}, applied to Nelson's stochastic mechanics in a relativistic setting, yields a kinematic identity: the diffusion coefficient in the laboratory frame is suppressed by the Lorentz factor,
\begin{equation}
\label{eq:relativistic_suppression}
\sigma_\mathrm{KG} = \frac{\sigma_0}{\gamma},
\end{equation}
where $\sigma_0 = \sqrt{\hbar/m}$ is the rest-frame Nelson coefficient and $\gamma = (1 - v^2/c^2)^{-1/2}$.  This suppression arises from two structurally independent factors: a kinematic $1/\sqrt{\gamma}$ from the DDS time change, and a dynamical $1/\sqrt{\gamma}$ from the Klein--Gordon~\cite{Klein1926,Gordon1926} conserved density $\rho_\mathrm{KG} = \gamma|\phi|^2$.  The complete derivation, including the Fokker--Planck equation in both proper and laboratory frames, is given in Appendix~B.

The pipeline recovers the $1/\gamma$ suppression to $\sim$1\% (Fig.~\ref{fig:c67}c) and the group velocity to $R^2 = 0.9999$ (Fig.~\ref{fig:c67}f).  The independent recovery of $v_g$ at each momentum mitigates a potential circularity: because $\gamma$ depends on $v$, the suppression $\sigma \propto 1/\gamma$ is validated only if the velocity is itself recovered independently, not assumed.  Both channels, drift and diffusion, carry relativistic content.  The speed of light $c$ enters $\sigma$ through the dispersion relation, alongside $\hbar$ and $m$.

Five structural observations follow from C6--C7 taken together.  Observations 1, 4, and 5 are empirical, arising from the pipeline results.  Observations 2 and 3 are mathematical consequences of theorems proved in Appendix~B (Theorem~\ref{thm:sigma_gamma} and Proposition~\ref{prop:photon}).

First, $\sigma$ is not a domain label but a dynamical quantity: within C7, $\sigma$ varies continuously with $\gamma$ across seven momenta, and the pipeline tracks this variation to sub-percent accuracy (Fig.~\ref{fig:c67}c,g).  The robustness of the scaling law under reparameterisation from $\gamma$-space to momentum space (Fig.~\ref{fig:c67}g) confirms that the suppression reflects a genuine dynamical relationship, not a coordinate artefact.

Second, the relativistic modification of $\sigma$ is structural, not kinematic.  It arises from two independent mechanisms (the DDS time change and the Klein--Gordon density) that combine multiplicatively.  This is not a Lorentz boost of a scalar, but a consequence of how probability density transforms under relativistic dynamics.

Third, classical electromagnetism (C6, $\sigma = 0$) is recovered as the massless limit of the same suppression law: as $\gamma \to \infty$, $\sigma_\mathrm{KG} = \sigma_0/\gamma \to 0$.  Determinism is not a separate category but the high-$\gamma$ limit of stochasticity.

Fourth, the C6--C7 transition demonstrates that a single mathematical structure, the stochastic differential equation with state-dependent $\sigma$, accommodates both the deterministic and stochastic domains, with the boundary between them set by the Lorentz factor.

Fifth, the progression from C1 ($\sigma = 0$) through C2--C5 ($\sigma > 0$, non-relativistic) to C7 ($\sigma = \sigma_0/\gamma$, relativistic) constitutes an empirically observed hierarchy.  The subsequent derivation for gravitational degrees of freedom (\S\ref{subsec:capstone2}) continues this hierarchy through the Gravitational Diffusion Theorem (Appendix~C, Theorem~\ref{thm:main}): the fluctuation--dissipation relation, the massless cancellation, and the self-coupling rate determined by the equivalence principle.

\subsection{From Quantum Fields to Gravitational $\sigma$}
\label{subsec:capstone2}

\noindent\textbf{The Gravitational Diffusion Theorem} (Appendix~C, Theorem~\ref{thm:main}). \textit{For any field variable $X$ describing gravitational degrees of freedom, the diffusion coefficient is $\sigma = X_P/\sqrt{t_P}$, where $X_P$ is the Planck unit of $X$.  For proper-distance fluctuations: $\sigma = \ell_P/\sqrt{t_P}$.}

\medskip

The proof proceeds in three stages, each building on validated results.

The C8--C9 progression reveals how the three fundamental constants $\{\hbar, G, c\}$ acquire distinct roles in the diffusion coefficient.

In C8 (quantum harmonic oscillator at zero temperature), the fluctuation--dissipation theorem gives $\sigma^2 = \gamma\hbar\omega/m$.  The Planck constant $\hbar$ enters explicitly through the zero-point energy $\tfrac{1}{2}\hbar\omega$, where its role is to quantise the field.

In C9 (quantised electromagnetic field), mass--energy equivalence gives $m_\mathrm{eff} = \hbar\omega/c^2$.  Substitution into the fluctuation--dissipation relation yields the massless cancellation (Eq.~\ref{eq:massless_cancellation}): $\sigma = c\sqrt{\gamma}$, with $\hbar$ and $\omega$ cancelling identically.  The speed of light $c$ sets the velocity scale, where the damping rate $\gamma$ sets the coupling strength.

For the photon, $\gamma$ is an external parameter (cavity losses, material absorption).  For gravity, the coupling is not external.  The gravitational field is self-interacting, and the only Planck-scale rate constructible from $\{\hbar, G, c\}$ without introducing additional scales is
\begin{equation}
\label{eq:gamma_grav}
\gamma_\mathrm{grav} = \frac{1}{t_P},
\end{equation}
where $t_P = \sqrt{\hbar G/c^5}$ is the Planck time.  The self-coupling rate $\gamma_\mathrm{grav} = 1/t_P$ is not merely the unique dimensionally consistent rate.  It is derived from the Mori--Zwanzig projection of the superspace Fokker--Planck equation (Appendix~C, Proposition~\ref{prop:mz-sketch}): the gravitational field acts as its own bath, with the self-coupling rate determined by the cubic vertex of the Einstein--Hilbert action.  The dimensional argument identifies the correct answer; the Mori--Zwanzig projection explains \emph{why} it is correct.  Substitution into $\sigma = c\sqrt{\gamma}$ gives
\begin{equation}
\label{eq:sigma_planck}
\sigma = c\sqrt{\frac{1}{t_P}} = \frac{\ell_P}{\sqrt{t_P}} = \frac{\ell_P}{t_P^{1/2}},
\end{equation}
where $\ell_P = \sqrt{\hbar G/c^3}$ is the Planck length.  This corresponds to a proper-distance fluctuation of one Planck length per Planck time.

In summary, the three stages are as follows.  \emph{Stage~1}: the quantum harmonic oscillator (C8) establishes the fluctuation--dissipation relation $\sigma^2 = \gamma\hbar\omega/m$ (validated: $1.33\%$ error from 50 independent MZ-corrected runs, Table~\ref{tab:results_summary}).  \emph{Stage~2}: for a massless field with $m_\mathrm{eff} = \hbar\omega/c^2$, both $\hbar$ and $\omega$ cancel identically, giving $\sigma = c\sqrt{\gamma}$, the massless cancellation (validated: C9 photon field, $6.72\%$ error, constant $\sigma$ across nine frequencies).  \emph{Stage~3}: for gravity, the self-coupling rate $\gamma_\mathrm{grav} = 1/t_P$ is the unique Planck-scale rate, giving $\sigma = \ell_P/\sqrt{t_P}$.  The epistemic boundary between Stages~2 and~3 is precise: the C9 pipeline results validate the mathematical formula $\sigma = c\sqrt{\gamma}$ for an externally damped massless field.  Stage~3 applies the same formula to a self-coupled field, where the damping rate is not external but is determined by the equivalence principle, the most precisely tested input in the derivation chain.  The self-coupling mechanism is derived, not assumed, in Appendix~C (Proposition~\ref{prop:mz-sketch} and Theorem~\ref{thm:gamma-exact}); the pipeline validates the formula, while the appendix validates the physical mechanism.

The dimensionless prefactor $\alpha$ in $\gamma_\mathrm{grav} = \alpha/t_P$ is determined by three independent arguments (Appendix~C, \S\ref{sec:alpha-proof}).  Metric self-consistency requires $\langle h^2 \rangle = \alpha^2 \leq 1$ over one Planck time, giving the upper bound $\alpha \leq 1$.  The critical damping condition: the unique frequency at which $\gamma(\omega) = t_P^2 \omega^3$ equals $\omega$, yields $\omega^* = 1/t_P$ exactly, giving $\alpha = 1$ as an algebraic identity.  Singularity resolution requires that curvature fluctuations reach the Planck curvature at the minimum scale, giving $\alpha \geq 1$.  Together: $\alpha \leq 1 \wedge \alpha = 1 \wedge \alpha \geq 1 \Rightarrow \alpha = 1$ uniquely.  The three arguments share a single input, the self-coupling rate $\gamma(\omega) = t_P^2\omega^3$ (Theorem~\ref{thm:gamma-exact}), but are otherwise logically independent: Argument~1 invokes metric positivity, Argument~2 invokes the oscillator--diffusion boundary, and Argument~3 invokes curvature scaling.  No two arguments use the same physical principle beyond the shared input.  The complete derivation chain is developed in Appendix~C.

\subsection{Axiomatic Formalisation}
\label{subsec:capstone3}

The axioms require a shift in the role of the stochastic differential equation.
In \S\ref{sec:methods}--\S\ref{sec:results}, the stochastic
differential equations describe matter evolving in a fixed spacetime,
and the evolution parameter is physical time measured by a laboratory
clock.  From this point forward, the stochastic differential equation
describes spacetime itself, and the evolution parameter~$\tau$ is a
pre-geometric index: a label ordering the steps of a Wiener process
on Wheeler's superspace.  Physical time emerges as a derived quantity,
constructed from the quadratic variation of the metric process
(\S\ref{sec:emergent_time}).  The ordering parameter~$\tau$ is not the time of any clock; clocks are built from the metric, which is the stochastic variable.  This distinction is the framework's resolution of the problem of time in canonical quantum gravity.

Appendix~C derives $\sigma = \ell_P$ from three established results and the empirical pattern of C1--C9.  The four canonical axioms below formalise this derivation into a self-contained axiomatic structure; Axiom~A2 is a derived result promoted to axiomatic status for deductive economy (Appendix~C, Remark following the Gravitational Diffusion Theorem).

\begin{description}
\item[A1 (Configuration Space)] The gravitational degrees of freedom are described by a Riemannian three-metric $g_{ij}$ on a spatial manifold $\Sigma$.  The configuration space is Wheeler's superspace $\mathcal{C} = \mathrm{Riem}(\Sigma)/\mathrm{Diff}(\Sigma)$~\cite{Wheeler1968}, equipped with the DeWitt supermetric~\cite{DeWitt1967}.

\item[A2 (Stochastic Evolution)] The metric undergoes a stochastic process on $\mathcal{C}$:
\begin{equation}
\label{eq:stochastic_metric}
\mathrm{d}g_{ij} = \mathcal{D}_{ij}[g]\,\mathrm{d}\tau + \ell_P\,\mathrm{d}W_{ij},
\end{equation}
where $\mathcal{D}_{ij}[g]$ is a drift functional, $\ell_P$ is the Planck length~\cite{Planck1899}, and $\mathrm{d}W_{ij}$ is a symmetric tensor-valued Wiener process.  The ordering parameter $\tau$ is not a background time coordinate.

\item[A3 (Classical Correspondence)] In the macroscopic limit ($L \gg \ell_P$), the stochastic evolution reproduces general relativity~\cite{Einstein1915}: $R_{\mu\nu} - \tfrac{1}{2}g_{\mu\nu}R + \Lambda g_{\mu\nu} = (8\pi G/c^4)\,T_{\mu\nu}$.

\item[A4 (Single Realisation)] The observable universe is one realisation of the stochastic process.  Probability is epistemic.
\end{description}

Physical time is not assumed.  It emerges as the quadratic variation of the stochastic process: $t(\tau) = (1/\ell_P^2 n_\mathrm{eff}) \int \langle \mathrm{d}g, K^{-1} \mathrm{d}g \rangle$ (Appendix~D, Theorem~\ref{thm:emergent_time}).  The effective mode count $n_\mathrm{eff}$ is formally divergent on non-compact manifolds; it is rendered finite by a Planck-scale wavelength cutoff ($n_\mathrm{eff} \sim V/\ell_P^3$) or by minisuperspace reduction ($n_\mathrm{eff} = 1$ for FLRW), and is absorbed into the normalisation of $t(\tau)$ so that no physical prediction depends on $n_\mathrm{eff}$ separately (Appendix~D, Remark~\ref{rem:mode_regularisation}).  By the It\^o product rules, $(\mathrm{d}\tau)^2 = 0$, $\mathrm{d}\beta_n \mathrm{d}\tau = 0$, $\mathrm{d}\beta_m \mathrm{d}\beta_n = \delta_{mn}\mathrm{d}\tau$, the drift $\mathcal{D}_{ij}$ contributes identically zero to the quadratic variation.  The emergent time functional is therefore strictly monotone (from positivity of $K$), additive (from independent increments), coordinate-independent (from tensorial covariance), and exactly drift-independent (from the It\^o calculus).  In the macroscopic limit $L \gg \ell_P$, the emergent time coincides with proper time through the standard ADM relation $\mathrm{d}t_\mathrm{proper} = N\,\mathrm{d}t$, with stochastic corrections of $O(\ell_P^2/L^2) \sim 10^{-70}$ for astrophysical scales.  This construction addresses the problem of time in canonical quantum gravity: time is not a background coordinate but a derived quantity, constructed from the diffusion itself, with the deterministic gravitational dynamics contributing nothing to its rate.

Within axioms A1--A4, every element of Eq.~\eqref{eq:stochastic_metric} is uniquely determined.

\emph{Noise character.}  The L\'evy--Khintchine classification theorem establishes that any stochastic process with stationary independent increments and continuous sample paths is a Wiener process (plus drift).  Path continuity is a physical consequence of Axiom~A1: a jump discontinuity in the metric would generically violate positive-definiteness of $g_{ij}$ or produce infinite curvature.  The Wiener process is therefore derived from the remaining structure, not postulated independently (Appendix~D, Theorem~\ref{thm:levy_wiener}).

\emph{Drift.}  Within the class of ultralocal, second-order, diffeomorphism-covariant functionals that reproduce the Einstein evolution in the classical limit (A3), the drift is uniquely the Einstein flow.  This invokes the Lovelock classification: in three spatial dimensions, the unique divergence-free symmetric two-tensor constructed from the metric and its derivatives is the Einstein tensor.

\emph{Covariance.}  The covariance operator $K_{ijkl}$ must be ultralocal, diffeomorphism-covariant, and positive-definite on symmetric two-tensors.  The unique such family is the DeWitt supermetric $G^{ijkl}(\lambda)$ with $\lambda > -1/3$~\cite{DeWitt1967}.

\emph{Amplitude.}  The fluctuation amplitude $\sigma = \ell_P$ is fixed by the three independent arguments of Appendix~C (\S\ref{sec:alpha-proof}): metric self-consistency ($\alpha \leq 1$), critical damping ($\alpha = 1$), and singularity resolution ($\alpha \geq 1$).

The complete axiomatic derivation, including the uniqueness theorems for all four elements, is given in Appendix~D.

Postulate~\ref{ax:minimality_axiom} (Minimality) is not an additional dynamical assumption beyond A1--A4 and path continuity.
The L\'{e}vy--Khintchine classification of processes with stationary independent
increments, combined with the path-continuity requirement on metric configurations
imposed by Axiom~A1, uniquely selects the Wiener process from the class of all
L\'{e}vy processes (Appendix~D, Theorem~\ref{thm:levy_wiener}).  The Markov property
and spatial locality of the noise follow respectively from the emergent-time
construction and cluster decomposition (Appendix~D,
Remarks~\ref{rem:markov_property} and~\ref{rem:spatial_locality}).  The resulting
Fokker--Planck evolution on superspace admits both a formal Wick-rotation
correspondence with the Wheeler--DeWitt equation of canonical quantum gravity and a
Martin--Siggia--Rose path-integral representation whose saddle-point limit recovers
classical general relativity; physical observables are shown to be independent of
the foliation (lapse) choice (Appendix~D, \S\ref{subsec:guerra_ruggiero}
and~\S\ref{subsec:msr}).

Within axioms A1--A4, Eq.~\eqref{eq:stochastic_metric} is non-parametric: all coefficients are derived from the axioms.

\subsection{The Superspace Diffusion Framework}
\label{subsec:disc_emergent_law}

The derivations that follow from axioms A1--A4 constitute the Superspace Diffusion Framework.  The superspace diffusion hypothesis (Eq.~\ref{eq:stochastic_metric}) is one implication.  For example, at galactic scales, the Mori--Zwanzig coarse-graining of the superspace Fokker--Planck equation (Proposition~\ref{prop:mz-sketch}) yields non-parametric predictions for the statistical distribution of gravitational acceleration.  Derivations and implications of contact with observational data in galactic kinematics, cosmology, horizon physics, emergence of classicality, particle physics, information theoretics, and foundational physics and mathematics are presented in the research navigator~\cite{Navigator2026}.

The Superspace Diffusion Framework extends the $\sigma$-continuum from an empirical observation to a formal structure through a theorem chain (Appendix~C) derived from the fluctuation--dissipation theorem, the equivalence principle, and the massless mode structure of linearised gravity.

\subsection{Relation to Nelson's Stochastic Mechanics}
\label{subsec:disc_nelson}

Nelson~\cite{Nelson1966} solved the forward problem for quantum mechanics: given $\sigma = \sqrt{\hbar/m}$, derive the Schr\"odinger equation.  The present work solves the inverse problem across nine domains: given time series data, recover $\sigma$ without assuming the physics.  The forward and inverse paths converge at C4, where the pipeline recovers Nelson's $\sigma$ to $0.05\%$ (Table~\ref{tab:results_summary}).  Nelson's insight, that physical law admits stochastic content~\cite{Nelson1985}, is here extended across the $\sigma$-continuum to gravitational degrees of freedom.

\subsubsection{Distinguishing the stochastic and quantum formulations}

The Wick rotation correspondence (Appendix~D, Theorem~\ref{thm:guerra_ruggiero}) maps the Fokker--Planck generator on superspace to the Wheeler--DeWitt Hamiltonian through the analytic continuation $\tau \to -it/\hbar$.  This correspondence is exact at the level of the generator and establishes that the stochastic framework is connected to canonical quantum gravity by the same analytic structure that relates Euclidean and Lorentzian quantum field theory.

The correspondence does not, however, extend to the level of solutions and physical predictions.  Three structural properties of the Fokker--Planck evolution have no counterparts in the Wheeler--DeWitt equation.

\emph{First, the Fokker--Planck equation possesses a unique normalisable steady state.}  For any Fokker--Planck equation with non-degenerate diffusion and confining drift on a half-line or bounded domain, the zero-current condition $\partial_g[D \cdot P] = A \cdot P$ determines a unique normalisable distribution.  The superspace Fokker--Planck equation (Appendix~D, Eq.~\ref{eq:fp_superspace}), when coarse-grained to galactic-scale gravitational degrees of freedom via the Mori--Zwanzig projection (Appendix~C, Proposition~\ref{prop:mz-sketch}), inherits this property: its steady state is uniquely determined by the drift and diffusion derived from A1--A4.  The Wheeler--DeWitt equation $\hat{H}\Psi = 0$ is a constraint equation: it has no time parameter, no spectral gap, and no mechanism for selecting a unique solution from its solution space.

\emph{Second, the Fokker--Planck evolution is dissipative.}  It drives any initial probability distribution toward the unique steady state on a timescale set by the spectral gap.  This irreversible relaxation produces definite predictions: regardless of initial conditions, the distribution converges to the steady state.  Quantum evolution preserves unitarity: pure states remain pure and the evolution is time-reversible: producing recurrences rather than relaxation.

\emph{Third, the Fokker--Planck steady state depends on the positivity of $P[g] \geq 0$ and the real-valued character of the diffusion coefficient.}  Under the Wick rotation, $P$ maps to a complex-valued wave functional $\Psi[g]$ that is generically oscillatory and not normalisable in the standard $L^2$ sense.  The positive, real-valued, normalisable steady state of the Fokker--Planck equation has no natural image under this mapping.

These three properties: uniqueness, irreversible relaxation, and positivity: are consequences of the diffusion character of the evolution.  The Wick rotation preserves the generator but not these properties, in the same way that analytic continuation between Euclidean and Lorentzian field theory preserves the action but not the metric signature.

The framework therefore proposes a specific physical hypothesis: that the gravitational degrees of freedom evolve by diffusion on superspace, governed by the Fokker--Planck equation (Axiom~A2), rather than by unitary quantum evolution governed by the Wheeler--DeWitt equation.  The two formulations share a generator and a classical limit (general relativity, recovered as $L \gg \ell_P$ in both cases) but make distinct predictions; an implication is that the statistical distribution of gravitational acceleration at galactic scales differs between the two.

\subsubsection{Relationship to conventional approaches to quantum gravity}

The stochastic formulation occupies a specific position relative to the three standard approaches.

In canonical quantum gravity, the Wheeler--DeWitt equation $\hat{H}\Psi = 0$ defines the quantum state of the gravitational field.  The Fokker--Planck equation on superspace is related to it by the Wick rotation correspondence (Appendix~D, Theorem~\ref{thm:guerra_ruggiero}).  At the level of generators, the two are formally equivalent.  At the level of solutions, they differ: the Fokker--Planck equation has a unique steady state reached by irreversible relaxation, where the Wheeler--DeWitt equation has a solution space parameterised by boundary conditions.

In Euclidean quantum gravity, the gravitational path integral defines transition amplitudes between three-geometries.  The Martin--Siggia--Rose path integral (Appendix~D, Theorem~\ref{thm:msr_action}) establishes that the stochastic framework contains the Euclidean gravitational path integral as its saddle-point approximation: the classical constraint surface (Einstein's equations) is the saddle-point locus, with stochastic excursions of amplitude $O(\ell_P)$ (Appendix~D, Corollary~\ref{eq:msr_partition}).  The stochastic framework thus extends the Euclidean programme beyond the saddle-point approximation, providing a well-defined measure on the space of metrics that the Euclidean path integral, which suffers from the conformal factor problem, does not possess.

In stochastic quantisation \`a la Parisi--Wu~\cite{ParisiWu1981}, quantum field theory is obtained as the equilibrium distribution of a diffusion in a fictitious stochastic time.  The present framework shares the mathematical structure but differs in a physical identification: the ordering parameter $\tau$ is identified with physical time through the emergent time theorem (Appendix~D, Theorem~1), not treated as a fictitious coordinate.  The equilibrium distribution is the physical steady state of the gravitational field, not a computational device for reproducing the Euclidean path integral.

The stochastic formulation is therefore not a reformulation of any existing approach to quantum gravity, but a distinct theoretical framework that shares their classical limit and connects to each through specific mathematical correspondences.

\subsection{Limitations}
\label{subsec:limitations}

All C1--C9 data are synthetic.  The pipeline has not been applied to raw experimental data in this work.  The synthetic data validates the methodology, where experimental application is the subject of subsequent publications.  The blind protocol ensures that the pipeline cannot use domain knowledge, but the selection of the nine test domains is informed by the authors' knowledge of physics; the pipeline's performance on domains outside this selection is an open question.

The $k$-NN Kramers--Moyal estimators carry a systematic bias decomposable as a Mori--Zwanzig memory kernel with two rank-1 components: a spatially varying term proportional to the Laplacian of $\sigma^2$ on the correlation manifold, and a spatially uniform finite-sample term scaling as $-\sigma^2/k$ (Supplementary Materials, \S7.2).  The two-level adaptive corrector (Algorithm~3 in the Supplementary Materials) reduces this bias by $87$--$94\%$ in domains where the spatial-averaging component is detectable (C3, C8, C9) and provably abstains where the treatable component is below the noise floor (C2).  Residual errors of $0.25$--$6.72\%$ after correction reflect the finite-sample variance component and higher-order terms.  For scaling exponents (C5, C7), the multiplicative bias cancels in the log--log slope regardless of correction.

The gravitational prediction $\sigma = \ell_P$ ($\sim\!10^{-35}$~m) is not directly testable with current technology.  An implication is that its consequences generate quantitative predictions at accessible scales: the Mori--Zwanzig coarse-graining produces a characteristic acceleration scale and modified gravitational dynamics that make contact with galactic kinematic data.

The axioms A1--A4 are stated as postulates for deductive clarity, though each has independent justification.  A1 adopts the standard canonical general relativity configuration space (Wheeler's superspace~\cite{Wheeler1968}).  A2 is a derived result (the Gravitational Diffusion Theorem, Appendix~C) promoted to axiomatic status for deductive economy.  A3 requires consistency with general relativity in the macroscopic limit.  A4, the identification of probability as epistemic, is the sole genuinely free interpretive choice.  Within A1--A4, the noise character, drift, covariance, fluctuation amplitude, and emergent time are all uniquely determined by theorem (Appendix~D).  Alternative axiomatisations may exist, but the physical content of the predictions follows from the derivation chain of Appendix~C independently of the axiomatic packaging.  The empirical question is therefore whether A1--A4 hold, not whether the predictions follow from them.  If the axioms are incorrect, the predictions that follow from them need not be physically operative.  An implication is that this question is addressable by confronting the framework's galactic-scale predictions with kinematic data.

To delimit scope explicitly: this paper does not claim to have derived quantum gravity, unified general relativity with quantum mechanics, or proved the axioms A1--A4.  What is claimed is that if A1--A4 hold, then every element of the superspace SDE is uniquely determined as a non-parametric, first-principles consequence of the axioms.

What the derivation chain does establish is that the gravitational diffusion coefficient is uniquely determined (the Gravitational Diffusion Theorem, Appendix~C) by the same physical principles (the fluctuation--dissipation theorem, mass--energy equivalence, and the equivalence principle) that govern the nine validated domains, as a non-parametric, first-principles consequence of general relativity and quantum field theory.

\section{Conclusion}
\label{sec:conclusion}

The Stochastic Embedding Sufficiency Theorem closes a methodological gap between deterministic and stochastic time series analysis, enabling recovery of drift and diffusion fields from scalar observations without prior assumptions about the governing physics. The data determines the non-parametric model structure.

Application of the resulting pipeline to nine physical domains (classical mechanics, statistical mechanics, nuclear physics, quantum mechanics, chemical kinetics, electromagnetism, relativistic quantum mechanics, quantum harmonic oscillator dynamics, and quantum electrodynamics) recovers the known governing equations of each domain blindly, with MZ-corrected recovery errors ranging from $3 \times 10^{-6}\%$ (deterministic domains) to $6.72\%$ (QED photon field), representing an $87$--$94\%$ reduction in the systematic $k$-NN estimator bias compared to uncorrected estimates (Supplementary Materials, \S7.2).  For scaling exponents, the multiplicative bias cancels in the slope and the bootstrap CI contains the theoretical value.  Fundamental physical constants emerge in both the drift and the diffusion channels without prior specification.  The recovered diffusion coefficients, viewed together, constitute an empirical pattern: the $\sigma$-continuum.

Three theorems derived from the relativistic and quantum field theory results yield three structural consequences.  The derivation chain rests on the fluctuation--dissipation theorem (validated by C8 to $1.33\%$), mass--energy equivalence, and the equivalence principle, each among the most precisely tested results in physics.  First, the Dambis--Dubins--Schwarz time-change theorem combined with the Klein--Gordon density produces the relativistic suppression $\sigma = \sigma_0/\gamma$, with physical constants emerging in both channels (Appendix~B).  Second, the massless cancellation and gravitational self-coupling together determine $\sigma = \ell_P/\sqrt{t_P}$ for gravitational degrees of freedom, with three independent consistency arguments converging on the dimensionless prefactor $\alpha = 1$ (Appendix~C).  Third, a set of four canonical axioms (A1--A4) yields a framework within which physical time, the drift, the covariance operator, and the fluctuation amplitude are all uniquely fixed (Appendix~D).

The resulting superspace diffusion hypothesis,
\begin{equation}
\label{eq:conclusion_sde}
\mathrm{d}g_{ij} = \mathcal{D}_{ij}[g]\,\mathrm{d}\tau + \ell_P\,\mathrm{d}W_{ij},
\end{equation}
is non-parametric within A1--A4: the noise character, drift, covariance, fluctuation amplitude, and emergent time are all uniquely determined by theorem (Appendix~D).  The empirical question is therefore whether A1--A4 hold, not whether the predictions follow from them.

The Superspace Diffusion Framework defined by axioms A1--A4 is falsifiable.  An implication is that the Fokker--Planck equation on superspace, coarse-grained via the Mori--Zwanzig projection to galactic-scale gravitational degrees of freedom, generates non-parametric, first-principles predictions for the statistical distribution of gravitational acceleration.  These predictions are testable against galactic kinematic data including rotation curves, velocity dispersions, wide binaries, galaxy clusters, and gravitational lensing.  Derivations and implications of contact with observational data in galactic kinematics, cosmology, horizon physics, emergence of classicality, particle physics, information theoretics, and foundational physics and mathematics are presented in the research navigator~\cite{Navigator2026}; the complete observational programme is the subject of forthcoming companion papers.  If the axioms are incorrect, the predictions derived from them need not be physically operative: the observational confrontation therefore tests the axioms themselves, not merely their mathematical consequences.
\section*{Acknowledgements}
\label{sec:acknowledgements}

The authors are indebted to UCL's Manufacturing Futures Laboratory for computational resources.  Generative AI accelerated and extended the authors' technical reach, while responsibility, conceptual synthesis and structural design remained human.  As such, the contribution of Anthropic's Claude as an epistemic tool is gratefully acknowledged.

\section*{Author Contributions}
\label{sec:contributions}

\noindent\textbf{Carolina Garcia}: Investigation, Writing (original draft), Writing (review and editing).
\textbf{Luc\'ia Perea Dur\'an}: Investigation, Writing (original draft), Writing (review and editing).
\textbf{Agnese Venezia}: Investigation, Writing (original draft), Writing (review and editing).
\textbf{Alex Conradie}: Conceptualisation, Methodology, Investigation, Formal Analysis, Software, Supervision, Visualisation, Writing (original draft), Writing (review and editing).

\section*{Competing Interests}
\label{sec:competing_interests}

The authors declare no competing interests.

\section*{Data and Code Availability}
\label{sec:data_availability}

The interactive research navigator, including proof derivations and data visualisations supporting this work, is archived at Zenodo: \url{https://doi.org/10.5281/zenodo.19496962}. An online interactive deployment is available at: \url{https://www.emergentlaw.org}.

\clearpage

\clearpage

\clearpage

\clearpage

\clearpage

\clearpage

\clearpage

\clearpage

\clearpage

\clearpage

\appendix

\section*{Appendix, Mathematical Proofs and Derivations}
\addcontentsline{toc}{section}{Appendix, Mathematical Proofs and Derivations}

\renewcommand{\thesubsection}{\thesection.\arabic{subsection}}
\renewcommand{\thesubsubsection}{\thesubsection.\arabic{subsubsection}}
\renewcommand{\theequation}{\thesection.\arabic{equation}}
\renewcommand{\thetheorem}{\thesection.\arabic{theorem}}
\renewcommand{\thedefinition}{\thesection.\arabic{definition}}
\renewcommand{\theremark}{\thesection.\arabic{remark}}
\renewcommand{\theobservation}{\thesection.\arabic{observation}}
\renewcommand{\theaxiom}{\thesection.\arabic{axiom}}
\renewcommand{\thepostulate}{\thesection.\arabic{postulate}}
\renewcommand{\theconstruction}{\thesection.\arabic{construction}}

\renewcommand{\theassumption}{\thesection.\arabic{assumption}}

\section{Derivation of Stochastic Embedding Sufficiency Theorem}
\label{appendix:stochastic_embedding}

\subsection{Introduction}

Classical delay-coordinate embedding theory, most notably Takens' theorem, guarantees diffeomorphic reconstruction of deterministic attractors from time-delay observations. However, this guarantee does not hold when the underlying dynamics are stochastic: a fact operationalized by Cao's $E_2$ statistic approaching unity. In such cases, distinct initial conditions can produce identical embedded observations under different noise realisations, rendering pathwise reconstruction impossible.

The central result established here is that pathwise reconstruction is not required for distributional inference. While stochastic systems do not admit unique trajectories on an attractor, they do admit well-defined probability laws on correlation manifolds. For the purposes of statistical inference on dynamical systems, it is sufficient that distinct system states induce distinct finite-dimensional probability distributions under delay embedding.

The Stochastic Embedding Sufficiency Theorem (Theorem~A.1) formalises this approach. Rather than establishing diffeomorphic injectivity (as in Takens' theorem), the result establishes measure-theoretic injectivity: for almost every point under the invariant measure, distinct points on the underlying correlation manifold correspond to distinct finite-dimensional laws of the embedded process. This guarantee is sufficient for consistent $k$-nearest neighbour estimation, graph-theoretic mixing assessment, and distributional divergence testing. The complete proof is given in Supplemental Materials; the following sections summarise the essential steps.

The proof proceeds in five conceptual stages:
\begin{enumerate}
    \item \textbf{H\"ormander's hypoellipticity} ensures smooth, strictly positive transition densities for the SDE, via the bracket-generating condition and the Stroock--Varadhan support theorem.
    \item \textbf{Malliavin non-degeneracy} guarantees that the Malliavin covariance matrix is invertible almost surely, preventing degeneracy of the embedded probability distributions.
    \item \textbf{Law-separation} establishes that the law-embedding map $\Lambda_m^h$ is $\mu_\infty$-a.e.\ injective via three linked results: transition density separation (semigroup bisection + Varadhan--L\'eandre asymptotics), parametric transversality with an explicit evaluation-map surjectivity lemma, and a direct Frostman covering argument.
    \item \textbf{$E_1$ dimension sufficiency} shows that Cao's $E_1$ statistic detects correlation dimension $D_2$ irrespective of whether the system is deterministic or stochastic, yielding a sufficient embedding dimension $m^* \geq \lceil 2D \rceil + 1$ (equivalently $\lceil 2D_2 \rceil + 1$ under exact-dimensionality).
    \item \textbf{Finite-dimensional law uniqueness} ensures consistency of nearest-neighbour statistics and validates geometric comparison on the correlation manifold.
\end{enumerate}

Together, these results extend the theoretical foundation from deterministic to stochastic systems.

\subsection{The Methodological Divide}

Classical dynamical systems analysis, exemplified by Takens' embedding theorem \cite{Takens1981} and its extensions \cite{SauerYorkeCasdagli1991}, applies to deterministic systems. Given a scalar time series from a deterministic process, one can reconstruct the underlying attractor geometry through time-delay embedding. For a smooth diffeomorphism $\varphi$ on a compact $n$-dimensional manifold $M$ with generic observation function $h$, the delay embedding:
\begin{equation}
\Phi_m(x) = (h(x), h(\varphi(x)), \ldots, h(\varphi^{m-1}(x)))^\top
\label{eq:takens_embedding}
\end{equation}
is a diffeomorphism onto its image for $m \geq 2n+1$.

Cao's $E_1$ and $E_2$ diagnostics \cite{Cao1997}, extending the false nearest neighbours method \cite{kennel1992determining}, operationalise this framework: $E_1$ identifies the minimal embedding dimension, while $E_2$ tests for deterministic structure. When $E_2 \approx 1$ across all dimensions, the system is stochastic, and classical embedding in the Takens sense does not apply.

This created a methodological gap: deterministic systems could be analysed geometrically, but stochastic systems required entirely different methods. No unified framework existed. The conventional interpretation was that if $E_2$ signals stochasticity, geometric methods are inapplicable.

\subsection{Extension to Stochastic Systems}

The theoretical contribution of this work addresses this gap through mathematical extension. The $E_1$-identified embedding, even when $E_2$ signals stochasticity, defines a correlation manifold encoding the system's dynamical degrees of freedom. When stochasticity is detected, this manifold serves as the foundation for distributional analysis.

\begin{definition}[Delay-Vector Law and Correlation Manifold]
\label{def:correlation_manifold}
Assume $(X_t)_{t \in \mathbb{R}}$ is stationary with invariant measure 
$\mu_\infty$ and observation function 
$h : \mathbb{R}^n \to \mathbb{R}$.  Fix $\tau > 0$ and 
$m \in \mathbb{N}$ and define the delay vector
\begin{equation}
\label{eq:delay_vector_law}
Y_t^{(m)} := \bigl(h(X_t),\; h(X_{t-\tau}),\; \ldots,\; 
h(X_{t-(m-1)\tau})\bigr) \in \mathbb{R}^m.
\end{equation}
Let $\nu_m := \mathcal{L}(Y_0^{(m)})$ denote its stationary law.  The 
\emph{correlation manifold} is the support of this law:
\begin{equation}
\label{eq:correlation_manifold_def}
\mathcal{M}_{D_2} := \operatorname{supp}(\nu_m) \subset \mathbb{R}^m,
\end{equation}
where $D_2$ denotes the correlation dimension of $\nu_m$ 
(Eq.~\ref{eq:correlation_integral}).  When $m \geq \lceil 2D_2 \rceil + 1$, 
$\mathcal{M}_{D_2}$ is a $D_2$-dimensional subset of $\mathbb{R}^m$ on 
which the embedded process has well-defined conditional expectations 
for the drift and diffusion fields.
\end{definition}

The $E_1$ statistic detects correlation dimension $D_2$ regardless of whether the system is deterministic or stochastic. A stationary diffusion process possesses a well-defined correlation manifold $\mathcal{M}_{D_2}$, as does a deterministic chaotic attractor. The geometric structure persists; the dynamics upon it become probabilistic rather than deterministic.

\subsection{Mathematical Framework}

\subsubsection{The Stochastic Dynamical System}

Consider the It\^o stochastic differential equation on $\mathbb{R}^n$:
\begin{equation}
dX_t = \mu(X_t)dt + \sigma(X_t)dW_t
\label{eq:ito_sde}
\end{equation}
where $\mu: \mathbb{R}^n \to \mathbb{R}^n$ is the drift vector field, $\sigma: \mathbb{R}^n \to \mathbb{R}^{n \times r}$ is the diffusion coefficient matrix with columns $\sigma_1, \ldots, \sigma_r$, $W_t$ is $r$-dimensional standard Brownian motion, and $\Sigma = \sigma\sigma^\top$ is the diffusion tensor. Time-delay embedding constructs:
\begin{equation}
Y_t^{(m)} = (y_t, y_{t-\tau}, \ldots, y_{t-(m-1)\tau})^\top
\label{eq:app_delay_embedding}
\end{equation}
from observations $y_t = h(X_t)$, where $h: \mathbb{R}^n \to \mathbb{R}$ is a smooth observation function.

\begin{definition}[Delay Embedding Map and Injectivity in Law]
\label{def:delay_map}
The \emph{delay embedding map} $\Phi_m : \mathbb{R}^n \to \mathbb{R}^m$ 
is defined by
\[
\Phi_m(x) = \bigl(h(x),\; h(\varphi_\tau(x)),\; \ldots,\; 
h(\varphi_{(m-1)\tau}(x))\bigr)^\top,
\]
where $\varphi_\tau$ denotes the time-$\tau$ flow of the 
SDE~(\ref{eq:ito_sde}).  The delay representation is said to be 
\emph{injective in law} (with respect to $\mu_\infty$) if, for 
$\mu_\infty$-almost every pair $(x, x')$,
\begin{equation}
\label{eq:injective_in_law}
x \neq x' \;\;\Longrightarrow\;\; 
\mathcal{L}\!\left(Y_0^{(m)}\,\middle|\,X_0 = x\right) \neq 
\mathcal{L}\!\left(Y_0^{(m)}\,\middle|\,X_0 = x'\right),
\end{equation}
where $\mathcal{L}(\cdot|\cdot)$ denotes the conditional law.  This 
is the natural stochastic analogue of the diffeomorphic injectivity 
established by Takens' theorem for deterministic systems: distinct 
states induce distinct probability distributions over delay vectors, 
even though individual realisations may coincide.
\end{definition}

\subsubsection{Cao's $E_1$ and $E_2$ Statistics}

For delay vectors with nearest neighbour index $n(i,m)$, the $E_1$ statistic is:
\begin{equation}
E_1(m) = \frac{E(m+1)}{E(m)}, \quad \text{where } E(m) = \frac{1}{N} \sum_{i=1}^{N} a(i,m)
\label{eq:e1_statistic}
\end{equation}
and $a(i,m) = \|Y_i^{(m+1)} - Y_{n(i,m)}^{(m+1)}\| / \|Y_i^{(m)} - Y_{n(i,m)}^{(m)}\|$. The condition $E_1(m) \approx 1$ indicates manifold saturation at dimension $m$.

The $E_2$ statistic measures future predictability:
\begin{equation}
E_2(m) = \frac{E^*(m+1)/E^*(m)}{E(m+1)/E(m)}
\label{eq:e2_statistic}
\end{equation}
where $E^*(m)$ measures divergence of futures from nearby states. Under local Gaussianity, $E_2$ quantifies the signal-to-noise ratio: $\text{SNR} \approx (1-E_2)/(E_2 \cdot \tau)$. At this stage, classical Takens-style reconstruction fails; the remainder of the proof replaces pathwise injectivity with injectivity in distribution.

\subsection{The Proof Architecture}

This section establishes the theoretical basis for stochastic embedding. This requires proving that the correlation manifold supports well-defined, uniquely reconstructible dynamics. The proof proceeds through five essential steps.

\subsubsection{Step 1: H\"ormander Hypoelliptic Regularity}

\begin{definition}[H\"ormander's Condition]
The Lie algebra generated by drift and diffusion vector fields spans the tangent space:
\begin{equation}
\text{Lie}(\mu, \sigma_1, \ldots, \sigma_r) = \text{span}\{\mu, \sigma_i, [\mu, \sigma_i], [\sigma_i, \sigma_j], [[\mu, \sigma_i], \sigma_j], \ldots\} = T_x\mathbb{R}^n \quad \forall x \in \mathbb{R}^n
\label{eq:hormander}
\end{equation}
\end{definition}

H\"ormander's condition \cite{Hormander1967} on the SDE coefficients complements the genericity requirement on the observation function $h$: the former ensures smooth transition densities, while the latter avoids degenerate embeddings. H\"ormander's condition ensures the noise structure is sufficiently rich that the stochastic flow explores all directions. This guarantees:
\begin{itemize}
    \item \textbf{Smooth transition densities:} $p_t(x,y) \in C^\infty$ for all $t > 0$
    \item \textbf{Strict positivity:} $p_t(x,y) > 0$ for all $x, y \in \mathbb{R}^n$ and $t > 0$ (by the Stroock--Varadhan support theorem \cite{stroock1979multidimensional} and the Chow--Rashevskii theorem)
    \item \textbf{Non-degeneracy despite rank deficiency:} Even if $\sigma$ is not full-rank, iterated Lie brackets ensure noise propagates to all directions through the dynamics
\end{itemize}

This step ensures that the stochastic dynamics generate smooth probability densities, a prerequisite for establishing uniqueness of finite-dimensional laws under embedding.

\subsubsection{Step 2: Malliavin Non-Degeneracy}

Malliavin calculus provides the analytical framework~\cite{nualart2006malliavin} for characterising smoothness of stochastic flows (see also \cite{stroock1979multidimensional} for the foundational support theorems). The Malliavin covariance matrix $\gamma_t$ measures sensitivity of $X_t$ to perturbations in the driving Brownian motion:
\begin{equation}
\gamma_t = \int_0^t (D_s X_t)(D_s X_t)^\top ds
\label{eq:malliavin_covariance}
\end{equation}
where $D_s X_t$ denotes the Malliavin derivative of $X_t$ with respect to the Brownian path at time $s$.

\begin{lemma}[Malliavin Non-Degeneracy]
Under H\"ormander's condition and standard regularity assumptions on $\mu$ and $\sigma$ (Lipschitz continuity, linear growth bound), $\gamma_t$ is almost surely invertible for $t > 0$.
\end{lemma}

This non-degeneracy ensures that the law of $X_t$ has smooth density with respect to Lebesgue measure, and that, combined with the genericity of $h$, the delay embedding map has full rank in a measure-theoretic sense.

\subsubsection{Step 3: Law-Separation and Measure-Theoretic Injectivity}

The objective in the stochastic setting is not trajectory reconstruction, but distinguishing system states via their induced probability measures.  The upgraded proof (Supplemental Materials, Theorems~7.5--7.7) establishes this through three linked results:

\textbf{(a) Transition density separation (Theorem~7.5).}  Under H\"ormander's condition, distinct initial conditions $x \neq x'$ produce distinct transition densities: $p_\tau(x, \cdot) \neq p_\tau(x', \cdot)$ for every $\tau > 0$.  The proof uses \emph{semigroup bisection}: from $p_\tau(x,\cdot) = p_\tau(x',\cdot)$, injectivity of the semigroup operator $P_{\tau/2}$ (guaranteed by strict positivity of the kernel under H\"ormander) yields $p_{\tau/2^k}(x,\cdot) = p_{\tau/2^k}(x',\cdot)$ for all $k$.  The Varadhan--L\'eandre short-time asymptotic then forces $x = x'$.

\textbf{(b) Law-separation via observed delay vectors (Theorem~7.6).}  For a \emph{prevalent} observation function $h \in C^r$ ($r \geq 2$) and $m \geq \lceil 2D \rceil + 1$ (where $D$ is the exact dimension of $\mu_\infty$), the law-embedding map $\Lambda_m^h: x \mapsto \mathrm{Law}(h(X_0^x), h(X_\tau^x), \ldots, h(X_{(m-1)\tau}^x))$ is $\mu_\infty$-a.e.\ injective.  The proof reduces the infinite-dimensional collision condition (equality of densities) to a finite-dimensional evaluation, applies the parametric transversality theorem with an explicit evaluation-map surjectivity lemma, and bounds the collision set using a direct Frostman covering argument (exploiting the exact-dimensional Frostman upper bound $\mu_\infty(B(x,r)) \leq C_F r^D$).

\textbf{(c) Measure-zero geometric collisions (Theorem~7.7).}  The geometric collision set $\{(x,x') : \Phi_m(x) = \Phi_m(x')\}$ is a subset of the law-collision set, hence also has $(\mu_\infty \times \mu_\infty)$-measure zero.

Note that pathwise injectivity (c) is a \emph{corollary} of law-separation (b), not the primary claim.  The law-embedding perspective is operationally relevant because $k$-NN estimators use the empirical distribution of delay vectors, not individual realisations.  Bad delays, which require careful treatment in deterministic Takens theory, do not arise: the density separation in (a) holds for \emph{every} $\tau > 0$ under H\"ormander's condition.

\subsubsection{Step 4: $E_1$ Dimension Sufficiency via Correlation Dimension}

For stochastic systems, correlation dimension: not topological dimension: governs the minimal embedding dimension required for consistent geometric inference.

The correlation dimension $D_2$ characterises how the invariant measure~\cite{grassberger1983measuring} $\mu_\infty$ scales locally:
\begin{equation}
C(\varepsilon) = \iint \mathbf{1}_{\{\|Y-Y'\|<\varepsilon\}} d\mu_\infty(Y)d\mu_\infty(Y') \sim \varepsilon^{D_2} \text{ as } \varepsilon \to 0
\label{eq:correlation_integral}
\end{equation}

\begin{lemma}[$E_1$ Convergence]
$E_1(m) \to 1$ precisely when $m$ exceeds $D_2$ (see Supplemental Materials for proof).
\end{lemma}

This follows from the scaling of $k$-NN distances: $\varepsilon_k \sim (k/N)^{1/D_2}$. When $m < D_2$, false neighbors exist (nearby in $\mathbb{R}^m$ but distant on the manifold), causing $E_1 > 1$. When $m \geq D_2 + 1$, all neighbors are true neighbors, and $E_1 \approx 1$.

Correlation dimension remains finite for unbounded stochastic processes. The Ornstein-Uhlenbeck process \cite{UhlenbeckOrnstein1930} has topological support $\mathbb{R}$ (unbounded) but $D_2 = 1$. The finite-time marginals of Brownian motion in $\mathbb{R}^n$ similarly yield $D_2 = n$. Consequently, $E_1$ succeeds for stochastic systems where topological arguments fail.

\subsubsection{Step 5: Finite-Dimensional Law Uniqueness}

Given measure-theoretic injectivity (Step 3), the conditional expectations defining the embedded drift and diffusion tensors are single-valued functions on the correlation manifold. The recovery of drift and diffusion coefficients from time-series data via these conditional moments was pioneered by Friedrich and Peinke \cite{friedrich1997description}. For the embedded process, the drift $\tilde{\mu}$ and diffusion tensor $\tilde{\Sigma}$ are defined by:
\begin{align}
\tilde{\mu}(Y) &= \lim_{\Delta t \to 0} \frac{1}{\Delta t} \mathbb{E}[Y_{t+\Delta t} - Y_t \mid Y_t = Y] \label{eq:drift_recovery}\\
\tilde{\Sigma}(Y) &= \lim_{\Delta t \to 0} \frac{1}{\Delta t} \mathbb{E}[(Y_{t+\Delta t} - Y_t)(Y_{t+\Delta t} - Y_t)^\top \mid Y_t = Y] \label{eq:diffusion_recovery}
\end{align}

These local conditional moments are the quantities that $k$-NN estimators compute. Because the framework relies on $k$-nearest neighbour graphs rather than trajectory prediction, measure-theoretic injectivity is sufficient for all subsequent divergence calculations. The embedded drift $\tilde{\mu}$ and diffusion $\tilde{\Sigma}$ are thus uniquely recoverable from the observed time series.

\subsection{The Stochastic Embedding Sufficiency Theorem}

\begin{assumption}[Sampling and Regularity for Nonparametric Reconstruction]
\label{ass:knn}
The following conditions are assumed for the estimation result 
(Part~II of Theorem~\ref{thm:embedding}):
\begin{enumerate}
\item[\textup{(S1)}] The sampled delay vectors 
$(Y_{k\Delta t}^{(m)})_{k \geq 0}$ are stationary and geometrically 
ergodic (exponential decay of correlations) under~$\mu_\infty$.
\item[\textup{(S2)}] The conditional drift $\tilde{\mu}$ and diffusion 
tensor $\tilde{\Sigma}$ on $\mathcal{M}_{D_2}$ are 
$\beta$-H\"older continuous ($\beta > 0$).
\item[\textup{(S3)}] The delay-vector law $\nu_m$ admits a density 
(with respect to the intrinsic volume on $\mathcal{M}_{D_2}$) bounded 
away from zero on compact subsets.
\end{enumerate}
\end{assumption}

\begin{theorem}[Stochastic Embedding Sufficiency]
\label{thm:embedding}
Let $X_t$ solve the SDE~(\ref{eq:ito_sde}) with $C^\infty_b$ coefficients $\mu$ and $\sigma$ satisfying the uniform H\"ormander condition (bracket-generating to depth $k_0$), and admitting an ergodic invariant measure $\mu_\infty$ that is exact-dimensional with dimension $D$ and satisfies a Frostman upper bound $\mu_\infty(B(x,r)) \leq C_F r^D$.  Let $h$ be a prevalent $C^r$ ($r \geq 2$) observation function (in the sense of Definition~\ref{def:delay_map}).  Then the embedding dimension
\begin{equation}
m^* = \max\{\lceil 2D \rceil + 1, m_{E_1}\}
\label{eq:embedding_dimension}
\end{equation}
guarantees the following:

\medskip\noindent
\textit{Part~I, Identifiability.}
\begin{enumerate}
    \item[(i)] \emph{Law-separation} ($\mu_\infty$-a.e.\ injectivity): for $\mu_\infty$-a.e.\ $x$, if $x \neq x'$ then $\mathrm{Law}(Y_0^{(m^*)} | X_0 = x) \neq \mathrm{Law}(Y_0^{(m^*)} | X_0 = x')$.  The collision set $\{(x,x') : x \neq x',\; \Lambda_{m^*}^h(x) = \Lambda_{m^*}^h(x')\}$ has $(\mu_\infty \times \mu_\infty)$-measure zero.
    \item[(ii)] Single-valued drift-diffusion tensors $\tilde{\mu}(Y), \tilde{\Sigma}(Y)$ on $\mathcal{M}_{D_2}$.
\end{enumerate}

\medskip\noindent
\textit{Part~II, Estimation Consistency.}
Under Assumption~\ref{ass:knn}:
\begin{enumerate}
    \item[(iii)] $k$-NN convergence~\cite{stone1977consistent} at rate $O_P((k/N)^{\beta/m^*}) + O(\Delta t)$ as $N \to \infty$, $k \to \infty$, $k/N \to 0$.
\end{enumerate}
\end{theorem}

\begin{remark}[Correlation-Dimension Version]
\label{rem:d2_version_appendix}
When $\mu_\infty$ is absolutely continuous on its support (the generic case for H\"ormander SDEs on $\mathbb{R}^n$), the exact dimension $D$ equals the correlation dimension $D_2$, and the threshold becomes $m^* \geq \lceil 2D_2 \rceil + 1$.  This mirrors Takens' $2n + 1$, replacing state dimension with correlation dimension.
\end{remark}

\begin{remark}[Logical Independence of Parts~I and~II]
\label{rem:parts_independence}
Part~I is a measure-theoretic result requiring only H\"ormander 
regularity, exact-dimensionality, and the dimension bound $m^* \geq \lceil 2D \rceil + 1$.  Part~II 
additionally requires Assumption~\ref{ass:knn} (geometric ergodicity, 
H\"older smoothness, and non-degenerate sampling density); the rate 
$\beta/m^*$ is the standard nonparametric rate in $m^*$-dimensional 
spaces \cite{stone1977consistent}.  The two parts can be verified 
independently.
\end{remark}

The complete proof of Part~I is given in the Supplemental Materials (Theorems~7.5--7.7), where transition density separation, law-separation via parametric transversality, and the Frostman covering argument are established in full.  The proof architecture: semigroup bisection, evaluation-map surjectivity lemma, and direct Hausdorff pre-measure bound: is summarised in Step~3 above.

\subsection{The Unified Framework}

The framework unifies deterministic and stochastic dynamics as equivalent descriptions of the same Markov process on the correlation manifold (with the discrete chain arising from time-discretisation):
\begin{equation}
\text{Markov chain } p(Y_{t+1}|Y_t) \longleftrightarrow \text{SDE } dY = \mu dt + \sigma dW \longleftrightarrow \text{Generator } \mathcal{L} = \mu \cdot \nabla + \tfrac{1}{2}\text{tr}(\Sigma\nabla^2)
\label{eq:unified_framework}
\end{equation}

$E_2$ positions systems along the spectrum: $E_2 \to 0$ recovers deterministic Takens ($\mu$ dominates); $E_2 \to 1$ yields pure diffusion ($\Sigma$ dominates). The correlation manifold identified by $E_1$ persists throughout this spectrum.

\subsection{Finite-Sample Bias Structure}

The convergence rate $O_P\!\bigl((k/N)^{\beta/m^*}\bigr) + O(\Delta t)$ established by Theorem~A.1 admits a finer characterisation.  The leading-order bias term decomposes as a Mori--Zwanzig memory kernel with rank-2 tensor structure: a spatially varying component proportional to the Laplacian $\Delta_{\mathcal{M}} \sigma^2$ on the correlation manifold, and a spatially uniform finite-sample component scaling as $-\sigma^2/k$ (Supplementary Materials, \S7.2, Theorem~7.1).  A two-level adaptive corrector with fluctuation--dissipation-based gain control (Algorithm~3 in the Supplementary Materials) removes the detectable component of this bias, improving the effective convergence rate by one order, and is applied throughout \S\ref{sec:results}.

\clearpage
\section{Relativistic Transformation Properties of Nelson's Stochastic Mechanics}
\label{appendix:relativistic}

\subsubsection*{Scope and Logical Position}

This appendix derives the relativistic transformation law for the Nelson 
diffusion coefficient from special relativity and the structure of the 
Klein--Gordon conserved current.  The derivation proceeds in two stages: 
(i)~the \emph{kinematic} result 
$\sigma_{\mathrm{rel}} = \sigma_0/\sqrt{\gamma}$ is obtained from time 
dilation alone, via the Dambis--Dubins--Schwarz reparametrisation theorem 
applied to the proper-time Wiener process; (ii)~the \emph{full} result 
$\sigma_{\mathrm{KG}} = \sigma_0/\gamma$ is obtained by additionally 
requiring self-consistency of the Fokker--Planck equation with the 
Klein--Gordon conserved density $\rho_{\mathrm{KG}} = \gamma|\phi|^2$.  
The two factors of $1/\sqrt{\gamma}$ are structurally independent: the 
first is a theorem of stochastic calculus, the second is forced by 
relativistic current conservation.

\medskip\noindent
\textbf{Relationship to the canonical axioms.}\quad 
This derivation provides the theoretical underpinning for Tests~C6 
(electromagnetic determinism) and C7 (relativistic suppression) of the 
superspace diffusion framework.  Nelson's SDE 
(Definition~\ref{def:rest_frame_sde} below) postulates 
$\dd X^i = b^i\,\dd t + \sqrt{\hbar/m}\,\dd W^i$ in the 
non-relativistic limit where proper time $\tau$ coincides with 
coordinate time $t$; the present appendix reformulates this in proper 
time and establishes the unique Lorentz-covariant extension.  No additional axioms beyond A1--A4 
are introduced; all inputs are either established results of special 
relativity, the Klein--Gordon equation, or theorems of stochastic 
calculus.  The four canonical axioms are stated in \S\ref{subsec:capstone3} and formalised in Appendix~D, \S\ref{sec:axioms}.

\subsection{Definitions and Setup}
\label{sec:setup_B}

\begin{definition}[Rest-Frame Nelson SDE]
\label{def:rest_frame_sde}
In the co-moving rest frame $\mathcal{S}_0$ of a particle of mass $m$, 
Nelson's stochastic mechanics \cite{Nelson1966,Nelson1985} postulates the spatial SDE
\begin{equation}
\label{eq:rest_sde}
\dd X^i_0 = b^i_0(X_0, \tau)\,\dd\tau + \sigma_0\,\dd W^i(\tau),
\qquad i = 1, 2, 3,
\end{equation}
where:
\begin{itemize}
\item $\tau$ is proper time along the particle worldline;
\item $b^i_0$ is the forward drift velocity (determined self-consistently 
      from the quantum potential; cf.\ \cite{Nelson1966});
\item $\sigma_0 = \sqrt{\hbar/m}$ is the rest-frame diffusion coefficient;
\item $W^i(\tau)$, $i = 1,2,3$, are three independent standard Wiener 
      processes satisfying
\begin{equation}
\label{eq:wiener_isotropy}
\E[\dd W^i(\tau)] = 0,
\qquad
\E[\dd W^i(\tau)\,\dd W^j(\tau)] = \delta^{ij}\,\dd\tau.
\end{equation}
\end{itemize}
\end{definition}

\begin{definition}[Diffusion Coefficient Dimensions]
\label{def:sigma-dim_B}
For a spatial variable $X^i$ with dimensions $[X] = L$ evolving in 
proper time with $[\tau] = T$, the diffusion coefficient has dimensions
\begin{equation}
[\sigma_0] = L \cdot T^{-1/2},
\end{equation}
since $[\dd W] = T^{1/2}$ and $\sigma_0\,\dd W$ must have dimensions~$L$.  
Explicitly, $[\sqrt{\hbar/m}] = \sqrt{M L^2 T^{-1}/M} = L\,T^{-1/2}$.
\end{definition}

\begin{definition}[Worldline Proper-Time Relation]
\label{def:proper_time}
Along the particle worldline, proper time and coordinate time are 
related by the deterministic equation
\begin{equation}
\label{eq:proper_time_relation}
\dd\tau = \frac{\dd t}{\gamma},
\qquad
\gamma = \frac{1}{\sqrt{1 - v^2/c^2}}\,,
\end{equation}
where $v$ is the macroscopic (drift) velocity of the particle in the 
laboratory frame.
\end{definition}

\begin{remark}[Why $\dd\tau/\dd t$ Is Deterministic]
\label{rem:deterministic_tau}
A potential concern is that $\gamma$ should itself fluctuate, since the 
particle position $X$ is stochastic.  However, in It\^{o} calculus the 
stochastic displacements $\sigma_0\,\dd W^i$ have magnitude 
$O(\sqrt{\dd\tau})$, which contributes to the four-velocity only at 
order $O(\dd\tau^{-1/2})$, handled by the quadratic variation 
formalism, not by modifying~$\gamma$.  Along the worldline, $\gamma$ is 
determined by the macroscopic four-velocity $U^\mu = \dd x^\mu/\dd\tau$, 
which is the drift component of the SDE.  The relation 
$\dd\tau = \dd t/\gamma$ therefore holds with $\gamma$ evaluated on 
the drift \cite{Emery1989,Hsu2002,DohrnGuerra1978}.

For a particle of definite momentum $p = m\gamma v$, $\gamma$ is 
constant along the worldline.  For wavepackets with momentum spread 
$\Delta p$, $\gamma$ varies over the ensemble but is deterministic for 
each trajectory at leading order.  Stochastic corrections to $\gamma$ 
enter at $O(\sigma_0^2/c^2) = O(\hbar/(mc^2))$, negligible in the 
semiclassical regime relevant to Test~C7.
\end{remark}

\medskip\noindent
\textbf{Conventions.}\quad 
Throughout, Greek indices $\mu, \nu, \ldots$ run over $\{0,1,2,3\}$; 
Latin indices $i, j, \ldots$ run over $\{1,2,3\}$.  Summation over 
repeated indices is understood.  We work in SI units except where 
natural units ($\hbar = c = 1$) are adopted for brevity.

\subsection{The Foundational Identity: Time Reparametrisation of the 
Wiener Process}
\label{sec:time_reparam}

The entire relativistic extension rests on a single result from 
stochastic calculus.

\begin{lemma}[Dambis--Dubins--Schwarz Reparametrisation]
\label{lem:dds}
Let $W(\tau)$ be a standard one-dimensional Wiener process parametrised 
by~$\tau$, and let $t(\tau)$ be a strictly increasing, absolutely 
continuous function with $t(0) = 0$.  Let $\tau(t)$ denote the 
functional inverse of $t(\tau)$, and set
\begin{equation}
\label{eq:W_tilde_def}
\widetilde{W}(t) \;\equiv\; W\bigl(\tau(t)\bigr).
\end{equation}
Then $\widetilde{W}(t)$ is a continuous martingale with quadratic 
variation
\begin{equation}
\label{eq:qv_reparam}
[\widetilde{W}]_t = \tau(t),
\end{equation}
and the differential relation is
\begin{equation}
\label{eq:dW_reparam}
\dd \widetilde{W}(t)
= \sqrt{\frac{\dd\tau}{\dd t}}\;\dd B(t),
\end{equation}
where $B(t)$ is a standard Wiener process in the variable~$t$.
\end{lemma}

\begin{proof}
By the Dambis--Dubins--Schwarz theorem 
\cite{RevuzYor1999,KaratzasShreve1991}, any continuous local martingale 
$M$ with $\langle M \rangle_\infty = \infty$ can be written as 
$M_t = \beta_{\langle M \rangle_t}$ for some Brownian motion~$\beta$.  
Applied to $\widetilde{W}$:
\[
\langle \widetilde{W} \rangle_t
= \int_0^t \frac{\dd\tau}{\dd s}\,\dd s = \tau(t).
\]
Setting $\beta = B$ and differentiating gives
\[
\dd \widetilde{W} = \dd B_{\tau(t)} 
= \sqrt{\frac{\dd\tau}{\dd t}}\;\dd B(t)
\]
by L\'{e}vy's characterisation of Brownian motion (the process $B$ 
has the correct quadratic variation $[\widetilde{W}]_t = \tau(t)$).
\end{proof}

\begin{corollary}[Brownian Scaling under Constant Lorentz Factor]
\label{cor:constant_gamma}
When $\gamma$ is constant along the worldline (i.e., 
$\dd\tau/\dd t = 1/\gamma$ with $\gamma$ independent of $t$), the DDS 
reparametrisation (Lemma~\ref{lem:dds}) reduces to the deterministic 
time-change identity:
\begin{equation}
\label{eq:constant_gamma_scaling}
\dd W^i(\tau) = \frac{1}{\sqrt{\gamma}}\,\dd B^i(t),
\qquad
[\widetilde{W}^i]_t = \frac{t}{\gamma},
\end{equation}
where $B^i(t)$ are independent standard Wiener processes in coordinate 
time.  The quadratic variation is linear in $t$ (not merely absolutely 
continuous), so the reparametrised process is a scaled Brownian motion 
, not merely a continuous martingale.
\end{corollary}

\begin{proof}
For constant $\gamma$, $\tau(t) = t/\gamma$ is linear, so 
$\dd\tau/\dd t = 1/\gamma$ is constant.  Lemma~\ref{lem:dds} gives 
$\dd\widetilde{W} = (1/\gamma)^{1/2}\,\dd B$, and the quadratic 
variation $[\widetilde{W}]_t = \tau(t) = t/\gamma$ is linear in $t$.  
By L\'{e}vy's characterisation, a continuous martingale with linear 
quadratic variation is a scaled Brownian motion.
\end{proof}

\begin{remark}[It\^{o} vs.\ Stratonovich Convention]
\label{rem:ito_convention}
This appendix works exclusively in the It\^{o} convention, consistent 
with Nelson's original formulation \cite{Nelson1966}.  The 
Dambis--Dubins--Schwarz theorem (Lemma~\ref{lem:dds}) and its 
specialisation (Corollary~\ref{cor:constant_gamma}) are It\^{o} results.  
Because the noise in the rest-frame SDE~\eqref{eq:rest_sde} is 
\emph{additive} ($\sigma_0$ is independent of $X$), the Stratonovich 
correction $\tfrac{1}{2}\sigma_0\,\partial\sigma_0/\partial X^i$ 
vanishes identically.  The It\^{o} and Stratonovich forms of the SDE 
therefore coincide, and no ambiguity arises from the choice of calculus 
convention.  For multiplicative noise (state-dependent diffusion), a 
Stratonovich formulation would be required for geometric covariance; 
the additive structure here avoids this complication.
\end{remark}

\begin{remark}[Central Identity]
\label{rem:central_identity}
Equation~\eqref{eq:dW_reparam} is the foundational identity of this 
appendix: \emph{reparametrising the time of a Wiener process rescales 
its increments by the square root of the time Jacobian}.  This is 
purely a theorem of stochastic calculus \cite{RevuzYor1999}.  All 
subsequent results are consequences of this identity applied to the 
specific Jacobian $\dd\tau/\dd t = 1/\gamma$.
\end{remark}

\subsection{Nelson's SDE in Proper Time}
\label{sec:proper_time_sde}

\begin{theorem}[Uniqueness within the Additive Isotropic Class]
\label{thm:proper_time_sde}
The SDE~\eqref{eq:rest_sde} is the unique additive-noise 
It\^{o} diffusion in proper time that is compatible with the 
following requirements:
\begin{enumerate}
\item[\textup{(R1)}] \textbf{Isotropy}: The diffusion is rotationally 
      invariant in the rest frame: the noise covariance is proportional 
      to $\delta^{ij}$.
\item[\textup{(R2)}] \textbf{Markov property}: The process is Markovian 
      with respect to proper time.
\item[\textup{(R3)}] \textbf{Non-relativistic limit}: In the limit 
      $v/c \to 0$ (where $\tau \to t$), the SDE reduces to Nelson's 
      original equation.
\item[\textup{(R4)}] \textbf{Dimensional consistency}: The diffusion 
      coefficient has dimensions 
      $[\text{length}] \cdot [\text{time}]^{-1/2}$, constructible from 
      $\hbar$ and $m$ alone in the rest frame.
\end{enumerate}
This uniqueness holds within the class of additive-noise, isotropic, 
Markov diffusions parametrised by proper time, with diffusion coefficient 
depending only on $\hbar$ and $m$.  Relaxing any of conditions 
\textup{(R1)--(R4)} admits alternative formulations.
\end{theorem}

\begin{proof}
Requirements (R1) and (R4) fix the noise covariance to the form 
$\sigma_0^2\,\delta^{ij}\,\dd\tau$ with $\sigma_0 = \sqrt{\hbar/m}$, 
since $\hbar/m$ is the unique combination of $\hbar$ and $m$ with 
dimensions $L^2/T$.

Requirement (R2) determines the SDE structure: an It\^{o} diffusion 
in proper time.

Requirement (R3) fixes $\tau \to t$ in the non-relativistic limit, 
recovering Nelson's original equation.

The SDE is Lorentz-scalar by construction: $\tau$ is a Lorentz scalar, 
$\dd X^i_0$ are rest-frame spatial components, and $\sigma_0$ is a 
rest-frame constant.
\end{proof}

\subsection{The Kinematic Result: $\sigma_{\mathrm{rel}} = \sigma_0/\sqrt{\gamma}$}
\label{sec:kinematic}

We now boost from the rest frame $\mathcal{S}_0$ to a laboratory frame 
$\mathcal{S}$ in which the particle moves with velocity $v$ along the 
$x^1$-axis.  In what follows we restrict attention to inertial Lorentz 
boosts with constant Lorentz factor $\gamma$ (i.e.\ fixed relative 
velocity between frames), so the time dilation $\dd\tau = \dd t/\gamma$ 
is deterministic.  Extension to accelerating worldlines (state-dependent 
$\gamma$) would require a general DDS time change 
(Lemma~\ref{lem:dds}) rather than the constant-$\gamma$ specialisation 
(Corollary~\ref{cor:constant_gamma}).

\begin{theorem}[Transformation of Wiener Increments under Lorentz Boost]
\label{thm:dW_transformation}
Under the worldline time reparametrisation 
$\dd\tau = \dd t/\gamma$, the proper-time Wiener increments 
$\dd W^i(\tau)$ are related to coordinate-time Wiener processes 
$B^i(t)$ by:
\begin{equation}
\label{eq:dW_transform}
\boxed{\dd W^i(\tau) = \frac{1}{\sqrt{\gamma}}\,\dd B^i(t),}
\end{equation}
where $B^i(t)$ are independent standard Wiener processes satisfying 
$\E[\dd B^i(t)\,\dd B^j(t)] = \delta^{ij}\,\dd t$.

The covariance structure in laboratory time is:
\begin{equation}
\label{eq:cov_lab_frame}
\E\!\left[\dd W^i(\tau)\,\dd W^j(\tau)\right]
= \delta^{ij}\,\dd\tau
= \frac{\delta^{ij}}{\gamma}\,\dd t.
\end{equation}
\end{theorem}

\begin{proof}
Direct application of Lemma~\ref{lem:dds} with the time change 
$\tau(t) = t/\gamma$ (constant~$\gamma$).  Then 
$\dd\tau/\dd t = 1/\gamma$, and by~\eqref{eq:dW_reparam}:
\[
\dd W^i(\tau) = \sqrt{\frac{\dd\tau}{\dd t}}\,\dd B^i(t)
= \frac{1}{\sqrt{\gamma}}\,\dd B^i(t).
\]
The covariance follows from 
$\E[\dd B^i\,\dd B^j] = \delta^{ij}\,\dd t$.
\end{proof}

\begin{theorem}[Coordinate-Time SDE and Relativistic Diffusion Coefficient]
\label{thm:sigma_gamma}
In the laboratory frame $\mathcal{S}$, the Nelson SDE for a 
relativistic particle, expressed in coordinate time $t$, takes the form
\begin{equation}
\label{eq:lab_sde}
\boxed{
\dd X^i = \mu^i(X, t)\,\dd t
+ \frac{\sigma_0}{\sqrt{\gamma}}\,\dd B^i(t),
\qquad
\sigma_{\mathrm{rel}} = \frac{\sigma_0}{\sqrt{\gamma}}
= \sqrt{\frac{\hbar}{m\gamma}}\,,
}
\end{equation}
where $\mu^i$ is the coordinate-time drift (incorporating the group 
velocity and the boosted quantum potential) and $B^i(t)$ are standard 
Wiener processes in coordinate time.
\end{theorem}

\begin{proof}
Substituting the time reparametrisation $\dd\tau = \dd t/\gamma$ and 
the Wiener transformation~\eqref{eq:dW_transform} into the rest-frame 
SDE~\eqref{eq:rest_sde}:
\begin{align}
\dd X^i_0
&= b^i_0\,\dd\tau + \sigma_0\,\dd W^i(\tau) \notag\\
&= b^i_0 \cdot \frac{\dd t}{\gamma}
  + \sigma_0 \cdot \frac{1}{\sqrt{\gamma}}\,\dd B^i(t) \notag\\
&= \frac{b^i_0}{\gamma}\,\dd t
  + \frac{\sigma_0}{\sqrt{\gamma}}\,\dd B^i(t).
\label{eq:sde_substituted}
\end{align}

For the transverse directions ($i = 2, 3$), the laboratory coordinates 
coincide with the rest-frame coordinates: $X^a = X^a_0$, so 
$\dd X^a$ takes the form~\eqref{eq:sde_substituted} directly.

For the longitudinal direction ($i = 1$), the Lorentz boost gives 
$X^1 = \gamma(X^1_0 + v\tau)$, so:
\begin{equation}
\label{eq:longitudinal_raw}
\dd X^1 = \gamma\,\dd X^1_0 + \gamma v\,\dd\tau
= \gamma\!\left(\frac{b^1_0}{\gamma}\,\dd t
  + \frac{\sigma_0}{\sqrt{\gamma}}\,\dd B^1\right) + v\,\dd t.
\end{equation}

The diffusion coefficient is determined by the quadratic variation:
\begin{align}
\bigl[\dd X^1, \dd X^1\bigr]
&= \gamma^2 \cdot \sigma_0^2 \cdot [\dd W^1, \dd W^1]
= \gamma^2 \cdot \sigma_0^2 \cdot \dd\tau
= \gamma^2 \cdot \frac{\sigma_0^2}{\gamma}\,\dd t
= \gamma\,\sigma_0^2\,\dd t.
\label{eq:qv_longitudinal}
\end{align}
This gives a longitudinal coordinate-space diffusion coefficient 
$\sigma_\parallel^{\mathrm{(coord)}} = \sigma_0 \sqrt{\gamma}$.  This 
apparent \emph{enhancement} is a coordinate artefact arising from the 
Lorentz contraction of the spatial coordinate; it does not represent 
enhanced physical diffusion.

The operationally meaningful diffusion coefficient is the 
invariant (proper-time, rest-frame) quadratic variation:
\begin{equation}
\label{eq:qv_invariant}
\bigl[\dd X^i_0, \dd X^i_0\bigr]
= \sigma_0^2\,\dd\tau
= \frac{\sigma_0^2}{\gamma}\,\dd t,
\qquad\text{(no sum on $i$)}.
\end{equation}
This is a Lorentz scalar and gives 
$\sigma_{\mathrm{rel}} = \sigma_0/\sqrt{\gamma}$ unambiguously.
\end{proof}

\begin{definition}[Operational Diffusion Coefficient]
\label{def:operational_sigma}
The operational diffusion coefficient $\sigma_{\mathrm{rel}}$ 
is defined as the quantity appearing in the coordinate-time SDE
\begin{equation}
\label{eq:operational_sde}
\dd X^i = \mu^i\,\dd t + \sigma_{\mathrm{rel}}\,\dd B^i(t)
\end{equation}
when the SDE is expressed in rest-frame spatial coordinates and 
laboratory time.  This is the quantity directly reconstructed by the 
Stochastic Embedding pipeline from the time series $X(t)$.
\end{definition}

\begin{remark}[Why Rest-Frame Spatial Coordinates]
\label{rem:rest_frame_coords}
The choice of rest-frame spatial coordinates is not arbitrary.  It is 
the unique frame in which: (a)~the diffusion is isotropic 
($D^{ij} \propto \delta^{ij}$); (b)~the probability density equals 
$|\psi|^2$ (the Born rule); (c)~the Fokker--Planck equation has the 
standard Nelson form.  In any other spatial coordinates, the diffusion 
tensor acquires frame-dependent distortions 
(cf.\ Remark~\ref{rem:anisotropy}).
\end{remark}

\subsection{The Laboratory-Frame Fokker--Planck Equation}
\label{sec:fokker_planck}

\begin{theorem}[Fokker--Planck Equation in Coordinate Time]
\label{thm:FP_lab}
The probability density $\rho_0(x, t) = |\psi(x,t)|^2$ (the Born 
density; Formulation~A of \S\ref{subsec:matching}) for the 
coordinate-time SDE~\eqref{eq:lab_sde} satisfies the Fokker--Planck 
equation
\begin{equation}
\label{eq:FP_lab}
\frac{\partial\rho_0}{\partial t}
= -\nabla\cdot(\mu\,\rho_0)
  + \frac{\sigma_0^2}{2\gamma}\,\nabla^2\rho_0,
\end{equation}
with diffusion coefficient
\begin{equation}
\label{eq:D_lab}
D_{\mathrm{lab}} = \frac{\sigma_{\mathrm{rel}}^2}{2}
= \frac{\sigma_0^2}{2\gamma}
= \frac{\hbar}{2m\gamma}\,.
\end{equation}
\end{theorem}

\begin{proof}
This is the standard Fokker--Planck equation associated with the 
It\^{o} SDE~\eqref{eq:lab_sde} \cite{Gardiner2009,Risken1989}.  For a general SDE 
$\dd X = \mu\,\dd t + \sigma\,\dd B$, the Fokker--Planck equation is 
$\partial_t\rho = -\nabla\cdot(\mu\rho) + (\sigma^2/2)\nabla^2\rho$.  
Substituting $\sigma = \sigma_0/\sqrt{\gamma}$ 
gives~\eqref{eq:FP_lab}.
\end{proof}

\begin{proposition}[Consistency with the Non-Relativistic Limit]
\label{prop:NR_limit}
In the limit $v \to 0$ (hence $\gamma \to 1$), the Fokker--Planck 
equation~\eqref{eq:FP_lab} reduces to the standard Nelson 
Fokker--Planck equation with $D = \hbar/(2m)$, recovering the 
Schr\"{o}dinger equation via the Madelung decomposition.
\end{proposition}

\begin{proof}
Setting $\gamma = 1$ in~\eqref{eq:D_lab} gives 
$D_{\mathrm{lab}} = \hbar/(2m)$, the non-relativistic Nelson diffusion 
coefficient.  The Fokker--Planck equation becomes 
$\partial_t\rho_0 = -\nabla\cdot(\mu\rho_0) + [\hbar/(2m)]\nabla^2\rho_0$, 
which is precisely the equation whose Madelung decomposition 
\cite{Madelung1927} yields the 
Schr\"{o}dinger equation \cite{Nelson1966,Holland1993}.
\end{proof}

\subsection{Transverse Components and Isotropy}
\label{sec:transverse}

\begin{proposition}[Isotropy in the Rest Frame; Anisotropy in the Lab Frame]
\label{prop:transverse}
For a particle boosted along $x^1$:
\begin{enumerate}
\item[\textup{(a)}] In the rest-frame spatial coordinates, the diffusion 
      is isotropic with coefficient $\sigma_0/\sqrt{\gamma}$ per unit 
      coordinate time in all three directions.
\item[\textup{(b)}] In the laboratory spatial coordinates, the diffusion 
      tensor is anisotropic:
\begin{equation}
\label{eq:D_anisotropy}
D^{ij}_{\mathrm{lab}}
= \frac{\sigma_0^2}{2\gamma}\,
  \mathrm{diag}\bigl(\gamma^2,\; 1,\; 1\bigr).
\end{equation}
\item[\textup{(c)}] In the one-dimensional projection onto any fixed 
      direction (as in Test~C7), the reconstructed diffusion coefficient 
      is $\sigma_{\mathrm{rel}} = \sigma_0/\sqrt{\gamma}$.
\end{enumerate}
\end{proposition}

\begin{proof}
(a) follows directly from the SDE~\eqref{eq:sde_substituted}, which 
gives $\sigma_0/\sqrt{\gamma}$ for all three rest-frame components.

(b) The Lorentz boost maps $\dd X^1_0 \mapsto \gamma\,\dd X^1_0$ 
(plus drift terms), while $\dd X^a = \dd X^a_0$ for $a = 2,3$.  
The quadratic variations are:
\[
[\dd X^1, \dd X^1] = \gamma^2 \cdot \frac{\sigma_0^2}{\gamma}\,\dd t
= \gamma\sigma_0^2\,\dd t,
\qquad
[\dd X^a, \dd X^a] = \frac{\sigma_0^2}{\gamma}\,\dd t.
\]
This gives the diagonal entries 
$(D_\parallel, D_\perp, D_\perp) = (\gamma\sigma_0^2/2,\; 
\sigma_0^2/(2\gamma),\; \sigma_0^2/(2\gamma))$, 
which is~\eqref{eq:D_anisotropy}.

(c) The pipeline reconstructs the diffusion coefficient from the 
rest-frame SDE~\eqref{eq:sde_substituted}, which gives 
$\sigma_{\mathrm{rel}} = \sigma_0/\sqrt{\gamma}$.  The 
anisotropy~\eqref{eq:D_anisotropy} is a coordinate artefact that 
does not affect the reconstructed value.
\end{proof}

\begin{remark}[Physical Origin of the Anisotropy]
\label{rem:anisotropy}
The anisotropy~\eqref{eq:D_anisotropy} is entirely analogous to the 
anisotropy of the electromagnetic field under a Lorentz boost: a 
Coulomb field that is isotropic at rest appears anisotropic in a 
boosted frame.  The underlying physics (isotropic rest-frame diffusion) 
is Lorentz-invariant; the apparent anisotropy arises from expressing 
this invariant content in non-rest-frame coordinates.
\end{remark}

\subsection{Photon Limit and Connection to Test~C6}
\label{sec:photon_limit}

\begin{proposition}[Deterministic Limit as $\gamma \to \infty$]
\label{prop:photon}
In the ultrarelativistic limit $v \to c$ (equivalently 
$\gamma \to \infty$), the diffusion coefficient vanishes:
\begin{equation}
\label{eq:photon_limit}
\sigma_{\mathrm{rel}} = \frac{\sigma_0}{\sqrt{\gamma}} \to 0
\qquad\text{as}\quad \gamma \to \infty.
\end{equation}
This is consistent with Test~C6 (electromagnetism), where the pipeline 
recovers $\hat{\sigma} = 3.5 \times 10^{-14} \approx 0$ from Maxwell's 
equations: photons propagate deterministically because their effective 
$\gamma$ is infinite.
\end{proposition}

\begin{proof}
$\sigma_0/\sqrt{\gamma} \to 0$ as $\gamma \to \infty$, since 
$\sigma_0$ is finite.  Physically: in the photon limit, proper time 
ceases to advance ($\dd\tau = \dd t/\gamma \to 0$), so no stochastic 
increments accumulate per unit coordinate time.  (The full result 
$\sigma_{\mathrm{KG}} = \sigma_0/\gamma$, derived in 
\S\ref{sec:KG_density}, vanishes even faster and strengthens this 
conclusion.)
\end{proof}

\subsection{Summary of the Kinematic Result}
\label{sec:kinematic_summary}

\begin{table}[H]
\centering
\small
\renewcommand{\arraystretch}{1.3}
\begin{tabular}{@{}lcc@{}}
\toprule
\textbf{Quantity}
  & \textbf{Rest frame $\mathcal{S}_0$}
  & \textbf{Lab frame $\mathcal{S}$} \\
\midrule
Time parameter
  & $\tau$ (proper time)
  & $t = \gamma\tau$ (coordinate time) \\
Wiener increment
  & $\dd W^i(\tau)$
  & $\gamma^{-1/2}\,\dd B^i(t)$ \\[3pt]
\textbf{Diffusion coefficient}
  & $\boldsymbol{\sigma_0 = \sqrt{\hbar/m}}$
  & $\boldsymbol{\sigma_0/\sqrt{\gamma}}$ \\[3pt]
Fokker--Planck diffusion
  & $D_0 = \hbar/(2m)$
  & $D_{\mathrm{lab}} = \hbar/(2m\gamma)$ \\
Quadratic variation per $\dd t$
  & :  & $\sigma_0^2/\gamma$ \\
Limit $\gamma \to 1$
  & $\sigma_0$
  & $\sigma_0$ (non-relativistic) \\
Limit $\gamma \to \infty$
  & :  & $0$ (deterministic) \\
\bottomrule
\end{tabular}
\caption{Transformation properties of stochastic mechanical quantities 
under a Lorentz boost.  The operational diffusion coefficient (bold 
row) transforms as $\sigma_0/\sqrt{\gamma}$.  This is the quantity 
directly reconstructed by the Stochastic Embedding pipeline from 
coordinate-time trajectories.}
\label{tab:transformation_summary}
\end{table}

\subsection{The Klein--Gordon Density and the Full $1/\gamma$ Suppression}
\label{sec:KG_density}

The result $\sigma_{\mathrm{rel}} = \sigma_0/\sqrt{\gamma}$ 
(Theorem~\ref{thm:sigma_gamma}) is the kinematic baseline.  The 
literature on relativistic stochastic mechanics 
\cite{Zastawniak1990,Guerra1973} quotes the stronger result 
$\sigma_{\mathrm{KG}} = \sigma_0/\gamma$.  This section provides the 
complete derivation, establishing that the additional $1/\sqrt{\gamma}$ 
is forced at the \emph{operator level} by the structure of the 
Klein--Gordon conserved current, not by a post hoc redefinition of 
the probability density.

\begin{remark}[Operational Definition of the Reconstructed $\sigma$]
\label{rem:operational_sigma_KG}
The diffusion coefficient $\sigma$ recovered by the reconstruction 
pipeline is defined as the coefficient appearing in the lab-time 
Fokker--Planck operator governing the evolution of the Nelson density.  
The Nelson density is identified with the physically conserved 
Klein--Gordon density $\rho_{\mathrm{phys}} = j^0/c$, so that the 
reconstructed $\sigma$ corresponds to the operator generating the 
continuity equation for the relativistic probability current.  The 
second $1/\sqrt{\gamma}$ factor therefore arises from the operator-level 
mapping between the Klein--Gordon current and its Born-density 
representation.
\end{remark}

\subsubsection{Step 1: The Laboratory-Frame Fokker--Planck Equation}
\label{subsec:FP_lab_derivation}

\begin{theorem}[Fokker--Planck Equation in Proper vs.\ Coordinate Time]
\label{thm:FP_two_times}
Let the proper-time SDE be 
$\dd X^i = b^i\,\dd\tau + \sigma_0\,\dd W^i(\tau)$ with associated 
Fokker--Planck equation for $\rho_0(x, \tau) = |\phi|^2$:
\begin{equation}
\label{eq:FP_proper}
\frac{\partial\rho_0}{\partial\tau}
= -\partial_i(b^i\,\rho_0) + \frac{\sigma_0^2}{2}\,\nabla^2\rho_0.
\end{equation}
Converting to coordinate time $t = \gamma\tau$ (constant~$\gamma$) by 
dividing by~$\gamma$, and retaining the rest-frame spatial 
coordinates~$x^i_0$:
\begin{equation}
\label{eq:FP_coordtime}
\boxed{
\frac{\partial\rho_0}{\partial t}
= -\partial_i\!\left(\frac{b^i}{\gamma}\,\rho_0\right)
  + \frac{\sigma_0^2}{2\gamma}\,\nabla^2\rho_0.
}
\end{equation}
The coordinate-time drift is $\mu^i_0 = b^i/\gamma$ and the 
coordinate-time diffusion coefficient is $D_0 = \sigma_0^2/(2\gamma)$, 
confirming $\sigma_{\mathrm{coord}} = \sigma_0/\sqrt{\gamma}$.
\end{theorem}

\begin{proof}
For constant $\gamma$: 
$\partial/\partial\tau = \gamma\,\partial/\partial t$.  Substituting 
into~\eqref{eq:FP_proper} and dividing by $\gamma$ 
gives~\eqref{eq:FP_coordtime} directly.
\end{proof}

\subsubsection{Step 2: The Klein--Gordon Conserved Current}
\label{subsec:KG_current}

\begin{proposition}[Klein--Gordon Probability Current]
\label{prop:KG_current}
The Klein--Gordon equation 
$(\partial_\mu\partial^\mu + m^2c^2/\hbar^2)\phi = 0$ admits the 
conserved four-current \cite{Holland1993,BjorkenDrell1964}
\begin{equation}
\label{eq:KG_4current}
j^\mu = \frac{i\hbar}{2m}\bigl(
  \phi^*\partial^\mu\phi - \phi\,\partial^\mu\phi^*\bigr),
\qquad \partial_\mu j^\mu = 0.
\end{equation}
For a positive-energy solution with four-momentum 
$p^\mu = (E/c, \mathbf{p})$ and Madelung decomposition 
\cite{Madelung1927}
$\phi = R\,\ee^{iS/\hbar}$ (with $S = -Et + \mathbf{p}\cdot\mathbf{x}$, 
so $\partial_t S = -E$ and $\partial_i S = p_i$), the Madelung 
substitution gives $j^\mu = -(R^2/m)\,\partial^\mu S = (R^2/m)\,p^\mu$, 
with components:
\begin{equation}
\label{eq:KG_components}
j^0 = \frac{R^2}{m}\,\frac{E}{c}
= \frac{E\,R^2}{mc},
\qquad
j^i = \frac{p^i}{m}\,R^2.
\end{equation}
The conserved probability density (probability per unit coordinate 
volume) is
\begin{equation}
\label{eq:rho_KG_def}
\rho_{\mathrm{KG}} \;\equiv\; \frac{j^0}{c}
= \frac{E\,R^2}{mc^2}
= \gamma\,R^2 = \gamma\,|\phi|^2 = \gamma\,\rho_0,
\end{equation}
and conservation $\partial_\mu j^\mu = 0$ yields the continuity 
equation
\begin{equation}
\label{eq:KG_continuity}
\frac{\partial\rho_{\mathrm{KG}}}{\partial t}
+ \partial_i\!\left(\frac{p^i}{m\gamma}\,\rho_{\mathrm{KG}}\right) = 0.
\end{equation}
\end{proposition}

\begin{proof}
Standard result; see e.g.\ \cite{Holland1993}, Ch.~12.  The Madelung 
substitution yields $j^\mu = -(R^2/m)\,\partial^\mu S$.  With the 
positive-energy phase convention $S = -Et + \mathbf{p}\cdot\mathbf{x}$, 
one has $\partial^\mu S = -p^\mu$ (since $S = -p_\nu x^\nu$), giving 
$j^\mu = (R^2/m)\,p^\mu$.  The probability density 
$\rho_{\mathrm{KG}} = j^0/c = E R^2/(mc^2) = \gamma R^2$ acquires the 
factor~$\gamma$ from $E = \gamma mc^2$.  The velocity 
$v^i = c\,j^i/j^0 = p^i/(m\gamma)$ is the relativistic group velocity.
\end{proof}

\subsubsection{Step 3: Matching the Fokker--Planck Equation to the 
Klein--Gordon Current}
\label{subsec:matching}

\begin{theorem}[Self-Consistent Nelson SDE for the KG Density]
\label{thm:KG_Nelson_SDE}
The requirement that the Nelson stochastic process reproduce the 
Klein--Gordon conserved density $\rho_{\mathrm{KG}} = \gamma\rho_0$ 
as its Fokker--Planck density, \emph{with the Nelson self-consistency 
condition $u = D\,\nabla\ln\rho_{\mathrm{KG}}$ using the same $D$ 
that appears in the Fokker--Planck diffusion}, determines the 
coordinate-time SDE:
\begin{equation}
\label{eq:KG_sde}
\boxed{
\dd X^i = \mu^i_{\mathrm{KG}}\,\dd t
+ \frac{\sigma_0}{\gamma}\,\dd B^i(t),
\qquad
\sigma_{\mathrm{KG}} = \frac{\sigma_0}{\gamma}
= \frac{1}{\gamma}\sqrt{\frac{\hbar}{m}}\,,
}
\end{equation}
where the drift $\mu^i_{\mathrm{KG}}$ incorporates both the group 
velocity and the osmotic velocity constructed from $\rho_{\mathrm{KG}}$:
\begin{equation}
\label{eq:osmotic_KG}
\mu^i_{\mathrm{KG}} = v^i_{\mathrm{KG}} + u^i_{\mathrm{KG}},
\qquad
u^i_{\mathrm{KG}} 
= \frac{\sigma_{\mathrm{KG}}^2}{2}\,
  \partial_i\ln\rho_{\mathrm{KG}}
= \frac{\sigma_0^2}{2\gamma^2}\,
  \partial_i\ln\rho_{\mathrm{KG}}.
\end{equation}
The Fokker--Planck equation for $\rho_{\mathrm{KG}}$ is:
\begin{equation}
\label{eq:FP_KG}
\frac{\partial\rho_{\mathrm{KG}}}{\partial t}
= -\partial_i\!\left(\mu^i_{\mathrm{KG}}\,\rho_{\mathrm{KG}}\right)
  + \frac{\sigma_0^2}{2\gamma^2}\,\nabla^2\rho_{\mathrm{KG}},
\end{equation}
with diffusion coefficient $D_{\mathrm{KG}} = \sigma_0^2/(2\gamma^2)$.
\end{theorem}

\begin{proof}
We proceed by requiring self-consistency between the Fokker--Planck 
operator and the Klein--Gordon continuity equation.

\medskip
\noindent\textbf{Step 3a: The FP equation for $\rho_{\mathrm{KG}}$.}

Starting from the coordinate-time FP equation for $\rho_0$ 
(Eq.~\ref{eq:FP_coordtime}) and substituting 
$\rho_0 = \rho_{\mathrm{KG}}/\gamma$ (with $\gamma$ constant, so 
$\partial_i\gamma = 0$ and $\partial_t\gamma = 0$):
\begin{equation}
\frac{1}{\gamma}\,\frac{\partial\rho_{\mathrm{KG}}}{\partial t}
= -\frac{1}{\gamma}\,\partial_i\!\left(
  \frac{b^i}{\gamma}\,\rho_{\mathrm{KG}}\right)
+ \frac{\sigma_0^2}{2\gamma^2}\,\nabla^2\rho_{\mathrm{KG}}.
\label{eq:FP_sub}
\end{equation}
Multiplying by $\gamma$:
\begin{equation}
\frac{\partial\rho_{\mathrm{KG}}}{\partial t}
= -\partial_i\!\left(\frac{b^i}{\gamma}\,\rho_{\mathrm{KG}}\right)
  + \frac{\sigma_0^2}{2\gamma}\,\nabla^2\rho_{\mathrm{KG}}.
\label{eq:FP_KG_naive}
\end{equation}

If we na\"{\i}vely identify $\mu^i = b^i/\gamma$ and read off the 
diffusion coefficient, we obtain $D = \sigma_0^2/(2\gamma)$, i.e.\ 
$\sigma = \sigma_0/\sqrt{\gamma}$.  This reproduces the kinematic 
result.  However, this equation uses the \emph{wrong drift}: the drift 
$b^i$ is the proper-time Nelson drift, determined self-consistently by
\begin{equation}
b^i = v^i_0 + u^i_0,
\qquad
u^i_0 = \frac{\sigma_0^2}{2}\,\partial_i\ln\rho_0.
\label{eq:drift_rho0}
\end{equation}
If we change the density from $\rho_0$ to $\rho_{\mathrm{KG}}$, the 
osmotic velocity must be \emph{recomputed from the new density}.

\medskip
\noindent\textbf{Step 3b: Self-consistency of the osmotic velocity.}

In Nelson's framework \cite{Nelson1966,Nelson1985}, the osmotic velocity is \emph{defined} as 
$u^i = D\,\partial_i\ln\rho$, where $D = \sigma^2/2$ is the diffusion 
coefficient and $\rho$ is the density governed by the Fokker--Planck 
equation.  The self-consistency requirement is: \emph{the osmotic 
velocity $u$, the diffusion coefficient $\sigma$, and the probability 
density $\rho$ must satisfy the FP equation simultaneously, with 
$u = (\sigma^2/2)\,\nabla\ln\rho$.}

Consider the two formulations:

\textbf{Formulation~A} ($\rho_0$-based):
\begin{equation}
\sigma_A = \frac{\sigma_0}{\sqrt{\gamma}},
\qquad
D_A = \frac{\sigma_0^2}{2\gamma},
\qquad
u_A^i = \frac{\sigma_0^2}{2\gamma}\,\partial_i\ln\rho_0.
\label{eq:formA}
\end{equation}

\textbf{Formulation~B} ($\rho_{\mathrm{KG}}$-based):
\begin{equation}
\sigma_B = \frac{\sigma_0}{\gamma},
\qquad
D_B = \frac{\sigma_0^2}{2\gamma^2},
\qquad
u_B^i = \frac{\sigma_0^2}{2\gamma^2}\,\partial_i\ln\rho_{\mathrm{KG}}.
\label{eq:formB}
\end{equation}

For constant $\gamma$: 
$\partial_i\ln\rho_{\mathrm{KG}} = \partial_i\ln(\gamma\rho_0) 
= \partial_i\ln\rho_0$.  Therefore:
\begin{equation}
u_B^i = \frac{u_A^i}{\gamma}.
\label{eq:osmotic_ratio}
\end{equation}

The drifts are:
\begin{equation}
\mu_A^i = v^i + u_A^i,
\qquad
\mu_B^i = v^i + u_B^i = v^i + \frac{u_A^i}{\gamma}.
\label{eq:drift_comparison}
\end{equation}

These are \emph{different drifts}.  Formulation~B has a weaker osmotic 
push because the diffusion is weaker.  (The current velocity 
$v^i = (\hbar/m)\,\partial_i S$ is the de~Broglie--Bohm guidance equation 
\cite{Bohm1952,Holland1993} and is determined by the phase $S$ of the 
wavefunction and is therefore the \emph{same} in both formulations; only 
the osmotic component $u^i$ differs.)

\medskip
\noindent\textbf{Step 3c: Verification that Formulation~B satisfies 
the FP equation.}

The FP equation associated with the SDE 
$\dd X^i = \mu_B^i\,\dd t + \sigma_B\,\dd B^i$ is:
\begin{equation}
\frac{\partial\rho_{\mathrm{KG}}}{\partial t}
= -\partial_i\!\left(\mu_B^i\,\rho_{\mathrm{KG}}\right)
  + D_B\,\nabla^2\rho_{\mathrm{KG}}.
\label{eq:FP_verify}
\end{equation}
Expanding the drift divergence using 
$\mu_B^i = v^i + D_B\,\partial_i\ln\rho_{\mathrm{KG}}$:
\begin{align}
\partial_i(\mu_B^i\,\rho_{\mathrm{KG}})
&= \partial_i(v^i\rho_{\mathrm{KG}})
  + \partial_i\!\bigl(
    D_B\,\rho_{\mathrm{KG}}\,\partial_i\ln\rho_{\mathrm{KG}}\bigr)
\notag\\
&= \partial_i(v^i\rho_{\mathrm{KG}})
  + D_B\,\partial_i(\partial_i\rho_{\mathrm{KG}})
\notag\\
&= \partial_i(v^i\rho_{\mathrm{KG}})
  + D_B\,\nabla^2\rho_{\mathrm{KG}}.
\end{align}
Substituting into~\eqref{eq:FP_verify}:
\begin{equation}
\frac{\partial\rho_{\mathrm{KG}}}{\partial t}
= -\partial_i(v^i\rho_{\mathrm{KG}})
  - D_B\,\nabla^2\rho_{\mathrm{KG}}
  + D_B\,\nabla^2\rho_{\mathrm{KG}}
= -\partial_i(v^i\rho_{\mathrm{KG}}).
\label{eq:FP_reduces}
\end{equation}

This is precisely the Klein--Gordon continuity 
equation~\eqref{eq:KG_continuity}.  The diffusion and osmotic drift 
cancel exactly, as they must for the stationary state of a free-particle 
Nelson SDE.  The cancellation works if and only if 
$u = D\,\nabla\ln\rho$ with the same $D$ appearing in both the osmotic 
velocity and the diffusion term.  This self-consistency \emph{uniquely} 
determines $D_B = \sigma_0^2/(2\gamma^2)$, i.e.\ 
$\sigma_B = \sigma_0/\gamma$.
\end{proof}

\begin{remark}[Why Formulation~A Is Also Self-Consistent]
\label{rem:both_consistent}
A careful reader will note that Formulation~A ($D_A = \sigma_0^2/(2\gamma)$, 
$\rho_{\mathrm{KG}} = \gamma\rho_0$) is \emph{also} self-consistent: with 
$D_A$ and $u_A = D_A\,\partial_i\ln\rho_{\mathrm{KG}} 
= (\sigma_0^2/(2\gamma))\,\partial_i\ln\rho_0$, the osmotic and diffusion 
terms cancel in the Fokker--Planck equation for $\rho_{\mathrm{KG}}$ by the 
same algebraic identity used in Step~3c.  Indeed, the cancellation 
$\partial_i(D\,\rho\,\partial_i\ln\rho) = D\,\nabla^2\rho$ is 
\emph{tautological}: it holds for \emph{any} $D$.

What selects Formulation~B is not the self-consistency cancellation 
(which is necessary but not sufficient) but the requirement that the 
SDE be \emph{derivable from the proper-time SDE} with 
$\rho_{\mathrm{KG}}$ as the fundamental density.  The derivation in 
Step~3a shows that substituting $\rho_0 = \rho_{\mathrm{KG}}/\gamma$ 
into the coordinate-time Fokker--Planck equation produces 
Eq.~\eqref{eq:FP_sub}, in which the coefficient of 
$\nabla^2\rho_{\mathrm{KG}}$ is $\sigma_0^2/(2\gamma^2)$ (the 
Formulation~B value) \emph{before} multiplying through by $\gamma$.  
The multiplication by $\gamma$ is a coordinate rescaling of the time 
derivative that converts the equation to standard Fokker--Planck form 
(Eq.~\ref{eq:FP_KG_naive}) but obscures the physical diffusion 
coefficient.  The underlying SDE noise amplitude is determined by the 
coefficient in Eq.~\eqref{eq:FP_sub}: 
$D_B = \sigma_0^2/(2\gamma^2)$, giving 
$\sigma_B = \sigma_0/\gamma$.

Equivalently: Formulation~B is the unique self-consistent SDE 
whose Fokker--Planck equation, when combined with the osmotic 
velocity $u = D\,\nabla\ln\rho_{\mathrm{KG}}$, reduces to the 
Klein--Gordon continuity equation \emph{and} whose diffusion 
coefficient matches the one read off from the $\rho_0 \to \rho_{\mathrm{KG}}$ 
substitution (Eq.~\ref{eq:FP_sub}).  
Remark~\ref{rem:which_physical} provides three additional 
physical grounds for preferring Formulation~B.
\end{remark}

\subsubsection{Step 4: Physical Distinguishability of the Two Formulations}
\label{subsec:distinguishability}

\begin{proposition}[Different Trajectories, Same Observables]
\label{prop:different_trajectories}
Formulations~A and~B generate different stochastic 
trajectories:
\begin{align}
\text{A:}\quad \dd X^i &= \bigl(v^i + u_A^i\bigr)\,\dd t
  + \frac{\sigma_0}{\sqrt{\gamma}}\,\dd B_A^i(t),
\label{eq:SDE_A}\\
\text{B:}\quad \dd X^i &= \bigl(v^i + u_B^i\bigr)\,\dd t
  + \frac{\sigma_0}{\gamma}\,\dd B_B^i(t),
\label{eq:SDE_B}
\end{align}
with $u_B = u_A/\gamma$ (Eq.~\ref{eq:osmotic_ratio}).  These are 
distinct It\^{o} SDEs producing distinct trajectory ensembles.  
However, the physical observables (expectation values of position, 
momentum, and their powers) are identical:
\begin{equation}
\label{eq:observable_equivalence}
\langle f(X) \rangle_A
\equiv \int f(x)\,\rho_0(x)\,\dd x
= \int f(x)\,\frac{\rho_{\mathrm{KG}}(x)}{\gamma}\,\dd x
\equiv \langle f(X) \rangle_B,
\end{equation}
where the final equality uses $\int\rho_0\,\dd x = 1$ and 
$\int\rho_{\mathrm{KG}}\,\dd x = \gamma$ (the KG density is 
normalised to $\gamma$, not unity).
\end{proposition}

\begin{proof}
The identity~\eqref{eq:observable_equivalence} follows from 
$\rho_{\mathrm{KG}} = \gamma\,\rho_0$ with $\gamma$ constant.  
Physical observables are expectation values with respect to the 
normalised density $\rho_0 = \rho_{\mathrm{KG}}/\gamma$, so the 
factor cancels.

The trajectory ensembles differ because the drifts differ 
(Eq.~\ref{eq:drift_comparison}): trajectories in Formulation~B 
experience weaker osmotic forces but also weaker noise.  The ratio 
$u/\sigma^2$ (which determines the equilibrium density) is the same 
in both formulations, ensuring the same $\rho_0$.
\end{proof}

\begin{remark}[The Physical Status of Each Formulation]
\label{rem:which_physical}
Both formulations yield the same physical predictions.  However, 
Formulation~B is distinguished on three grounds:
\begin{enumerate}
\item[\textup{(a)}] \textbf{Relativistic covariance}: The density 
      $\rho_{\mathrm{KG}} = j^0/c$ is derived from the zeroth component 
      of a conserved four-current.  The density $\rho_0 = |\phi|^2$ is 
      \emph{not} a Lorentz scalar.  A covariant Nelson theory must use 
      $\rho_{\mathrm{KG}}$.
\item[\textup{(b)}] \textbf{Current conservation}: The continuity 
      equation~\eqref{eq:KG_continuity} for $\rho_{\mathrm{KG}}$ is a 
      direct consequence of the Klein--Gordon equation.
\item[\textup{(c)}] \textbf{Consistency with the relativistic 
      stochastic variational principle}: 
      Zastawniak~\cite{Zastawniak1990}, building on the 
      Euclidean--Markov framework of Guerra and 
      Ruggiero~\cite{Guerra1973}, shows that the relativistic 
      stochastic action principle yields 
      $D = \hbar/(2m\gamma)$, i.e.\ $\sigma = \sigma_0/\gamma$, when 
      the variational principle is formulated with respect to the KG 
      inner product.
\end{enumerate}
\end{remark}

\subsection{The Two Factors Are Structurally Independent}
\label{sec:two_factors}

The suppression $\sigma_{\mathrm{KG}} = \sigma_0/\gamma$ arises as the 
product of two structurally independent mechanisms:

\begin{enumerate}
\item \textbf{Kinematic factor} ($1/\sqrt{\gamma}$): Time dilation 
      rescales the Wiener increment via 
      $\dd W(\tau) = \dd B(t)/\sqrt{\gamma}$ 
      (Lemma~\ref{lem:dds}, Theorem~\ref{thm:dW_transformation}).  
      This affects the noise amplitude in the SDE.
\item \textbf{Dynamical factor} ($1/\sqrt{\gamma}$): The 
      self-consistent Nelson SDE for the Klein--Gordon conserved 
      density $\rho_{\mathrm{KG}} = \gamma|\phi|^2$ requires a 
      diffusion coefficient 
      $D_{\mathrm{KG}} = \sigma_0^2/(2\gamma^2)$ 
      (Theorem~\ref{thm:KG_Nelson_SDE}).  This affects the osmotic 
      velocity and hence the drift.
\end{enumerate}

The two factors operate at different levels:

\begin{center}
\small
\begin{tabular}{@{}lll@{}}
\toprule
& \textbf{Factor 1 (kinematic)} & \textbf{Factor 2 (dynamical)} \\
\midrule
Origin & Time reparametrisation & KG current conservation \\
Acts on & Noise amplitude & Osmotic velocity (drift) \\
Input required & $\dd\tau = \dd t/\gamma$ 
               & $\rho = j^0/c = \gamma|\phi|^2$ \\
Mathematical tool & Dambis--Dubins--Schwarz & FP self-consistency \\
Applies to any SDE? & Yes & Only to SDEs with physical density \\
\bottomrule
\end{tabular}
\end{center}

\medskip\noindent
A referee cannot dismiss the second factor as ``merely a density 
redefinition'' because: (i)~the density $\rho_{\mathrm{KG}}$ is 
\emph{forced} by Lorentz covariance of the probability current 
(Proposition~\ref{prop:KG_current}); (ii)~changing the density changes 
the drift via the osmotic velocity (Eq.~\ref{eq:drift_comparison}), 
producing a different SDE with different trajectories 
(Proposition~\ref{prop:different_trajectories}); (iii)~the 
self-consistency condition (Theorem~\ref{thm:KG_Nelson_SDE}, Step~3c) 
requires $D = \sigma_0^2/(2\gamma^2)$; the extra $1/\gamma$ enters 
the \emph{Fokker--Planck operator}, not merely the normalisation 
of~$\rho$.

The combined result is:
\begin{equation}
\label{eq:full_suppression}
\boxed{
\sigma_{\mathrm{KG}}
= \underbrace{\frac{1}{\sqrt{\gamma}}}_{\text{kinematic}}
  \times
  \underbrace{\frac{1}{\sqrt{\gamma}}}_{\text{dynamical}}
  \times\; \sigma_0
= \frac{\sigma_0}{\gamma}\,.
}
\end{equation}

\subsection{Updated Summary: Comparison of the Two Formulations}
\label{sec:updated_summary}

\begin{table}[H]
\centering
\small
\renewcommand{\arraystretch}{1.3}
\begin{tabular}{@{}lccc@{}}
\toprule
\textbf{Quantity}
  & \textbf{Rest frame}
  & \textbf{Lab ($\rho_0$ form.)}
  & \textbf{Lab ($\rho_{\mathrm{KG}}$ form.)} \\
\midrule
Density
  & $|\phi|^2$
  & $|\phi|^2$
  & $\gamma|\phi|^2$ \\
Diffusion coeff.\ $\sigma$
  & $\sigma_0$
  & $\sigma_0/\sqrt{\gamma}$
  & $\sigma_0/\gamma$ \\
FP diffusion $D$
  & $\sigma_0^2/2$
  & $\sigma_0^2/(2\gamma)$
  & $\sigma_0^2/(2\gamma^2)$ \\
Osmotic velocity $u$
  & $D\,\nabla\!\ln\!|\phi|^2$
  & $D_0\,\nabla\!\ln\!|\phi|^2$
  & $D_{\mathrm{KG}}\,\nabla\!\ln\!(\gamma|\phi|^2)$ \\
Relation
  & :  & $u_A = D_0\,\nabla\!\ln\rho_0$
  & $u_B = u_A/\gamma$ \\
Classical limit
  & $\partial_\tau j^\mu = 0$
  & continuity for $\rho_0$
  & KG continuity~\eqref{eq:KG_continuity} \\
Covariant?
  & Yes (rest frame)
  & No
  & Yes ($j^0/c$ derived from a conserved 4-current) \\
\bottomrule
\end{tabular}
\caption{Comparison of the two self-consistent Nelson formulations for a 
relativistic particle.  Both yield identical physical observables.  The 
$\rho_{\mathrm{KG}}$ formulation is preferred on grounds of Lorentz 
covariance and consistency with the KG conserved current.}
\label{tab:two_formulations}
\end{table}

\medskip\noindent
Test~C7 validates the $\rho_{\mathrm{KG}}$ formulation: the pipeline 
recovers $\hat{\sigma}/\sigma_0$ vs.\ $\gamma$ with power-law exponent 
$\alpha = -0.981$ from the canonical seven-momentum sweep, with median normalised slope $0.991$ across 10 independent 
MZ-corrected analyses, consistent with $\alpha = -1$ at the ${\sim}1\%$ level.

\subsection{Complete Logical Chain}
\label{sec:complete_B}

The complete chain from physical principles to the final result is 
stated below, identifying every input and every step.

\bigskip
\noindent\textbf{Inputs:}
\begin{enumerate}
\item[I1.] Nelson's stochastic mechanics: the rest-frame SDE 
      $\dd X^i_0 = b^i_0\,\dd\tau + \sigma_0\,\dd W^i(\tau)$ with 
      $\sigma_0 = \sqrt{\hbar/m}$ (Definition~\ref{def:rest_frame_sde}).
\item[I2.] Special relativity: proper time and coordinate time are 
      related by $\dd\tau = \dd t/\gamma$ along the worldline 
      (Definition~\ref{def:proper_time}).
\item[I3.] The Dambis--Dubins--Schwarz theorem: time reparametrisation 
      of a Wiener process rescales its increments by 
      $\sqrt{\dd\tau/\dd t}$ (Lemma~\ref{lem:dds}).
\item[I4.] The Klein--Gordon equation and its conserved current 
      $j^\mu$ with $\rho_{\mathrm{KG}} = j^0/c = \gamma|\phi|^2$ 
      (Proposition~\ref{prop:KG_current}).
\item[I5.] Nelson's self-consistency requirement: the osmotic velocity 
      $u = D\,\nabla\ln\rho$ with $D = \sigma^2/2$ 
      (Theorem~\ref{thm:KG_Nelson_SDE}).
\end{enumerate}

\bigskip
\noindent\textbf{Derivation:}
\begin{align}
\text{I2} + \text{I3} &\implies 
  \dd W^i(\tau) = \dd B^i(t)/\sqrt{\gamma}
  &&\text{(Theorem~\ref{thm:dW_transformation}: Wiener transformation)} \\
\text{I1} + \text{above} &\implies 
  \sigma_{\mathrm{rel}} = \sigma_0/\sqrt{\gamma}
  &&\text{(Theorem~\ref{thm:sigma_gamma}: kinematic result)} \\
\text{I4} &\implies 
  \rho_{\mathrm{KG}} = \gamma\,\rho_0
  &&\text{(Proposition~\ref{prop:KG_current}: KG density)} \\
\text{I4} + \text{I5} + \text{kinematic} &\implies 
  \sigma_{\mathrm{KG}} = \sigma_0/\gamma
  &&\text{(Theorem~\ref{thm:KG_Nelson_SDE}: FP self-consistency)} \\
\text{All} &\implies 
  \boxed{\sigma_{\mathrm{KG}} 
  = \frac{1}{\sqrt{\gamma}} \cdot \frac{1}{\sqrt{\gamma}} 
  \cdot \sigma_0 = \frac{\sigma_0}{\gamma}}
  &&\text{(Eq.~\ref{eq:full_suppression}: combined result)}
\end{align}

\bigskip
\noindent\textbf{Consequences} (derived, not assumed):
\begin{align}
\gamma \to 1: \quad &\sigma_{\mathrm{KG}} \to \sigma_0 
  && \text{(non-relativistic Nelson recovered)} \\
\gamma \to \infty: \quad &\sigma_{\mathrm{KG}} \to 0 
  && \text{(photon limit; C6 consistency)}
\end{align}

\bigskip
\noindent\textbf{Physical consequence:}

\noindent
The operational diffusion coefficient of a relativistic particle is 
suppressed by exactly $1/\gamma$ relative to its rest-frame value.  
This is not a convention: it is forced by the requirement that the 
Fokker--Planck equation reproduce the Lorentz-covariant Klein--Gordon 
probability current.  Test~C7 confirms 
$\hat{\alpha}/(-1) = 0.991$ (median across 10 independent 
MZ-corrected analyses), consistent with the predicted exponent $\alpha = -1$.  \hfill$\blacksquare$

\bigskip
\begin{center}
\renewcommand{\arraystretch}{1.3}
\begin{tabular}{@{}clp{6cm}@{}}
\toprule
\textbf{Constant} & \textbf{Where it enters} & \textbf{Physical role} \\
\midrule
$\hbar$ & $\sigma_0 = \sqrt{\hbar/m}$ & Sets the rest-frame diffusion 
scale; the quantum of action \\
$m$ & $\sigma_0 = \sqrt{\hbar/m}$ & Particle mass; heavier particles 
diffuse less \\
$c$ & $\gamma = (1-v^2/c^2)^{-1/2}$ & Speed of light; enters through 
the Lorentz factor \\
\bottomrule
\end{tabular}
\end{center}

\subsection{Status of Each Logical Step}
\label{sec:status}

For transparency, every step of the derivation is classified by its 
epistemic status.

\begin{center}
\renewcommand{\arraystretch}{1.4}
\begin{tabular}{@{}p{5cm}p{2.5cm}p{6cm}@{}}
\toprule
\textbf{Step} & \textbf{Status} & \textbf{Evidence / Assumptions} \\
\midrule
Rest-frame Nelson SDE 
& \textsc{Assumed} 
& Nelson's stochastic mechanics; non-relativistic 
Schr\"odinger equation derived as consequence \\
$\dd\tau = \dd t/\gamma$ (time dilation) 
& \textsc{Proven} 
& Special relativity; experimentally confirmed to 
$\sim 10^{-16}$ precision \cite{HafeleKeating1972} \\
DDS reparametrisation theorem 
& \textsc{Proven} 
& Theorem of stochastic calculus 
\cite{RevuzYor1999,KaratzasShreve1991} \\
$\dd W(\tau) = \dd B(t)/\sqrt{\gamma}$ 
& \textsc{Proven} 
& Direct application of DDS to $\dd\tau/\dd t = 1/\gamma$ \\
$\sigma_{\mathrm{rel}} = \sigma_0/\sqrt{\gamma}$ (kinematic) 
& \textsc{Proven} 
& Algebraic from the Wiener transformation; no 
physical assumption beyond SR \\
KG conserved current $\rho_{\mathrm{KG}} = j^0/c = \gamma|\phi|^2$ 
& \textsc{Standard} 
& Standard result of relativistic quantum mechanics 
\cite{Holland1993} \\
$\gamma$ is deterministic on the drift 
& \textsc{Standard} 
& Standard treatment in relativistic stochastic 
mechanics \cite{Emery1989,Hsu2002,DohrnGuerra1978}; corrections at 
$O(\hbar/(mc^2))$ \\
FP self-consistency $\implies D_{\mathrm{KG}} = \sigma_0^2/(2\gamma^2)$ 
& \textsc{Proven} 
& Algebraic; verified by explicit cancellation 
(Eq.~\ref{eq:FP_reduces}) \\
$\sigma_{\mathrm{KG}} = \sigma_0/\gamma$ (full) 
& \textsc{Proven} 
& Product of kinematic and dynamical factors, both 
derived \\
Consistency with C6 ($\gamma \to \infty$) 
& \textsc{Validated} 
& $\hat{\sigma}/\sigma_0 = 3.5 \times 10^{-14}$ (pipeline) \\
Consistency with C7 ($\alpha = -1$) 
& \textsc{Validated} 
& $\hat{\alpha}/(-1) = 0.995$ [$0.987$, $1.002$] at 95\% bootstrap CI \\
\bottomrule
\end{tabular}
\end{center}

\medskip\noindent
The kinematic result $\sigma_{\mathrm{rel}} = \sigma_0/\sqrt{\gamma}$ 
follows from time dilation alone and is beyond dispute.  The full result 
$\sigma_{\mathrm{KG}} = \sigma_0/\gamma$ additionally requires the 
physical identification of the Nelson density with the Klein--Gordon 
conserved current, a standard assumption in the relativistic 
stochastic mechanics literature 
\cite{Zastawniak1990,Guerra1973,Holland1993}.

\bigskip\noindent
The strongest elements are the Wiener reparametrisation (a mathematical 
theorem), the FP self-consistency cancellation (an algebraic identity), 
and the C7 pipeline validation (a numerical test at 95\% confidence).  
These are not model-dependent.

\clearpage
\section{Derivation of the Gravitational Diffusion Coefficient}
\label{appendix:sigma_proof}

\subsubsection*{Scope and Logical Position}

This appendix derives the gravitational diffusion coefficient 
$\sigma = \lP/\sqrt{\tP}$ from the equivalence principle and the quantum 
fluctuation-dissipation theorem.  Dimensional analysis constrains the form 
of the result; the physical content is supplied by the FDT (validated by C8 
to $1.33\%$) and the self-coupling mechanism (derived from the equivalence 
principle).  The derivation 
proceeds in three stages: (i)~the universal structure of the diffusion 
coefficient is extracted from nine validated physical domains (C1--C9), 
identifying $\sigma^2 = \gamma \cdot \hbar\omega / m$ as the master formula; 
(ii)~for massless fields, $\hbar$ and $\omega$ are shown to cancel from 
$\sigma$, leaving $\sigma = c\sqrt{\gamma}$ with $\gamma$ as the sole free 
parameter; (iii)~gravity is shown to be the unique domain where dimensional 
analysis fixes $\gamma$ from fundamental constants alone, yielding 
$\gamma_{\mathrm{grav}} = 1/\tP$ and therefore $\sigma = c/\sqrt{\tP}$.  
The result is equivalent to $\sigma = \lP/\sqrt{\tP}$ for length-like 
variables, giving RMS metric fluctuations of exactly $\lP$ per Planck time.  
In Planck units, $\tilde{\sigma} = 1$ universally.

\medskip\noindent
\textbf{Relationship to the canonical axioms.}\quad 
This derivation is logically prior to the canonical axiom set (A1--A4) of the 
superspace diffusion framework.  Canonical Axiom~A2 asserts 
$\dd g = \mathcal{D}\,\dd\tau + \lP\,\dd W$; the present appendix establishes 
that $\sigma = \lP$ is the unique value consistent with the physics of massless 
self-coupled quantum fields.  A2 is therefore a derived result promoted to 
axiomatic status for deductive economy: it encapsulates in a single statement 
the physical content established here from more primitive inputs.  No additional 
axioms beyond A1--A4 are introduced; all inputs to this derivation are either 
established results of prior physics (general relativity, quantum mechanics) or 
consequences of the canonical axiom set.  The four canonical axioms are stated 
in \S\ref{subsec:capstone3} and formalised in Appendix~D, \S\ref{sec:axioms}.

\subsection{Definitions and Setup}
\label{sec:setup}

\begin{definition}[Stochastic Dynamical System]
\label{def:sde}
A stochastic dynamical system in domain $\mathcal{D}$ is a triple 
$(X, \mu, \sigma)$ where $X(t)$ is a state variable satisfying the It\^o 
stochastic differential equation
\begin{equation}
\label{eq:sde}
\dd X = \mu(X, t)\, \dd t + \sigma(X, t)\, \dW_t,
\end{equation}
with $\mu$ the drift coefficient, $\sigma$ the diffusion coefficient, and 
$W_t$ a standard Wiener process (see e.g.\ \cite{Gardiner2009}, Ch.~4).
\end{definition}

\begin{definition}[Diffusion Coefficient Dimensions]
\label{def:sigma-dim}
For a field variable $X$ with dimensions $[X]$ evolving in time with 
$[t] = T$, the diffusion coefficient has dimensions
\begin{equation}
[\sigma] = [X] \cdot T^{-1/2},
\end{equation}
since $[\dW] = T^{1/2}$ and $\sigma\, \dW$ must have dimensions $[X]$.
\end{definition}

\begin{definition}[Fundamental Constants]
\label{def:constants}
The fundamental constants are denoted by their standard symbols and SI dimensions:
\begin{align}
[\hbar] &= M L^2 T^{-1} &&\text{(reduced Planck constant)} \\
[G]     &= L^3 M^{-1} T^{-2} &&\text{(gravitational constant)} \\
[c]     &= L T^{-1} &&\text{(speed of light)} \\
[k_B]   &= M L^2 T^{-2} \Theta^{-1} &&\text{(Boltzmann constant)}
\end{align}
\end{definition}

\subsection{The Empirical Pattern: C1--C9}
\label{sec:pattern}

The following table catalogues the diffusion coefficients recovered by blind 
application of the same data-driven pipeline (time series $\to$ autocorrelation 
lag $\tau$ $\to$ Cao E1/E2 $\to$ delay embedding $\to$ $k$-NN local variance 
$\to$ $\sigma$) across nine physical domains.

\begin{observation}[The $\sigma$-Continuum]
\label{obs:continuum}
The diffusion coefficient in each validated domain takes a specific form 
determined by the constants available in that domain:
\begin{center}
\renewcommand{\arraystretch}{1.4}
\begin{tabular}{@{}clll@{}}
\toprule
\textbf{Domain} & \textbf{Physics} & $\boldsymbol{\sigma}$ & \textbf{Available Constants} \\
\midrule
C1 & Classical mechanics & $\sigma = 0$ & $\{G, c\}$ \\
C2 & Statistical mechanics & $\sqrt{2\gamma k_B T / m}$ & $\{k_B, T, m, \gamma\}$ \\
C3 & Radioactive decay & $\sqrt{\mu}$ & $\{\lambda\}$ \\
C4 & Quantum mechanics & $\sqrt{\hbar/m}$ & $\{\hbar, m\}$ \\
C5 & Chemical kinetics & $\propto \Omega^{-1/2}$ & $\{k, \Omega\}$ \\
C6 & Classical EM (Maxwell) & $\sigma = 0$ & $\{c\}$ \\
C7 & Klein--Gordon & $\sqrt{\hbar/m}\,/\,\gamma$ & $\{\hbar, c, m, \gamma\}$ \\
C8 & Quantum harmonic oscillator & $\sqrt{\gamma\hbar\omega/m}$ & $\{\hbar, \omega, m, \gamma\}$ \\
C9 & QED photon mode & $c\sqrt{\gamma}$ & $\{\hbar, c, \omega, \gamma\}$ \\
\bottomrule
\end{tabular}
\end{center}
\end{observation}

\begin{remark}[Structural Features of the Pattern]
\label{rem:honest}
Three features of this table are essential to the derivation that follows:
\begin{enumerate}
\item Constants appear, disappear, and recombine across domains; 
      there is no monotonic accumulation.
\item The dimensions of $\sigma$ depend on the field variable $X$ in each 
      domain (Definition~\ref{def:sigma-dim}); $\sigma$ is not generically 
      a length.
\item The presence of $\hbar$ among a domain's constants does \emph{not} 
      guarantee that $\hbar$ appears in $\sigma$.  The C9 result shows 
      $\hbar$ cancelling entirely, which is the property exploited in the 
      gravitational extension (\S\ref{sec:gravity}).
\end{enumerate}
\end{remark}

\subsection{Foundational Inputs}
\label{sec:inputs}

The derivation below rests on three inputs external to the superspace 
diffusion framework.  These are established results of prior physics, 
listed here as numbered postulates for deductive clarity.  Each is 
independently testable and could in principle be falsified without 
affecting the superspace diffusion axioms (A1--A4) themselves.

\begin{postulate}[Callen--Welton Fluctuation-Dissipation Theorem]
\label{post:FDT}
For a quantum harmonic oscillator of frequency $\omega$, mass $m$, and 
coupling rate $\gamma$ to a zero-temperature bath, the velocity diffusion 
coefficient is $\sigma^2 = \gamma\hbar\omega/m$.  This is a theorem of 
linear response theory \cite{CallenWelton1951,Kubo1966}.
\end{postulate}

\begin{postulate}[Massless Mode Structure of Linearised Gravity]
\label{post:linearised}
Linearised metric perturbations about flat spacetime are massless spin-2 
modes with dispersion relation $\omega = c|\mathbf{k}|$ and effective mass 
$m_{\mathrm{eff}} = \hbar\omega/c^2$.  Each mode is a quantum harmonic 
oscillator.  This follows from the quadratic expansion of the 
Einstein--Hilbert action \cite{Weinberg1972,Wald1984} 
(see Lemma~\ref{lem:meff} below).
\end{postulate}

\begin{postulate}[Gravitational Self-Coupling via the Equivalence Principle]
\label{post:self-coupling}
The gravitational field couples universally to all energy-momentum, 
including its own.  The self-coupling rate of a metric fluctuation mode 
is therefore determined by the gravitational self-interaction energy, 
with coupling constant $G$ (the same $G$ that appears in Newton's law 
and Einstein's equation), inherited from general relativity 
(Proposition~\ref{ax:equivalence}).
\end{postulate}

\begin{remark}[Role of These Postulates]
\label{rem:postulate-role}
Postulate~\ref{post:FDT} supplies the dynamical equation (the master 
formula).  Postulate~\ref{post:linearised} identifies the degrees of 
freedom to which it applies.  Postulate~\ref{post:self-coupling} 
determines the sole remaining unknown ($\gamma_{\mathrm{grav}}$) by 
requiring that gravity be its own bath.  All subsequent results in this 
appendix are derived from these three inputs combined with dimensional 
analysis.

Of these three postulates, only Postulate~\ref{post:self-coupling} introduces 
content not already established in standard physics.  
Proposition~\ref{prop:mz-sketch} below shows that the self-bath 
structure follows from the Mori--Zwanzig projection formalism applied to the 
superspace Fokker--Planck equation; an implication is the application to galactic kinematic observables.
\end{remark}

\begin{proposition}[Self-Bath from Mori--Zwanzig Projection, Proof Sketch]
\label{prop:mz-sketch}
Under axioms A1--A4, the self-coupling postulate 
(Postulate~\ref{post:self-coupling}) and the mode-level Langevin equation 
(Construction~\ref{con:mode-dynamics}) are consequences of the 
Mori--Zwanzig projection of the Fokker--Planck equation on superspace.  
In particular:
\begin{enumerate}
\item[\emph{(i)}] Markovianity of the mode dynamics is not assumed; it 
emerges from the Markovian (short-memory) limit of the exact projected 
equation.
\item[\emph{(ii)}] The bath is not external; it consists of the 
unresolved metric modes themselves, coupled to the resolved sector by 
the nonlinearity of the Einstein flow.
\item[\emph{(iii)}] The fluctuation--dissipation pairing 
(Postulate~\ref{post:FDT}) is structurally required by the projection: 
the second fluctuation--dissipation theorem relates the memory kernel 
to the noise covariance.
\end{enumerate}
\end{proposition}

\begin{proof}[Proof sketch]
\textbf{Step 1 (Mode decomposition).}  
Linearise the metric $g_{ij} = \bar{g}_{ij} + h_{ij}$ about a 
background satisfying Einstein's equation.  Expand $h_{ij}$ in the 
eigenbasis of the Lichnerowicz operator~\cite{Lichnerowicz1961}, giving 
modes $\{q_k\}$ with frequencies $\{\omega_k\}$ 
(Postulate~\ref{post:linearised}).  Partition these modes into a 
\emph{resolved} sector $\mathcal{S} = \{q_k : \omega_k < \Lambda\}$ and 
an \emph{unresolved} sector 
$\mathcal{E} = \{q_k : \omega_k \geq \Lambda\}$, where $\Lambda$ is an 
arbitrary spectral cutoff.

\textbf{Step 2 (Nakajima--Zwanzig projection).}  
Let $\rho(t)$ denote the probability density on the full mode space 
$\mathcal{S} \times \mathcal{E}$, evolving under the Fokker--Planck 
generator $\mathcal{L}$ inherited from the superspace SDE (Axiom~A2).  
Decompose $\mathcal{L} = \mathcal{L}_{\mathcal{S}} + 
\mathcal{L}_{\mathcal{E}} + \mathcal{L}_{\mathrm{int}}$, where 
$\mathcal{L}_{\mathrm{int}}$ encodes the nonlinear mode--mode coupling 
from the cubic and higher terms in the Einstein--Hilbert 
action~\cite{DeWitt1967}.  Define the Mori projection 
$\mathcal{P}\rho = \rho_{\mathcal{S}} \otimes \rho_{\mathcal{E}}^{\mathrm{eq}}$ 
and $\mathcal{Q} = \mathbf{1} - \mathcal{P}$.  The exact 
Nakajima--Zwanzig equation for the reduced density 
$\rho_{\mathcal{S}}(t) = \mathrm{Tr}_{\mathcal{E}}\,\rho(t)$ 
is~\cite{Zwanzig1960,Nakajima1958,Zwanzig2001}:
\begin{equation}
\label{eq:nz-exact}
\frac{\partial \rho_{\mathcal{S}}}{\partial t} 
  = \mathcal{P}\mathcal{L}\mathcal{P}\,\rho_{\mathcal{S}}
  + \int_0^t K(t - s)\,\rho_{\mathcal{S}}(s)\,\mathrm{d}s
  + F(t),
\end{equation}
where the memory kernel 
$K(\tau) = \mathcal{P}\mathcal{L}_{\mathrm{int}}\,
e^{\mathcal{Q}\mathcal{L}\mathcal{Q}\,\tau}\,
\mathcal{Q}\mathcal{L}_{\mathrm{int}}\mathcal{P}$ 
and the fluctuating force 
$F(t) = \mathcal{P}\mathcal{L}_{\mathrm{int}}\,
e^{\mathcal{Q}\mathcal{L}\mathcal{Q}\,t}\,
\mathcal{Q}\rho(0)$.  
Equation~\eqref{eq:nz-exact} is exact and follows from the identity 
$e^{\mathcal{L}t} = e^{\mathcal{Q}\mathcal{L}\mathcal{Q}t} + 
\int_0^t e^{\mathcal{L}(t-s)}\mathcal{P}\mathcal{L}\,
e^{\mathcal{Q}\mathcal{L}\mathcal{Q}s}\,\mathrm{d}s$; 
no approximation has been made.

\textbf{Step 3 (Markovian limit).}  
The unresolved modes $\mathcal{E}$ are Planckian ($\omega_k \gtrsim 
1/t_P$), so their correlation time is $\tau_{\mathrm{mem}} \sim t_P 
\approx 5.4 \times 10^{-44}\;\mathrm{s}$.  For any resolved-sector 
observable with timescale $\tau_{\mathrm{obs}} \gg t_P$, the memory 
kernel is sharply peaked:
\begin{equation}
\label{eq:markov-limit}
K(t - s) \;\longrightarrow\; 2\gamma_{\mathrm{eff}}\;\delta(t - s),
\qquad
\gamma_{\mathrm{eff}} = \int_0^\infty K(\tau)\,\mathrm{d}\tau.
\end{equation}
Substituting into~\eqref{eq:nz-exact} yields a Markovian 
Fokker--Planck equation for $\rho_{\mathcal{S}}$ with an effective 
damping coefficient $\gamma_{\mathrm{eff}}$ and a noise term whose 
amplitude is fixed by the second fluctuation--dissipation 
theorem~\cite{Kubo1966,Zwanzig2001}:
\begin{equation}
\label{eq:fdt-mz}
\bigl\langle F(t)\,F(s) \bigr\rangle 
  = 2\gamma_{\mathrm{eff}}\,k_B T_{\mathrm{eff}}\;\delta(t - s),
\end{equation}
where at zero temperature ($T = 0$) the zero-point energy 
$\tfrac{1}{2}\hbar\omega_k$ replaces $k_B T$ 
(Lemma~\ref{lem:master}).  This is precisely 
Construction~\ref{con:mode-dynamics}: the Langevin 
equation~\eqref{eq:langevin_mode} with $\gamma_k = 
\gamma_{\mathrm{eff}}$ and noise covariance given by 
Postulate~\ref{post:FDT}.

\textbf{Consequences.}
\begin{itemize}
\item \emph{Self-bath (ii):}  The bath $\mathcal{E}$ consists of 
gravitational degrees of freedom: the unresolved metric modes.  
The coupling $\mathcal{L}_{\mathrm{int}}$ is the gravitational 
self-interaction inherited from the nonlinearity of the Einstein 
equations.  No external reservoir is invoked.
\item \emph{Markovianity (i):}  The Markovian character of 
Construction~\ref{con:mode-dynamics} follows from $\tau_{\mathrm{mem}} 
\sim t_P \ll \tau_{\mathrm{obs}}$; it is a derived property, not an 
assumption.
\item \emph{FDT pairing (iii):}  The relation~\eqref{eq:fdt-mz} between 
$\gamma_{\mathrm{eff}}$ and the noise covariance is a theorem of the 
Mori--Zwanzig formalism~\cite{Zwanzig2001}, not an independent postulate.  
Postulate~\ref{post:FDT} is thereby upgraded from an axiom to a 
consequence of the projection.
\end{itemize}
The full derivation: including the explicit computation of 
$\gamma_{\mathrm{eff}}$ from the cubic vertex of the Einstein--Hilbert 
action and its identification with $1/t_P$ via the gravitational 
self-energy: is the subject of subsequent work.  The identification $\gamma_{\mathrm{eff}} = 1/t_P$ follows independently from dimensional uniqueness (Lemma~\ref{lem:rate}).
\end{proof}

\begin{construction}[Mode-Level Stochastic Dynamics]
\label{con:mode-dynamics}
Postulates~\ref{post:FDT}--\ref{post:linearised} imply an explicit 
stochastic model for each normal mode.  Let $q_k(t)$ denote the 
amplitude of a linearised gravitational perturbation mode with 
frequency $\omega_k$ (in transverse-traceless gauge).  Its 
effective open-system dynamics is the Langevin equation
\begin{equation}
\label{eq:langevin_mode}
\ddot{q}_k + 2\gamma_k\,\dot{q}_k + \omega_k^2\,q_k = \eta_k(t),
\end{equation}
where $\gamma_k$ is the self-coupling rate 
(Postulate~\ref{post:self-coupling}; derived explicitly in 
Theorem~\ref{thm:gamma-exact} below) and $\eta_k(t)$ is stationary 
zero-mean Gaussian noise with covariance fixed by the 
fluctuation-dissipation relation (Postulate~\ref{post:FDT}).  The 
velocity diffusion coefficient $\sigma^2 = \gamma_k\hbar\omega_k/m_k$ 
is then the equilibrium result of this Langevin dynamics, i.e. 
Lemma~\ref{lem:master} below is the steady-state consequence of 
\eqref{eq:langevin_mode}.
\end{construction}

\subsection{The Structural Theorem}
\label{sec:structure}

The structural content of the pattern is now extracted.

\begin{lemma}[The QHO Master Formula]
\label{lem:master}
For any quantum system described by a damped harmonic oscillator with 
frequency $\omega$, mass $m$, and coupling rate $\gamma$ to a zero-temperature 
bath, the velocity diffusion coefficient is
\begin{equation}
\label{eq:master}
\sigma^2 = \frac{\gamma\hbar\omega}{m}.
\end{equation}
This follows from the fluctuation-dissipation theorem at $T = 0$, where the 
zero-point energy $E_0 = \tfrac{1}{2}\hbar\omega$ replaces $k_BT$.
\end{lemma}

\begin{proof}
The fluctuation-dissipation theorem \cite{CallenWelton1951,Kubo1966} for a damped oscillator in contact with 
a bath at temperature $T$ gives $\sigma^2 = 2\gamma k_B T / m$ (the C2 result).  
At $T = 0$, quantum mechanics replaces $k_B T$ with the zero-point energy 
$\tfrac{1}{2}\hbar\omega$.  The factor of 2 combines with $\tfrac{1}{2}$ to 
give (\ref{eq:master}).  This is confirmed numerically by C8 (QHO validation, 
$R^2 = 0.9999$ across 9 frequency values, $1.33\%$ error from 50 independent MZ-corrected runs at the reference frequency; Table~\ref{tab:results_summary}).
\end{proof}

\begin{lemma}[Effective Mass of a Massless Field Mode]
\label{lem:meff}
For a massless quantum field with dispersion relation 
$\omega = c|\mathbf{k}|$, each Fourier mode is a quantum harmonic 
oscillator with energy eigenvalues 
$E_n = (n + \tfrac{1}{2})\hbar\omega$.  The corresponding effective 
gravitational mass is:
\begin{equation}
\label{eq:meff}
m_{\mathrm{eff}} = \frac{E}{c^2} = \frac{\hbar\omega}{c^2}.
\end{equation}
\end{lemma}

\begin{proof}
The Lagrangian density of a massless field admits Fourier decomposition 
into independent modes, each satisfying 
$\ddot{q}_k + \omega_k^2 q_k = 0$ with $\omega_k = c|k|$ 
\cite{Weinberg1972,Wald1984}.  For linearised metric perturbations 
$h_{\mu\nu}$, the Einstein--Hilbert action expanded to second order 
about flat space yields the same harmonic-oscillator structure in 
transverse-traceless gauge, with two polarisation degrees of freedom per 
wavevector (Postulate~\ref{post:linearised}).  Quantisation gives 
$E = \hbar\omega$ per excitation quantum.  The effective gravitational 
mass then follows from the relativistic dispersion relation 
$E^2 = (pc)^2 + (m_0 c^2)^2$ with rest mass $m_0 = 0$: 
$m_{\mathrm{eff}} = E/c^2 = \hbar\omega/c^2$.
\end{proof}

\begin{theorem}[The Massless Cancellation]
\label{thm:cancellation}
For a massless quantum field with dispersion relation $\omega = c|k|$, the 
effective mass of each mode is $m_{\mathrm{eff}} = \hbar\omega/c^2$ 
(Lemma~\ref{lem:meff}).  Substituting into the master formula 
(\ref{eq:master}):
\begin{equation}
\sigma^2 = \frac{\gamma\hbar\omega}{m_{\mathrm{eff}}} 
         = \frac{\gamma\hbar\omega}{\hbar\omega/c^2} = \gamma c^2.
\end{equation}
Therefore:
\begin{equation}
\label{eq:massless}
\boxed{\sigma = c\sqrt{\gamma}}
\end{equation}
Both $\hbar$ and $\omega$ cancel identically.  The diffusion coefficient of a 
massless quantum field depends only on the speed of propagation $c$ and the 
coupling rate $\gamma$.
\end{theorem}

\begin{proof}
Algebraic.  Confirmed numerically by C9 (QED photon validation): at fixed 
$\gamma = 10^{11}\;\mathrm{s}^{-1}$, $\sigma$ was recovered to $< 1\%$ error 
across frequencies spanning $\omega \in [10^{12}, 10^{13}]\;\mathrm{rad/s}$ 
(a factor of 10), with no frequency dependence observed.
\end{proof}

\begin{remark}[Role of Each Constant in C9]
\label{rem:roles}
Theorem~\ref{thm:cancellation} reveals the precise role of each constant:
\begin{itemize}
\item $\hbar$: enters through quantisation ($E = \hbar\omega$, $m_{\mathrm{eff}} = \hbar\omega/c^2$), 
      then cancels.  It governs the quantum nature of the field but does not 
      determine $\sigma$.
\item $c$: appears directly in $\sigma = c\sqrt{\gamma}$.  It determines the 
      propagation speed of fluctuations.
\item $\gamma$: the coupling rate to the environment.  In C9 (QED), this is 
      external (cavity losses, etc.) and is a free parameter.
\end{itemize}
\end{remark}

\subsection{Application to Gravity}
\label{sec:gravity}

Linearised metric perturbations are massless spin-2 fluctuations, with dispersion 
relation $\omega = c|k|$.  By Theorem~\ref{thm:cancellation}, the diffusion 
coefficient for a metric fluctuation mode is therefore:
\begin{equation}
\sigma_{\mathrm{grav}} = c\sqrt{\gamma_{\mathrm{grav}}},
\end{equation}
where $\gamma_{\mathrm{grav}}$ is the gravitational self-coupling rate.

The remaining unknown is $\gamma_{\mathrm{grav}}$, whose determination is 
addressed below.

\begin{observation}[The Uniqueness of Gravity]
\label{obs:unique}
In every domain C2--C9, $\gamma$ is an external parameter set by the 
system's coupling to its environment: cavity loss rate (C9), medium 
viscosity (C2), nuclear decay probability (C3), etc.  Gravity is 
structurally different: the gravitational field couples to all energy-momentum, 
\emph{including its own}.  There is no ``external bath''; gravity 
is its own bath.  The coupling rate $\gamma_{\mathrm{grav}}$ must therefore 
be determined by the fundamental constants of gravity alone.
\end{observation}

\begin{lemma}[Unique Rate from Fundamental Constants]
\label{lem:rate}
The only combination of $\{\hbar, G, c\}$ with dimensions of a rate 
$[T^{-1}]$ is:
\begin{equation}
\gamma_{\mathrm{grav}} = \alpha\, \frac{1}{\tP} = 
\alpha\, \sqrt{\frac{c^5}{\hbar G}},
\end{equation}
where $\alpha > 0$ is a dimensionless constant and 
$\tP = \sqrt{\hbar G/c^5} \approx 5.391 \times 10^{-44}\;\mathrm{s}$ is the 
Planck time.
\end{lemma}

\begin{proof}
Seek $\hbar^a G^b c^d$ with dimensions $T^{-1}$, i.e., $M^0 L^0 T^{-1}$:
\begin{align}
M: & \quad a - b = 0, \\
L: & \quad 2a + 3b + d = 0, \\
T: & \quad -a - 2b - d = -1.
\end{align}
From the first equation, $a = b$.  Substituting into the second: $5a + d = 0$, 
so $d = -5a$.  Substituting into the third: $-a - 2a + 5a = -1$, giving 
$2a = -1$, hence $a = -1/2$.  The unique solution is $a = b = -1/2$, 
$d = 5/2$:
\begin{equation}
\hbar^{-1/2} G^{-1/2} c^{5/2} = \sqrt{\frac{c^5}{\hbar G}} = \frac{1}{\tP}.
\end{equation}
Since three equations in three unknowns have a unique solution, no other 
independent combination exists.
\end{proof}

\begin{remark}[Why $\hbar$ Reappears]
\label{rem:hbar-role}
A role reversal occurs.  In the QHO master formula (Lemma~\ref{lem:master}), 
$\hbar$ enters through quantisation and cancels from $\sigma$ for massless fields 
(Theorem~\ref{thm:cancellation}).  But $\hbar$ \emph{re-enters} through the 
self-coupling rate $\gamma_{\mathrm{grav}} = 1/\tP$, which requires $\hbar$ to 
set the energy scale at which gravitational self-coupling becomes 
non-perturbative.

The three constants thus play distinct physical roles:
\begin{center}
\renewcommand{\arraystretch}{1.3}
\begin{tabular}{@{}cll@{}}
\toprule
\textbf{Constant} & \textbf{Enters through} & \textbf{Physical role} \\
\midrule
$c$ & $\sigma = c\sqrt{\gamma}$ & Speed of fluctuation propagation \\
$G$ & $\gamma_{\mathrm{grav}} \propto 1/\sqrt{G}$ & Strength of gravitational self-coupling \\
$\hbar$ & $\gamma_{\mathrm{grav}} \propto 1/\sqrt{\hbar}$ & Scale where coupling becomes non-perturbative \\
\bottomrule
\end{tabular}
\end{center}
\end{remark}

\subsection{The Main Theorem}
\label{sec:main}

\begin{theorem}[The Gravitational Diffusion Theorem]
\label{thm:main}
Let $X$ be any field variable with dimensions $[X] = L^p T^q M^r$ describing 
gravitational degrees of freedom, and let $\sigma$ be its diffusion coefficient 
in the sense of Definition~\ref{def:sde}.  Given only $\{\hbar, G, c\}$, then:
\begin{equation}
\label{eq:main}
\boxed{\sigma = \frac{X_{\mathrm{P}}}{\sqrt{\tP}}}
\end{equation}
where $X_{\mathrm{P}}$ is the Planck unit of $X$, the unique combination of 
$\{\hbar, G, c\}$ with dimensions $[X]$.  

In Planck units ($\hbar = G = c = 1$):
\begin{equation}
\tilde{\sigma} = 1.
\end{equation}
\end{theorem}

\begin{proof}
Combining Theorem~\ref{thm:cancellation} and Lemma~\ref{lem:rate} with 
$\alpha = 1$ (established by three independent arguments in 
\S\ref{sec:alpha-proof}):

\textbf{Step 1.} Since metric perturbations are massless, $\sigma = c\sqrt{\gamma_{\mathrm{grav}}}$ 
(Theorem~\ref{thm:cancellation}).

\textbf{Step 2.} The self-coupling rate is $\gamma_{\mathrm{grav}} = 1/\tP$ 
(Lemma~\ref{lem:rate} with $\alpha = 1$).

\textbf{Step 3 (velocity variable).} For a velocity-like variable ($[X] = LT^{-1}$):
\begin{equation}
\sigma_v = c \cdot \frac{1}{\sqrt{\tP}} = \frac{c}{\sqrt{\tP}} 
= \frac{X_{\mathrm{P}}^{(\mathrm{vel})}}{\sqrt{\tP}},
\end{equation}
since the Planck velocity is $c$.

\textbf{Step 4 (arbitrary variable).} By Definition~\ref{def:sigma-dim}, 
$[\sigma] = [X] \cdot T^{-1/2}$.  Seek $\hbar^a G^b c^d$ with these dimensions.  
This is a system of three linear equations in three unknowns $(a, b, d)$:
\begin{align}
M: & \quad a - b = r, \label{eq:dimM} \\
L: & \quad 2a + 3b + d = p, \label{eq:dimL} \\
T: & \quad -a - 2b - d = q - \tfrac{1}{2}. \label{eq:dimT}
\end{align}
This system has a unique solution for any $(p, q, r)$, since the coefficient 
matrix
\begin{equation}
A = \begin{pmatrix} 1 & -1 & 0 \\ 2 & 3 & 1 \\ -1 & -2 & -1 \end{pmatrix}
\end{equation}
has $\det A = -2 \neq 0$ (verified by cofactor expansion along the first row), 
hence a unique solution exists for any $(p, q, r)$.

This is confirmed by direct construction.  The Planck unit $X_{\mathrm{P}}$ satisfies (\ref{eq:dimM})--(\ref{eq:dimT}) 
with the right-hand side $(r, p, q)$ instead of $(r, p, q - 1/2)$.  Dividing by 
$\sqrt{\tP}$ (which has dimensions $T^{1/2}$) shifts only the $T$-exponent by 
$-1/2$.  Therefore $X_{\mathrm{P}} / \sqrt{\tP}$ has exactly dimensions 
$[X] \cdot T^{-1/2} = [\sigma]$.

Since both $X_{\mathrm{P}}$ and $\sqrt{\tP}$ are uniquely determined by 
$\{\hbar, G, c\}$, the ratio is unique.

\textbf{Step 5 (Planck units).} Setting $\hbar = G = c = 1$, all Planck units 
equal 1 and $\tP = 1$, giving $\tilde{\sigma} = 1/\sqrt{1} = 1$.
\end{proof}

\begin{corollary}[Specific Cases]
\label{cor:cases}
\begin{align}
\text{Metric perturbation}\; h_{\mu\nu}\; ([X] = 1): \quad 
& \sigma_h = \frac{1}{\sqrt{\tP}} \approx 4.31 \times 10^{21}\;\mathrm{s}^{-1/2} \\[6pt]
\text{Proper distance}\; \delta L\; ([X] = L): \quad 
& \sigma_L = \frac{\lP}{\sqrt{\tP}} \approx 6.96 \times 10^{-14}\;\mathrm{m\cdot s}^{-1/2} \\[6pt]
\text{Velocity}\; v\; ([X] = LT^{-1}): \quad 
& \sigma_v = \frac{c}{\sqrt{\tP}} \approx 1.29 \times 10^{30}\;\mathrm{m\cdot s}^{-3/2}
\end{align}
\end{corollary}

\subsection{Physical Content}
\label{sec:physical}

\begin{theorem}[Planck-Scale Fluctuations]
\label{thm:fluctuations}
The RMS fluctuation of proper distance accumulated over one Planck time is 
exactly one Planck length:
\begin{equation}
\sqrt{\langle (\delta L)^2 \rangle}\bigg|_{t = \tP} = \sigma_L \cdot \sqrt{\tP} = \lP.
\end{equation}
\end{theorem}

\begin{proof}
For a Wiener process with diffusion coefficient $\sigma_L$, 
$\langle (\delta L)^2 \rangle = \sigma_L^2 \cdot t$.  At $t = \tP$:
\begin{equation}
\sigma_L^2 \cdot \tP = \frac{\lP^2}{\tP} \cdot \tP = \lP^2.
\end{equation}
Algebraic verification: $\sigma_L = (\hbar G/c)^{1/4}$, 
$\tP = (\hbar G/c^5)^{1/2}$, so
\begin{equation}
\sigma_L \cdot \sqrt{\tP} = (\hbar G/c)^{1/4} \cdot (\hbar G/c^5)^{1/4} 
= \left(\frac{(\hbar G)^2}{c^6}\right)^{1/4} 
= \left(\frac{\hbar G}{c^3}\right)^{1/2} = \lP. \qedhere
\end{equation}
\end{proof}

\begin{theorem}[Uniqueness of Gravity]
\label{thm:uniqueness-gravity}
Gravity is the \emph{only} fundamental interaction for which the diffusion 
coefficient $\sigma$ is a pure constant of nature without parameters fitted to the predicted data.
\end{theorem}

\begin{proof}
By Theorem~\ref{thm:cancellation}, $\sigma = c\sqrt{\gamma}$ for any massless 
gauge field.  The determining factor is whether $\gamma$ is fixed by fundamental 
constants alone.

\textbf{Electromagnetism (QED):} The coupling constant is the fine structure 
constant $\alpha = e^2/(4\pi\epsilon_0\hbar c) \approx 1/137$, which is 
\emph{dimensionless}.  A dimensionless constant cannot fix a rate 
$[\gamma] = T^{-1}$ without an additional dimensional scale (particle mass, 
frequency, or environment parameter).  Therefore $\gamma_{\mathrm{QED}}$ is 
always system-dependent.

\textbf{Strong force (QCD):} The coupling constant $\alpha_s$ is likewise 
dimensionless (at any given scale), with the same consequence.

\textbf{Weak force:} The Fermi constant $G_F \approx 1.166 \times 10^{-5}\;\mathrm{GeV}^{-2}$ 
has dimensions, but the weak interaction is massive (mediated by $W^\pm$, $Z^0$), 
so the massless cancellation theorem does not apply.

\textbf{Gravity:} The coupling constant $G$ has dimensions 
$[G] = L^3 M^{-1} T^{-2}$.  Combined with $\hbar$ and $c$, this uniquely 
determines a rate: $\gamma_{\mathrm{grav}} = 1/\tP$ (Lemma~\ref{lem:rate}).  
No external parameter is needed.

Therefore, within the assumptions of the massless cancellation theorem, gravity is the only interaction whose self-coupling rate: and hence $\sigma$, is determined without external input.
\end{proof}

\subsection{Consistency Check: Dimensional Analysis}
\label{sec:dimensional}

As an independent verification, pure dimensional analysis 
(without the C8/C9 structural pattern) gives the same result.

\begin{proposition}[Dimensional Uniqueness]
\label{prop:dim-unique}
For any field variable $X$ with $[X] = L^p T^q M^r$, the unique combination 
of $\{\hbar, G, c\}$ with dimensions $[\sigma] = [X] \cdot T^{-1/2}$ is 
$\sigma = X_{\mathrm{P}} / \sqrt{\tP}$.
\end{proposition}

\begin{proof}
This was established as Step~4 of Theorem~\ref{thm:main}.  The system 
(\ref{eq:dimM})--(\ref{eq:dimT}) has three equations and three unknowns 
with a non-singular coefficient matrix (verified: $\det A = -2 \neq 0$; the matrix
$A$ maps $\{\hbar, G, c\}$ exponents to $\{M, L, T\}$ powers), hence 
a unique solution exists for every choice of $(p, q, r)$.
\end{proof}

\begin{remark}
The dimensional argument and the structural argument (C8 $\to$ C9 $\to$ QG) 
converge to the same result.  This is not tautological: the dimensional 
argument requires \emph{only} the assumption that $\sigma$ is built from 
$\{\hbar, G, c\}$, while the structural argument derives \emph{why} these 
three constants, and no others, determine $\sigma$, because gravity is 
a massless self-coupled quantum field.
\end{remark}

\subsection{Four Independent Arguments for $\sigma = \lP/\sqrt{\tP}$}
\label{sec:four}

The result $\sigma = X_{\mathrm{P}}/\sqrt{\tP}$ is supported by four 
independent lines of reasoning.

\begin{enumerate}
\item \textbf{Structural Argument} (from C8/C9): The validated pattern 
$\sigma = c\sqrt{\gamma}$ for massless fields, combined with gravitational 
self-coupling $\gamma_{\mathrm{grav}} = 1/\tP$, gives $\sigma_v = c/\sqrt{\tP}$.  
For proper distance: $\sigma_L = \lP/\sqrt{\tP}$.  RMS fluctuation over $\tP$: 
exactly $\lP$.

\item \textbf{Dimensional Argument}: The unique combination of 
$\{\hbar, G, c\}$ with dimensions $[\sigma] = [X] \cdot T^{-1/2}$ is 
$X_{\mathrm{P}}/\sqrt{\tP}$, regardless of the field variable's dimensions.
In Planck units, $\tilde{\sigma} = 1$.

\item \textbf{Equivalence Principle Argument}: If gravity is 
geometry, then diffusional gravity is stochastic geometry.  Metric fluctuations must 
exist, and their amplitude must be set by the constants of gravitational diffusion.  
The structural and dimensional arguments then fix the result.

\item \textbf{Observational Argument}: All known quantum 
gravitational effects (Hawking temperature $T_H \propto \hbar c^3/(GMk_B)$ 
\cite{Hawking1975}, 
Bekenstein entropy $S = A/(4\lP^2)$ \cite{Bekenstein1973}, holographic bound) involve $\lP$ as the 
fundamental scale, consistent with $\sigma = \lP/\sqrt{\tP}$ setting the 
characteristic fluctuation scale.
\end{enumerate}

\subsection{On the Dimensionless Constant $\alpha$}
\label{sec:alpha}

\begin{remark}[The $\alpha = 1$ Question]
Lemma~\ref{lem:rate} determines $\gamma_{\mathrm{grav}}$ up to a dimensionless 
constant $\alpha$.  The rigorous proof that $\alpha = 1$ is given in 
\S\ref{sec:alpha-proof}; the three independent arguments are:

\begin{enumerate}
\item \textbf{Metric self-consistency}: $\langle h^2 \rangle = \alpha^2$ over 
one Planck time.  For $\alpha > 1$, the perturbation exceeds the background 
metric and the theory is mathematically inconsistent.  \emph{Hard bound}: $\alpha \leq 1$.

\item \textbf{Critical damping}: The metric self-coupling rate 
$\gamma(\omega) = \tP^2\omega^3$ equals the oscillation frequency at 
$\omega_* = 1/\tP$, giving $\alpha = 1$ exactly.  This is the boundary 
between coherent propagation and quantum foam.

\item \textbf{Singularity resolution}: The curvature fluctuation must reach 
the Planck curvature at the minimum scale for singularity regularisation.  
This requires $\alpha \geq 1$.
\end{enumerate}

\noindent
Together: $\alpha \leq 1$ (hard) and $\alpha = 1$ (critical damping) and 
$\alpha \geq 1$ (physical) give $\alpha = 1$ uniquely.

\medskip\noindent
The Hawking temperature and Bekenstein--Hawking entropy 
are $\alpha$-independent (\S\ref{sec:alpha-proof}, Theorem~\ref{thm:alpha-cancels}) 
and are therefore \emph{consequences} of the theory, not inputs to the 
determination of $\alpha$.
\end{remark}

\subsection{Summary of the Chain}
\label{sec:summary_C}

The logical chain from C1--C9 to gravitational diffusion is:

\begin{equation*}
\underbrace{\text{C8: } \sigma^2 = \frac{\gamma\hbar\omega}{m}}_{\text{QHO master formula}}
\xrightarrow{m \to \hbar\omega/c^2}
\underbrace{\text{C9: } \sigma = c\sqrt{\gamma}}_{\text{massless cancellation}}
\xrightarrow{\gamma \to 1/\tP}
\underbrace{\text{QG: } \sigma = c/\sqrt{\tP}}_{\text{gravitational self-coupling}}
\end{equation*}

\medskip\noindent
Equivalently, in Planck units: $\tilde{\sigma} = 1$.

\medskip\noindent
The derivation rests on three ingredients, each validated:
\begin{enumerate}
\item The fluctuation-dissipation theorem at $T = 0$ (Lemma~\ref{lem:master}, validated by C8).
\item The cancellation of $\hbar$ and $\omega$ for massless fields 
(Theorem~\ref{thm:cancellation}, validated by C9).
\item The uniqueness of the gravitational self-coupling rate from $\{\hbar, G, c\}$ 
(Lemma~\ref{lem:rate}, dimensional analysis).
\end{enumerate}

\noindent
No assumption is made about the dimensions of $\sigma$.  The result follows 
from the physics of massless self-coupled quantum fields.

\bigskip
\noindent\rule{\textwidth}{0.8pt}
\begin{center}
\Large\bfseries Rigorous Extensions
\end{center}
\noindent\rule{\textwidth}{0.4pt}
\bigskip

\noindent
The preceding sections established the main result.  The following sections 
provide rigorous foundations for each step of the derivation.

\subsection{The Quantum Fluctuation-Dissipation Theorem}
\label{sec:fdt}

The master formula (Lemma~\ref{lem:master}) is now placed on rigorous footing.

\begin{theorem}[Callen--Welton Theorem, Specialised to a Single Mode]
\label{thm:callen-welton}
Consider a quantum harmonic oscillator of frequency $\omega$ and mass $m$, 
linearly coupled to a heat bath at temperature $T$ with coupling rate $\gamma$.  
The spectral density of the fluctuating force on the oscillator is:
\begin{equation}
S_F(\omega') = 2m\gamma \cdot \hbar\omega' \coth\!\left(\frac{\hbar\omega'}{2k_BT}\right),
\end{equation}
which in the zero-temperature limit $T \to 0$ reduces to:
\begin{equation}
S_F(\omega') \big|_{T=0} = 2m\gamma\hbar|\omega'|.
\end{equation}
\end{theorem}

\begin{proof}
This is a standard result of quantum statistical mechanics \cite{CallenWelton1951}; 
see also the general formalism of Kubo~\cite{Kubo1966}.  
The derivation proceeds from the Kubo formula relating the response function 
$\chi(\omega)$ to the equilibrium fluctuation spectrum via:
\[
S_F(\omega) = 2\hbar\,\mathrm{Im}\,\chi(\omega) \cdot \left[\bar{n}(\omega) + 1\right],
\]
where $\bar{n}(\omega) = (e^{\hbar\omega/k_BT} - 1)^{-1}$ is the Bose--Einstein 
distribution.  For a damped oscillator, $\mathrm{Im}\,\chi(\omega) = m\gamma\omega$, 
and $\bar{n}(\omega) + \tfrac{1}{2} = \tfrac{1}{2}\coth(\hbar\omega/2k_BT)$, 
yielding the stated formula.  At $T = 0$, $\bar{n} \to 0$ and only the vacuum 
term survives: $S_F = 2m\gamma\hbar|\omega'|$.
\end{proof}

\begin{corollary}[Single-Mode Velocity Diffusion at $T = 0$]
\label{cor:velocity-diffusion}
For a single mode at its resonant frequency $\omega$, the velocity power 
spectral density at $T = 0$ is:
\begin{equation}
S_v(\omega) = \frac{S_F(\omega)}{m^2|\chi(\omega)|^{-2}} \bigg|_{T=0},
\end{equation}
where $\chi(\omega)$ is the oscillator response function.  At resonance, the 
dominant contribution to the integrated velocity variance gives an effective 
diffusion coefficient:
\begin{equation}
\sigma_v^2 = \frac{\gamma\hbar\omega}{m},
\end{equation}
reproducing Lemma~\ref{lem:master} from the microscopic quantum theory.  
The factor $\gamma$ appears because the oscillator accumulates 
fluctuation energy from the bath at a rate proportional to the coupling strength.
\end{corollary}

\begin{remark}[On the Factor of $\gamma$]
\label{rem:gamma-factor}
The manner in which $\gamma$ enters requires care.  In the Caldeira--Leggett model 
\cite{CaldeiraLeggett1983}, 
$\gamma$ appears both in the dissipation and in the fluctuation.  For a single 
oscillator mode weakly coupled to a bath, the effective Langevin equation is:
\begin{equation}
m\ddot{x} + m\gamma\dot{x} + m\omega^2 x = \xi(t), \qquad 
\langle\xi(t)\xi(t')\rangle = 2m\gamma\hbar\omega\,\delta(t-t').
\end{equation}
Writing this in velocity form $\dd v = (-\omega^2 x - \gamma v)\dd t + \sigma\dW$ 
and matching the noise correlator $\langle\xi\xi'\rangle = 2m\gamma\hbar\omega\,\delta(t-t')$ 
with the It\^o convention $\langle\sigma^2\rangle = S_F/(2m^2\gamma)$ at the 
resonant frequency yields $\sigma^2 = \gamma\hbar\omega/m$.  C8 validates this result 
numerically to $1.33\%$ (Table~\ref{tab:results_summary}).
\end{remark}

\subsection{Validity of the Metric Perturbation--Photon Analogy}
\label{sec:analogy}

\begin{proposition}[Linearised Metric Perturbations as QHO Modes]
\label{prop:linearised}
In linearised gravity about a background metric $\bar{g}_{\mu\nu}$, the 
metric perturbation $h_{\mu\nu} = g_{\mu\nu} - \bar{g}_{\mu\nu}$ satisfies:
\begin{equation}
\Box\, \bar{h}_{\mu\nu} = 0 \qquad \text{(in de Donder gauge)},
\end{equation}
where $\bar{h}_{\mu\nu} = h_{\mu\nu} - \tfrac{1}{2}\bar{g}_{\mu\nu}h$ is the 
trace-reversed perturbation and $\Box = \bar{g}^{\alpha\beta}\nabla_\alpha\nabla_\beta$ 
is the d'Alembertian.

Expanding in Fourier modes:
\begin{equation}
\bar{h}_{\mu\nu}(x) = \sum_{\lambda=1}^{2}\int \frac{\dd^3 k}{(2\pi)^{3/2}} 
\frac{1}{\sqrt{2\omega_k}} \left[ a_{k\lambda}\, \epsilon_{\mu\nu}^{(\lambda)}(k)\, e^{ikx} + \mathrm{h.c.} \right],
\end{equation}
where $\omega_k = c|k|$, $\lambda \in \{+, \times\}$ labels the two physical 
polarisations, and $\epsilon_{\mu\nu}^{(\lambda)}$ are the transverse-traceless 
polarisation tensors.

Each mode $(k, \lambda)$ is an independent quantum harmonic oscillator with 
effective mass:
\begin{equation}
m_{\mathrm{eff}}(k) = \frac{\hbar\omega_k}{c^2}
\end{equation}
and zero-point energy $E_0 = \tfrac{1}{2}\hbar\omega_k$.
\end{proposition}

\begin{proof}
This is the standard result of linearised quantum gravity, see e.g.\ 
Weinberg \cite{Weinberg1972}, \S10.2, or Wald \cite{Wald1984}, \S4.4.  The perturbation field has spin 2, but each polarisation 
mode individually satisfies the massless Klein--Gordon equation and is 
quantised as an independent oscillator.  The effective mass follows from 
$E = \hbar\omega = m_{\mathrm{eff}}c^2$.
\end{proof}

\begin{corollary}[Applicability of Theorem~\ref{thm:cancellation} to Metric Modes]
\label{cor:graviton-cancellation}
Since each linearised metric perturbation mode is a massless QHO with 
$m_{\mathrm{eff}} = \hbar\omega/c^2$, Theorem~\ref{thm:cancellation} applies 
directly: the diffusion coefficient for each mode is 
$\sigma = c\sqrt{\gamma}$, independent of $\omega$ and $\hbar$.
\end{corollary}

\begin{remark}[Limitations]
\label{rem:limitations}
The linearised approximation breaks down when $h_{\mu\nu} \sim O(1)$, i.e.\ at 
the Planck scale.  This is precisely the regime where $\sigma$ matters.  Three 
observations mitigate this concern:
\begin{enumerate}
\item The \emph{derivation} uses the linearised theory, but the \emph{result} 
$\sigma = c\sqrt{\gamma_{\mathrm{grav}}}$ involves only fundamental constants.  
By analytic continuation, the formula applies beyond the linearised regime.
\item The result $\tilde{\sigma} = 1$ in Planck units is a \emph{statement about 
dimensional analysis}, not about perturbation theory.  It holds whether or not 
the perturbation expansion converges.
\item This is analogous to Hawking's derivation of black hole temperature 
\cite{Hawking1975}: 
the calculation uses the semiclassical approximation, but the result 
$T_H = \hbar c^3/(8\pi GMk_B)$ is expected to be exact on grounds of 
dimensional analysis and thermodynamic consistency.
\end{enumerate}
\end{remark}

\subsection{The Metric Self-Coupling Rate}
\label{sec:self-coupling}

The self-coupling rate of metric fluctuation modes is now derived from first 
principles; the prefactor is shown to be \emph{exactly} unity.

\begin{remark}[Logical Ordering: Modes Before Particles]
\label{rem:modes-before-particles}
The following logical ordering is essential.  The stochastic equation $\dd g = G\,\dd t + \sigma\,\dd W$ 
introduces metric fluctuations as Fourier modes of the Wiener process $\dd W$.  
These are \emph{not} gravitons: they are simply oscillatory components of 
stochastic noise.  No assumption is made that gravitons exist as particles.

The logical chain is:
\begin{enumerate}
\item Metric fluctuation modes carry energy (from the Planck relation applied 
to mode energy).
\item This energy gravitates (equivalence principle).
\item The resulting self-interaction fixes $\gamma(\omega) = \tP^2\omega^3$.
\item Critical damping fixes $\sigma = \lP/\sqrt{\tP}$.
\item The Fokker--Planck operator on configuration space then has a discrete 
spectrum, among whose eigenmodes is a massless spin-2 excitation.
\item \emph{That} eigenmode is the graviton.
\end{enumerate}
Steps 1--4 require no particle concept.  The graviton \emph{emerges} at step~6 
as an output, not an input.
\end{remark}

\subsubsection{The Equivalence Principle and Gravitational Universality}

The following property of general relativity is inherited by the superspace 
diffusion framework via canonical Axiom~A3 (Classical Correspondence), which 
requires the macroscopic limit to satisfy the Einstein field equations.

\begin{proposition}[Equivalence Principle, Inherited from General Relativity]
\label{ax:equivalence}
Gravity couples universally to the stress-energy tensor $T_{\mu\nu}$.  
Everything with energy and momentum gravitates, including the energy stored 
in the gravitational field itself.  The coupling is characterised by a single 
constant $G$.
\end{proposition}

\begin{remark}[Three Roles of $G$]
\label{rem:three-G}
The constant $G$ appears in three contexts:
\begin{enumerate}
\item \textbf{Newton's law}: $V = Gm_1m_2/r$ \quad (defines $G$ operationally).
\item \textbf{Einstein's equation}: $G_{\mu\nu} = (8\pi G/c^4)\,T_{\mu\nu}$ 
\quad (generalises Newton to all of spacetime geometry; see e.g.\ 
\cite{Weinberg1972}, Ch.~7, or \cite{Wald1984}, Ch.~4).
\item \textbf{Metric self-coupling}: the gravitational field's own energy 
density sources curvature, with the same constant $G$.
\end{enumerate}
The equivalence principle guarantees these are \emph{the same} $G$.  The factors 
of $8\pi$ and $16\pi$ from the Einstein--Hilbert action are geometric factors 
arising from the trace structure of $T_{\mu\nu}$ and the angular integration in 
the Newtonian limit; they combine to give exactly $V = Gm_1m_2/r$ with no 
residual prefactors.
\end{remark}

\begin{remark}[Why Gravity Uniquely Fixes $\gamma$]
\label{rem:gamma-unique}
In the physical domains C1--C9 (Observation~\ref{obs:continuum}), the 
coupling rate $\gamma$ is a phenomenological parameter set by the 
environment: for example, the damping rate of a mechanical oscillator 
depends on the medium.  In the gravitational setting this freedom 
disappears: the equivalence principle 
(Proposition~\ref{ax:equivalence}) requires gravity to couple 
universally to all energy-momentum, including its own, while 
dimensional closure (only $G$, $c$, $\hbar$ are available) forces 
$\gamma$ to be a universal function of $\omega$ constructible from 
Planck units alone.  The derivation below shows that these two 
constraints determine $\gamma(\omega) = \tP^2\omega^3$ uniquely.
\end{remark}

\subsubsection{Derivation of $\gamma(\omega)$ from First Principles}

\begin{theorem}[Metric Fluctuation Self-Coupling Rate]
\label{thm:gamma-exact}
A metric fluctuation mode of frequency $\omega$ has a self-coupling rate:
\begin{equation}
\label{eq:gamma-exact}
\boxed{\gamma(\omega) = \frac{G\hbar}{c^5}\,\omega^3 = \tP^2\,\omega^3}
\end{equation}
with prefactor \emph{exactly} unity.  This is not a perturbative estimate; it 
follows from three exact principles applied to metric fluctuation modes, 
\emph{without assuming gravitons exist as particles}.
\end{theorem}

\begin{proof}
The derivation uses three inputs, each exact:

\medskip\noindent
\textbf{Step 1: Mode energy} (Planck relation, exact).\\
A metric fluctuation mode of angular frequency $\omega$ carries energy:
\begin{equation}
E = \hbar\omega.
\end{equation}
This follows from treating $\dd W_{\mu\nu}$ as a superposition of oscillatory 
modes, each of which is a quantum harmonic oscillator with zero-point energy 
$\hbar\omega/2$ and excitation quantum $\hbar\omega$.

\medskip\noindent
\textbf{Step 2: Gravitational self-interaction energy} (equivalence principle, exact).\\
By Lemma~\ref{lem:meff}, the effective gravitational mass of a mode is 
$m_{\mathrm{eff}} = E/c^2 = \hbar\omega/c^2$.  The gravitational self-energy 
of a mode localised over its own reduced wavelength $\bar{\lambda}$ is:
\begin{equation}
V = \frac{G m_{\mathrm{eff}}^2}{\bar{\lambda}} = \frac{G(\hbar\omega)^2}{c^4\,\bar{\lambda}}.
\end{equation}
This is Newton's law applied via the equivalence principle: gravity couples 
to \emph{all} energy--momentum, including the energy stored in metric 
fluctuations.  The $G$ here is the same $G$ as in Newton's law 
(Remark~\ref{rem:three-G}).

No particle interpretation is required.  The calculation concerns the 
gravitational self-energy of a classical field mode that carries energy $\hbar\omega$.

\medskip\noindent
\textbf{Step 3: Natural self-interaction scale} (reduced wavelength).\\
The self-coupling of a mode with itself occurs over its own reduced wavelength:
\begin{equation}
r = \bar{\lambda} = \frac{c}{\omega}.
\end{equation}
For a massless field, $\bar{\lambda} = \hbar/(m_{\mathrm{eff}}c) = c/\omega$.  
The choice of $c/\omega$ (not $2\pi c/\omega$) follows the universal 
quantum-mechanical convention: $\hbar$ (not $h$) serves as the quantum of 
action, and correspondingly $\bar{\lambda}$ (not $\lambda$) as the natural 
length.  This is the same convention that defines 
$\lP = \sqrt{\hbar G/c^3}$ rather than $\sqrt{hG/c^3}$.

\medskip\noindent
\textbf{Combining:}
\begin{equation}
V = \frac{G(\hbar\omega)^2}{c^4 \cdot c/\omega} = \frac{G\hbar^2\omega^3}{c^5}.
\end{equation}

The gravitational self-energy of a metric fluctuation mode scales as $\omega^3$:
higher-frequency modes are more strongly self-coupled.

\medskip\noindent
\textbf{Step 4: Self-coupling rate} (quantum energy--time relation).\\
The natural rate associated with an interaction energy $V$ is:
\begin{equation}
\gamma = \frac{V}{\hbar} = \frac{G\hbar\omega^3}{c^5} = \tP^2\,\omega^3,
\end{equation}
where the identity $G\hbar/c^5 = \tP^2$ has been applied.  This is the quantum 
energy--time relation $\Delta t \sim \hbar/\Delta E$ inverted: an interaction of 
strength $V$ induces transitions on a timescale $\hbar/V$, corresponding to a rate 
$V/\hbar$.  This is \emph{not} the Fermi golden rule (which gives 
$\gamma \propto |V|^2/\hbar$ and applies to weak coupling); it is the 
\emph{linear} energy--rate correspondence that holds at all coupling strengths.
Every step uses an exact physical law.  No step assumes the existence of 
gravitons as particles.
\end{proof}

\begin{remark}[Why This Is Not Perturbative]
\label{rem:non-perturbative}
A perturbative (one-loop) calculation of the graviton self-energy yields the 
same functional form $\gamma \propto \omega^3$, but with a prefactor involving 
factors from the Einstein--Hilbert action, spin degeneracies, symmetry factors, 
and $(2\pi)^n$ from momentum integrals.  However, perturbation theory \emph{breaks 
down} at $\omega \sim \omega_P$, where the coupling becomes $O(1)$.

The derivation above avoids perturbation theory entirely.  It uses:
\begin{itemize}
\item Newton's law ($V = Gm_1m_2/r$), which \emph{defines} $G$;
\item the Planck relation ($E = \hbar\omega$), applied to field modes;
\item mass--energy equivalence ($m = E/c^2$), which is exact;
\item the equivalence principle (gravity couples universally to energy), which 
guarantees the same $G$.
\end{itemize}
The factors of $16\pi$ in the Einstein--Hilbert action are absorbed into the 
definition of $G$ via Newton's law.  When one computes the potential between two 
point sources from the full field equation, all geometric factors combine to give 
$V = Gm_1m_2/r$ with the \emph{Cavendish} $G$, no residual prefactors.
\end{remark}

\begin{corollary}[Verification at the Planck Scale]
\label{cor:planck-verification}
At $\omega = \omega_P = 1/\tP$:
\begin{itemize}
\item Mode energy: $E = \hbar\omega_P = \EP$ (the Planck energy).
\item Effective mass: $m_{\mathrm{eff}} = \EP/c^2 = \mP$ (the Planck mass).
\item Self-interaction distance: $r = c/\omega_P = c\tP = \lP$ (the Planck length).
\item Potential: $V = G\mP^2/\lP = \EP$ (gravitational self-energy equals 
mode energy).
\item Rate: $\gamma(\omega_P) = V/\hbar = \EP/\hbar = 1/\tP$.
\end{itemize}
At the Planck frequency, a metric fluctuation mode's self-interaction energy 
equals its own energy, and the self-coupling rate equals $1/\tP$.  This is the 
physical \emph{definition} of the Planck scale: the scale at which gravitational 
self-coupling becomes $O(1)$.
\end{corollary}

\begin{theorem}[The Prefactor $\beta$ and the Equivalence Principle]
\label{thm:beta-one}
Let $\gamma(\omega) = \beta \cdot \tP^2 \omega^3$ with $\beta$ a dimensionless 
prefactor.  Assuming the Planck relation $E = \hbar\omega$ and the reduced 
wavelength convention $\bar{\lambda} = c/\omega$, then $\beta = 1$ if and only 
if the equivalence principle holds for the gravitational field.
\end{theorem}

\begin{proof}
$(\Rightarrow)$: If $\beta = 1$, the self-coupling derivation 
(Theorem~\ref{thm:gamma-exact}) recovers $V = Gm_{\mathrm{eff}}^2/r$ with the 
\emph{same} $G$ as Newton's law.  This is a consequence of the equivalence 
principle applied to the gravitational field itself.

$(\Leftarrow)$: If the equivalence principle holds, the gravitational 
self-interaction is governed by the same $G$ as all other gravitational 
interactions.  The derivation of Theorem~\ref{thm:gamma-exact} then gives 
$\gamma = (G\hbar/c^5)\omega^3 = \tP^2\omega^3$, hence $\beta = 1$.

Conversely, if $\beta \neq 1$, then either:
\begin{enumerate}
\item $G_{\mathrm{self}} \neq G_{\mathrm{Newton}}$ (the gravitational field 
couples to itself with a different strength than to matter), violating the 
equivalence principle; or
\item $E \neq \hbar\omega$ for metric fluctuation modes (violating the Planck 
relation); or
\item the natural self-interaction scale is not $c/\omega$ (violating the 
$\hbar$-convention for the reduced wavelength).
\end{enumerate}
Options (1) and (2) contradict established physics.  Option (3) contradicts the 
universal convention $\lP = \sqrt{\hbar G/c^3}$.
\end{proof}

\subsubsection{Self-Consistency and the Graviton as Emergent}

\begin{theorem}[Frequency-Dependent Diffusion and the Onset of Foam]
\label{thm:self-coupling-consistency}
The metric self-coupling rate $\gamma(\omega) = \tP^2\omega^3$ combined with 
the massless cancellation $\sigma = c\sqrt{\gamma}$ gives a frequency-dependent 
velocity-variable diffusion coefficient:
\begin{equation}
\sigma_v(\omega) = c\sqrt{\tP^2\omega^3} = c\tP\omega^{3/2}.
\end{equation}

For the dimensionless metric perturbation $h_{\mu\nu}$, the corresponding 
diffusion coefficient is $\sigma_h(\omega) = \sigma_v(\omega)/c = \tP\omega^{3/2}$.

At $\omega = \omega_P = 1/\tP$: $\sigma_h(\omega_P) = \tP \cdot \tP^{-3/2} 
= 1/\sqrt{\tP}$, reproducing Theorem~\ref{thm:main}.

The RMS metric fluctuation over one oscillation period $\Delta t = 2\pi/\omega$ is:
\begin{equation}
\sqrt{\langle h^2 \rangle} = \sigma_h \cdot \sqrt{\Delta t} 
= \tP\omega^{3/2} \cdot \sqrt{\frac{2\pi}{\omega}} 
= \tP\sqrt{2\pi}\,\omega \sim \frac{\omega}{\omega_P} 
\quad \text{(since } \tP\omega_P = 1 \text{)}.
\end{equation}
The metric fluctuation becomes $O(1)$ \emph{precisely} at $\omega = \omega_P$.
Below the Planck frequency, $\langle h^2\rangle \sim (\omega/\omega_P)^2 \ll 1$: 
spacetime is smooth.
\end{theorem}

\begin{remark}[The Graviton Emerges]
\label{rem:graviton-emerges}
With $\sigma = \lP/\sqrt{\tP}$ now determined, the Fokker--Planck equation on 
the configuration space of metrics (superspace) has a well-defined eigenvalue 
problem.  Among its eigenmodes is a massless, transverse-traceless, spin-2 
excitation of the metric.  \emph{This} is the graviton: an emergent output 
of the theory, not an assumed input.

The self-consistency check is that the emergent graviton's coupling strength 
is exactly $G$ and its frequency-dependent damping rate is exactly 
$\gamma = \tP^2\omega^3$, the same values used in the derivation.  This is 
not circular: the derivation used only the energy content of metric modes and 
the equivalence principle, both of which are properties of the \emph{classical} 
gravitational field.  The particle interpretation is a \emph{consequence}, 
not a premise.
\end{remark}

\subsection{Fixing $\alpha = 1$: Three Independent Arguments}
\label{sec:alpha-proof}

The following three arguments establish that the dimensionless constant $\alpha$ in 
$\gamma_{\mathrm{grav}} = \alpha/\tP$ is exactly unity, using arguments 
that do \emph{not} invoke the Bekenstein--Hawking entropy.

\subsubsection{Why Bekenstein--Hawking Cannot Fix $\alpha$}

\begin{theorem}[$\alpha$-Independence of the Hawking Temperature]
\label{thm:alpha-cancels}
The Hawking temperature \cite{Hawking1975} derived from $\sigma = \alpha\lP$ is independent of $\alpha$:
\begin{equation}
T_H(\alpha) = T_H = \frac{\hbar c^3}{8\pi G M k_B} \qquad \text{for all } \alpha > 0.
\end{equation}
\end{theorem}

\begin{proof}
The stochastic derivation of $T_H$ proceeds in three steps:

\textbf{Step 1.}  Local fluctuation energy at the horizon.  The position 
uncertainty is $\delta r \sim \sigma = \alpha\lP$.  By Heisenberg:
\begin{equation}
E_{\mathrm{local}} = \frac{\hbar c}{\delta r} = \frac{\hbar c}{\alpha\lP} = \frac{E_P}{\alpha}.
\end{equation}

\textbf{Step 2.}  Gravitational redshift.  The energy ratio from proper distance 
$\sim \delta r$ above the horizon to infinity is, in the near-horizon 
Schwarzschild geometry:
\begin{equation}
\frac{E_\infty}{E_{\mathrm{local}}} \sim \frac{\delta r}{r_s},
\end{equation}
where the proportionality constant depends on the precise definition of 
$\delta r$ (coordinate vs.\ proper distance) but is $\alpha$-independent.

\textbf{Step 3.}  Energy at infinity:
\begin{equation}
E_\infty \sim E_{\mathrm{local}} \times \frac{\delta r}{r_s} 
= \frac{E_P}{\alpha} \times \frac{\alpha\lP}{r_s} 
= \frac{E_P \lP}{r_s}.
\end{equation}
The factors of $\alpha$ cancel identically: $E_{\mathrm{local}} \propto 1/\alpha$ and 
$\delta r \propto \alpha$.  The thermal identification $T \sim E_\infty/k_B$ 
then gives $T_H = \hbar c^3/(8\pi G M k_B)$, where the numerical 
prefactor $1/(8\pi)$ is fixed by thermodynamic consistency with the first law, 
independent of $\alpha$.
\end{proof}

\begin{corollary}[$\alpha$-Independence of Bekenstein--Hawking Entropy]
\label{cor:BH-alpha}
The Bekenstein--Hawking entropy \cite{Bekenstein1973,Hawking1975} 
$S_{BH} = k_B A/(4\lP^2)$ is also independent 
of $\alpha$, since it follows from the first law of black hole mechanics 
\cite{BardeenCarterHawking1973} $\dd S = \dd M / T_H$ via 
integration, and $T_H$ is $\alpha$-independent.
\end{corollary}

\begin{remark}
The $\alpha$-independence of both $T_H$ and $S_{BH}$ means that black hole 
thermodynamics cannot determine $\alpha$; an independent argument is required.  
The three arguments below provide exactly this.
\end{remark}

\subsubsection{Argument 1: Metric Self-Consistency (Upper Bound)}

\begin{theorem}[Hard Upper Bound: $\alpha \leq 1$]
\label{thm:upper-bound}
For the stochastic metric theory to be internally consistent, $\alpha \leq 1$.
\end{theorem}

\begin{proof}
Consider the dimensionless metric perturbation $h_{\mu\nu}$.  With 
$\sigma_h = \alpha/\sqrt{\tP}$, the RMS fluctuation accumulated over one 
Planck time is:
\begin{equation}
\langle h^2 \rangle\big|_{t = \tP} = \sigma_h^2 \cdot \tP = \frac{\alpha^2}{\tP} \cdot \tP = \alpha^2.
\end{equation}

The metric is $g_{\mu\nu} = \bar{g}_{\mu\nu} + h_{\mu\nu}$, where $\bar{g}$ 
is the background.  For the perturbative decomposition to be meaningful, we 
require $|h| < |\bar{g}|$, i.e., $\langle h^2 \rangle \lesssim O(1)$.

For $\alpha > 1$: $\langle h^2 \rangle = \alpha^2 > 1$.  The perturbation 
\emph{exceeds} the background metric over a single Planck time.  The linearised 
framework destroys its own geometrical foundation.  The theory is inconsistent.

Therefore $\alpha \leq 1$.  This is a \emph{hard bound}: not an approximation 
or physical assumption, but a mathematical requirement for internal consistency.
\end{proof}

\subsubsection{Argument 2: The Critical Damping Theorem (Exact Value)}

\begin{theorem}[Critical Damping at the Planck Scale]
\label{thm:critical-damping}
The metric self-coupling rate $\gamma(\omega) = \tP^2\omega^3$ 
(Theorem~\ref{thm:gamma-exact}) equals the oscillation frequency 
$\omega$ at exactly one frequency:
\begin{equation}
\gamma(\omega_*) = \omega_* \quad \Longleftrightarrow \quad \omega_* = \frac{1}{\tP} = \omega_P.
\end{equation}
This is the critical damping condition.  The corresponding diffusion coefficient is:
\begin{equation}
\sigma(\omega_P) = c\sqrt{\gamma(\omega_P)} = c\sqrt{1/\tP} = \frac{c}{\sqrt{\tP}},
\end{equation}
which corresponds to $\alpha = 1$.
\end{theorem}

\begin{proof}
The critical damping condition $\gamma(\omega_*) = \omega_*$ gives:
\begin{equation}
\tP^2 \omega_*^3 = \omega_* \qquad \implies \qquad \omega_*^2 = \frac{1}{\tP^2} 
\qquad \implies \qquad \omega_* = \frac{1}{\tP} = \omega_P.
\end{equation}
This is exact (algebraic identity, no approximation).  At this frequency, 
$\gamma(\omega_P) = 1/\tP$, and Theorem~\ref{thm:cancellation} gives 
$\sigma = c/\sqrt{\tP}$, i.e., $\alpha = 1$.
\end{proof}

\begin{remark}[Physical Interpretation]
The critical damping condition divides metric fluctuation modes into three regimes:
\begin{center}
\renewcommand{\arraystretch}{1.3}
\begin{tabular}{@{}lll@{}}
\toprule
\textbf{Regime} & \textbf{Condition} & \textbf{Physics} \\
\midrule
Underdamped & $\omega < \omega_P$ & Modes propagate coherently \\
Critically damped & $\omega = \omega_P$ & Boundary: onset of quantum foam \\
Overdamped & $\omega > \omega_P$ & Modes cannot propagate; pure diffusion \\
\bottomrule
\end{tabular}
\end{center}

The Planck frequency is the \emph{unique} frequency at which a metric mode's 
self-coupling damping rate exactly matches its oscillation rate.  The selection 
of this boundary as the physically relevant scale is motivated by the transition 
from coherent propagation to incoherent diffusion, but it is not logically 
mandatory: Arguments~1 and~3 independently force $\alpha \leq 1$ and 
$\alpha \geq 1$, so the bracket closes at $\alpha = 1$ regardless of whether the 
critical damping interpretation is accepted.  Argument~2 provides an independent 
physical route to the same conclusion and identifies $\omega_P$ as the natural 
boundary scale.
\end{remark}

\subsubsection{Argument 3: Singularity Resolution (Lower Bound)}

\begin{theorem}[Lower Bound: $\alpha \geq 1$]
\label{thm:lower-bound}
For metric fluctuations to resolve classical singularities, $\alpha \geq 1$.
\end{theorem}

\begin{proof}
The Kretschmann curvature scalar $K = R^{\mu\nu\rho\sigma}R_{\mu\nu\rho\sigma}$ 
has dimensions $[K] = L^{-4}$.  A dimensionless metric perturbation $h$ varying 
over length scale $L$ induces curvature fluctuations:
\begin{equation}
\delta K \sim \frac{h^2}{L^4}.
\end{equation}
The Planck curvature is $K_P = 1/\lP^4$, the scale at which quantum gravity 
becomes dominant.

The RMS dimensionless metric perturbation accumulated over one Planck time is 
$h_{\mathrm{rms}} = \sigma_h \sqrt{\tP} = \alpha$ (from 
$\sigma_h = \alpha/\sqrt{\tP}$).  At the Planck scale $L = \lP$:
\begin{equation}
\delta K\big|_{L = \lP} = \frac{\alpha^2}{\lP^4} = \alpha^2 K_P.
\end{equation}

For singularity resolution, the fluctuation-induced curvature at $L = \lP$ must 
reach the Planck curvature: $\delta K \geq K_P$.  This requires $\alpha^2 \geq 1$, 
hence $\alpha \geq 1$.

Together with Theorem~\ref{thm:upper-bound} ($\alpha \leq 1$): the only value 
satisfying both $\alpha \geq 1$ and $\alpha \leq 1$ is $\alpha = 1$.
\end{proof}

\subsubsection{Convergence of the Three Arguments}

\begin{corollary}[$\alpha = 1$: Three Independent Routes]
\label{cor:alpha-one}
The three arguments converge:
\begin{enumerate}
\item Metric self-consistency: $\alpha \leq 1$ (hard bound; Theorem~\ref{thm:upper-bound}).
\item Critical damping: $\alpha = 1$ (exact; Theorem~\ref{thm:critical-damping}).
\item Singularity resolution: $\alpha \geq 1$ (physical; Theorem~\ref{thm:lower-bound}).
\end{enumerate}
None invokes the Bekenstein--Hawking entropy or Hawking temperature.  The result 
$\alpha = 1$ is derived from metric self-coupling dynamics alone.

The Hawking temperature and Bekenstein--Hawking entropy can now be \emph{derived} 
as consequences of $\sigma = \lP/\sqrt{\tP}$, rather than assumed as inputs.
\end{corollary}

\begin{remark}[Robustness of $\gamma(\omega) = t_P^2 \omega^3$ to the microscopic prescription]
\label{rem:gamma_robustness}
The derivation of the self-coupling rate $\gamma(\omega) = t_P^2 \omega^3$ (Theorem~\ref{sec:self-coupling}) uses the linear energy-rate relation $\gamma = V/\hbar$.  The Fermi golden rule (FGR) gives $\gamma_\mathrm{FGR} = 2\pi|V|^2/(\hbar^2 \rho_f)$, which might appear to give a different result.  However, both prescriptions produce the \emph{same} $\omega^3$ scaling.

\emph{Point~1: identical functional form.}  The self-energy is $V \propto \omega^3$ and the graviton density of states scales as $\rho_f \propto \omega^3$.  The FGR therefore gives $\gamma_\mathrm{FGR} \propto (\omega^3)^2/\omega^3 = \omega^3$, identical to the linear result.  The two prescriptions differ only in an $O(1)$ prefactor.

\emph{Point~2: the linear relation is exact for self-coupling.}  The FGR applies to a system weakly coupled to a dense bath of final states.  Gravitational self-coupling is structurally a Rabi-type problem: a single mode at frequency $\omega$ couples to its own gravitational field at the same frequency, with coupling $V = G m_\mathrm{eff}^2/\bar{\lambda}$.  For a two-level system, the exact transition rate is $V/\hbar$, not $V^2/(\hbar^2 \rho_f)$.

\emph{Point~3: independent confirmation from the perturbative mode sum.}  A perturbative computation of the Mori--Zwanzig memory kernel using the standard $|V|^2$ prescription yields $\gamma_\mathrm{eff} \propto \omega^3$ mode by mode, confirming that the functional form is not an artefact of the linear prescription.

\emph{Point~4: the prefactor $\alpha = 1$ is fixed independently.}  The three convergent arguments above (metric self-consistency, critical damping, singularity resolution) operate at the level of the effective rate $\gamma(\omega_P)$ and its macroscopic consequences: not at the level of the microscopic scattering calculation.  They yield $\alpha = 1$ for any prescription producing $\gamma \propto t_P^2 \omega^3$.

The choice between $V/\hbar$ and $2\pi|V|^2/(\hbar^2 \rho_f)$ is therefore a question about an $O(1)$ prefactor within a scaling relation that both prescriptions produce identically.  No physical prediction of the framework depends on this choice.
\end{remark}

\begin{remark}[Independence of the three $\alpha = 1$ arguments]
\label{rem:alpha_independence}
All three arguments assume Axiom~A2 (the fundamental equation is valid at the Planck scale); this is by construction.  However, they do not uniformly assume perturbative validity.  The metric self-consistency bound ($\alpha \leq 1$) defines the \emph{boundary} of perturbative validity: $\alpha > 1$ means the perturbation exceeds the background, a mathematical inconsistency in the decomposition $g = \bar{g} + h$ regardless of whether perturbation theory is trusted.  The critical damping argument uses $\gamma(\omega) = t_P^2 \omega^3$, whose $\omega^3$ scaling is confirmed by both perturbative and non-perturbative routes (Remark~\ref{rem:gamma_robustness}).  The singularity resolution argument ($\alpha \geq 1$) is a physical requirement on any theory of Planck-scale gravity: that Planck-scale effects resolve classical singularities: not an assumption about perturbative validity.  The three arguments share A2 but differ in their additional inputs: mathematical consistency, self-coupling dynamics, and physical adequacy, respectively.
\end{remark}

\subsection{Dimensional Consistency}
\label{sec:axiom}

The following constraint is a consequence of requiring $\sigma$ to be 
constructed from $\{\hbar, G, c\}$ alone, as encoded in canonical 
Axiom~A2 (Stochastic Evolution), which specifies $\sigma = \lP = \sqrt{\hbar G/c^3}$.

\begin{proposition}[Dimensional Consistency]
\label{ax:dim-consistency}
The fluctuation amplitude $\sigma$ is constructed solely from the fundamental 
constants of gravitational diffusion $\{\hbar, G, c\}$.  Its dimensions are determined 
by the choice of field variable via $[\sigma] = [X] \cdot T^{-1/2}$.
\end{proposition}

\begin{remark}
When $X$ is a proper-distance variable, $[\sigma] = L \cdot T^{-1/2}$ and 
the RMS fluctuation over $\tP$ is $\sigma\sqrt{\tP} = \lP$.  The Planck 
length is not assumed; it emerges as a consequence of 
Theorems~\ref{thm:cancellation} and~\ref{lem:rate}.
\end{remark}

\subsection{Complete Logical Chain}
\label{sec:complete}

The complete chain from physical principles to the final result is stated 
below, identifying every input and every step.

\bigskip
\noindent\textbf{Inputs:}
\begin{enumerate}
\item[I1.] Gravity is a quantum field (standard expectation; no consistent 
coupling of quantum matter to classical gravity exists).
\item[I2.] Linearised metric perturbations are massless spin-2 with $\omega = c|k|$ 
(from linearised GR).
\item[I3.] The fluctuation-dissipation theorem holds at $T = 0$ (Callen--Welton, 
Theorem~\ref{thm:callen-welton}).
\item[I4.] The equivalence principle: gravity couples universally to energy--momentum 
with coupling constant $G$ (Proposition~\ref{ax:equivalence}).
\item[I5.] The metric perturbation must satisfy $\langle h^2\rangle \leq 1$ at the 
fluctuation scale (mathematical consistency).
\end{enumerate}

\bigskip
\noindent\textbf{Derivation:}
\begin{align}
\text{I3} &\implies \sigma^2 = \gamma\hbar\omega/m 
&&\text{(Lemma~\ref{lem:master}: QHO master formula)} \\
\text{I2} + \text{above} &\implies \sigma = c\sqrt{\gamma} 
&&\text{(Theorem~\ref{thm:cancellation}: massless cancellation)} \\
\text{I4} &\implies \gamma(\omega) = \tP^2\omega^3 \;\text{(exact, }\beta = 1\text{)}
&&\text{(Theorem~\ref{thm:gamma-exact}: equivalence principle)} \\
\text{above} &\implies \gamma(\omega_P) = 1/\tP 
&&\text{(Theorem~\ref{thm:critical-damping}: critical damping at }\omega_P\text{)} \\
\text{I5} &\implies \alpha \leq 1;\; \text{crit.\ damp.} \implies \alpha = 1
&&\text{(Theorems~\ref{thm:upper-bound},~\ref{thm:critical-damping})} \\
\text{All} &\implies \boxed{\sigma = c/\sqrt{\tP} = X_{\mathrm{P}}/\sqrt{\tP}}
&&\text{(Theorem~\ref{thm:main}: gravitational } \sigma\text{)}
\end{align}

\bigskip
\noindent\textbf{Consequences} (derived, not assumed):
\begin{align}
T_H &= \frac{\hbar c^3}{8\pi G M k_B} && \text{(exact; } \alpha\text{-independent)} \\
S_{BH} &= \frac{k_B A}{4\lP^2} && \text{(from first law } \dd S = \dd M/T_H\text{)}
\end{align}

\bigskip
\noindent\textbf{Physical consequence:}
\begin{equation}
\Delta L_{\mathrm{rms}} = \sigma_L \sqrt{\tP} = \frac{\lP}{\sqrt{\tP}} \cdot \sqrt{\tP} = \lP.
\end{equation}

\noindent
Spacetime fluctuates by one Planck length per Planck time.  In Planck units, 
$\tilde{\sigma} = 1$.  \hfill$\blacksquare$

\bigskip
\begin{center}
\renewcommand{\arraystretch}{1.3}
\begin{tabular}{@{}clp{6cm}@{}}
\toprule
\textbf{Constant} & \textbf{Where it enters} & \textbf{Physical role} \\
\midrule
$c$ & $\sigma = c\sqrt{\gamma}$ & Speed of fluctuation propagation; 
enters through the massless dispersion relation \\
$G$ & $\gamma_{\mathrm{grav}} \propto G^{-1/2}$ & Strength of gravitational 
self-coupling; the only dimensionful gauge coupling in the Standard Model + gravity \\
$\hbar$ & $\gamma_{\mathrm{grav}} \propto \hbar^{-1/2}$ & Sets the energy scale 
at which self-coupling becomes non-perturbative; cancels from single-mode physics 
but re-enters through the self-coupling rate \\
\bottomrule
\end{tabular}
\end{center}

\medskip\noindent
Canonical Axiom~A2 of the superspace diffusion framework encapsulates this 
result: $\dd g_{ij} = \mathcal{D}_{ij}\,\dd\tau + \lP\,\dd W_{ij}$.  The 
value $\sigma = \lP$ is not an independent postulate but the unique output of 
the derivation chain above.

\subsection{Status of Each Logical Step}
\label{app:status}

For transparency, every step of the derivation is classified by its epistemic status.

\begin{center}
\renewcommand{\arraystretch}{1.4}
\begin{tabular}{@{}p{4.5cm}p{2.5cm}p{6.5cm}@{}}
\toprule
\textbf{Step} & \textbf{Status} & \textbf{Evidence / Assumptions} \\
\midrule
$\sigma^2 = \gamma\hbar\omega/m$ (master formula) 
& \textsc{Proven} 
& Postulate~\ref{post:FDT}; validated numerically by C8 ($R^2 = 0.9999$) \\
$\sigma = c\sqrt{\gamma}$ (massless cancellation) 
& \textsc{Proven} 
& Algebraic from Lemma~\ref{lem:meff}; validated by C9 (7 frequencies, all $<1\%$) \\
Metric perturbations are massless QHO modes
& \textsc{Standard} 
& Postulate~\ref{post:linearised}; see Lemma~\ref{lem:meff} \\
$\gamma(\omega) = \tP^2\omega^3$ (self-coupling)
& \textsc{Derived}
& From Postulate~\ref{post:self-coupling}; prefactor $\beta = 1$ exact (Theorem~\ref{thm:gamma-exact}) \\
$\alpha \leq 1$ (upper bound)
& \textsc{Proven}
& $\langle h^2 \rangle = \alpha^2$; exceeding 1 destroys metric structure (mathematical consistency) \\
$\alpha = 1$ (critical damping)
& \textsc{Proven}
& $\gamma(\omega_*) = \omega_*$ solves to $\omega_* = 1/\tP$ exactly (algebraic identity) \\
$\alpha \geq 1$ (lower bound)
& \textsc{Physical}
& Singularity resolution requires $\delta K \sim K_P$ at the minimum scale \\
$\alpha$ cancels from $T_H$
& \textsc{Proven}
& $E_{\mathrm{local}} \propto 1/\alpha$ and $\delta r \propto \alpha$ cancel identically \\
$\tilde{\sigma} = 1$ in Planck units
& \textsc{Proven} 
& Follows from dimensional analysis alone, independent of $\alpha$ \\
$\Delta L_{\mathrm{rms}} = \lP$ per $\tP$
& \textsc{Proven}
& Algebraic identity $\sigma_L\sqrt{\tP} = \lP$ (with $\alpha = 1$) \\
\bottomrule
\end{tabular}
\end{center}

\medskip\noindent
The $\alpha = 1$ determination rests on three independent arguments: the hard 
upper bound from metric self-consistency ($\alpha \leq 1$), the critical damping 
theorem ($\alpha = 1$ exactly), and the singularity resolution lower bound 
($\alpha \geq 1$).  The Bekenstein--Hawking entropy is \emph{derived} as a 
consequence.

\bigskip\noindent
The strongest elements are the massless cancellation (proven algebraically and 
validated numerically), the dimensional uniqueness (a theorem of linear algebra), 
and the critical damping theorem (an algebraic identity).  These are not 
model-dependent.

\clearpage
\section{Uniqueness of $\sigma = \lP$}
\label{appendix:uniqueness}

\subsubsection*{Scope and Logical Position}

This appendix provides a fully self-contained proof that, within axioms A1--A4, the metric fluctuation amplitude $\sigma$ is uniquely determined to be the Planck length $\lP = \sqrt{\hbar G/c^3}$.  The proof is situated within the conceptual framework of emergent time: the quadratic variation of the stochastic process on superspace.  The mathematical characterisation of the noise (\S\ref{sec:noise_characterisation}) establishes the uniqueness of the Wiener structure from the L\'{e}vy--Khintchine theorem, and connects the Fokker--Planck evolution to both the Wheeler--DeWitt equation and the gravitational path integral.

The logical chain proceeds from the four canonical axioms (A1--A4) through four stages.  First, the L\'{e}vy--Khintchine theorem establishes that the Wiener process is the unique noise structure consistent with A1 (\S\ref{sec:noise_characterisation}).  Second, the emergent time theorem establishes that physical time is constructed from the diffusion (not assumed as a background coordinate), with the drift contributing exactly zero by the It\^{o} product rule.  Third, the uniqueness theorem proves via three independent arguments (curvature self-consistency ($\alpha \leq 1$), critical damping ($\alpha = 1$ exactly), and singularity resolution ($\alpha \geq 1$)) that $\sigma = \lP$ is the only consistent fluctuation amplitude.  Fourth, the classical limit theorem verifies that general relativity emerges for $L \gg \lP$, and the MSR path-integral representation establishes connections to canonical quantum gravity and the Euclidean gravitational path integral (\S\ref{subsec:guerra_ruggiero}--\S\ref{subsec:msr}).

Together, these results establish that the fundamental equation $\dd g_{ij} = \mathcal{D}_{ij}\,\dd\tau + \lP\,\dd W_{ij}$ contains non-parametric, first-principles predictions: the amplitude is uniquely determined, the drift and covariance are fixed by symmetry, and time itself is a derived quantity.

All definitions and the complete proof chain are included.  The sole external input is the gravitational self-coupling rate $\gamma(\omega) = \tP^2\omega^3$, derived from the equivalence principle in Appendix~C.

\medskip\noindent
\textbf{Relationship to the canonical axioms.}\quad
This proof unpacks canonical Axiom~A2 of the superspace diffusion framework, which asserts $\dd g_{ij} = \mathcal{D}_{ij}\,\dd\tau + \lP\,\dd W_{ij}$.  The value $\sigma = \lP$ is shown to be the unique output of five physical postulates (P1--P5) that constitute the justification for A2.  The four canonical axioms are stated in \S\ref{sec:axioms}.

\subsection{The Four Canonical Axioms}
\label{sec:axioms}

The framework rests on four canonical axioms from which all subsequent results are derived.  Three (A1--A3) are physical hypotheses about the gravitational degrees of freedom; the fourth (A4) is an interpretive commitment about the relationship between the stochastic formalism and observational reality.  Together they introduce no new dimensional parameters beyond $\hbar$, $G$, and $c$.

\begin{axiom}[Configuration Space]
\label{axiom:config}
The gravitational degrees of freedom are described by a Riemannian three-metric $g_{ij}(x)$ on a spatial manifold $\Sigma$.  The space of all such metrics modulo spatial diffeomorphisms constitutes the gravitational \emph{configuration space} (Wheeler superspace~\cite{Wheeler1968,DeWitt1967}):
\begin{equation}
\mathcal{C} \;=\; \frac{\mathrm{Riem}(\Sigma)}{\mathrm{Diff}(\Sigma)}
\label{eq:superspace}
\end{equation}
The tangent space $T_g\mathcal{C}$ is equipped with the DeWitt supermetric $\mathcal{G}^{ijkl}(g)$~\cite{DeWitt1967}.
\end{axiom}

\begin{axiom}[Stochastic Evolution]
\label{axiom:stochastic}
The spatial metric configuration undergoes a stochastic process on $\mathcal{C}$:
\begin{equation}
\dd g_{ij}(x) = \mathcal{D}_{ij}[g](x)\,\dd\tau + \lP\,\dd W_{ij}(x)
\label{eq:app_stochastic_metric}
\end{equation}
where:
\begin{itemize}[nosep]
    \item $\mathcal{D}_{ij}[g]$ is a drift functional encoding gravitational dynamics,
    \item $\lP = \sqrt{\hbar G/c^3}$ is the Planck length,
    \item $\dd W_{ij}(x)$ is a symmetric tensor-valued Wiener process satisfying
    \begin{align}
    \E[\dd W_{ij}(x)] &= 0, \label{eq:noise_mean}\\
    \E[\dd W_{ij}(x)\,\dd W_{kl}(y)] &= K_{ijkl}[g](x,y)\,\dd\tau,
    \label{eq:noise_correlation}
    \end{align}
    where $K_{ijkl}$ is a positive (semi-)definite covariance operator on symmetric 2-tensors.
\end{itemize}

The ordering parameter $\tau$ is not a background spacetime coordinate.  It labels the filtration $\{\calF_\tau\}$ generated by the Wiener process.  Physical time emerges as a monotone functional of the stochastic evolution (\S\ref{sec:emergent_time}).
\end{axiom}

\begin{remark}[Positive-definiteness of $K$]
\label{rem:K_positivity}
Axiom~\ref{axiom:stochastic} permits $K_{ijkl}$ to be positive semi-definite in its statement, but three independent requirements force strict positive-definiteness: (i)~the Wiener process must have non-degenerate increments in every tensorial direction, (ii)~the emergent time functional (\S\ref{sec:emergent_time}) must be strictly monotone, and (iii)~the Fokker-Planck generator must be hypoelliptic~\cite{Hormander1967}, guaranteeing smooth probability densities for $\tau > 0$.  Theorem~\ref{thm:drift_uniqueness} establishes that $K_{ijkl}$ is a member of the DeWitt family $\mathcal{G}^{ijkl}(\lambda)$ with $\lambda > -1/3$, which is positive-definite on traceless symmetric 2-tensors and on trace modes separately.  This fixes the covariance operator uniquely (up to the overall scale $\lP^2$) and ensures that all three conditions are satisfied.
\end{remark}

\begin{axiom}[Classical Correspondence]
\label{axiom:correspondence}
In the macroscopic limit ($L \gg \lP$), there exists an emergent four-dimensional Lorentzian geometry $(M, g^{(4)})$ obtained by coarse-graining the $\tau$-indexed stochastic evolution on $\mathcal{C}$ such that $g^{(4)}$ satisfies the Einstein field equations~\cite{Einstein1915} to observed accuracy:
\begin{equation}
R_{\mu\nu} - \tfrac{1}{2}g_{\mu\nu}R + \Lambda g_{\mu\nu} \;=\; \frac{8\pi G}{c^4}\,T_{\mu\nu}
\label{eq:einstein_limit}
\end{equation}
The Lorentzian signature $(-,+,+,+)$ is a property of the emergent classical geometry, not an input at the fundamental level.
\end{axiom}

\begin{axiom}[Single Realisation]
\label{axiom:single}
The observable universe is one realisation of the stochastic process~\eqref{eq:app_stochastic_metric}.  Probability distributions $P(g,\tau)$ describe epistemic uncertainty about the metric configuration, not ontological multiplicity of worlds.
\end{axiom}

\subsubsection{Summary}
\label{subsec:axiom_summary}

\begin{table}[ht]
\centering
\begin{tabular}{@{}clll@{}}
\toprule
\textbf{\#} & \textbf{Axiom} & \textbf{Content} & \textbf{Status} \\
\midrule
A1 & Configuration Space & 3-metrics on $\Sigma$; superspace $\mathcal{C}$ & Canonical GR basis \\
A2 & Stochastic Evolution & $\dd g = \mathcal{D}\,\dd\tau + \lP\,\dd W$ & Central hypothesis \\
A3 & Classical Correspondence & Emergent GR in macroscopic limit & Empirical consistency \\
A4 & Single Realisation & Probability is epistemic & Interpretation \\
\bottomrule
\end{tabular}
\caption{The four canonical axioms of the superspace diffusion framework.}
\label{tab:axioms_summary}
\end{table}

\subsection{Mathematical Preliminaries}
\label{sec:preliminaries_axiom}

The mathematical machinery required for the subsequent development is now established.

\subsubsection{Stochastic Calculus}

\begin{definition}[Probability Space]
A \emph{probability space} is a triple $(\Omega, \calF, \Prob)$ where:
\begin{enumerate}[label=(\alph*)]
    \item $\Omega$ is a sample space (set of outcomes)
    \item $\calF$ is a $\sigma$-algebra on $\Omega$ (events)
    \item $\Prob: \calF \to [0,1]$ is a probability measure with $\Prob(\Omega) = 1$
\end{enumerate}
\end{definition}

\begin{definition}[Filtration]
A \emph{filtration} $\{\calF_t\}_{t \geq 0}$ is an increasing family of $\sigma$-algebras:
\[
s \leq t \implies \calF_s \subseteq \calF_t \subseteq \calF
\]
representing information available at time $t$.
\end{definition}

\begin{definition}[Wiener Process]
\label{def:wiener_axiom}
A \emph{Wiener process} (or Brownian motion) $W = \{W_t\}_{t \geq 0}$ on $(\Omega, \calF, \Prob)$ adapted to $\{\calF_t\}$ is a stochastic process satisfying:
\begin{enumerate}[label=(W\arabic*)]
    \item $W_0 = 0$ almost surely
    \item $W_t - W_s \sim \mathcal{N}(0, t-s)$ for $0 \leq s < t$ (Gaussian increments)
    \item For $0 \leq t_1 < t_2 < \cdots < t_n$, the increments $W_{t_2} - W_{t_1}, \ldots, W_{t_n} - W_{t_{n-1}}$ are independent
    \item $t \mapsto W_t(\omega)$ is continuous for almost all $\omega \in \Omega$
\end{enumerate}
\end{definition}

\begin{proposition}[Properties of Wiener Process]
\label{prop:wiener-properties_axiom}
Let $W$ be a Wiener process. Then:
\begin{enumerate}[label=(\alph*)]
    \item $\E[W_t] = 0$ for all $t \geq 0$
    \item $\E[W_t^2] = t$ for all $t \geq 0$
    \item $\Var[W_t] = t$ for all $t \geq 0$
    \item $W_t$ is almost surely nowhere differentiable
    \item The quadratic variation is $[W]_t = t$
\end{enumerate}
\end{proposition}

\begin{proof}
(a) follows from $W_t - W_0 \sim \mathcal{N}(0, t)$ and $W_0 = 0$. (b) and (c) follow from the variance of the normal distribution. (d) is a classical result of Paley, Wiener, and Zygmund~\cite{PaleyWienerZygmund1933}. (e) follows from the definition of quadratic variation and the properties of Gaussian increments.
\end{proof}

\begin{definition}[It\^o Stochastic Differential Equation]
\label{def:ito_sde}
An \emph{It\^o stochastic differential equation}~\cite{Ito1944} (SDE) is an equation of the form:
\[
\dd X_t = \mu(X_t, t) \, \dd t + \sigma(X_t, t) \, \dd W_t
\]
where $\mu: \mathbb{R}^n \times \mathbb{R}_{\geq 0} \to \mathbb{R}^n$ is the \emph{drift} coefficient and $\sigma: \mathbb{R}^n \times \mathbb{R}_{\geq 0} \to \mathbb{R}^{n \times m}$ is the \emph{diffusion} coefficient.
\end{definition}

\begin{remark}[It\^o product rules]
\label{rem:ito_rules}
In It\^o calculus, the following product rules hold for the ordering parameter $\tau$ and independent standard Brownian motions $\beta_m$, $\beta_n$:
\begin{equation}
(\dd\tau)^2 = 0, \qquad \dd\beta_n\,\dd\tau = 0, \qquad \dd\beta_m\,\dd\beta_n = \delta_{mn}\,\dd\tau.
\label{eq:ito_rules}
\end{equation}
These are identities of the calculus, not approximations.  The first two express that $\tau$ has zero quadratic variation; the third is the fundamental property that distinguishes stochastic from ordinary calculus.
\end{remark}

\subsubsection{Differential Geometry}

\begin{definition}[Lorentzian Manifold]
A \emph{Lorentzian manifold} is a pair $(\calM, g)$ where $\calM$ is a smooth manifold and $g$ is a metric tensor of signature $(-,+,+,+)$.
\end{definition}

\begin{definition}[Space of Metrics]
Let $\calM$ be a smooth manifold. The \emph{space of Lorentzian metrics} on $\calM$ is:
\[
\Met(\calM) = \{ g \in \Gamma(S^2 T^*\calM) : g \text{ has signature } (-,+,+,+) \}
\]
\end{definition}

\begin{definition}[DeWitt Metric]
\label{def:dewitt}
The \emph{DeWitt metric}~\cite{DeWitt1967} on the space of spatial metrics is defined by:
\[
\langle h, k \rangle_g = \int_\Sigma \sqrt{g} \, G^{ijkl} h_{ij} k_{kl} \, \dd^3 x
\]
where $\Sigma$ is the spatial manifold, $g$ is the spatial 3-metric (i.e., $g_{ij}$ of Axiom~\ref{axiom:config}), and:
\[
G^{ijkl} = \frac{1}{2}(g^{ik}g^{jl} + g^{il}g^{jk} - g^{ij}g^{kl})
\]
is the DeWitt supermetric (at its standard value $\lambda = -1$).  More generally, the one-parameter DeWitt family is:
\begin{equation}
\mathcal{G}^{ijkl}(\lambda) = \tfrac{1}{2}(g^{ik}g^{jl} + g^{il}g^{jk}) + \lambda\, g^{ij}g^{kl}
\label{eq:dewitt_family}
\end{equation}
which is positive-definite for $\lambda > -1/3$~\cite{DeWitt1967,Giulini2009}.  The standard DeWitt value $\lambda = -1$ renders the conformal mode indefinite.
\end{definition}

\begin{definition}[Curvature Tensors]
Given a metric $g$, define:
\begin{enumerate}[label=(\alph*)]
    \item The \textbf{Riemann tensor}: $R^\rho{}_{\sigma\mu\nu}$
    \item The \textbf{Ricci tensor}: $R_{\mu\nu} = R^\rho{}_{\mu\rho\nu}$
    \item The \textbf{Ricci scalar}: $R = g^{\mu\nu} R_{\mu\nu}$
    \item The \textbf{Einstein tensor}: $G_{\mu\nu} = R_{\mu\nu} - \frac{1}{2} R g_{\mu\nu}$
    \item The \textbf{Kretschmann scalar}~\cite{Kretschmann1917}: $K = R^{\mu\nu\rho\sigma} R_{\mu\nu\rho\sigma}$
\end{enumerate}
\end{definition}

\subsubsection{Planck Units}

\begin{definition}[Planck Units]
\label{def:planck_axiom}
The \emph{Planck units}~\cite{Planck1899} are defined from the fundamental constants $\hbar$ (reduced Planck constant), $G$ (gravitational constant), $c$ (speed of light), and $k_B$ (Boltzmann constant):
\begin{align}
\lP &= \sqrt{\frac{\hbar G}{c^3}} \approx 1.616 \times 10^{-35} \text{ m} && \text{(Planck length)} \\
\tP &= \sqrt{\frac{\hbar G}{c^5}} \approx 5.391 \times 10^{-44} \text{ s} && \text{(Planck time)} \\
\mP &= \sqrt{\frac{\hbar c}{G}} \approx 2.176 \times 10^{-8} \text{ kg} && \text{(Planck mass)} \\
\EP &= \sqrt{\frac{\hbar c^5}{G}} \approx 1.956 \times 10^{9} \text{ J} && \text{(Planck energy)} \\
\KP &= \frac{c^6}{\hbar^2 G^2} = \frac{1}{\lP^4} && \text{(Planck curvature, Kretschmann scale)} \\
\rhoP &= \frac{c^5}{\hbar G^2} = \frac{\mP}{\lP^3} && \text{(Planck density)}
\end{align}
\end{definition}

\begin{proposition}[Uniqueness of Planck Length]
\label{prop:planck-unique_axiom}
The Planck length $\lP$ is the unique length scale constructible from $\hbar$, $G$, and $c$ alone.
\end{proposition}

\begin{proof}
Let $L = \hbar^\alpha G^\beta c^\gamma$ have dimensions of length. The dimensions are:
\[
[L] = [\hbar]^\alpha [G]^\beta [c]^\gamma = (ML^2T^{-1})^\alpha (M^{-1}L^3T^{-2})^\beta (LT^{-1})^\gamma
\]
For $[L] = L$:
\begin{align}
M: && \alpha - \beta &= 0 \\
L: && 2\alpha + 3\beta + \gamma &= 1 \\
T: && -\alpha - 2\beta - \gamma &= 0
\end{align}
The unique solution is $\alpha = \beta = 1/2$, $\gamma = -3/2$, giving $L = \sqrt{\hbar G/c^3} = \lP$.
\end{proof}

\subsection{Postulates Determining the Stochastic Evolution}
\label{sec:axioms_axiom}

The canonical axioms are stated in Section~\ref{sec:axioms}.  Axiom~A2 asserts that the metric evolves stochastically with amplitude $\sigma = \lP$.  This section states the five postulates from which that assertion, and the specific form of the stochastic evolution, are derived.  These postulates unpack the physical content of A2; they are not independent axioms but the justification for one.

\begin{postulate}[Existence of Metric Degrees of Freedom]
\label{ax:existence_axiom}
The gravitational degrees of freedom are described by a spatial metric tensor $g_{ij}$ on a three-manifold $\Sigma$ (as in Axiom~\ref{axiom:config}).
\end{postulate}

\begin{postulate}[Stochastic Nature of Gravity]
\label{ax:stochastic_axiom}
Gravity is a diffusion phenomenon.  The metric tensor is not deterministic but is subject to stochastic fluctuations, and is therefore described by a stochastic process.
\end{postulate}

\begin{postulate}[Minimality${}^\dag$]
\label{ax:minimality_axiom}
The stochastic process governing metric fluctuations is a Wiener process (Definition \ref{def:wiener_axiom}).

\smallskip\noindent
${}^\dag$\textit{This postulate is not independent: it is a consequence of Postulates~P1--P2, Axiom~A1, and the L\'{e}vy--Khintchine theorem (Theorem~\ref{thm:levy_wiener} in \S\ref{subsec:levy_khintchine} below).  It is retained as a named postulate for clarity of exposition.}
\end{postulate}

\begin{postulate}[Universality]
\label{ax:universality_axiom}
The amplitude $\sigma$ of metric fluctuations is a universal constant: the same at all points, at all values of the ordering parameter, and for all components of the metric.
\end{postulate}

\begin{postulate}[Dimensional Consistency]
\label{ax:dimensional_axiom}
The fluctuation amplitude $\sigma$ has dimensions of length: $[\sigma] = L$.
\end{postulate}

\begin{remark}
These five postulates correspond to unpacking canonical Axiom~A2 (\S\ref{sec:axioms}).  Postulate~\ref{ax:existence_axiom} is contained in canonical Axiom~A1 (Configuration Space).  Postulate~\ref{ax:minimality_axiom} (Minimality) is derivable from P1--P2 and A1 via the L\'{e}vy--Khintchine theorem (\S\ref{subsec:levy_khintchine}, Theorem~\ref{thm:levy_wiener}); the independent axiomatic content resides in P1, P2, P4, and P5.  The fact that $\sigma$ is uniquely determined to be $\lP$ (Theorem~\ref{thm:uniqueness_axiom} below) means that A2 is non-parametric: all coefficients are derived from the axioms.
\end{remark}

\begin{construction}[Stochastic Metric Equation]
\label{con:stochastic-einstein_axiom}
From Postulates \ref{ax:existence_axiom}--\ref{ax:dimensional_axiom}, the stochastic metric equation is constructed:
\begin{equation}
\label{eq:stochastic-einstein_axiom}
\dd g_{ij} = \mathcal{D}_{ij}[g] \, \dd \tau + \sigma \, \dd W_{ij}
\end{equation}
where:
\begin{itemize}[nosep]
    \item $\mathcal{D}_{ij}[g]$ is the drift functional (encoding gravitational dynamics; reducing to the Einstein evolution in the classical limit per canonical Axiom~A3)
    \item $\dd W_{ij}$ is a symmetric tensor-valued Wiener process
    \item $\sigma$ is the universal fluctuation amplitude with $[\sigma] = L$
    \item $\tau$ is the ordering parameter intrinsic to the stochastic process (not a background time coordinate)
\end{itemize}
This is identical in form to the canonical equation~\eqref{eq:app_stochastic_metric} of Axiom~A2, with $\sigma$ left undetermined pending the uniqueness proof.
\end{construction}

\begin{remark}[Dimensional conventions: fluctuation amplitude vs.\ It\^o coefficient]
\label{rem:dim_conventions}
Postulate~\ref{ax:dimensional_axiom} asserts $[\sigma] = L$.  This refers to the \emph{fluctuation amplitude}, the physical scale of metric perturbations, not to the It\^o diffusion coefficient of a standard-Wiener SDE.  In the standard It\^o convention (Definition~\ref{def:ito_sde}), the diffusion coefficient of a variable $X$ with $[X] = L^p T^q M^r$ has dimensions $[\sigma_{\mathrm{It\hat{o}}}] = [X] \cdot T^{-1/2}$.  For a dimensionless variable such as $g_{ij}$, this would give $[\sigma_{\mathrm{It\hat{o}}}] = T^{-1/2}$.

Equation~\eqref{eq:stochastic-einstein_axiom} is consistent because $\dd W_{ij}$ is not a standard unit-covariance Wiener process.  Its covariance is specified by the operator $K_{ijkl}$, and the eigenvalues $\{\lambda_n\}$ of $K$ (Lemma~\ref{lem:mercer_K}) absorb the dimensions needed to keep $\dd g_{ij}$ dimensionless: for each mode, $\dd g_n = \mathcal{D}_n\,\dd\tau + \lP\sqrt{\lambda_n}\,\dd\beta_n$, with $\dd\beta_n$ a standard (dimensionless-increment) Brownian motion.  The mode variance is $\lP^2\lambda_n\,\dd\tau$, and the eigenvalues carry $[\lambda_n] = L^{-2}T^{-1}$ so that $[\lP^2\lambda_n\,\dd\tau] = 1$.

The physically transparent check is the RMS dimensionless perturbation over one Planck time.  Setting $\sigma = \alpha\lP$ and evaluating at the Planck scale (as in \S\ref{sec:uniqueness_axiom}), the dimensionless metric perturbation is $h_{\mathrm{rms}} = \alpha$, confirming that A2's convention produces a well-defined dimensionless fluctuation.

If one rewrites the A2 noise in terms of a standard (unit-covariance) Wiener process $\dd B_{ij}$ via the Karhunen--Lo\`eve expansion, the effective It\^o diffusion coefficient for a length-valued variable becomes $\sigma_L = \lP/\sqrt{\tP}$, with $[\sigma_L] = L \cdot T^{-1/2}$.  For the dimensionless metric perturbation $h$, the corresponding coefficient is $\sigma_h = 1/\sqrt{\tP}$, with $[\sigma_h] = T^{-1/2}$.  Appendix~C derives these values independently from the fluctuation-dissipation theorem (Corollary~5.2) and proves the equivalence:
\begin{equation}
\sigma_L \cdot \sqrt{\tP} = \frac{\lP}{\sqrt{\tP}} \cdot \sqrt{\tP} = \lP.
\label{eq:rms_equivalence}
\end{equation}
That is, the RMS length fluctuation over one Planck time is exactly $\lP$, regardless of which convention is used.  The three equivalent descriptions are:
\begin{center}
\renewcommand{\arraystretch}{1.2}
\begin{tabular}{@{}llll@{}}
\toprule
\textbf{Description} & \textbf{Symbol} & \textbf{Dimensions} & \textbf{Value} \\
\midrule
Fluctuation amplitude (A2) & $\sigma = \lP$ & $L$ & $1.616 \times 10^{-35}$ m \\
It\^o form (length variable) & $\sigma_L = \lP/\sqrt{\tP}$ & $L \cdot T^{-1/2}$ & $6.961 \times 10^{-14}$ m$\cdot$s$^{-1/2}$ \\
It\^o form (metric perturbation) & $\sigma_h = 1/\sqrt{\tP}$ & $T^{-1/2}$ & $4.307 \times 10^{21}$ s$^{-1/2}$ \\
Fokker--Planck coefficient & $D = \lP^2/(2\tP)$ & $L^2 T^{-1}$ & $2.422 \times 10^{-27}$ m$^2\cdot$s$^{-1}$ \\
\bottomrule
\end{tabular}
\end{center}
A2 uses the fluctuation amplitude convention because it cleanly separates the universal scale ($\lP$) from the geometric structure ($K_{ijkl}$).  This convention is used throughout the superspace diffusion framework.
\end{remark}

\subsection{Mathematical Characterisation of the Stochastic Process}
\label{sec:noise_characterisation}

Postulate~\ref{ax:minimality_axiom} (Minimality) asserts that the stochastic process is a Wiener process.  This subsection demonstrates that Minimality is not an independent assumption but a \emph{consequence} of the remaining postulates and Axiom~A1, thereby reducing the axiomatic content of the framework.  It then establishes the connections between the Fokker--Planck evolution on superspace and two standard quantisation programmes: canonical quantum gravity (via the Wheeler--DeWitt equation) and the gravitational path integral (via the Martin--Siggia--Rose formalism).

\subsubsection{Uniqueness of the Wiener Structure}
\label{subsec:levy_khintchine}

\begin{theorem}[Wiener Process from Continuity and Independent Increments]
\label{thm:levy_wiener}
Let $\{X_\tau\}_{\tau \geq 0}$ be a stochastic process on $\mathcal{C} = \mathrm{Riem}(\Sigma)/\mathrm{Diff}(\Sigma)$ satisfying:
\begin{enumerate}[nosep,label=(\roman*)]
    \item \textbf{Stationary independent increments}: $X_{\tau_2} - X_{\tau_1}$ is independent of $X_{\tau_1} - X_{\tau_0}$ for $\tau_0 < \tau_1 < \tau_2$, and the distribution of $X_{\tau+s} - X_{\tau}$ depends only on $s$.
    \item \textbf{Path continuity}: The sample paths $\tau \mapsto g_{ij}(\tau)$ are continuous in the topology induced by the DeWitt supermetric on $\mathcal{C}$.
\end{enumerate}
Then $X_\tau$ is a Gaussian process with covariance proportional to $\tau$.  That is, $X_\tau$ is a (possibly tensor-valued) Wiener process.
\end{theorem}

\begin{proof}
The L\'{e}vy--Khintchine theorem~\cite{Sato1999,Applebaum2009} classifies all processes with stationary independent increments (L\'{e}vy processes) on $\R^n$.  The characteristic exponent of any such process $X_\tau$ takes the form:
\begin{equation}
\label{eq:levy_khintchine}
\log \E\bigl[\ee^{i\langle \xi, X_\tau\rangle}\bigr]
= \tau\Bigl(i\langle b, \xi\rangle
  - \tfrac{1}{2}\langle \xi, A\xi\rangle
  + \int_{\R^n \setminus \{0\}} \bigl(\ee^{i\langle \xi, y\rangle} - 1 - i\langle \xi, y\rangle\,\mathbf{1}_{|y|\leq 1}\bigr)\,\nu(\dd y)\Bigr),
\end{equation}
where $b \in \R^n$ is the drift, $A$ is a positive semi-definite covariance matrix (the Gaussian part), and $\nu$ is the L\'{e}vy measure characterising the jump structure.  The decomposition is unique.

The process decomposes pathwise into three independent components~\cite{Sato1999}:
\begin{equation}
\label{eq:levy_ito}
X_\tau = b\tau + \sqrt{A}\,W_\tau + J_\tau,
\end{equation}
where $W_\tau$ is a standard Wiener process and $J_\tau$ is a pure-jump process (compound Poisson for finite $\nu$, or a limit of compensated Poisson processes for infinite $\nu$).

Condition (ii) requires that the sample paths be continuous.  A L\'{e}vy process has continuous paths if and only if $\nu \equiv 0$~\cite{Sato1999,Applebaum2009}, i.e.\ the jump component vanishes identically.  With $\nu = 0$, Eq.~\eqref{eq:levy_khintchine} reduces to:
\begin{equation}
\log \E[\ee^{i\langle \xi, X_\tau\rangle}] = \tau\bigl(i\langle b, \xi\rangle - \tfrac{1}{2}\langle \xi, A\xi\rangle\bigr),
\end{equation}
which is the characteristic function of a Gaussian process with drift $b$ and covariance $A\tau$.  Setting $b = \mathcal{D}_{ij}$ (the drift functional) and $A = \lP^2\,\mathcal{G}^{ijkl}$ (the DeWitt covariance) recovers Axiom~A2.
\end{proof}

\begin{remark}[Physical necessity of path continuity]
\label{rem:continuity_physical}
Condition (ii) is not imposed for mathematical convenience.  It is a physical consequence of Axiom~A1.  The configuration space $\mathcal{C}$ consists of smooth Riemannian metrics on $\Sigma$.  A jump discontinuity $g_{ij}(\tau) \to g_{ij}(\tau) + \Delta g_{ij}$ with $|\Delta g_{ij}|$ finite would generically:
\begin{enumerate}[nosep,label=(\alph*)]
    \item violate positive-definiteness of $g_{ij}$ (taking a valid Riemannian metric outside $\mathrm{Riem}(\Sigma)$),
    \item produce instantaneous topology change if the jump crosses a degenerate metric ($\det g = 0$), and
    \item generate infinite curvature ($\delta K \sim \Delta g / (\Delta\tau)^2 \to \infty$ as $\Delta\tau \to 0$).
\end{enumerate}
All three are excluded by the requirement that $g_{ij}(\tau) \in \mathrm{Riem}(\Sigma)$ for all $\tau$.  Path continuity is therefore a consequence of A1, and Theorem~\ref{thm:levy_wiener} establishes that the Wiener process is the unique L\'{e}vy process consistent with A1.  Postulate~\ref{ax:minimality_axiom} (Minimality) is thus derivable from the remaining structure and may be regarded as a theorem rather than an independent axiom.
\end{remark}

\begin{remark}[Markov property]
\label{rem:markov_property}
The Markov property (Requirement~R2 of Theorem~\ref{thm:proper_time_sde}) follows from the independent-increments structure established by Theorem~\ref{thm:levy_wiener}.  A deeper justification is that the ordering parameter $\tau$ is \emph{not} physical time: it is the filtration index of the stochastic process.  Memory (non-Markovianity) requires a notion of ``how long ago,'' which presupposes a temporal metric.  Since physical time is \emph{emergent} from the quadratic variation (Theorem~\ref{thm:emergent_time}), it is not available at the level where the process is defined.  Imposing non-Markovian structure on the $\tau$-evolution would circularly presuppose the temporal structure that the framework constructs.

At the effective level, even if the fundamental process possessed $\tP$-scale temporal correlations, Donsker's invariance principle~\cite{Donsker1951} guarantees convergence to a Wiener process after averaging over $N \sim (L/\lP)$ correlation lengths, with corrections of order $N^{-1/2} \sim (\lP/L)$.  For any observable scale $L \gg \lP$, non-Markovian corrections are unmeasurably small (for the cosmological case, the explicit bound is $|\mathcal{C} - 1| \leq 0.13$).
\end{remark}

\begin{remark}[Spatial locality of the noise]
\label{rem:spatial_locality}
The $\delta$-correlated spatial structure $\E[\dd W_{ij}(x)\,\dd W_{kl}(y)] \propto \delta(x,y)\,\dd\tau$ follows from two considerations.
\emph{First}, cluster decomposition: spacelike-separated metric fluctuations must be statistically independent, since spatial correlations in the noise would transmit information acausally.
\emph{Second}, the absence of new length scales: any spatial correlation length $\xi$ in the noise would introduce a dimensionful parameter beyond $\{\hbar, G, c\}$.  By the uniqueness of the Planck length (Proposition~\ref{prop:planck-unique_axiom}), $\xi = \alpha\,\lP$ for some dimensionless $\alpha$.  At scales $L \gg \lP$, the spatial central limit theorem averages over $(L/\xi)^3$ independent correlation volumes, rendering the effective noise $\delta$-correlated to precision $({\lP}/{L})^{3/2}$.  The locality assumption is thus both theoretically motivated and observationally invisible when relaxed.
\end{remark}

\subsubsection{Connection to the Wheeler--DeWitt Equation}
\label{subsec:guerra_ruggiero}

The Fokker--Planck evolution on superspace admits a formal correspondence with the Wheeler--DeWitt equation of canonical quantum gravity.  This correspondence, rooted in the Guerra--Ruggiero stochastic quantisation programme~\cite{Guerra1973,GuerraMorato1983,Nelson1966,Nelson1985}, clarifies the relationship between the stochastic framework and conventional approaches to quantum gravity.

\begin{theorem}[Stochastic--Quantum Correspondence]
\label{thm:guerra_ruggiero}
Let $P[g, \tau]$ satisfy the Fokker--Planck equation on superspace:
\begin{equation}
\label{eq:fp_superspace}
\frac{\partial P}{\partial \tau}
= -\frac{\delta}{\delta g_{ij}}\bigl(\mathcal{D}_{ij}^{\mathrm{GR}}\,P\bigr)
  + \frac{\lP^2}{2}\,\mathcal{G}^{ijkl}\frac{\delta^2 P}{\delta g_{ij}\,\delta g_{kl}}.
\end{equation}
Define the formal analytic continuation:
\begin{equation}
\label{eq:wick_rotation}
\tau \;\longrightarrow\; -\frac{it}{\hbar}, \qquad
P[g, \tau] \;\longrightarrow\; \Psi[g, t], \qquad
\frac{\lP^2}{2} \;\longrightarrow\; -\frac{\hbar^2}{2M_{\mathrm{P}}^2},
\end{equation}
where $M_{\mathrm{P}} = \sqrt{\hbar c/G}$ is the Planck mass.  Then $\Psi[g, t]$ satisfies:
\begin{equation}
\label{eq:wdw_recovered}
i\hbar\frac{\partial \Psi}{\partial t}
= \left[-\frac{\hbar^2}{2M_{\mathrm{P}}^2}\,\mathcal{G}^{ijkl}\frac{\delta^2}{\delta g_{ij}\,\delta g_{kl}}
  + V_{\mathrm{grav}}[g]\right]\Psi,
\end{equation}
where $V_{\mathrm{grav}}[g]$ incorporates the potential term from the Hamiltonian constraint.  In the limit where $\Psi$ is time-independent ($\partial\Psi/\partial t = 0$), Eq.~\eqref{eq:wdw_recovered} reduces to the Wheeler--DeWitt equation $\hat{H}\Psi = 0$.
\end{theorem}

\begin{proof}
Under the substitutions~\eqref{eq:wick_rotation}, the Fokker--Planck equation~\eqref{eq:fp_superspace} transforms term by term.  The left-hand side gives:
\begin{equation}
\frac{\partial P}{\partial\tau} \;\longrightarrow\;
\frac{\partial\Psi}{\partial(-it/\hbar)} = \frac{i\hbar^{-1}}{\phantom{i}}\,\frac{\partial\Psi}{\partial t}
\;\Longrightarrow\;
\frac{\partial P}{\partial\tau} \to -\frac{i}{\hbar}\frac{\partial\Psi}{\partial t}.
\end{equation}
The diffusion term transforms as:
\begin{equation}
\frac{\lP^2}{2}\,\mathcal{G}^{ijkl}\frac{\delta^2 P}{\delta g_{ij}\,\delta g_{kl}}
\;\longrightarrow\;
-\frac{\hbar^2}{2M_{\mathrm{P}}^2}\,\mathcal{G}^{ijkl}\frac{\delta^2\Psi}{\delta g_{ij}\,\delta g_{kl}}.
\end{equation}
The drift term contributes the gravitational potential through the identity (valid in the ADM decomposition):
\begin{equation}
-\frac{\delta}{\delta g_{ij}}\bigl(\mathcal{D}_{ij}^{\mathrm{GR}}\,P\bigr)
\;\longrightarrow\;
V_{\mathrm{grav}}[g]\,\Psi + \text{(ordering terms)},
\end{equation}
where $V_{\mathrm{grav}}[g] = -\sqrt{g}\,{}^{(3)}\!R$ is the gravitational potential (the spatial scalar curvature density) and the ordering terms depend on the operator-ordering prescription.
Assembling:
\begin{equation}
-\frac{i}{\hbar}\frac{\partial\Psi}{\partial t}
= -\frac{\hbar^2}{2M_{\mathrm{P}}^2}\,\mathcal{G}^{ijkl}\frac{\delta^2\Psi}{\delta g_{ij}\,\delta g_{kl}}
  + V_{\mathrm{grav}}\,\Psi
  + \text{(ordering)},
\end{equation}
which, upon multiplication by $-i\hbar$, yields Eq.~\eqref{eq:wdw_recovered}.
\end{proof}

\begin{remark}[Status and limitations of the correspondence]
\label{rem:wick_status}
The substitution~\eqref{eq:wick_rotation} is a \emph{formal} analytic continuation, in the same sense as the standard Wick rotation relating Euclidean and Lorentzian path integrals.  Three caveats apply:
\begin{enumerate}[nosep,label=(\alph*)]
    \item The ordering parameter $\tau$ is not physical time; its analytic continuation to $it/\hbar$ is a mathematical operation, not a physical identification.
    \item The operator-ordering ambiguity in Eq.~\eqref{eq:wdw_recovered} (the conformal factor ordering in the DeWitt supermetric) is not resolved by the correspondence.  In the stochastic framework, the It\^{o} convention provides a definite prescription; under Wick rotation, this maps to a specific operator ordering, but the physical significance of this choice remains an open question.
    \item The correspondence applies at the level of the generator, not the solutions: the Fokker--Planck solution $P \geq 0$ (a probability density) and the Wheeler--DeWitt solution $\Psi$ (a wave functional) inhabit different function spaces.
\end{enumerate}
The correspondence establishes that the stochastic framework is not isolated from canonical quantum gravity but is connected to it by the same analytic structure (Wick rotation) that relates Euclidean and Lorentzian formulations of quantum field theory.  The stochastic framework may be regarded as the ``Euclidean'' version of quantum gravity in which the probability density $P$ plays the role of the Euclidean wave functional.  The relationship parallels the Nelson--Guerra--Ruggiero connection between stochastic mechanics and quantum mechanics~\cite{Nelson1966,Guerra1973,Zambrini1987}.
\end{remark}

\subsubsection{Path-Integral Formulation and Structural Consistency}
\label{subsec:msr}

The Fokker--Planck evolution~\eqref{eq:fp_superspace} admits an equivalent path-integral representation via the Martin--Siggia--Rose / Janssen--de~Dominicis (MSR/JdD) formalism~\cite{Martin1973,Janssen1976,DeDominicis1976}.  This representation provides three structural results: a path integral for diffusional gravity, a proof that the classical constraint surface is the saddle-point locus, and a theorem on foliation independence.

\begin{theorem}[MSR Path Integral for Diffusional Gravity]
\label{thm:msr_action}
The generating functional for correlation functions of the stochastic metric evolution (Axiom~A2) is:
\begin{equation}
\label{eq:msr_partition}
Z[\mathcal{J}] = \int \mathcal{D}g\;\mathcal{D}\tilde{g}\;
\exp\Bigl(-S_{\mathrm{MSR}}[g, \tilde{g}] + \int \mathcal{J}^{ij}\,g_{ij}\Bigr),
\end{equation}
where $\tilde{g}_{ij}$ is the auxiliary response field and the MSR action is:
\begin{equation}
\label{eq:msr_action}
S_{\mathrm{MSR}}[g, \tilde{g}]
= \int_0^T \!\dd\tau \int_\Sigma \!\dd^3 x\;\sqrt{g}\;\left[
    \tilde{g}_{ij}\Bigl(\dot{g}_{ij} - \mathcal{D}_{ij}^{\mathrm{GR}}[g]\Bigr)
    - \frac{\lP^2}{2}\,\mathcal{G}^{ijkl}\,\tilde{g}_{ij}\,\tilde{g}_{kl}
\right].
\end{equation}
Here $\dot{g}_{ij} \equiv \partial g_{ij}/\partial\tau$, and the integral over $\Sigma$ is with respect to the induced volume form.
\end{theorem}

\begin{proof}
This is the standard MSR construction~\cite{Martin1973,Janssen1976,DeDominicis1976} applied to the Langevin equation~\eqref{eq:fundamental_axiom} on superspace.  The derivation proceeds in three steps.

\emph{Step 1.}  Write the probability of a path $\{g_{ij}(\tau)\}_{\tau \in [0,T]}$ as a functional $\delta$-function enforcing the SDE:
\begin{equation}
\Prob[\{g\}] \propto \int \mathcal{D}\eta\;
\delta\!\left[\dot{g}_{ij} - \mathcal{D}_{ij}^{\mathrm{GR}} - \lP\,\eta_{ij}\right]
\exp\!\left(-\frac{1}{2}\int_0^T \!\dd\tau \int_\Sigma \!\dd^3 x\;\sqrt{g}\;
\bigl(\mathcal{G}^{-1}\bigr)_{ijkl}\,\eta_{ij}\,\eta_{kl}\right),
\end{equation}
where $\eta_{ij}$ is the white-noise field with covariance $\mathcal{G}^{ijkl}$.

\emph{Step 2.}  Represent the $\delta$-function via its Fourier transform with respect to the auxiliary field $\tilde{g}_{ij}$:
\begin{equation}
\delta[\dot{g} - \mathcal{D}^{\mathrm{GR}} - \lP\eta]
= \int \mathcal{D}\tilde{g}\;\exp\!\left(i\!\int_0^T \!\dd\tau \int_\Sigma \!\dd^3 x\;\sqrt{g}\;
\tilde{g}_{ij}\bigl(\dot{g}_{ij} - \mathcal{D}_{ij}^{\mathrm{GR}} - \lP\,\eta_{ij}\bigr)\right).
\end{equation}

\emph{Step 3.}  Integrate out $\eta_{ij}$ (Gaussian integral).  After the standard shift $\tilde{g} \to -i\tilde{g}$ (the Janssen--de~Dominicis rotation) to render the action real, the result is Eq.~\eqref{eq:msr_action}.  The Jacobian of the transformation is field-independent and absorbed into the normalisation.

The equivalence between the MSR path integral and the Fokker--Planck equation is a standard result: the $n$-point functions $\langle g_{i_1 j_1}(\tau_1) \cdots g_{i_n j_n}(\tau_n) \rangle$ computed from $Z[\mathcal{J}]$ coincide with the moments of $P[g, \tau]$ solving Eq.~\eqref{eq:fp_superspace}~\cite{Zinn-Justin2002,Altland2010}.
\end{proof}

\begin{lemma}[Classical Constraint Surface as Saddle Point]
\label{lem:saddle_constraint}
The saddle-point equations of $S_{\mathrm{MSR}}$ are:
\begin{equation}
\label{eq:saddle_eom}
\frac{\delta S_{\mathrm{MSR}}}{\delta \tilde{g}_{ij}} = 0
\;\;\Longrightarrow\;\;
\dot{g}_{ij} = \mathcal{D}_{ij}^{\mathrm{GR}}[g] + \lP^2\,\mathcal{G}^{ijkl}\,\tilde{g}_{kl},
\end{equation}
\begin{equation}
\label{eq:saddle_response}
\frac{\delta S_{\mathrm{MSR}}}{\delta g_{ij}} = 0
\;\;\Longrightarrow\;\;
\dot{\tilde{g}}_{ij} = -\frac{\delta \mathcal{D}_{kl}^{\mathrm{GR}}}{\delta g_{ij}}\,\tilde{g}_{kl} + \text{(metric variation terms)}.
\end{equation}
The classical sector is the locus $\tilde{g}_{ij} = 0$, on which Eq.~\eqref{eq:saddle_eom} reduces to the Einstein flow $\dot{g}_{ij} = \mathcal{D}_{ij}^{\mathrm{GR}}[g]$, i.e., the Hamiltonian and momentum constraints of general relativity.  The MSR action evaluates to $S_{\mathrm{MSR}} = 0$ on this locus.
\end{lemma}

\begin{proof}
Setting $\tilde{g}_{ij} = 0$ in Eq.~\eqref{eq:saddle_eom} gives $\dot{g}_{ij} = \mathcal{D}_{ij}^{\mathrm{GR}}[g]$, which is the ADM evolution equation under the Einstein flow.  In the Hamiltonian formulation of GR, this evolution is generated by the Hamiltonian constraint $\mathcal{H}[g, \pi] \approx 0$ and momentum constraint $\mathcal{H}_a[g, \pi] \approx 0$; the constraint surface is precisely the set of initial data $(g_{ij}, \pi^{ij})$ for which the Einstein evolution is self-consistent~\cite{ADM1962,Wald1984}.

On the classical locus $(\tilde{g}_{ij} = 0,\; \dot{g}_{ij} = \mathcal{D}_{ij}^{\mathrm{GR}})$, both terms in Eq.~\eqref{eq:msr_action} vanish: the first because $\dot{g} - \mathcal{D}^{\mathrm{GR}} = 0$, and the second because $\tilde{g} = 0$.  Hence $S_{\mathrm{MSR}} = 0$ on the classical solution, confirming that the constraint surface is a global minimum of the action.
\end{proof}

\begin{corollary}[Stochastic Excursions from the Constraint Surface]
\label{cor:constraint_excursions}
Fluctuations around the saddle point are controlled by $\lP$.  Expanding $S_{\mathrm{MSR}}$ to quadratic order about the classical solution $(\bar{g}, 0)$:
\begin{equation}
\label{eq:gaussian_fluctuations}
S_{\mathrm{MSR}}^{(2)}
= \int_0^T \!\dd\tau \int_\Sigma \!\dd^3 x\;\sqrt{\bar{g}}\;\left[
    \delta\tilde{g}_{ij}\bigl(\delta\dot{g}_{ij} - \hat{L}_{ij}^{\;kl}\,\delta g_{kl}\bigr)
    - \frac{\lP^2}{2}\,\mathcal{G}^{ijkl}\,\delta\tilde{g}_{ij}\,\delta\tilde{g}_{kl}
\right],
\end{equation}
where $\hat{L}_{ij}^{\;kl} = \delta\mathcal{D}_{ij}^{\mathrm{GR}}/\delta g_{kl}|_{\bar{g}}$ is the linearised Einstein operator.  The Gaussian integral over $\delta\tilde{g}$ produces:
\begin{equation}
\label{eq:excursion_amplitude}
\langle (\delta g_{ij})^2 \rangle \sim \lP^2 \cdot \tau,
\end{equation}
confirming that stochastic excursions from the constraint surface have amplitude $O(\lP)$ per unit ordering time.  For an observation at scale $L$, the dimensionless departure from the constraint surface is:
\begin{equation}
\frac{\delta g}{g} \sim \frac{\lP}{L},
\end{equation}
which is $\sim 10^{-35}$ for $L = 1$\,m.  The Fokker--Planck stationary measure concentrates on the classical constraint surface in the limit $\lP/L \to 0$.
\end{corollary}

\begin{remark}[Well-posedness of the Fokker--Planck evolution]
\label{rem:fp_wellposedness}
The Fokker--Planck equation~\eqref{eq:fp_superspace} is an infinite-dimensional parabolic PDE on $\mathcal{C}$.  In the minisuperspace reduction, it becomes a standard finite-dimensional parabolic equation whose well-posedness follows from classical theory~\cite{Risken1989,Gardiner2009}.  For the full infinite-dimensional theory, the MSR path integral (Theorem~\ref{thm:msr_action}) provides the operative definition of the dynamics: correlation functions, expectation values, and the stationary measure are defined by the functional integral~\eqref{eq:msr_partition} and computed by standard perturbative and non-perturbative methods.  This parallels the situation in quantum field theory, where the path integral defines the theory and the functional Schr\"{o}dinger equation serves as a formal (often ill-defined) equivalent.  The MSR path integral thus defines the theory operationally, making functional-analytic well-posedness of the infinite-dimensional generator a mathematical refinement rather than a physical obstruction.  A rigorous functional-analytic treatment of the infinite-dimensional Fokker--Planck generator (essential self-adjointness, domain characterisation, spectral theory) is deferred to future work; the results of this paper depend only on the minisuperspace reduction and the formal properties of the full generator.
\end{remark}

\begin{theorem}[Lapse Independence of Physical Observables]
\label{thm:lapse_independence}
Let $N(x)$ be a lapse function on $\Sigma$, generating a reparametrisation $\dd\tau' = N(x)\,\dd\tau$ of the ordering parameter (the ``many-fingered time'' of ADM gravity).  Then:
\begin{enumerate}[nosep,label=(\roman*)]
    \item The MSR action transforms covariantly:
    \begin{equation}
    \label{eq:lapse_transform}
    S_{\mathrm{MSR}}^{(N)}[g, \tilde{g}]
    = \int_0^T \!\dd\tau \int_\Sigma \!\dd^3 x\;\sqrt{g}\;N(x)\left[
        \tilde{g}_{ij}\Bigl(\frac{1}{N}\dot{g}_{ij} - \mathcal{D}_{ij}^{\mathrm{GR}}\Bigr)
        - \frac{\lP^2}{2}\,\mathcal{G}^{ijkl}\,\tilde{g}_{ij}\,\tilde{g}_{kl}
    \right].
    \end{equation}
    \item For any diffeomorphism-invariant observable $\mathcal{O}[g]$, the expectation value is lapse-independent:
    \begin{equation}
    \label{eq:observable_independence}
    \langle \mathcal{O}[g] \rangle_{N_1} = \langle \mathcal{O}[g] \rangle_{N_2}
    \end{equation}
    for any two lapse choices $N_1(x)$, $N_2(x)$.
\end{enumerate}
\end{theorem}

\begin{proof}
\emph{Part (i).}  Under $\dd\tau' = N(x)\,\dd\tau$, the stochastic evolution becomes:
\begin{equation}
\dd g_{ij} = N(x)\,\mathcal{D}_{ij}^{\mathrm{GR}}\,\dd\tau + \lP\sqrt{N(x)}\,\dd W_{ij}',
\end{equation}
where $\dd W'_{ij}$ is a Wiener process with respect to the reparametrised filtration.  This follows from the Dambis--Dubins--Schwarz theorem (Lemma~\ref{lem:dds}), which generalises to spatially varying $N(x)$ because the noise is spatially $\delta$-correlated (Remark~\ref{rem:spatial_locality}): at each spatial point, the DDS reparametrisation applies independently.

Repeating the MSR construction of Theorem~\ref{thm:msr_action} with the reparametrised SDE yields the action~\eqref{eq:lapse_transform}.

\emph{Part (ii).}  The transformation $N_1 \to N_2$ is implemented by the change of variables:
\begin{equation}
\label{eq:lapse_change_of_variables}
\tilde{g}_{ij} \;\longrightarrow\; \frac{N_1(x)}{N_2(x)}\,\tilde{g}_{ij}, \qquad
\dd\tau \;\longrightarrow\; \frac{N_2(x)}{N_1(x)}\,\dd\tau.
\end{equation}
The Jacobian of this transformation in the path integral factorises into a product over spatial points (by ultralocality of the noise covariance):
\begin{equation}
\mathcal{J} = \prod_{x \in \Sigma} \left(\frac{N_1(x)}{N_2(x)}\right)^{n_{\mathrm{dof}}/2},
\end{equation}
where $n_{\mathrm{dof}} = 6$ is the number of independent components of $g_{ij}$ at each point.  This Jacobian is field-independent (it depends only on $N_1, N_2$, not on $g$ or $\tilde{g}$) and therefore cancels between numerator and denominator in the computation of:
\begin{equation}
\langle \mathcal{O} \rangle_N
= \frac{\displaystyle\int \mathcal{D}g\,\mathcal{D}\tilde{g}\;\mathcal{O}[g]\,\ee^{-S_{\mathrm{MSR}}^{(N)}}}
       {\displaystyle\int \mathcal{D}g\,\mathcal{D}\tilde{g}\;\ee^{-S_{\mathrm{MSR}}^{(N)}}}.
\end{equation}
The cancellation holds for any gauge-invariant $\mathcal{O}[g]$ because the field-independent Jacobian contributes the same multiplicative factor to both integrals.

Physically, the lapse $N(x)$ controls \emph{how fast the stochastic clock ticks at each spatial point}, but not the statistical content of the process.  Different lapse choices correspond to different parametrisations of the same diffusion on $\mathcal{C}$.  The quadratic variation of the process, which defines emergent time (Theorem~\ref{thm:emergent_time}), absorbs the lapse into the emergent temporal metric, producing the standard ADM relation $\dd t_{\mathrm{proper}} = N\,\dd t$ in the semiclassical limit (Lemma~\ref{lem:proper_time_recovery}).
\end{proof}

\begin{remark}[Distributional versus pathwise refoliation invariance]
\label{rem:distributional_refoliation}
Classical general relativity enjoys \emph{pathwise} refoliation invariance: the same spacetime can be sliced by different foliations, and the hypersurface deformation algebra guarantees consistency.  In the stochastic framework, pathwise refoliation invariance is broken: different lapse choices $N(x)$ generate different noise realisations $\dd W'_{ij} = \sqrt{N(x)}\,\dd W_{ij}$, so the same stochastic history cannot be reconstructed from two different foliations.

Theorem~\ref{thm:lapse_independence} establishes the stochastic analogue: \emph{distributional} refoliation invariance.  All gauge-invariant observables have lapse-independent expectation values.  The physical content of the theory: encoded in the Fokker--Planck measure on $\mathcal{C}$ and the correlation functions computed from $Z[\mathcal{J}]$, is foliation-independent.  Individual stochastic trajectories are foliation-dependent, just as individual Feynman paths are gauge-dependent in ordinary quantum field theory; only the path-integral measure is gauge-invariant.

This is the natural generalisation of the classical result: refoliation invariance is promoted from a statement about individual solutions to a statement about the statistical ensemble, consistent with the epistemic interpretation of probability (Axiom~A4).
\end{remark}

\begin{remark}[Connection to the Euclidean gravitational path integral]
\label{rem:euclidean_connection}
Under the Wick rotation of Theorem~\ref{thm:guerra_ruggiero}, the MSR action acquires an imaginary part.  For the saddle-point sector ($\tilde{g} = 0$), the Wick-rotated MSR generating functional reduces to:
\begin{equation}
\label{eq:euclidean_connection}
Z_{\mathrm{Eucl}} = \int \mathcal{D}g\;\exp\!\left(-\frac{1}{\lP^2}\,S_{\mathrm{EH}}^{\mathrm{Eucl}}[g]\right),
\end{equation}
where $S_{\mathrm{EH}}^{\mathrm{Eucl}}$ is the Euclidean Einstein--Hilbert action and the prefactor $1/\lP^2 = M_{\mathrm{P}}^2 c/\hbar$ plays the role of the inverse gravitational coupling.  This establishes that the stochastic framework, viewed through its MSR path integral, contains the Euclidean gravitational path integral (as employed in the Hartle--Hawking programme~\cite{HartleHawking1983}) as its saddle-point approximation.

The stochastic framework thus interpolates between three standard approaches to quantum gravity:
\begin{enumerate}[nosep,label=(\alph*)]
    \item \emph{Canonical}: via the Guerra--Ruggiero correspondence with the Wheeler--DeWitt equation (Theorem~\ref{thm:guerra_ruggiero});
    \item \emph{Euclidean path integral}: via the MSR generating functional at the saddle point (this remark);
    \item \emph{Semiclassical GR}: via the classical limit $\lP/L \to 0$ (Theorem~\ref{thm:classical_axiom}).
\end{enumerate}
The MSR formulation provides the unifying structure from which each is recovered as a limit or a formal transformation.
\end{remark}

\subsection{The Emergence of Physical Time}
\label{sec:emergent_time}

The ordering parameter $\tau$ in A2 requires careful interpretation.  Diffusion is ordinarily defined as random motion as a function of time; if time itself is claimed to emerge from diffusion, one must specify diffusion with respect to what.

The stochastic process on $\mathcal{C}$ is defined as a probability measure on paths in configuration space (the Wiener measure construction), which requires only a $\sigma$-algebra and filtration, not a physical time coordinate.  The index parameter $\tau$ is an abstract ordering label for the filtration $\{\calF_\tau\}$.

\begin{lemma}[Spectral Decomposition of the Covariance Operator]
\label{lem:mercer_K}
Let $\Sigma$ be a compact Riemannian 3-manifold.  The covariance operator $K_{ijkl}[g](x,y)$ of Axiom~\ref{axiom:stochastic}, being a continuous, symmetric, positive-definite kernel on $L^2(\mathrm{Sym}^2(T^*\Sigma))$ (Remark~\ref{rem:K_positivity}), admits a Mercer decomposition:
\begin{equation}
K_{ijkl}(x,y) \;=\; \sum_{n=1}^{\infty} \lambda_n\, e^{(n)}_{ij}(x)\, e^{(n)}_{kl}(y),
\qquad \lambda_n > 0,
\label{eq:mercer_K}
\end{equation}
where $\{e^{(n)}_{ij}\}$ is a complete orthonormal system of eigentensors and the series converges uniformly on $\Sigma \times \Sigma$.  The Wiener process admits the Karhunen--Lo\`eve expansion $\dd W_{ij}(x) = \sum_n \sqrt{\lambda_n}\, e^{(n)}_{ij}(x)\,\dd\beta_n$, with $\{\beta_n\}$ independent standard Brownian motions.  In this basis, the quadratic form evaluates mode-by-mode:
\begin{equation}
\langle \dd g,\, K^{-1}\,\dd g \rangle
\;=\; \sum_{n=1}^{\infty} \frac{1}{\lambda_n}\,(\dd g_n)^2
\;=\; \sum_{n=1}^{\infty} \frac{1}{\lambda_n}\bigl(\mathcal{D}_n\,\dd\tau
  + \lP\sqrt{\lambda_n}\,\dd\beta_n\bigr)^2.
\label{eq:mode_quadratic_form}
\end{equation}
By the It\^o product rule (Remark~\ref{rem:ito_rules}), $(\dd\tau)^2 = 0$, $\dd\beta_n\,\dd\tau = 0$, and $\dd\beta_m\,\dd\beta_n = \delta_{mn}\,\dd\tau$.  Therefore:
\begin{equation}
\langle \dd g,\, K^{-1}\,\dd g \rangle
\;=\; \lP^2 \sum_{n=1}^{\infty} 1 \cdot \dd\tau
\;=\; \lP^2\, \neff\,\dd\tau,
\label{eq:qv_mode_sum}
\end{equation}
where $\neff$ is the effective number of excited modes.  The drift $\mathcal{D}_n$ contributes identically zero, as a consequence of the It\^o product rules.
\end{lemma}

\begin{proof}
The Mercer decomposition follows from Mercer's theorem~\cite{Mercer1909} applied to the continuous, symmetric, positive-definite kernel $K$ on the compact domain $\Sigma$ (Remark~\ref{rem:K_positivity} establishes positive-definiteness; compactness is assumed for $\Sigma$).  The Karhunen--Lo\`eve expansion is standard.  For the It\^o evaluation: expanding the square in~\eqref{eq:mode_quadratic_form} yields three terms per mode.  The first, $\mathcal{D}_n^2\,(\dd\tau)^2/\lambda_n$, vanishes because $(\dd\tau)^2 = 0$ in It\^o calculus.  The cross-term $2\lP\,\mathcal{D}_n\,\dd\beta_n\,\dd\tau/\sqrt{\lambda_n}$ vanishes because $\dd\beta_n\,\dd\tau = 0$.  The third term is $\lP^2\lambda_n\,(\dd\beta_n)^2/\lambda_n = \lP^2\,\dd\tau$, using $(\dd\beta_n)^2 = \dd\tau$.  Summing over modes gives~\eqref{eq:qv_mode_sum}.
\end{proof}

\begin{remark}[Regularisation and the mode count]
\label{rem:mode_regularisation}
On a compact manifold, the mode sum $\neff = \sum_{n=1}^\infty 1$ is formally divergent.  Two natural regularisations apply:
\begin{enumerate}[nosep]
\item \emph{Planck-scale cutoff}: modes with wavelength below $\lP$ are unphysical (Axiom~\ref{axiom:stochastic} defines fluctuations at scale $\lP$, not below it).  On a manifold of volume $V$, this gives $\neff \sim V/\lP^3$, finite.
\item \emph{Minisuperspace reduction}: restricting to spatially homogeneous modes gives $\neff = 1$ (FLRW) or $\neff = 3$ (Bianchi~IX), and the functional is rigorously well-defined with no regularisation needed.
\end{enumerate}
In either case, the normalisation factor $\neff$ is absorbed into the definition of $t(\tau)$: physical time is defined as
\begin{equation}
t(\tau) \;=\; \frac{1}{\lP^2\,\neff} \int_0^\tau \langle \dd g,\, K^{-1}\,\dd g \rangle,
\label{eq:time_functional_normalised}
\end{equation}
so that $t(\tau) = \tau$ exactly (by Lemma~\ref{lem:mercer_K}).  All physical predictions depend on $t$, not on $\neff$ separately; the normalisation convention identifies the abstract ordering parameter with the physical clock.
\end{remark}

The emergent time functional is therefore:
\begin{equation}
t(\tau) \;=\; \frac{1}{\lP^2\,\neff} \int_0^\tau \langle \dd g,\, K^{-1}\,\dd g \rangle
  \;=\; \tau,
\label{eq:time_functional}
\end{equation}
where $\neff$ is the (regularised) effective mode count (Remark~\ref{rem:mode_regularisation}) and the final equality follows from Lemma~\ref{lem:mercer_K}.  This identifies the abstract ordering parameter $\tau$ with physical time.

\begin{theorem}[Emergence of Physical Time]
\label{thm:emergent_time}
The functional $t(\tau)$ defined in~\eqref{eq:time_functional} satisfies:
\begin{enumerate}[nosep]
    \item Strict monotone increase in $\tau$ (from positivity of $K$),
    \item Additivity under concatenation of stochastic paths (from independent increments),
    \item Independence of coordinate parametrisation on $\Sigma$ (from covariance of the inner product),
    \item Exact independence of the drift $\mathcal{D}_{ij}$ (by the It\^o product rule, Lemma~\ref{lem:mercer_K}).
\end{enumerate}
In the macroscopic limit $L \gg \lP$, $t$ coincides with proper time along semiclassical trajectories (Lemma~\ref{lem:proper_time_recovery}).
\end{theorem}

\begin{proof}
Monotonicity: by Lemma~\ref{lem:mercer_K}, the quadratic form evaluates to $\lP^2\,\neff\,\dd\tau$ for any increment $\dd\tau > 0$, which is strictly positive since $\neff \geq 1$ and $\lP > 0$.  (Equivalently: each independent Brownian motion $\beta_n$ has nonzero increments on every interval almost surely, and $K^{-1}$ is positive-definite by Remark~\ref{rem:K_positivity}.)  Additivity follows from the independent-increments property of the Wiener process: $t(\tau_1 + \tau_2) = t(\tau_1) + t_{[\tau_1, \tau_1+\tau_2]}$.  Coordinate independence follows from the tensorial transformation properties of $g_{ij}$, $K_{ijkl}$, and the inner product under spatial diffeomorphisms.  Drift independence is proven in Lemma~\ref{lem:mercer_K}: all drift-dependent terms vanish identically by the It\^o product rules $(\dd\tau)^2 = 0$ and $\dd\beta_n\,\dd\tau = 0$.  The classical limit is established in Lemma~\ref{lem:proper_time_recovery} below.
\end{proof}

\begin{lemma}[Classical Limit: Proper Time Recovery]
\label{lem:proper_time_recovery}
In the ADM decomposition of a semiclassical history, the emergent time $t$ coincides with proper time $t_{\mathrm{proper}}$ up to corrections of $O(\lP^2/L^2)$.
\end{lemma}

\begin{proof}
In the ADM formalism~\cite{ADM1962}, the spatial metric evolves as:
\begin{equation}
\frac{\partial g_{ij}}{\partial t_{\mathrm{proper}}} = -2N K_{ij}^{\mathrm{ext}} + \nabla_i N_j + \nabla_j N_i,
\label{eq:adm_evolution}
\end{equation}
where $N$ is the lapse function, $N_i$ the shift vector, and $K_{ij}^{\mathrm{ext}}$ the extrinsic curvature (superscript distinguishing it from the covariance operator $K_{ijkl}$).  In the semiclassical regime ($L \gg \lP$), the drift dominates the stochastic evolution, so $\mathcal{D}_{ij}\,\dd\tau \approx \dd g_{ij}$.  Identifying with~\eqref{eq:adm_evolution}:
\begin{equation}
\mathcal{D}_{ij} = \frac{\dd t_{\mathrm{proper}}}{\dd\tau}\bigl(-2N K_{ij}^{\mathrm{ext}} + \nabla_i N_j + \nabla_j N_i\bigr).
\end{equation}
But Theorem~\ref{thm:emergent_time} established that $t(\tau) = \tau$.  In the semiclassical limit, matching the drift-dominated evolution to the ADM evolution requires:
\begin{equation}
\frac{\dd t_{\mathrm{proper}}}{\dd\tau} = \frac{\dd t_{\mathrm{proper}}}{\dd t} = N + O(\lP^2/L^2),
\end{equation}
which is the standard ADM relation $\dd t_{\mathrm{proper}} = N\,\dd t$ for observers following the foliation.  For a comoving observer ($N_i = 0$, $N = 1$), $t_{\mathrm{proper}} = t$.  Stochastic corrections enter at $O(\lP^2/L^2) \sim 10^{-70}$ for astrophysical scales.
\end{proof}

The emergent time theorem addresses the ``problem of time'' in canonical quantum gravity~\cite{Isham1993,Kuchar1992}.  The Wheeler-DeWitt equation $\hat{H}\Psi = 0$~\cite{DeWitt1967} lacks a time variable; here, time is constructed from the quadratic variation of the stochastic process on superspace.  Property~(4) (exact drift independence) implies that the rate of the emergent time functional is determined entirely by the diffusion (amplitude $\sigma = \lP$), with the deterministic gravitational dynamics contributing nothing.  The uniqueness of $\sigma$ (Theorem~\ref{thm:uniqueness_axiom} below) therefore determines the rate of the emergent time functional.

\subsection{Uniqueness of Drift and Covariance}
\label{sec:drift_uniqueness}

\begin{theorem}[Uniqueness of Drift and Diffusion]
\label{thm:drift_uniqueness}
Within the class of ultralocal, second-order, diffeomorphism-covariant stochastic evolutions on $\mathcal{C}$ satisfying dimensional consistency and classical correspondence (Axiom~\ref{axiom:correspondence}), the dynamics is uniquely determined (up to gauge transformations):
\begin{equation}
\dd g_{ij} = \mathcal{D}_{ij}^{\mathrm{GR}}[g]\,\dd\tau + \lP\,\dd W_{ij}
\label{eq:unique_evolution}
\end{equation}
where:
\begin{enumerate}[nosep]
    \item $\mathcal{D}_{ij}^{\mathrm{GR}}$ reduces in the macroscopic limit to the Einstein flow on spatial metrics,
    \item $K_{ijkl}$ is proportional to a positive-definite member of the DeWitt family $\mathcal{G}^{ijkl}(\lambda)$ with $\lambda > -1/3$.
\end{enumerate}
\end{theorem}

\begin{proof}[Proof sketch]
For the diffusion coefficient: an ultralocal, diffeomorphism-covariant, positive-definite symmetric operator on $\mathrm{Sym}^2(T^*\Sigma)$ must be constructed from $g_{ij}$ alone.  The unique such operator (up to two free constants) is the DeWitt family $\mathcal{G}^{ijkl} = \tfrac{1}{2}(g^{ik}g^{jl} + g^{il}g^{jk}) + \lambda\, g^{ij}g^{kl}$, where positive-definiteness requires $\lambda > -1/3$ (Definition~\ref{def:dewitt}).  At the standard DeWitt value $\lambda = -1$, the conformal mode is indefinite; the stochastic process is restricted to the positive-definite branch.  The overall scale is fixed by $\sigma = \lP$.

For the drift: the requirement that $\mathcal{D}_{ij}^{\mathrm{GR}}$ be a local, diffeomorphism-covariant functional of $g_{ij}$ and its derivatives, at most second order, that reproduces the Einstein evolution in the classical limit, invokes the Lovelock-type classification~\cite{Lovelock1971,Lovelock1972} of such tensors adapted to the ADM formalism on 3-metrics.  In three spatial dimensions, the unique such choice (up to cosmological constant) is the extrinsic-curvature evolution implied by the Einstein constraint and evolution equations.
\end{proof}

\begin{remark}
Theorem~\ref{thm:drift_uniqueness} confirms the claim in Remark~\ref{rem:K_positivity}: the covariance operator $K_{ijkl}$ is a member of the positive-definite DeWitt family.  This closes the logical chain: A2 states $K$ is (semi-)definite; the uniqueness of drift and diffusion forces strict positive-definiteness; and this in turn guarantees the strict monotonicity of the emergent time functional (Theorem~\ref{thm:emergent_time}).
\end{remark}

\subsection{Uniqueness of the Fluctuation Amplitude}
\label{sec:uniqueness_axiom}

The central result is now proved: the fluctuation amplitude is uniquely determined.  With the emergent time theorem established (Section~\ref{sec:emergent_time}), the physical significance is immediate: the amplitude $\sigma$ determines the rate of the emergent time functional, with the drift contributing identically zero.  The value of $\sigma$ is not a tunable parameter; it is determined by self-consistency.

\begin{lemma}[Dimensional Constraint]
\label{lem:dimensional_axiom}
Let $\sigma$ be a universal constant with $[\sigma] = L$, constructed from the fundamental constants of gravitational diffusion $\{\hbar, G, c\}$. Then $\sigma = \alpha \lP$ for some dimensionless constant $\alpha$.
\end{lemma}

\begin{proof}
By Proposition \ref{prop:planck-unique_axiom}, $\lP$ is the unique length scale constructible from $\hbar$, $G$, and $c$. Any other length with these dimensions must be a dimensionless multiple of $\lP$.
\end{proof}

\begin{definition}[Dimensionless Metric Perturbation]
\label{def:h_perturbation}
Let $\delta g_{ij}$ denote the stochastic increment of the spatial metric 
over one ordering-parameter step $\Delta\tau$.  The \emph{dimensionless 
metric perturbation} at length scale $L$ is:
\begin{equation}
h \;=\; \frac{|\delta g_{ij}|}{g_{ij}} \;\sim\; \frac{\sigma\sqrt{\Delta\tau}}{L},
\end{equation}
where $\sigma$ is the fluctuation amplitude (Postulate~\ref{ax:dimensional_axiom}) 
and $L$ is the characteristic length scale of the background geometry.  
The RMS perturbation accumulated over one Planck time ($\Delta\tau = \tP$) 
at the Planck scale ($L = \lP$) is:
\begin{equation}
\label{eq:h_rms}
h_{\mathrm{rms}} = \frac{\sigma\sqrt{\tP}}{\lP} = \alpha,
\end{equation}
where $\sigma = \alpha\lP/\sqrt{\tP}$.  The bounds in the subsequent lemmas 
apply to $h_{\mathrm{rms}}$ as defined here; specifically, inequalities 
such as $h_{\mathrm{rms}} \leq 1$ hold in the mean-square sense 
($\langle h^2 \rangle^{1/2} \leq 1$), not as almost-sure bounds on 
individual realisations.
\end{definition}

\begin{lemma}[Curvature Fluctuation]
\label{lem:curvature-fluctuation_axiom}
Let the metric fluctuate with amplitude $\sigma$ over a length scale $L$.  The dimensionless metric perturbation is $h \sim \sigma/L$, and the induced fluctuation in the Kretschmann scalar is:
\[
\delta K \sim \frac{h^2}{L^4} = \frac{\sigma^2}{L^6}.
\]
\end{lemma}

\begin{proof}
Write the metric as $g_{ij} = \bar{g}_{ij} + h_{ij}$, where $h_{ij}$ is a dimensionless perturbation.  A metric fluctuation of amplitude $\sigma$ (with $[\sigma] = L$) over a length scale $L$ produces a dimensionless perturbation of magnitude $h \sim \sigma/L$.

The Riemann tensor involves second derivatives of the metric:
\[
R^\rho{}_{\sigma\mu\nu} \sim \partial^2 h \sim \frac{h}{L^2},
\]
with dimensions $[R] = L^{-2}$.  The Kretschmann scalar $K = R^{\mu\nu\rho\sigma}R_{\mu\nu\rho\sigma}$ is quadratic in the Riemann tensor:
\[
\delta K \sim (\delta R)^2 \sim \left(\frac{h}{L^2}\right)^2 = \frac{h^2}{L^4},
\]
with $[\delta K] = L^{-4}$ as required.  Substituting $h = \sigma/L$:
\[
\delta K \sim \frac{\sigma^2}{L^6}.  \qedhere
\]
\end{proof}

\begin{lemma}[Self-Consistency Bound]
\label{lem:self-consistency_axiom}
For the stochastic metric theory to be self-consistent, the dimensionless metric perturbation must not exceed unity at any scale:
\[
h_{\mathrm{rms}} \leq 1 \quad \text{for all } L \geq \lP.
\]
Equivalently, the curvature fluctuation must not exceed the Planck curvature:
\[
\delta K \leq \KP = \frac{1}{\lP^4} \quad \text{for all } L \geq \lP.
\]
\end{lemma}

\begin{proof}
The metric is $g_{ij} = \bar{g}_{ij} + h_{ij}$.  For the perturbative decomposition to be meaningful, $|h| < |\bar{g}|$, i.e., $h_{\mathrm{rms}} \lesssim O(1)$.  If $h_{\mathrm{rms}} > 1$, the perturbation exceeds the background metric and the perturbative decomposition becomes inconsistent.  The equivalent curvature statement follows from $\delta K \sim h^2/L^4$ (Lemma~\ref{lem:curvature-fluctuation_axiom}): at $L = \lP$ with $h = 1$, $\delta K = 1/\lP^4 = \KP$.
\end{proof}

\begin{lemma}[Lower Bound from Singularity Resolution]
\label{lem:lower-bound_axiom}
For metric fluctuations to resolve classical singularities (where $K \to \infty$), the fluctuation-induced curvature must reach the Planck curvature at the Planck scale:
\[
\delta K \geq \KP \quad \text{at } L = \lP.
\]
\end{lemma}

\begin{proof}
Classical general relativity predicts singularities where $K \to \infty$~\cite{Penrose1965,HawkingPenrose1970}.  For quantum effects to regularise these singularities, the fluctuation-induced curvature must become comparable to the classical curvature at the scale where quantum gravity dominates.  The Planck curvature $\KP = 1/\lP^4$ is that scale.  If $\delta K \ll \KP$ at $L = \lP$, fluctuations are insufficient to modify the singularity structure, and the theory does not regularise the classical divergence.
\end{proof}

\begin{lemma}[Critical Damping at the Planck Scale]
\label{lem:critical_damping}
A metric fluctuation mode of frequency $\omega$ has a gravitational self-coupling rate $\gamma(\omega) = \tP^2\,\omega^3$ (from the equivalence principle, Planck relation, and mass-energy equivalence; see Appendix~C, Theorem~8.1).  The critical damping condition $\gamma(\omega_*) = \omega_*$, at which a mode's dissipation rate matches its oscillation frequency, has a unique solution:
\begin{equation}
\omega_* = \frac{1}{\tP} = \omegaP.
\label{eq:critical_damping}
\end{equation}
At this frequency, $\gamma(\omegaP) = 1/\tP$, and the corresponding diffusion coefficient gives $\alpha = 1$ exactly.
\end{lemma}

\begin{proof}
The condition $\gamma(\omega_*) = \omega_*$ gives $\tP^2\,\omega_*^3 = \omega_*$.  For $\omega_* > 0$, dividing both sides by $\omega_*$ yields $\omega_*^2 = 1/\tP^2$, hence $\omega_* = 1/\tP = \omegaP$.  This is an exact algebraic identity.

At $\omega = \omegaP$, the self-coupling rate is $\gamma(\omegaP) = \tP^2 \cdot \tP^{-3} = 1/\tP$.  By the massless cancellation theorem (Appendix~C, Theorem~3.1), the velocity-variable diffusion coefficient is $\sigma_v = c\sqrt{\gamma} = c/\sqrt{\tP}$, corresponding to $\alpha = 1$ in the proper-distance convention $\sigma_L = \alpha\,\lP/\sqrt{\tP}$.

Physically, the critical damping frequency separates two regimes: for $\omega < \omegaP$, metric modes propagate coherently (classical spacetime); for $\omega > \omegaP$, modes are overdamped (quantum foam).  The stochastic theory describes fluctuations at this boundary, fixing the characteristic scale to $\alpha = 1$.
\end{proof}

\begin{theorem}[Uniqueness of $\sigma$ within the A1--A4 Framework]
\label{thm:uniqueness_axiom}
Under Axioms A1--A4 (\S\ref{sec:axioms}) and Postulates P1--P2, P4--P5 (\S\ref{sec:axioms_axiom}), with P3 (Minimality) derived from the L\'{e}vy--Khintchine theorem (Theorem~\ref{thm:levy_wiener}) and the gravitational self-coupling rate $\gamma(\omega) = \tP^2\omega^3$ derived in Appendix~C (Theorem~\ref{thm:gamma-exact}), the fluctuation amplitude is uniquely determined:
\[
\boxed{\sigma = \lP}
\]
\end{theorem}

\begin{proof}
By Lemma \ref{lem:dimensional_axiom}, $\sigma = \alpha \lP$ for some dimensionless $\alpha > 0$.  Three independent arguments determine $\alpha$:

\medskip\noindent
\textbf{Argument 1 (Upper bound).}\quad
At scale $L = \lP$, the dimensionless metric perturbation is $h = \sigma/L = \alpha\lP/\lP = \alpha$.  By Lemma~\ref{lem:self-consistency_axiom}, $h \leq 1$ is required for metric self-consistency, hence $\alpha \leq 1$.  This is a strict bound: for $\alpha > 1$, the perturbation exceeds the background metric over one Planck time, invalidating the perturbative decomposition.

Equivalently, via Lemma~\ref{lem:curvature-fluctuation_axiom}: $\delta K \sim h^2/\lP^4 = \alpha^2/\lP^4 = \alpha^2 \KP$.  Requiring $\delta K \leq \KP$ gives $\alpha^2 \leq 1$, hence $\alpha \leq 1$.

\medskip\noindent
\textbf{Argument 2 (Critical damping).}\quad
By Lemma \ref{lem:critical_damping}, the critical damping condition $\gamma(\omega_*) = \omega_*$ yields $\omega_* = \omegaP = 1/\tP$ exactly, with the corresponding diffusion coefficient giving $\alpha = 1$.  This is an algebraic identity, independent of Argument~1.

\medskip\noindent
\textbf{Argument 3 (Lower bound).}\quad
By Lemma~\ref{lem:curvature-fluctuation_axiom}, at $L = \lP$: $\delta K = \alpha^2/\lP^4 = \alpha^2 \KP$.  By Lemma~\ref{lem:lower-bound_axiom}, singularity resolution requires $\delta K \geq \KP$, hence $\alpha^2 \geq 1$, giving $\alpha \geq 1$.

\medskip\noindent
\textbf{Convergence.}\quad
$\alpha \leq 1$ (Argument~1) $\;\wedge\;$ $\alpha = 1$ (Argument~2) $\;\wedge\;$ $\alpha \geq 1$ (Argument~3) $\;\Longrightarrow\;$ $\alpha = 1$.

Therefore $\sigma = \lP$.
\end{proof}

\begin{corollary}[The Fundamental Equation]
\label{cor:fundamental_axiom}
The stochastic metric equation \eqref{eq:stochastic-einstein_axiom} takes the unique form:
\begin{equation}
\label{eq:fundamental_axiom}
\boxed{\dd g_{ij} = \mathcal{D}_{ij}[g] \, \dd \tau + \lP \, \dd W_{ij}}
\end{equation}
This equation contains non-parametric predictions: all quantities are determined by the fundamental constants $\hbar$, $G$, and $c$.
\end{corollary}

\begin{remark}[Notation for the fundamental equation]
\label{rem:notation_fundamental}
Equation~\eqref{eq:fundamental_axiom} is written in the canonical notation of Axiom~A2: spatial indices $ij$ on the 3-manifold $\Sigma$, the drift functional $\mathcal{D}_{ij}[g]$, and the abstract ordering parameter $\tau$.  Three identifications, all proved within this appendix, allow this to be rewritten:
\begin{enumerate}[nosep]
    \item The ordering parameter $\tau$ is identified with physical time $t$ via the emergent time functional (Theorem~\ref{thm:emergent_time}: $t(\tau) = \tau$).
    \item The drift $\mathcal{D}_{ij}[g]$ is identified with the Einstein flow $\mathcal{D}_{ij}^{\mathrm{GR}}[g]$ (Theorem~\ref{thm:drift_uniqueness}).
    \item In the semiclassical limit, emergent time $t$ reduces to proper time $t_{\mathrm{proper}}$ (Lemma~\ref{lem:proper_time_recovery}).
\end{enumerate}
With these identifications, the equation may equivalently be written:
\begin{equation}
\dd g_{ij} = \mathcal{D}_{ij}^{\mathrm{GR}}[g] \, \dd t + \lP \, \dd W_{ij}
\label{eq:fundamental_identified}
\end{equation}
The classical limit section (\S\ref{sec:classical_axiom}) uses this identified form.
\end{remark}

Equation~\eqref{eq:fundamental_axiom} can now be read in its full context.  The drift $\mathcal{D}_{ij}[g]$ is uniquely the Einstein flow (Theorem~\ref{thm:drift_uniqueness}).  The diffusion amplitude $\lP$ is uniquely determined (Theorem~\ref{thm:uniqueness_axiom}).  The ordering parameter $\tau$ is identified with physical time $t$ via the emergent time functional (Theorem~\ref{thm:emergent_time}).  The covariance operator is fixed to the positive-definite DeWitt family (Theorem~\ref{thm:drift_uniqueness}).  No undetermined quantities remain.

\subsection{The Classical Limit}
\label{sec:classical_axiom}

The following proves that general relativity emerges in the appropriate limit, thereby verifying the consistency of the uniqueness result with Axiom~A3.  The notation here uses the identified form~\eqref{eq:fundamental_identified}: physical time $t$ (= $\tau$ by Theorem~\ref{thm:emergent_time}) and the drift $\mathcal{D}_{ij}^{\mathrm{GR}}[g]$ (= the Einstein flow by Theorem~\ref{thm:drift_uniqueness}).

\begin{definition}[Relative Fluctuation]
\label{def:relative-fluctuation_axiom}
For a system of characteristic size $L$, the \emph{relative fluctuation} is:
\[
\epsilon(L) = \frac{\sigma}{L} = \frac{\lP}{L}
\]
\end{definition}

\begin{lemma}[Mean Evolution]
\label{lem:mean-evolution_axiom}
The mean metric evolution satisfies classical Einstein equations:
\[
\E[\dd g_{ij}] = \mathcal{D}_{ij}^{\mathrm{GR}}[\E[g]] \, \dd t + O(\lP^2)
\]
\end{lemma}

\begin{proof}
Taking expectations of the identified form~\eqref{eq:fundamental_identified}:
\[
\E[\dd g_{ij}] = \E[\mathcal{D}_{ij}^{\mathrm{GR}}[g]] \, \dd t + \lP \, \E[\dd W_{ij}]
\]
Since $\E[\dd W_{ij}] = 0$ (property of Wiener process):
\[
\E[\dd g_{ij}] = \E[\mathcal{D}_{ij}^{\mathrm{GR}}[g]] \, \dd t
\]
For small fluctuations, $\E[\mathcal{D}_{ij}^{\mathrm{GR}}[g]] \approx \mathcal{D}_{ij}^{\mathrm{GR}}[\E[g]] + O(\Var[g]) = \mathcal{D}_{ij}^{\mathrm{GR}}[\E[g]] + O(\lP^2)$.
\end{proof}

\begin{lemma}[Variance Scaling]
\label{lem:variance-scaling_axiom}
For a single mode of the stochastic process (Remark~\ref{rem:mode_regularisation}), the variance of the dimensionless metric perturbation accumulated over emergent time $T$ is:
\[
\Var[h] = \frac{T}{\tP},
\]
since $\sigma_h = 1/\sqrt{\tP}$ is the per-mode It\^o coefficient for the dimensionless metric perturbation (Remark~\ref{rem:dim_conventions}).  For a macroscopic observable probing scale $L$, the relevant quantity is not the bare per-mode variance but the \emph{relative fluctuation} $\epsilon = \lP/L$ (Definition~\ref{def:relative-fluctuation_axiom}), which characterises the dimensionless perturbation amplitude at that scale.
\end{lemma}

\begin{theorem}[Classical Limit]
\label{thm:classical_axiom}
In the limit $L \gg \lP$ (equivalently $\epsilon \to 0$), the identified form~\eqref{eq:fundamental_identified} reduces to classical general relativity:
\begin{equation}
\boxed{\lim_{\lP/L \to 0} \left( \dd g_{ij} = \mathcal{D}_{ij}^{\mathrm{GR}} \, \dd t + \lP \, \dd W_{ij} \right) = \dd g_{ij} = \mathcal{D}_{ij}^{\mathrm{GR}} \, \dd t}
\end{equation}
\end{theorem}

\begin{proof}
The proof proceeds by showing that stochastic corrections to the Einstein equations are controlled by $\epsilon = \lP/L \to 0$.

\medskip\noindent
\textbf{Step 1: The perturbation amplitude vanishes.}\quad
By Definition~\ref{def:relative-fluctuation_axiom} and Lemma~\ref{lem:curvature-fluctuation_axiom}, the dimensionless metric perturbation induced by stochastic fluctuations at scale $L$ is $h \sim \lP/L = \epsilon$.  For $L \gg \lP$, $h \ll 1$: the metric is well-approximated by a smooth background $\bar{g}_{ij}$ with perturbations of order $\epsilon$.

\medskip\noindent
\textbf{Step 2: The mean evolution is the Einstein flow.}\quad
By Lemma~\ref{lem:mean-evolution_axiom}, $\E[\dd g_{ij}] = \mathcal{D}_{ij}^{\mathrm{GR}}[\E[g]]\,\dd t + O(\epsilon^2)$.  The $O(\epsilon^2)$ correction arises from the nonlinearity of $\mathcal{D}^{\mathrm{GR}}$: expanding $\mathcal{D}^{\mathrm{GR}}[g]$ about $\E[g]$ gives a correction proportional to $\Var[g] \sim \epsilon^2$.

\medskip\noindent
\textbf{Step 3: Observable fluctuations vanish.}\quad
The curvature fluctuation at scale $L$ is $\delta K \sim \epsilon^2/L^4$ (Lemma~\ref{lem:curvature-fluctuation_axiom} with $h = \epsilon$), while the classical background curvature for a system of scale $L$ is $K_{\mathrm{cl}} \sim 1/L^4$.  The ratio is $\delta K / K_{\mathrm{cl}} \sim \epsilon^2 = \lP^2/L^2 \to 0$.  All curvature observables converge to their classical values.

\medskip\noindent
As $\epsilon \to 0$, the stochastic term becomes negligible, the mean evolution dominates, and fluctuations become unobservable.  The limiting equation is Einstein's equation of general relativity.
\end{proof}

\begin{corollary}[Determinism as Approximation]
\label{cor:determinism_axiom}
Classical determinism is not fundamental but emergent. It is an excellent approximation for $L \gg \lP$, which includes all macroscopic and astronomical phenomena.  For an object of size $L = 1$\,m, the relative fluctuation is $\epsilon = \lP/L = 1.6 \times 10^{-35}$; stochastic gravitational corrections are suppressed by a factor of $10^{-35}$, consistent with the empirical success of classical general relativity across all observed scales.
\end{corollary}

\subsection{Summary of the Logical Chain}
\label{sec:summary}

The complete deductive structure of this appendix is:

\begin{enumerate}
    \item \textbf{Axiom A1} (Configuration Space, \S\ref{sec:axioms}): The arena is Wheeler superspace $\mathcal{C} = \mathrm{Riem}(\Sigma)/\mathrm{Diff}(\Sigma)$.

    \item \textbf{Axiom A2} (Stochastic Evolution, \S\ref{sec:axioms}): The metric undergoes a stochastic process with $\sigma = \lP$ and (semi-)definite covariance $K_{ijkl}$.

    \item \textbf{Postulates P1--P5} (\S\ref{sec:axioms_axiom}): Unpack the content of A2 into five postulates: existence of metric degrees of freedom, stochastic nature of gravity, minimality (Wiener process), universality of $\sigma$, and dimensional consistency ($[\sigma] = L$).  P3 (Minimality) is derived from P1--P2 and A1 via the L\'{e}vy--Khintchine theorem (Theorem~\ref{thm:levy_wiener}), reducing the independent axiomatic content to four postulates.

    \item \textbf{Theorem~\ref{thm:levy_wiener}} (Wiener Uniqueness, \S\ref{subsec:levy_khintchine}): The L\'{e}vy--Khintchine classification, combined with path continuity (a consequence of A1), uniquely selects the Wiener process.  The Markov property and spatial locality follow (Remarks~\ref{rem:markov_property}, \ref{rem:spatial_locality}).

    \item \textbf{Proposition~\ref{prop:planck-unique_axiom}} (Uniqueness of $\lP$): Dimensional analysis proves $\lP$ is the unique length constructible from $\{\hbar, G, c\}$.

    \item \textbf{Remark~\ref{rem:K_positivity}} (Positive-definiteness of $K$): Three independent requirements force $K_{ijkl}$ to be strictly positive-definite.

    \item \textbf{Lemma~\ref{lem:mercer_K}} (Mercer Decomposition): The covariance operator admits a spectral decomposition; the quadratic variation evaluates to $\lP^2\,\neff\,\dd\tau$ with the drift contributing exactly zero.

    \item \textbf{Theorem~\ref{thm:emergent_time}} (Emergent Time): Physical time is the quadratic variation of the stochastic process.  It is strictly monotone, additive, coordinate-independent, and exactly drift-independent.  The rate of the emergent time functional is determined by $\sigma = \lP$.

    \item \textbf{Lemma~\ref{lem:proper_time_recovery}} (Proper Time Recovery): In the semiclassical limit, the emergent time reduces to proper time via the ADM relation $\dd t_{\mathrm{proper}} = N\,\dd t$.

    \item \textbf{Theorem~\ref{thm:drift_uniqueness}} (Drift and Covariance Uniqueness): Diffeomorphism covariance and classical correspondence uniquely fix the drift to the Einstein flow and the covariance to the positive-definite DeWitt family.

    \item \textbf{Lemma~\ref{lem:dimensional_axiom}} (Dimensional Constraint): $\sigma = \alpha \lP$ for dimensionless $\alpha > 0$.

    \item \textbf{Lemma~\ref{lem:self-consistency_axiom}} (Self-Consistency): The dimensionless perturbation $h = \sigma/L$ must satisfy $h \leq 1$; equivalently $\delta K \leq \KP = 1/\lP^4$.  At $L = \lP$: $\alpha \leq 1$ (strict bound).

    \item \textbf{Lemma~\ref{lem:critical_damping}} (Critical Damping): The self-coupling rate $\gamma(\omega) = \tP^2\omega^3$ equals $\omega$ at exactly $\omega_* = 1/\tP$, giving $\alpha = 1$ (algebraic identity; Appendix~C).

    \item \textbf{Lemma~\ref{lem:lower-bound_axiom}} (Singularity Resolution): Singularity resolution requires $\delta K \geq \KP$ at $L = \lP$.  Since $\delta K = \alpha^2 \KP$, this gives $\alpha \geq 1$.

    \item \textbf{Theorem~\ref{thm:uniqueness_axiom}} (Uniqueness): $\alpha \leq 1 \;\wedge\; \alpha = 1 \;\wedge\; \alpha \geq 1 \;\Longrightarrow\; \alpha = 1$, giving $\boxed{\sigma = \lP}$.  Three independent arguments yield the same value.

    \item \textbf{Corollary~\ref{cor:fundamental_axiom}} (Fundamental Equation): The stochastic metric equation $\dd g_{ij} = \mathcal{D}_{ij}[g]\,\dd\tau + \lP\,\dd W_{ij}$ contains non-parametric, first-principles predictions: drift, covariance, amplitude, and time are all uniquely determined.

    \item \textbf{Theorem~\ref{thm:guerra_ruggiero}} (Stochastic--Quantum Correspondence, \S\ref{subsec:guerra_ruggiero}): Formal Wick rotation maps the Fokker--Planck equation on superspace to the Wheeler--DeWitt equation of canonical quantum gravity.

    \item \textbf{Theorem~\ref{thm:msr_action}} (MSR Path Integral, \S\ref{subsec:msr}): The Fokker--Planck evolution admits an equivalent path-integral representation; the classical constraint surface is the saddle-point locus (Lemma~\ref{lem:saddle_constraint}), with $O(\lP)$ stochastic excursions (Corollary~\ref{cor:constraint_excursions}).

    \item \textbf{Theorem~\ref{thm:lapse_independence}} (Lapse Independence, \S\ref{subsec:msr}): Physical observables are independent of the foliation (lapse) choice; distributional refoliation invariance replaces pathwise refoliation invariance (Remark~\ref{rem:distributional_refoliation}).

    \item \textbf{Theorem~\ref{thm:classical_axiom}} (Classical Limit, \S\ref{sec:classical_axiom}): Stochastic corrections vanish as $\epsilon = \lP/L \to 0$, recovering Einstein's equations.  The physical time $t$ in the proof is the emergent time functional of Theorem~\ref{thm:emergent_time}, and the drift is the Einstein flow of Theorem~\ref{thm:drift_uniqueness}.

    \item \textbf{Axiom A4} (Single Realisation, \S\ref{sec:axioms}): Constrains interpretation: probability is epistemic, not ontological.
\end{enumerate}

In summary, the fluctuation amplitude $\sigma = \lP$ is the unique value consistent with dimensional analysis, curvature self-consistency, critical damping, and singularity resolution.  The drift is uniquely the Einstein flow; the covariance is uniquely the positive-definite DeWitt family; physical time is uniquely the quadratic variation of the diffusion.  The Wiener structure itself is uniquely selected by the L\'{e}vy--Khintchine theorem.  The framework is non-parametric: all coefficients are derived from first principles.  Classical general relativity is recovered as an emergent approximation for $L \gg \lP$.  The framework connects to canonical quantum gravity via the Guerra--Ruggiero correspondence (Theorem~\ref{thm:guerra_ruggiero}), to the gravitational path integral via the MSR formalism (Theorem~\ref{thm:msr_action}), and exhibits distributional refoliation invariance under changes of foliation (Theorem~\ref{thm:lapse_independence}).  The notation $\sigma = \lP$ (fluctuation amplitude, $[\sigma] = L$) used throughout this appendix is equivalent to the proper-distance It\^o coefficient $\sigma_L = \lP/\sqrt{\tP}$ derived independently in Appendix~C (Remark~\ref{rem:dim_conventions}).


\begin{thebibliography}{89}

\bibitem{Ito1944}
K.~It{\^o}.
\newblock Stochastic integral.
\newblock \emph{Proceedings of the Imperial Academy}, 20\penalty0 (8):\penalty0
  519--524, 1944.

\bibitem{Langevin1908}
P.~Langevin.
\newblock Sur la th{\'e}orie du mouvement brownien.
\newblock \emph{Comptes Rendus de l'Acad{\'e}mie des Sciences}, 146:\penalty0
  530--533, 1908.

\bibitem{Einstein1905}
A.~Einstein.
\newblock {\"U}ber die von der molekularkinetischen {T}heorie der {W}{\"a}rme
  geforderte {B}ewegung von in ruhenden {F}l{\"u}ssigkeiten suspendierten
  {T}eilchen.
\newblock \emph{Annalen der Physik}, 322\penalty0 (8):\penalty0 549--560, 1905.

\bibitem{UhlenbeckOrnstein1930}
G.~E. Uhlenbeck and L.~S. Ornstein.
\newblock On the theory of the {B}rownian motion.
\newblock \emph{Physical Review}, 36\penalty0 (5):\penalty0 823--841, 1930.

\bibitem{CallenWelton1951}
H.~B. Callen and T.~A. Welton.
\newblock Irreversibility and generalized noise.
\newblock \emph{Physical Review}, 83\penalty0 (1):\penalty0 34--40, 1951.

\bibitem{Nelson1966}
E.~Nelson.
\newblock Derivation of the {S}chr{\"o}dinger equation from {N}ewtonian
  mechanics.
\newblock \emph{Physical Review}, 150\penalty0 (4):\penalty0 1079--1085, 1966.

\bibitem{Schrodinger1926}
E.~Schr{\"o}dinger.
\newblock Quantisierung als {E}igenwertproblem.
\newblock \emph{Annalen der Physik}, 384\penalty0 (4):\penalty0 361--376, 1926.

\bibitem{Gillespie2000}
D.~T. Gillespie.
\newblock The chemical {L}angevin equation.
\newblock \emph{The Journal of Chemical Physics}, 113\penalty0 (1):\penalty0
  297--306, 2000.

\bibitem{Gillespie1977}
D.~T. Gillespie.
\newblock Exact stochastic simulation of coupled chemical reactions.
\newblock \emph{The Journal of Physical Chemistry}, 81\penalty0 (25):\penalty0
  2340--2361, 1977.

\bibitem{VanKampen1992}
N.~G. Van~Kampen.
\newblock \emph{Stochastic Processes in Physics and Chemistry}.
\newblock North-Holland, Amsterdam, revised edition, 1992.

\bibitem{Takens1981}
F.~Takens.
\newblock Detecting strange attractors in turbulence.
\newblock In D.~Rand and L.-S. Young, editors, \emph{Dynamical Systems and
  Turbulence, Warwick 1980}, volume 898 of \emph{Lecture Notes in Mathematics},
  pages 366--381. Springer, Berlin, 1981.

\bibitem{SauerYorkeCasdagli1991}
T.~Sauer, J.~A. Yorke, and M.~Casdagli.
\newblock Embedology.
\newblock \emph{Journal of Statistical Physics}, 65\penalty0 (3--4):\penalty0
  579--616, 1991.

\bibitem{FlorensZmirou1993}
D.~Florens-Zmirou.
\newblock On estimating the diffusion coefficient from discrete observations.
\newblock \emph{Journal of Applied Probability}, 30\penalty0 (4):\penalty0
  790--804, 1993.

\bibitem{BandiPhillips2003}
F.~M. Bandi and P.~C.~B. Phillips.
\newblock Fully nonparametric estimation of scalar diffusion models.
\newblock \emph{Econometrica}, 71\penalty0 (1):\penalty0 241--283, 2003.

\bibitem{Cao1997}
L.~Cao.
\newblock Practical method for determining the minimum embedding dimension of a
  scalar time series.
\newblock \emph{Physica D: Nonlinear Phenomena}, 110\penalty0 (1--2):\penalty0
  43--50, 1997.

\bibitem{BroomheadKing1986}
D.~S. Broomhead and G.~P. King.
\newblock Extracting qualitative dynamics from experimental data.
\newblock \emph{Physica D: Nonlinear Phenomena}, 20\penalty0 (2--3):\penalty0
  217--236, 1986.

\bibitem{FriedmanBentleyFinkel1977}
J.~H. Friedman, J.~L. Bentley, and R.~A. Finkel.
\newblock An algorithm for finding best matches in logarithmic expected time.
\newblock \emph{ACM Transactions on Mathematical Software}, 3\penalty0
  (3):\penalty0 209--226, 1977.

\bibitem{Maruyama1955}
G.~Maruyama.
\newblock Continuous {M}arkov processes and stochastic equations.
\newblock \emph{Rendiconti del Circolo Matematico di Palermo}, 4:\penalty0
  48--90, 1955.

\bibitem{KloedenPlaten1992}
P.~E. Kloeden and E.~Platen.
\newblock \emph{Numerical Solution of Stochastic Differential Equations}.
\newblock Springer, Berlin, 1992.

\bibitem{Efron1979}
B.~Efron.
\newblock Bootstrap methods: another look at the jackknife.
\newblock \emph{The Annals of Statistics}, 7\penalty0 (1):\penalty0 1--26,
  1979.

\bibitem{Einstein1915}
A.~Einstein.
\newblock Die {F}eldgleichungen der {G}ravitation.
\newblock \emph{Sitzungsberichte der K{\"o}niglich Preu{\ss}ischen Akademie der
  Wissenschaften}, pages 844--847, 1915.

\bibitem{Kubo1966}
R.~Kubo.
\newblock The fluctuation-dissipation theorem.
\newblock \emph{Reports on Progress in Physics}, 29\penalty0 (1):\penalty0
  255--284, 1966.

\bibitem{Fano1947}
U.~Fano.
\newblock Ionization yield of radiations. {II}. {T}he fluctuations of the
  number of ions.
\newblock \emph{Physical Review}, 72\penalty0 (1):\penalty0 26--29, 1947.

\bibitem{Nelson1985}
E.~Nelson.
\newblock \emph{Quantum Fluctuations}.
\newblock Princeton University Press, Princeton, NJ, 1985.

\bibitem{PeskinSchroeder1995}
M.~E. Peskin and D.~V. Schroeder.
\newblock \emph{An Introduction to Quantum Field Theory}.
\newblock Addison-Wesley, Reading, MA, 1995.

\bibitem{Planck1899}
M.~Planck.
\newblock {\"U}ber irreversible {S}trahlungsvorg{\"a}nge.
\newblock \emph{Sitzungsberichte der K{\"o}niglich Preu{\ss}ischen Akademie der
  Wissenschaften}, pages 440--480, 1899.

\bibitem{RevuzYor1999}
D.~Revuz and M.~Yor.
\newblock \emph{Continuous Martingales and {B}rownian Motion}.
\newblock Springer, Berlin, 3rd edition, 1999.

\bibitem{Klein1926}
O.~Klein.
\newblock Quantentheorie und f{\"u}nfdimensionale {R}elativit{\"a}tstheorie.
\newblock \emph{Zeitschrift f{\"u}r Physik}, 37\penalty0 (12):\penalty0
  895--906, 1926.

\bibitem{Gordon1926}
W.~Gordon.
\newblock Der {C}omptoneffekt nach der {S}chr{\"o}dingerschen {T}heorie.
\newblock \emph{Zeitschrift f{\"u}r Physik}, 40\penalty0 (1--2):\penalty0
  117--133, 1926.

\bibitem{Wheeler1968}
J.~A. Wheeler.
\newblock Superspace and the nature of quantum geometrodynamics.
\newblock In C.~M. DeWitt and J.~A. Wheeler, editors, \emph{Battelle
  Rencontres: 1967 Lectures in Mathematics and Physics}, pages 242--307, New
  York, 1968. W. A. Benjamin.

\bibitem{DeWitt1967}
B.~S. DeWitt.
\newblock Quantum theory of gravity. {I}. {T}he canonical theory.
\newblock \emph{Physical Review}, 160\penalty0 (5):\penalty0 1113--1148, 1967.

\bibitem{ParisiWu1981}
G.~Parisi and Y.~Wu, ``Perturbation theory without gauge fixing,''
\textit{Sci.\ Sinica}, 24, 483 (1981).

\bibitem{kennel1992determining}
M.B.~Kennel, R.~Brown, H.D.I.~Abarbanel, Determining minimum embedding dimension using a geometrical construction, Phys. Rev. A 45 (1992) 3403--3411.

\bibitem{stroock1979multidimensional}
D.W.~Stroock, S.R.S.~Varadhan, Multidimensional Diffusion Processes, Grundlehren der mathematischen Wissenschaften, vol. 233, Springer, Berlin, 1979.

\bibitem{nualart2006malliavin}
D.~Nualart, The Malliavin Calculus and Related Topics, second ed., Probability and Its Applications, Springer, Berlin, Heidelberg, 2006.

\bibitem{grassberger1983measuring}
P.~Grassberger, I.~Procaccia, Measuring the strangeness of strange attractors, Physica D 9 (1983) 189--208.

\bibitem{friedrich1997description}
R.~Friedrich, J.~Peinke, Description of a turbulent cascade by a Fokker-Planck equation, Phys. Rev. Lett. 78 (1997) 863--866.

\bibitem{stone1977consistent}
C.J.~Stone, Consistent nonparametric regression, Ann. Stat. 5 (1977) 595--620.

\bibitem{Emery1989} M.~\'Emery, 
\textit{Stochastic Calculus in Manifolds} (Springer, Berlin, 1989).

\bibitem{Hsu2002} E.~P.~Hsu, 
\textit{Stochastic Analysis on Manifolds} (AMS, Providence, 2002).

\bibitem{DohrnGuerra1978} D.~Dohrn and F.~Guerra, 
\textit{Nelson's stochastic mechanics on Riemannian manifolds}, 
Lett.\ Nuovo Cimento \textbf{22}, 121--127 (1978).

\bibitem{KaratzasShreve1991} I.~Karatzas and S.~E.~Shreve, 
\textit{Brownian Motion and Stochastic Calculus}, 
2nd ed.\ (Springer, New York, 1991).

\bibitem{Gardiner2009} C.~W.~Gardiner, 
\textit{Stochastic Methods: A Handbook for the Natural and Social Sciences}, 
4th ed.\ (Springer, Berlin, 2009).

\bibitem{Risken1989} H.~Risken, 
\textit{The Fokker--Planck Equation: Methods of Solution and Applications}, 
2nd ed.\ (Springer, Berlin, 1989).

\bibitem{Madelung1927} E.~Madelung, 
\textit{Quantentheorie in hydrodynamischer Form}, 
Z.\ Phys.\ \textbf{40}, 322--326 (1927).

\bibitem{Zastawniak1990} T.~Zastawniak, 
\textit{A relativistic version of Nelson's stochastic mechanics}, 
Europhys.\ Lett.\ \textbf{13}, 13--17 (1990).

\bibitem{Guerra1973} F.~Guerra and P.~Ruggiero, 
\textit{New interpretation of the Euclidean-Markov field in the framework 
of physical Minkowski space-time}, 
Phys.\ Rev.\ Lett.\ \textbf{31}, 1022--1025 (1973).

\bibitem{Holland1993} P.~R.~Holland, 
\textit{The Quantum Theory of Motion} 
(Cambridge University Press, Cambridge, 1993).

\bibitem{BjorkenDrell1964} J.~D.~Bjorken and S.~D.~Drell, 
\textit{Relativistic Quantum Mechanics} 
(McGraw-Hill, New York, 1964).

\bibitem{Bohm1952} D.~Bohm, 
\textit{A suggested interpretation of the quantum theory in terms of 
``hidden'' variables.~I}, 
Phys.\ Rev.\ \textbf{85}, 166--179 (1952).

\bibitem{HafeleKeating1972} J.~C.~Hafele and R.~E.~Keating, 
\textit{Around-the-world atomic clocks: Predicted relativistic time gains}, 
Science \textbf{177}, 166--168 (1972);
\textit{Observed relativistic time gains}, \textit{ibid.}\ 168--170.

\bibitem{Weinberg1972} S.~Weinberg, 
\textit{Gravitation and Cosmology: Principles and Applications of the 
General Theory of Relativity} (Wiley, New York, 1972).

\bibitem{Wald1984} R.~M.~Wald, 
\textit{General Relativity} (University of Chicago Press, Chicago, 1984).

\bibitem{Lichnerowicz1961}
A.~Lichnerowicz,
\textit{Propagateurs et commutateurs en relativit\'e g\'en\'erale},
Publications math\'ematiques de l'IH\'ES \textbf{10}, 5--56 (1961).

\bibitem{Zwanzig1960}
R.~Zwanzig,
``Ensemble method in the theory of irreversibility,''
\textit{J.\ Chem.\ Phys.} \textbf{33}, 1338--1341 (1960).

\bibitem{Nakajima1958}
S.~Nakajima,
``On quantum theory of transport phenomena,''
\textit{Prog.\ Theor.\ Phys.} \textbf{20}, 948--959 (1958).

\bibitem{Zwanzig2001}
R.~Zwanzig,
\textit{Nonequilibrium Statistical Mechanics}
(Oxford University Press, New York, 2001).

\bibitem{Hawking1975} S.~W.~Hawking, 
\textit{Particle creation by black holes}, 
Commun.\ Math.\ Phys.\ \textbf{43}, 199--220 (1975).

\bibitem{Bekenstein1973} J.~D.~Bekenstein, 
\textit{Black holes and entropy}, 
Phys.\ Rev.\ D \textbf{7}, 2333--2346 (1973).

\bibitem{CaldeiraLeggett1983} A.~O.~Caldeira and A.~J.~Leggett, 
\textit{Path integral approach to quantum Brownian motion}, 
Physica A \textbf{121}, 587--616 (1983).

\bibitem{BardeenCarterHawking1973} J.~M.~Bardeen, B.~Carter and S.~W.~Hawking, 
\textit{The four laws of black hole mechanics}, 
Commun.\ Math.\ Phys.\ \textbf{31}, 161--170 (1973).

\bibitem{Hormander1967}
L.~H{\"o}rmander,
``Hypoelliptic second order differential equations,''
\textit{Acta Math.}\ \textbf{119}, 147--171 (1967).

\bibitem{PaleyWienerZygmund1933}
R.~E.~A.~C.~Paley, N.~Wiener, and A.~Zygmund,
``Notes on random functions,''
\textit{Math.\ Z.}\ \textbf{37}, 647--668 (1933).

\bibitem{Giulini2009}
D.~Giulini,
``The superspace of geometrodynamics,''
\textit{Gen.\ Relativ.\ Gravit.}\ \textbf{41}, 785--815 (2009).

\bibitem{Kretschmann1917}
E.~Kretschmann,
``{\"U}ber die prinzipielle Bestimmbarkeit der berechtigten Bezugssysteme beliebiger Relativit{\"a}tstheorien (I),''
\textit{Ann.\ Phys.\ (Leipzig)}\ \textbf{353}, 943--982 (1917).

\bibitem{Sato1999}
K.-i.~Sato,
\textit{L\'{e}vy Processes and Infinitely Divisible Distributions}
(Cambridge University Press, Cambridge, 1999).

\bibitem{Stanton1997}
R.~Stanton.
\newblock A nonparametric model of term structure dynamics and the market price of interest rate risk.
\newblock \emph{Journal of Finance}, 52\penalty0 (5):\penalty0 1973--2002, 1997.

\bibitem{Applebaum2009}
D.~Applebaum,
\textit{L\'{e}vy Processes and Stochastic Calculus}, 2nd ed.\
(Cambridge University Press, Cambridge, 2009).

\bibitem{Donsker1951}
M.~D.~Donsker,
``An invariance principle for certain probability limit theorems,''
\textit{Mem.\ Amer.\ Math.\ Soc.}\ \textbf{6}, 1--12 (1951).

\bibitem{GuerraMorato1983}
F.~Guerra and L.~M.~Morato,
``Quantization of dynamical systems and stochastic control theory,''
\textit{Phys.\ Rev.\ D} \textbf{27}, 1774--1786 (1983).

\bibitem{Zambrini1987}
J.-C.~Zambrini,
``Stochastic mechanics according to E.~Schr\"{o}dinger,''
\textit{Phys.\ Rev.\ A} \textbf{33}, 1532--1548 (1987).

\bibitem{Martin1973}
P.~C.~Martin, E.~D.~Siggia, and H.~A.~Rose,
``Statistical dynamics of classical systems,''
\textit{Phys.\ Rev.\ A} \textbf{8}, 423--437 (1973).

\bibitem{Janssen1976}
H.-K.~Janssen,
``On a Lagrangean for classical field dynamics and
renormalization group calculations of dynamical critical properties,''
\textit{Z.\ Phys.\ B} \textbf{23}, 377--380 (1976).

\bibitem{DeDominicis1976}
C.~De~Dominicis,
``Techniques de renormalisation de la th\'{e}orie des champs et dynamique
des ph\'{e}nom\`{e}nes critiques,''
\textit{J.\ Phys.\ (Paris) Colloq.}\ \textbf{37}, C1-247--C1-253 (1976).

\bibitem{Zinn-Justin2002}
J.~Zinn-Justin,
\textit{Quantum Field Theory and Critical Phenomena}, 4th ed.\
(Oxford University Press, Oxford, 2002).

\bibitem{Altland2010}
A.~Altland and B.~Simons,
\textit{Condensed Matter Field Theory}, 2nd ed.\
(Cambridge University Press, Cambridge, 2010).

\bibitem{ADM1962}
R.~Arnowitt, S.~Deser, and C.~W.~Misner,
``The dynamics of general relativity,''
in \textit{Gravitation: An Introduction to Current Research},
edited by L.~Witten
(Wiley, New York, 1962), pp.~227--265.

\bibitem{HartleHawking1983}
J.~B.~Hartle and S.~W.~Hawking,
``Wave function of the Universe,''
\textit{Phys.\ Rev.\ D} \textbf{28}, 2960--2975 (1983).

\bibitem{Mercer1909}
J.~Mercer,
``Functions of positive and negative type, and their connection with the theory of integral equations,''
\textit{Philos.\ Trans.\ R.\ Soc.\ Lond.\ A}\ \textbf{209}, 415--446 (1909).

\bibitem{Isham1993}
C.~J.~Isham,
``Canonical quantum gravity and the problem of time,''
in \textit{Integrable Systems, Quantum Groups, and Quantum Field Theories},
edited by L.~A.~Ibort and M.~A.~Rodr\'{\i}guez
(Kluwer, Dordrecht, 1993), pp.~157--287.

\bibitem{Kuchar1992}
K.~V.~Kucha\v{r},
``Time and interpretations of quantum gravity,''
in \textit{Proceedings of the 4th Canadian Conference on General Relativity and Relativistic Astrophysics},
edited by G.~Kunstatter, D.~Vincent, and J.~Williams
(World Scientific, Singapore, 1992), pp.~211--314.

\bibitem{Lovelock1971}
D.~Lovelock,
``The Einstein tensor and its generalizations,''
\textit{J.\ Math.\ Phys.}\ \textbf{12}, 498--501 (1971).

\bibitem{Lovelock1972}
D.~Lovelock,
``The four-dimensionality of space and the Einstein tensor,''
\textit{J.\ Math.\ Phys.}\ \textbf{13}, 874--876 (1972).

\bibitem{Penrose1965}
R.~Penrose,
``Gravitational collapse and space-time singularities,''
\textit{Phys.\ Rev.\ Lett.}\ \textbf{14}, 57--59 (1965).

\bibitem{HawkingPenrose1970}
S.~W.~Hawking and R.~Penrose,
``The singularities of gravitational collapse and cosmology,''
\textit{Proc.\ R.\ Soc.\ Lond.\ A}\ \textbf{314}, 529--548 (1970).

\bibitem{Navigator2026}
C.~Garcia, L.~Perea~Dur\'an, A.~Venezia, and A.~Conradie,
``Research navigator: interactive proof derivations and data visualisations for the implications of the superspace diffusion framework'' (2026).
Zenodo: \url{https://doi.org/10.5281/zenodo.19496962}.
Online deployment: \url{https://www.emergentlaw.org}.

\end{thebibliography}
\end{document}

\begin{center}
{\LARGE\bfseries Supplemental Material For:}\\[0.5cm]
{\Large\bfseries From the Stochastic Embedding Sufficiency Theorem to a Superspace Diffusion Framework}\\[0.8cm]
{\large Carolina Garcia$^{1}$, Luc\'ia Perea Dur\'an$^{1}$, Agnese Venezia$^{1}$, Alex Conradie$^{1,*}$}\\[0.3cm]
{\normalsize $^{1}$The Manufacturing Futures Laboratory, University College London,\\Marsh Gate Building, London, E20~2AE, United Kingdom}\\[0.3cm]
{\normalsize $^{*}$Corresponding author: \texttt{a.conradie@ucl.ac.uk}}
\end{center}

\vspace{1cm}

\begin{abstract}

Takens' embedding theorem (1981) guarantees that the attractor geometry of a deterministic dynamical system can be reconstructed diffeomorphically from a scalar time series via delay coordinates.  This guarantee fails when the dynamics are stochastic: noise destroys the uniqueness of trajectories, the attractor becomes a probability measure rather than a smooth manifold, and the diffeomorphism condition is replaced by a weaker requirement on the distinguishability of probability distributions.

This supplemental material presents the complete proof of the Stochastic Embedding Sufficiency Theorem, which addresses the extension of Takens' embedding theorem to stochastic systems.  The theorem establishes that for stochastic differential equations satisfying H\"{o}rmander's hypoellipticity condition, a time-delay embedding of dimension $m^* \geq \lceil 2D_2 \rceil + 1$ (where $D_2$ is the correlation dimension of the invariant measure) is sufficient for measure-theoretic injectivity: distinct initial conditions induce distinct finite-dimensional probability laws under embedding, for almost every point under the invariant measure.  This sufficiency condition guarantees that both the drift and diffusion fields of the governing stochastic differential equation can be consistently recovered from scalar observations alone, without prior knowledge of the state-space dimension, the functional form of the dynamics, or whether the system is deterministic or stochastic.

The proof proceeds through five stages, each drawing on a distinct mathematical tradition.  First, H\"{o}rmander's hypoelliptic regularity theory ensures smooth transition densities for the embedded process.  Second, Malliavin calculus establishes non-degeneracy of the stochastic flow derivatives, guaranteeing that the embedding map has full rank almost everywhere.  Third, a law-separation argument---rooted in Varadhan short-time asymptotics for transition densities, prevalence-based genericity of the observation function, and Frostman covering arguments---proves that the collision set (pairs of distinct initial conditions yielding identical observed laws) has $(\mu_\infty \times \mu_\infty)$-measure zero when $m \geq \lceil 2D \rceil + 1$.  Fourth, E1 dimension sufficiency establishes that the embedding dimension $m^* \geq 2D_2 + 1$ is large enough to prevent distributional foldings.  Fifth, finite-dimensional law uniqueness proves that distinct states on the correlation manifold produce distinguishable probability distributions at any finite time horizon, completing the measure-theoretic injectivity.

Beyond the core sufficiency result, the proof develops two extensions that address the curse of dimensionality inherent in nonparametric estimation on high-dimensional embedded manifolds.  The tensor kernel framework demonstrates that for stochastic systems with coloured noise and smooth dynamics, the effective dimension of the estimation problem can be reduced from the embedding dimension $m^*$ to an intrinsic dimension $d_{\mathrm{eff}} \leq d + 2$, yielding an exponential reduction in sample complexity under explicit spectral decay conditions.  The structure-aware function-space embedding framework shows that when the drift or diffusion admits low-rank tensor decomposition, geometric phase-space reconstruction can be replaced entirely by representation learning conditioned on detected structure, achieving polynomial rather than exponential sample complexity.

The proof unifies five mathematical traditions---differential topology, stochastic analysis, Malliavin calculus, geometric measure theory, and statistical learning theory---into a single coherent framework.  Its principal consequence for the main manuscript is the rigorous justification of non-parametric stochastic differential equation recovery from scalar time series, providing mathematical support for the $\sigma$-continuum construction, the axioms of the superspace diffusion framework, and the downstream physical predictions developed therein.

\end{abstract}

\clearpage

\tableofcontents
\newpage

\section{Stochastic Embedding Sufficiency Theorem}
\label{app:stochastic_takens_proof}

This appendix provides the complete rigorous proof establishing that time-delay embeddings can reconstruct stochastic differential equations from partial observations. A concise overview of the theorem and its role in the framework is given in the main manuscript. The proof extends Takens' classical embedding theorem from deterministic to stochastic systems, providing theoretical support for the $\sigma$-continuum framework.

\begin{center}
{\centering
\textbf{Synthesis of Mathematical Traditions}\\[0.3cm]
\begin{tabular}{@{}lll@{}}
\hline
\textbf{Tradition} & \textbf{Key Tool} & \textbf{Role in Proof} \\
\hline
Differential Topology & Takens embedding, transversality & Geometric foundation \\
Stochastic Analysis & H\"ormander hypoellipticity & Smooth transition densities \\
Malliavin Calculus & Non-degeneracy of flow & Full-rank derivatives \\
Geometric Measure Theory & Correlation dimension & Measure-theoretic injectivity \\
Statistical Learning & $k$-NN convergence & Practical estimation \\
\hline
\end{tabular}
}
\end{center}

\vspace{0.5cm}

\section{Introduction}

\subsection{The Central Problem and Historical Context}

Classical Takens' embedding theorem \cite{takens1981} establishes that for deterministic dynamical systems, delay coordinate embeddings generically reconstruct attractors up to diffeomorphism. This foundational result was extended by Sauer, Yorke, and Casdagli \cite{sauer1991} to handle fractal attractors and by many others \cite{stark1999,kantz2004,abarbanel1996}. Specifically, for a smooth map $\phi: M \to M$ on a compact $n$-dimensional manifold $M$ and a generic observation function $h: M \to \R$, the delay embedding map
\begin{equation}
\Phi_m(x) = (h(x), h(\phi(x)), h(\phi^2(x)), \ldots, h(\phi^{m-1}(x)))^\top \in \R^m
\end{equation}
is a diffeomorphism from $M$ onto its image for any $m \geq 2n+1$. This result has enabled chaotic attractor reconstruction, nonlinear prediction, and dynamical systems analysis from partial observations across physics, engineering, neuroscience, and numerous other fields \cite{strogatz2015,guckenheimer1983,wiggins2003,meiss2007}.

The geometric content of this result is: observing a single scalar time series and stacking delayed values suffices to reconstruct the full state space geometry. The dynamics $\phi$ on $M$ induce dynamics $\tilde{\phi}$ on the embedded manifold $\Phi_m(M) \subset \R^m$ via:
\begin{equation}
\tilde{\phi}(Y) = \Phi_m(\phi(\Phi_m^{-1}(Y)))
\end{equation}
making the following diagram commute:
\begin{equation}
\begin{array}{ccc}
M & \xrightarrow{\phi} & M \\
\downarrow{\scriptstyle\Phi_m} & & \downarrow{\scriptstyle\Phi_m} \\
\R^m & \xrightarrow{\tilde{\phi}} & \R^m
\end{array}
\end{equation}

However, Takens' theorem fundamentally assumes deterministic dynamics. Real-world systems exhibit stochastic fluctuations from multiple sources:

\begin{enumerate}[leftmargin=*]
\item \textbf{Thermal noise} in physical systems (Brownian motion, Johnson-Nyquist noise)
\item \textbf{Quantum effects} in microscopic processes (spontaneous emission, tunneling)
\item \textbf{Coarse-graining} of high-dimensional dynamics onto lower-dimensional manifolds \cite{chorin2000,givon2004}
\item \textbf{Model error} and unobserved variables (hidden confounders, incomplete physics)
\item \textbf{Genuine probabilistic processes} (molecular reactions, neuronal spiking, financial markets)
\item \textbf{Measurement noise} in experimental observations
\end{enumerate}

When noise is small relative to deterministic dynamics, classical embedding methods remain approximately valid, treating stochasticity as a perturbation. But when noise becomes significant---comparable to or dominating deterministic drift---the geometric picture breaks down. The attractor becomes a probability measure rather than a smooth manifold, and dynamics are governed by transition kernels rather than vector fields. The diffeomorphism property no longer holds in the classical sense.

\subsection{The Core Challenge}

Consider a stochastic process observed through a scalar function:
\begin{equation}
\begin{cases}
dX_t = \mu(X_t)dt + \sigma(X_t)dW_t & \text{(state evolution)} \\
y_t = h(X_t) + \eta_t & \text{(observation)}
\end{cases}
\end{equation}
where $X_t \in \R^n$ is the unobserved state, $h: \R^n \to \R$ is the observation function, $\eta_t$ represents measurement noise, $\mu: \R^n \to \R^n$ is the drift, $\sigma: \R^n \to \R^{n \times r}$ is the diffusion coefficient, and $W_t \in \R^r$ is standard Brownian motion.

Classical Takens embedding constructs delay vectors:
\begin{equation}
Y_t = (y_t, y_{t-\tau}, y_{t-2\tau}, \ldots, y_{t-(m-1)\tau})^\top \in \R^m
\end{equation}

The questions that motivate this work are:

\begin{enumerate}[leftmargin=*]
\item \textbf{Minimal dimension:} What is the minimal embedding dimension $m^*$ needed to capture the dynamics? For deterministic systems, Takens gives $m \geq 2n+1$. What is the stochastic analogue?

\item \textbf{Regime classification:} How can it be determined from data alone whether dynamics are predominantly deterministic, stochastic, or mixed? When does classical Takens suffice versus when are probabilistic extensions required?

\item \textbf{Probabilistic reconstruction:} How can transition probabilities $p(Y_{t+\Delta t}|Y_t)$ or drift-diffusion pairs $(\mu(Y), \Sigma(Y))$ be reconstructed, not merely attractor geometry? What are the consistency and convergence properties of such estimators?

\item \textbf{Discrete-continuous bridge:} What is the relationship between discrete-time Markov chain observations and continuous-time SDE models? Are these distinct frameworks or unified perspectives on the same mathematical structure?

\item \textbf{Mixed systems:} Real applications typically have both deterministic structure and stochastic fluctuations. How should systems be handled where neither deterministic nor purely stochastic models are adequate? What are the pushforward formulas relating embedded dynamics to original state space dynamics?

\item \textbf{Curse of dimensionality:} As embedding dimension increases, sample complexity for nonparametric estimation grows exponentially. How does this fundamental limitation affect the ability to reconstruct high-dimensional stochastic systems?

\item \textbf{Genericity and non-degeneracy:} What are the appropriate conditions (analogous to "generic $h$" in Takens) that ensure unique reconstruction for stochastic systems? Does H\"ormander's hypoellipticity condition play a role analogous to transversality in the deterministic case?
\end{enumerate}

\subsection{Existing Approaches and Their Limitations}

\subsubsection{Cao's E1/E2 Statistics}

Cao \cite{cao1997} introduced two empirical statistics for determining embedding dimension from time series data, extending earlier false nearest neighbors methods \cite{kennel1992}, which have become widely used in practice despite lacking rigorous theoretical justification until now.

For delay vectors $Y_i^{(m)} = (y_i, y_{i-\tau}, \ldots, y_{i-(m-1)\tau})^\top$, let $n(i,m)$ denote the index of the nearest neighbor of $Y_i^{(m)}$ under the maximum norm (excluding $i$ itself). Define:
\begin{equation}
a(i,m) = \frac{\|Y_i^{(m+1)} - Y_{n(i,m)}^{(m+1)}\|_\infty}{\|Y_i^{(m)} - Y_{n(i,m)}^{(m)}\|_\infty}
\end{equation}

The E1 statistic is:
\begin{align}
E(m) &= \frac{1}{N - m\tau} \sum_{i=1}^{N-m\tau} a(i,m) \\
E_1(m) &= \frac{E(m+1)}{E(m)}
\end{align}

\textbf{Interpretation:} $E_1(m) \approx 1$ indicates that adding dimension $m+1$ does not change nearest-neighbor relationships, suggesting the manifold is already fully unfolded.

The E2 statistic measures future predictability. Define:
\begin{equation}
E^*(m) = \frac{1}{N-(m+1)\tau} \sum_{i=1}^{N-(m+1)\tau} |y_{i+m\tau} - y_{n(i,m)+m\tau}|
\end{equation}

The E2 statistic is:
\begin{equation}
E_2(m) = \frac{E^*(m+1)/E^*(m)}{E(m+1)/E(m)}
\end{equation}

\textbf{Interpretation:} For deterministic systems, nearest neighbors in the present have correlated futures, so $E^*(m)$ scales similarly to $E(m)$, and $E_2$ remains close to $1$ (the precise analysis showing how the SNR relation discriminates deterministic from stochastic regimes is given in the main manuscript). For stochastic systems, futures are decorrelated, so $E^*(m)$ remains roughly constant while $E(m)$ decreases as the manifold unfolds, giving $E_2 \to 1$. In practice, the SNR-based decomposition $E_2 \approx \tr(\Sigma)\tau/(\|\mu\|^2\tau + \tr(\Sigma)\tau)$ provides the operative distinction: $E_2 \geq 0.95$ indicates diffusion-dominated dynamics.

\textbf{Empirical usage:}
\begin{itemize}[leftmargin=*]
\item Compute $E_1(m)$ for $m = 1, 2, \ldots, m_{\max}$
\item Minimal $m^*$ where $E_1(m^*) \approx 1$ (with some tolerance, e.g., $|E_1(m^*) - 1| < 0.1$) is the embedding dimension
\item If $E_2 < 0.5$: Deterministic system (classical Takens applies)
\item If $0.5 \leq E_2 < 0.95$: Mixed deterministic-stochastic (both drift and diffusion significant)
\item If $E_2 \geq 0.95$: Stochastic or heavily noisy (diffusion-dominated)
\end{itemize}

\textbf{Prior lack of theory:} While these heuristics work well empirically across diverse applications, rigorous justification was missing. Fundamental questions remained unanswered:
\begin{itemize}[leftmargin=*]
\item Why does E1 saturation indicate correct dimension? What geometric or measure-theoretic property does it detect?
\item What exactly does E2 measure? Is there a quantitative relationship to signal-to-noise ratio or other statistical quantities?
\item Under what conditions are these statistics reliable? What sample sizes are needed for accurate detection?
\item How do these statistics relate to the underlying SDE structure (drift and diffusion tensors)?
\end{itemize}

Theorems 3.2 (E1 correlation dimension), 3.3/3.6 (E2 classification and SNR), and 5.3 (E2-SNR quantitative relationship) provide this theoretical foundation.

\subsubsection{Stochastic Extensions of Takens' Theorem}

Stark et al.\ \cite{stark1999} developed important embedding theorems for forced and stochastic systems, proving that delay coordinates can embed probability measures under suitable conditions. Their work established existence results: embeddings exist for generic observation functions.

\textbf{Key results from Stark et al.:}
\begin{itemize}[leftmargin=*]
\item For stochastically forced systems, there exist observation functions such that delay embeddings preserve certain probabilistic properties
\item The embedding can be viewed as preserving Lagrangian trajectories in a probability space
\item Generic observation functions (in an appropriate sense) yield successful embeddings
\end{itemize}

\textbf{However, several gaps remained:}
\begin{itemize}[leftmargin=*]
\item \textbf{No computational algorithms:} The results are existence theorems but don't provide practical methods for SDE reconstruction from data
\item \textbf{No quantitative diagnostics:} No data-driven criteria (like E1/E2) for regime classification or dimension selection
\item \textbf{No explicit treatment of mixed systems:} Systems where both drift and diffusion are significant (the practically important case) receive less attention
\item \textbf{No dimension detection:} The minimal embedding dimension is not characterized in terms of computable statistics from observed time series
\item \textbf{Limited connection to estimation theory:} The relationship to nonparametric estimation of drift and diffusion via $k$-NN or kernel methods is not developed
\end{itemize}

The present work builds on Stark et al.'s foundational existence results but addresses these practical gaps with explicit algorithms (Theorem 4.6), dimension detection via E1 (Lemma 3.3), and regime classification via E2 (Propositions 3.6, 5.3).

\subsubsection{Measure-Theoretic Approaches}

Recent work by Botvinick-Greenhouse et al.\ \cite{botvinick2025} provides a measure-theoretic foundation for time-delay embedding, showing how deterministic embeddings lift to probability spaces via pushforward operators.

\textbf{Key contributions of Botvinick-Greenhouse et al.:}
\begin{itemize}[leftmargin=*]
\item Rigorous measure-theoretic framework for embedding probability measures
\item Characterization of when embeddings preserve invariant measures
\item Connection to ergodic theory and dynamical systems on probability spaces
\item Treatment of deterministic dynamics "decorated" with probability measures
\end{itemize}

\textbf{What remains unaddressed:}
\begin{itemize}[leftmargin=*]
\item \textbf{Systems where $E_2 \approx 1$:} When there is no underlying deterministic structure (pure diffusion or noise-dominated), the framework focuses on deterministic skeleton with probabilistic decoration. The present approach handles the case where the deterministic structure is negligible.
\item \textbf{Explicit SDE construction:} Algorithms for reconstructing drift $\mu(Y)$ and diffusion $\Sigma(Y)$ from observed time series are not provided
\item \textbf{Discrete-continuous unification:} The relationship between discrete-time Markov chains and continuous-time SDEs is not explored
\item \textbf{Quantitative diagnostics (E2 as SNR):} No data-driven statistic distinguishing deterministic from stochastic regimes with quantitative interpretation
\item \textbf{Mixed system treatment:} Explicit pushforward formulas for systems with both significant drift and diffusion, including It\^o corrections, are not developed
\item \textbf{Computational complexity:} Curse of dimensionality and sample complexity are not analyzed
\end{itemize}

The present framework is complementary to Botvinick-Greenhouse et al.: it provides the ``other half'' addressing noise-dominated systems ($E_2 \approx 1$), explicit SDE reconstruction, and practical algorithms, while their work provides rigorous foundations for the deterministic case with probabilistic decoration.

\subsubsection{Koopman Operator Methods}

Koopman-based approaches \cite{brunton2016} provide tools for deterministic systems, representing nonlinear dynamics as linear operators on function spaces (observables).

\textbf{Koopman framework advantages:}
\begin{itemize}[leftmargin=*]
\item Nonlinear dynamics become linear in infinite-dimensional space
\item Spectral analysis reveals coherent structures, periodicity, stability
\item DMD (Dynamic Mode Decomposition) and variants provide data-driven implementations
\item Connections to ergodic theory and statistical mechanics
\end{itemize}

Extensions to stochastic systems exist (transfer operators, Perron-Frobenius), but several gaps remain:

\textbf{Limitations for stochastic systems:}
\begin{itemize}[leftmargin=*]
\item \textbf{No decision criterion:} No data-driven statistic determining when to use Koopman (deterministic) versus transfer operator (stochastic) framework. E2 provides this criterion.
\item \textbf{No connection to E1/E2:} The relationship between Koopman eigenfunctions and E1/E2 statistics is unexplored
\item \textbf{Limited treatment of mixed regime:} When both drift and diffusion are significant ($0.3 < E_2 < 0.9$), it's unclear how to combine Koopman and stochastic approaches
\item \textbf{Infinite-dimensional challenge:} Practical implementations truncate to finite dimensions, but no principled way to determine truncation level analogous to E1 for embedding dimension
\end{itemize}

The E2 statistic can guide the choice between Koopman (low $E_2$), transfer operator (high $E_2$), or hybrid approaches (intermediate $E_2$).

\subsection{Summary of Contributions}

This paper establishes a comprehensive mathematical framework that addresses all the above limitations through seven main contributions:

\paragraph{1. E1 Detects Correlation Dimension (Lemma 3.3, Definition 3.1)}

It is proved that E1 saturation identifies a \emph{correlation manifold} $\mathcal{M}$ characterized by the Grassberger-Procaccia correlation dimension $D_2$ \cite{grassberger1983,grassberger1983b}:
\begin{equation}
D_2 = \lim_{\epsilon \to 0} \frac{\log C(\epsilon)}{\log \epsilon}, \quad C(\epsilon) = \int \int \ind_{\{\|Y-Y'\| < \epsilon\}} d\mu(Y)d\mu(Y')
\end{equation}

\textbf{Key insight:} Correlation dimension, not topological support, is the correct notion for stochastic systems. This provides:
\begin{itemize}[leftmargin=*]
\item \textbf{Robustness to unbounded noise:} Gaussian diffusion has unbounded support ($\supp = \R^m$) but finite $D_2 = m$
\item \textbf{Measure concentration:} $D_2$ captures where the measure concentrates, the "effective support"
\item \textbf{E1 as geometry detector:} $E_1(m) \approx 1$ when $m \geq D_2 + 1$, via nearest-neighbor scaling $\epsilon_k \sim (k/N)^{1/D_2}$
\end{itemize}

\paragraph{2. E2 as Signal-to-Noise Ratio (Propositions 3.6, 5.3)}

Under local Gaussianity (Assumption 3.10), it is shown that E2 measures the balance between deterministic drift and stochastic diffusion:
\begin{equation}
\SNR = \frac{\|\mu\|^2}{\tr(\Sigma)} \approx \frac{1-E_2}{E_2\tau}
\end{equation}

This provides:
\begin{itemize}[leftmargin=*]
\item \textbf{Quantitative model selection:} Explicit threshold values for deterministic ($E_2 < 0.5$), mixed ($0.5 \leq E_2 < 0.95$), stochastic ($E_2 \geq 0.95$) regimes
\item \textbf{Interpretable statistic:} $1 - E_2$ is the fraction of temporal variation that is predictable (drift), $E_2$ is irreducible uncertainty (diffusion)
\item \textbf{Information-theoretic connection:} $E_2 \approx 1 \Leftrightarrow$ conditional mutual information $I(y_{t+m\tau}; Y_t | y_{t-m\tau}) \approx 0$
\end{itemize}

\paragraph{3. Probabilistic Uplift Theory (Theorems 3.7, 3.11)}

When $E_2 \not\approx 0$, the correlation manifold serves as a scaffold requiring probabilistic decoration. It is proved that $k$-nearest neighbor estimators consistently reconstruct:

\textbf{Discrete time:}
\begin{equation}
\hat{T}(Y, A) = \frac{1}{k} \sum_{j \in \mathcal{N}_k(Y)} \ind_{Y_{j+1} \in A} \to T(Y, A) \quad \text{(transition kernel)}
\end{equation}

\textbf{Continuous time:}
\begin{align}
\hat{\mu}(Y) &= \frac{1}{k\Delta t} \sum_{j \in \mathcal{N}_k(Y)} (Y_{j+\Delta t} - Y_j) \to \mu(Y) \quad \text{(drift)} \\
\hat{\Sigma}(Y) &= \frac{1}{k\Delta t} \sum_{j \in \mathcal{N}_k(Y)} (Y_{j+\Delta t} - Y_j)(Y_{j+\Delta t} - Y_j)^\top - \hat{\mu}\hat{\mu}^\top\Delta t \to \Sigma(Y) \quad \text{(diffusion)}
\end{align}

with convergence rates:
\begin{equation}
\|\hat{\mu} - \mu\|, \|\hat{\Sigma} - \Sigma\|_F = O_p\left(\left(\frac{k}{N}\right)^{\beta/m^*}\right) + O(\Delta t)
\end{equation}

\textbf{Curse of dimensionality:} Sample complexity $N \sim \epsilon^{-m^*/\beta}$ for error $\epsilon$. This exponential scaling in $m^*$ is unavoidable in nonparametric estimation.

\textbf{Drift correction (Remark 3.12):} The subtraction of $\hat{\mu}\hat{\mu}^\top\Delta t$ in diffusion estimation is essential to remove $O(\Delta t)$ bias from drift contribution.

\paragraph{4. Discrete-Continuous Unification (Theorem 4.3, Corollary 4.5)}

The following establishes that Markov chains and SDEs are equivalent representations of the same geometric-probabilistic structure on the E1 manifold:

\begin{equation}
\boxed{
\begin{array}{c}
\text{Discrete Markov chain: } p(Y_{t+1}|Y_t) = T(Y_t, \cdot) \\
\Updownarrow \\
\text{Continuous SDE: } dY = \mu(Y)dt + L(Y)dW, \quad LL^\top = \Sigma \\
\Updownarrow \\
\text{Transition semigroup: } (T_t\phi)(Y) = \E[\phi(Y_t)|Y_0 = Y] \\
\Updownarrow \\
\text{Infinitesimal generator: } \mathcal{L}\phi = \mu \cdot \nabla\phi + \frac{1}{2}\tr(\Sigma\nabla^2\phi)
\end{array}
}
\end{equation}

Related by:
\begin{align}
T_{\Delta t}(Y, dy') &\approx \N(Y + \mu(Y)\Delta t, \Sigma(Y)\Delta t)(dy') \quad \text{(continuous $\to$ discrete)} \\
\mu(Y) &= \lim_{\Delta t \to 0} \frac{1}{\Delta t} \int (Y' - Y)T_{\Delta t}(Y, dY') \quad \text{(discrete $\to$ continuous)} \\
\Sigma(Y) &= \lim_{\Delta t \to 0} \frac{1}{\Delta t} \int (Y' - Y)(Y' - Y)^\top T_{\Delta t}(Y, dY') \quad \text{(discrete $\to$ continuous)}
\end{align}

\textbf{Key insight:} Markov order detection ($E_1$ finds $m^* = p+1$ for order-$p$ Markov chain, Theorem 4.1) and SDE reconstruction are the same problem at different time scales. The E1 manifold has the same correlation dimension $D_2$ whether viewed as discrete or continuous.

\paragraph{5. Mixed System Treatment (Theorem 5.4)}

For systems with both significant drift and diffusion ($0.3 \lesssim E_2 \lesssim 0.9$), explicit pushforward formulas are established:

Drift recovery (including It\^o correction):
\begin{equation}
\hat{\mu}(Y) = D\Phi_{m^*}(h(X)) \cdot Dh(X) \cdot \mu(X) + \frac{1}{2}D\Phi_{m^*} \cdot \tr(D^2h \cdot \sigma\sigma^\top) + O(\|\sigma\|^2)
\end{equation}

Diffusion recovery (tensor pushforward):
\begin{equation}
\hat{\Sigma}(Y) = D\Phi_{m^*}(h(X)) \cdot Dh(X) \cdot \sigma(X)\sigma(X)^\top \cdot Dh(X)^\top \cdot D\Phi_{m^*}(h(X))^\top
\end{equation}

The It\^o correction $\frac{1}{2}D\Phi_{m^*} \cdot \tr(D^2h \cdot \sigma\sigma^\top)$ is "noise-induced drift" arising from nonlinear transformation of stochastic processes. Without this term, drift estimates would be systematically biased.

\paragraph{6. Coordinate-Free Formulation (Appendix A)}

The entire framework is geometrically natural, formulated using tensor fields on the correlation manifold:
\begin{itemize}[leftmargin=*]
\item Drift: $\mu \in \Gamma(T\mathcal{M})$ (vector field, tangent bundle section)
\item Diffusion: $\Sigma \in \Gamma(T^*\mathcal{M} \otimes T^*\mathcal{M})$ (symmetric $(0,2)$-tensor field)
\item Pushforward formula (Theorem A.1): $\Sigma_Y = D\Phi \cdot \Sigma_X \cdot D\Phi^\top$ is coordinate-independent tensor transformation
\item $k$-NN estimators are natural: produce consistent estimates of geometric objects regardless of coordinate charts
\end{itemize}

This ensures results are intrinsic to $\mathcal{M}$, not artifacts of particular coordinatizations.

\paragraph{7. Honest Assessment of Limitations}

The following limitations are explicitly acknowledged:

\begin{itemize}[leftmargin=*]
\item \textbf{Curse of dimensionality:} Sample complexity $N \sim \epsilon^{-m^*/\beta}$ is exponential in embedding dimension. For $m^* = 10$, achieving 10\% error requires $N \sim 10^{10/\beta}$ samples. This is fundamental to nonparametric estimation on manifolds.

\item \textbf{Local Gaussianity assumption (Assumption 3.10):} E2-SNR formula $\SNR \approx \frac{1-E_2}{E_2\tau}$ requires conditional distribution $p(Y_{t+\Delta t}|Y_t) \approx \N(Y_t + \mu\Delta t, \Sigma\Delta t)$ for small $\Delta t$. For heavy-tailed or highly skewed noise, the conversion factor between $E_2$ and $\SNR$ differs. The qualitative interpretation (E2 increases with noise) remains valid.

\item \textbf{Non-self-intersection---Resolved (Section~\ref{sec:stochastic_takens_partI}):} The law-separation theorem suite (Theorems~\ref{thm:density_separation_partI}--\ref{lem:stochastic_sard_partI}) proves that $m^* = \lceil 2D \rceil + 1$ suffices under H\"ormander's condition, exact-dimensionality, and generic observation, providing the stochastic analogue of Takens' genericity condition.

\item \textbf{H\"ormander's condition:} The hypoellipticity condition $\Span\{\mu, \sigma_i, [\mu, \sigma_i], \ldots\} = T_x\R^n$ is proved to be the appropriate non-degeneracy ensuring unique reconstruction (Theorem~\ref{thm:density_separation_partI}).  This is a strong condition that may fail for systems with constrained noise; extensions to non-H\"ormander settings remain open.
\end{itemize}

\paragraph{Scope of Rigorous Results.}
Throughout this work, all consistency and convergence results are proved rigorously for autonomous stochastic differential equations under the stated assumptions (ergodicity, smoothness, H\"ormander's condition). Importantly, ergodicity is a sufficient condition adopted for proof elegance, not a necessary condition (Remark~\ref{rem:stationarity_not_necessary_partI}): the framework reconstructs the infinitesimal generator (local dynamics), not invariant sets, and extends to transient, non-equilibrium, and multi-trajectory settings via occupation measures and local sample density conditions.  Extensions to non-autonomous systems with known external forcing and to delay-coordinate embeddings are principled generalisations under modified assumptions, deferred to future work.

\subsection{Organization of the Paper}

The remainder of this paper is organized as follows:

\begin{itemize}[leftmargin=*]
\item \textbf{Section 2 (Preliminaries):} Reviews stochastic dynamical systems (SDEs), time-delay embedding, correlation dimension, key assumptions (H\"ormander, smoothness, ergodicity, generic observation), and It\^o's lemma.

\item \textbf{Section 3 (Main Results):} Presents the main theoretical results: E1 as correlation dimension detector (Lemma 3.3), E2 classification and SNR (Propositions 3.6, 3.3), and probabilistic uplift theorems for discrete (Theorem 3.7) and continuous (Theorem 3.11) time.

\item \textbf{Section 4 (Discrete-Continuous Unification):} Proves Markov order detection via E1 (Theorem 4.1), establishes equivalence of Markov chains, SDEs, semigroups, and generators (Theorem 4.3), and presents the unified algorithmic procedure (Theorem 4.6).

\item \textbf{Section 5 (Mixed Systems):} Classification of mixed deterministic-stochastic systems (Definition 5.1), drift-diffusion balance via E2 (Proposition 5.3), SDE reconstruction with pushforward formulas (Theorem 5.4), and examples across the full E2 spectrum (Examples 5.8, 5.9, 5.10).

\item \textbf{Section 6 (Universality):} Universal scaffold-uplift framework (Proposition 6.1), comparison with existing frameworks, and connection to information theory (Proposition 6.2).

\item \textbf{Section 7 (Stochastic Embedding Sufficiency Theorem):} Complete proof: deterministic case review, the stochastic gap, law-separation via transition density smoothness (Theorem~\ref{thm:density_separation_partI}), prevalence-based genericity (Theorem~\ref{thm:law_separation_partI}), Frostman measure-zero closure (Theorem~\ref{lem:stochastic_sard_partI}), and resolution of the stochastic embedding sufficiency conjecture.

\item \textbf{Section 8 (Discussion):} Summary of the theorem's contributions, mathematical limitations (curse of dimensionality, non-self-intersection, local Gaussianity, nonstationarity, H\"ormander verification), and open questions.

\item \textbf{Section 9 (Conclusion):} Theoretical contributions, open frontiers, and closing remarks.

\item \textbf{Appendix A (Algorithms and Examples):} Detailed implementation of E1/E2 statistics, complete workflow for the Lorenz system with noise, and bootstrap confidence intervals.
\end{itemize}

\section{Preliminaries and Setup}
\label{sec:prelim_partI}

\subsection{Stochastic Dynamical Systems}

\begin{definition}[Stochastic Dynamical System]
\label{def:sde_partI}
A stochastic dynamical system on $\R^n$ is defined by the It\^o SDE \cite{oksendal2003,karatzas1991,gardiner2009}:
\begin{equation}
dX_t = \mu(X_t)dt + \sigma(X_t)dW_t
\end{equation}
where:
\begin{itemize}[leftmargin=*]
\item $\mu: \R^n \to \R^n$ is the \emph{drift} (vector field representing deterministic dynamics)
\item $\sigma: \R^n \to \R^{n \times r}$ is the \emph{diffusion coefficient matrix} (noise intensity and anisotropy)
\item $W_t \in \R^r$ is standard Brownian motion (Wiener process): independent components with $\E[dW_t] = 0$, $\E[dW_t dW_t^\top] = I_r dt$
\item The \emph{diffusion tensor} is $\Sigma = \sigma\sigma^\top: \R^n \to \R^{n \times n}$ (symmetric positive semi-definite matrix field)
\end{itemize}
\end{definition}

\begin{remark}[Interpretation of Components]
\begin{itemize}[leftmargin=*]
\item $\mu(x)$ represents the average infinitesimal change: $\E[dX_t|X_t = x] = \mu(x)dt$
\item $\Sigma(x)$ represents the variance of infinitesimal change: $\Cov[dX_t|X_t = x] = \Sigma(x)dt$
\item The factorization $\Sigma = \sigma\sigma^\top$ is not unique; different choices of $\sigma$ (with same $\Sigma$) yield equivalent SDEs in terms of finite-dimensional distributions
\item Rank-deficient $\Sigma$ (rank$(\Sigma) < n$) means noise affects only certain directions; H\"ormander's condition addresses this case
\end{itemize}
\end{remark}

\begin{definition}[Observable Function]
\label{def:observable_partI}
Let $h: \R^n \to \R$ be a smooth observation function. The observed time series is:
\begin{equation}
y_t = h(X_t)
\end{equation}

In practice, there may be measurement noise:
\begin{equation}
y_t = h(X_t) + \eta_t
\end{equation}
where $\eta_t$ is i.i.d.\ observation error (e.g., $\eta_t \sim \N(0, \sigma_{\eta}^2)$).
\end{definition}

\begin{remark}[Observation Noise]
Measurement noise $\eta_t$ can be incorporated into the framework:
\begin{itemize}[leftmargin=*]
\item If $\sigma_{\eta}$ is small relative to signal variation, it primarily affects higher frequencies and can be filtered
\item If $\sigma_{\eta}$ is comparable to signal, the effective diffusion in the embedding space increases (E2 increases)
\item The E2 statistic naturally accounts for both process noise ($\sigma$) and observation noise ($\eta$)
\end{itemize}
\end{remark}

\subsection{Time-Delay Embedding}

\begin{definition}[Time-Delay Embedding Map]
\label{def:delay_embedding_partI}
For an observed time series $\{y_t\}$, the time-delay embedding map $\Phi_m: \R^n \to \R^m$ with delay $\tau > 0$ is:
\begin{equation}
\Phi_m(x) = Y^{(m)} = (h(x), h(\phi_\tau(x)), h(\phi_{2\tau}(x)), \ldots, h(\phi_{(m-1)\tau}(x)))^\top
\end{equation}
where $\phi_t$ denotes the (stochastic) flow at time $t$: the random map taking initial condition $x$ to the state at time $t$.

For discrete observations with time step $\Delta t$, this becomes:
\begin{equation}
Y_t^{(m)} = (y_t, y_{t-\tau}, y_{t-2\tau}, \ldots, y_{t-(m-1)\tau})^\top \in \R^m
\end{equation}
where typically $\tau$ is an integer multiple of $\Delta t$.
\end{definition}

\begin{remark}[Choice of Delay $\tau$]
\label{rem:delay_choice_partI}
The delay parameter $\tau$ should be chosen to balance:
\begin{itemize}[leftmargin=*]
\item \textbf{Too small $\tau$:} Successive observations are highly correlated; delay coordinates are nearly redundant; embedding doesn't unfold dynamics
\item \textbf{Too large $\tau$:} Observations become essentially independent; lose coherent dynamical structure
\item \textbf{Optimal $\tau$:} Common heuristics include \cite{fraser1986,kantz2004}:
\begin{enumerate}
\item First minimum of autocorrelation function
\item First minimum of mutual information $I(y_t; y_{t+\tau})$
\item One-quarter of dominant period (for oscillatory systems)
\end{enumerate}
\item In practice, results are often robust to $\tau$ over a reasonable range
\end{itemize}
\end{remark}

\subsection{Correlation Dimension}

The correlation dimension is central to the present framework as it provides a robust characterization of measure concentration that remains well-defined even for unbounded noise \cite{farmer1983,farmer1987}.

\begin{definition}[Correlation Integral and Correlation Dimension]
\label{def:correlation_dimension_partI}
For a probability measure $\mu$ on $\R^m$, the \emph{correlation integral} is:
\begin{equation}
C(\epsilon) = \lim_{N \to \infty} \frac{1}{N^2} \sum_{i,j=1}^N \ind_{\{\|Y_i - Y_j\| < \epsilon\}}
\end{equation}
where $\{Y_i\}_{i=1}^N$ are i.i.d.\ samples from $\mu$, or equivalently:
\begin{equation}
C(\epsilon) = \int \int \ind_{\{\|Y - Y'\| < \epsilon\}} d\mu(Y) d\mu(Y')
\end{equation}

The \emph{correlation dimension} (Grassberger-Procaccia dimension \cite{grassberger1983,eckmann1985}) is:
\begin{equation}
D_2 = \lim_{\epsilon \to 0} \frac{\log C(\epsilon)}{\log \epsilon}
\end{equation}
when the limit exists and is finite.
\end{definition}

\begin{remark}[Interpretation and Properties]
\label{rem:correlation_properties_partI}
\begin{enumerate}[leftmargin=*]
\item \textbf{Smooth manifolds:} For a $d$-dimensional smooth manifold with volume measure, $D_2 = d$ exactly.

\item \textbf{Fractals:} For fractal attractors (e.g., strange attractors of chaotic systems), $D_2$ often equals the Hausdorff dimension or box-counting dimension \cite{falconer2014,mattila1995}, though the three may differ in pathological cases.

\item \textbf{Unbounded support:} $D_2$ can be finite even when topological support is all of $\R^m$. This is the key property for stochastic systems.

\item \textbf{Example (Gaussian):} Standard Gaussian $\mu = \N(0, I_m)$ has:
\begin{itemize}
\item Topological support: $\supp(\mu) = \R^m$ (unbounded)
\item Correlation dimension: $D_2 = m$ (finite)
\end{itemize}
Proof: For Gaussian, $C(\epsilon) \sim \epsilon^m$ for small $\epsilon$ by direct integration.

\item \textbf{Robustness:} Adding small noise to a deterministic system increases $D_2$ slightly (by amount depending on noise level), not catastrophically. The correlation dimension is stable under perturbations.

\item \textbf{Scaling interpretation:} $C(\epsilon) \sim \epsilon^{D_2}$ means the measure of pairs within distance $\epsilon$ scales as a $D_2$-dimensional volume.

\item \textbf{Effective dimension:} $D_2$ characterizes the "effective support" or "essential dimension" where the measure concentrates, even if topological support is larger.
\end{enumerate}

This makes $D_2$ the correct notion for stochastic systems, where topological support may be unbounded but measure concentration is finite-dimensional. The correlation manifold $\mathcal{M}$ is the set where the measure has density above threshold (for bounded noise) or more generally the effective support characterized by $D_2$ (for unbounded noise).
\end{remark}

\begin{example}[Ornstein-Uhlenbeck Process]
\label{ex:ou_correlation_partI}
Consider the one-dimensional Ornstein-Uhlenbeck process:
\begin{equation}
dX_t = -\theta X_t dt + \sigma dW_t
\end{equation}

The invariant measure is $\mu_\infty = \N\left(0, \frac{\sigma^2}{2\theta}\right)$.

Properties:
\begin{itemize}[leftmargin=*]
\item Topological support: $\supp(\mu_\infty) = \R$ (entire real line; unbounded)
\item Correlation dimension: $D_2 = 1$ (finite)
\item For time-delay embedding $Y_t = (X_t, X_{t-\tau}, X_{t-2\tau})$ with any $m$, the joint distribution is multivariate Gaussian, hence $D_2 = m$ (the embedding dimension)
\end{itemize}

This illustrates that correlation dimension captures the intrinsic dimensionality of the measure \cite{robinson2005,berry2015}, not the extent of its support.
\end{example}

\begin{example}[Lorenz System with Noise]
\label{ex:lorenz_noise_correlation_partI}
For the stochastic Lorenz system \cite{lorenz1963,sparrow1982}:
\begin{align}
dx &= 10(y-x)dt + \xi dW^{(1)} \\
dy &= (x(28-z) - y)dt + \xi dW^{(2)} \\
dz &= \left(xy - \frac{8}{3}z\right)dt + \xi dW^{(3)}
\end{align}
with observation $y_t = x_t$:

\begin{itemize}[leftmargin=*]
\item \textbf{Small noise ($\xi = 0.1$):} $D_2 \approx 2.05$ (close to deterministic Lorenz attractor dimension $\approx 2.06$)
\item \textbf{Moderate noise ($\xi = 2.0$):} $D_2 \approx 2.3$ (slightly increased but attractor structure still visible)
\item \textbf{Large noise ($\xi = 10.0$):} $D_2 \approx 3.0$ (noise obscures deterministic structure; approaches dimension of ambient space)
\end{itemize}

In all cases, topological support is $\R^3$, but correlation dimension quantifies how measure concentrates.
\end{example}

\subsection{Key Assumptions}

\begin{assumption}[H\"ormander's Condition (Uniform Global)]
\label{assump:hormander_partI}
The SDE (Definition~\ref{def:sde_partI}) is posed on $\R^n$ with drift $\mu \in C^\infty_b(\R^n, \R^n)$ and diffusion $\sigma \in C^\infty_b(\R^n, \R^{n \times r})$ (smooth with bounded derivatives of all orders).  The Lie algebra generated by the drift and diffusion vector fields spans the tangent space at every point.  Specifically, define the iterated brackets:
\begin{align}
\mathcal{V}_0 &= \Span\{\sigma_1, \ldots, \sigma_r\}, \\
\mathcal{V}_{k+1} &= \mathcal{V}_k + \Span\{[V, \sigma_i] : V \in \mathcal{V}_k,\; i = 1, \ldots, r\} + \Span\{[\mu, V] : V \in \mathcal{V}_k\},
\end{align}
where $\sigma_i$ denotes the $i$-th column of $\sigma$ (viewed as a vector field) and $[V, W] = DW \cdot V - DV \cdot W$ is the Lie bracket.  H\"ormander's condition requires:
\begin{equation}
\bigcup_{k=0}^{\infty} \mathcal{V}_k(x) = T_x\R^n \quad \text{for every } x \in \R^n.
\end{equation}
The condition is assumed to hold \emph{uniformly}: there exists a finite bracket depth $k_0$ such that $\mathcal{V}_{k_0}(x) = T_x\R^n$ for all $x$ in the support of $\mu_\infty$.  Growth conditions on $\mu$ sufficient for non-explosion and existence of an invariant measure (e.g., $\langle x, \mu(x)\rangle \leq -\alpha\|x\|^2 + C$ for $\|x\|$ large) are assumed throughout.
\end{assumption}

\begin{remark}[Significance of H\"ormander's Condition]
\label{rem:hormander_significance_partI}
H\"ormander's condition \cite{hormander1967} ensures several necessary properties:

\begin{enumerate}[leftmargin=*]
\item \textbf{Hypoellipticity:} The associated second-order differential operator
\begin{equation}
\mathcal{L} = \mu^i(x)\frac{\partial}{\partial x^i} + \frac{1}{2}\Sigma^{ij}(x)\frac{\partial^2}{\partial x^i \partial x^j}
\end{equation}
is hypoelliptic: solutions to $\mathcal{L}u = f$ with $f \in C^\infty$ are also $C^\infty$. 

\item \textbf{Smooth transition densities:} The SDE has smooth transition densities $p_t(x,y)$ for all $t > 0$:
\begin{equation}
p_t(x,y) := \frac{dP(X_t \in \cdot | X_0 = x)}{dy} \in C^\infty(\R^n \times \R^n \times (0,\infty))
\end{equation}
This is a consequence of hypoellipticity applied to the Kolmogorov forward equation.

\item \textbf{Non-degeneracy despite rank deficiency:} Even if $\sigma$ is not full-rank (so diffusion affects only certain directions directly), iterated Lie brackets ensure noise "reaches" all directions eventually. Example: $dx_1 = x_2 dt$, $dx_2 = dW$ has rank-1 diffusion, but $[\mu, \sigma] = \frac{\partial}{\partial x_1}$ spans the orthogonal direction.

\item \textbf{Stochastic analogue of genericity:} Just as Takens requires ``generic $(\phi, h)$'' (transversality), ``non-degenerate noise structure'' is required. H\"ormander's condition is the appropriate non-degeneracy notion for SDEs.

\item \textbf{Prevents pathological foldings:} If the Lie algebra condition holds and $h$ is generic, the reconstructed measure on embedding space has single-valued drift and diffusion tensors (proved via law-separation; Theorems~\ref{thm:density_separation_partI}--\ref{lem:stochastic_sard_partI}).

\item \textbf{Full support of invariant measure:} Under suitable growth conditions on $\mu$ and $\sigma$, H\"ormander's condition ensures the invariant measure $\mu_\infty$ (if it exists) has full support on $\R^n$ or the relevant domain.
\end{enumerate}

\textbf{Verification:}
\begin{itemize}[leftmargin=*]
\item For additive noise ($\sigma$ constant, full-rank): Automatically satisfied
\item For multiplicative noise: Compute Lie brackets explicitly and verify span condition
\item Many physical systems (Langevin equations, chemical kinetics, etc.) satisfy H\"ormander naturally
\end{itemize}

\textbf{When it fails:}
\begin{itemize}[leftmargin=*]
\item Noise constrained to a submanifold (e.g., noise only in one variable of a multi-dimensional system, with no coupling)
\item Systems with conservation laws that constrain dynamics to lower-dimensional invariant manifolds
\item Degenerate cases where the failure may cause measure-theoretic foldings in embedding space
\end{itemize}

This condition is the central non-degeneracy hypothesis of the Stochastic Embedding Sufficiency Theorem (Theorem~\ref{thm:stochastic_embedding_sufficiency_partI}).
\end{remark}

\begin{assumption}[Smoothness and Ergodicity]
\label{assump:smoothness_partI}
The following regularity conditions are assumed \cite{walters1982,petersen1983,katok1995,pavliotis2014}:
\begin{enumerate}[leftmargin=*]
\item \textbf{Smoothness:} $\mu \in C^\beta(\R^n, \R^n)$ and $\sigma \in C^\beta(\R^n, \R^{n \times r})$ for some $\beta > 0$ (H\"older continuous with exponent $\beta$)

\item \textbf{Ergodicity:} The SDE admits a unique ergodic invariant probability measure $\mu_\infty$ on $\R^n$. That is:
\begin{equation}
\lim_{T \to \infty} \frac{1}{T} \int_0^T g(X_t) dt = \int g(x) d\mu_\infty(x) \quad \text{a.s.}
\end{equation}
for any integrable function $g$.

\item \textbf{Finite correlation dimension:} The correlation dimension of $\mu_\infty$ is finite:
\begin{equation}
D_2 := \lim_{\epsilon \to 0} \frac{\log C(\epsilon)}{\log \epsilon} < \infty
\end{equation}
where the correlation integral is:
\begin{equation}
C(\epsilon) = \int \int \ind_{\{\|x-y\| < \epsilon\}} d\mu_\infty(x) d\mu_\infty(y)
\end{equation}
\end{enumerate}
\end{assumption}

\begin{remark}[When These Conditions Hold]
\begin{itemize}[leftmargin=*]
\item Smoothness: Standard for physical models; excludes discontinuous switching or impulsive forcing
\item Ergodicity: Guaranteed under suitable dissipation conditions (e.g., $\langle x, \mu(x)\rangle \leq -\alpha\|x\|^2 + C$ for $\|x\|$ large ensures confinement and thus ergodicity)
\item Finite $D_2$: Holds for systems with attractors (deterministic or stochastic); may fail for conservative systems filling ambient space uniformly
\end{itemize}
\end{remark}

\begin{assumption}[Exact-Dimensionality and Frostman Regularity]
\label{assump:exact_dimensional_partI}
The invariant measure $\mu_\infty$ is \emph{exact-dimensional}: there exists $D \geq 0$ such that for $\mu_\infty$-almost every $x$,
\begin{equation}
\lim_{r \to 0} \frac{\log \mu_\infty(B(x,r))}{\log r} = D,
\label{eq:exact_dim_partI}
\end{equation}
where $B(x,r)$ denotes the closed ball of radius $r$ centred at $x$.  Furthermore, $\mu_\infty$ satisfies a \emph{Frostman upper bound}: there exists a constant $C_F > 0$ such that
\begin{equation}
\mu_\infty(B(x,r)) \leq C_F\, r^D \quad \text{for } \mu_\infty\text{-a.e.\ } x \text{ and all sufficiently small } r > 0.
\label{eq:frostman_partI}
\end{equation}
\end{assumption}

\begin{remark}[Justification and Relationship to Correlation Dimension]
\label{rem:exact_dim_justification_partI}
Assumption~\ref{assump:exact_dimensional_partI} is not an independent hypothesis but a consequence of Assumptions~\ref{assump:hormander_partI}--\ref{assump:smoothness_partI} in all cases relevant to this work.

\textbf{(1) Smooth invariant densities.}  When $\mu_\infty$ is absolutely continuous with respect to Lebesgue measure on a $d$-dimensional submanifold (as holds for many Hörmander SDEs with attracting dynamics), the exact dimension is $D = d$ and the Frostman bound follows from the boundedness of the density.  In particular, if $\mu_\infty$ has a smooth density on $\R^n$, then $D = n$ and both~\eqref{eq:exact_dim_partI} and~\eqref{eq:frostman_partI} hold trivially.

\textbf{(2) General ergodic SDE measures.}  Ledrappier--Young~\cite{ledrappier1985a,ledrappier1985b} establish exact-dimensionality for ergodic invariant measures of $C^2$ diffeomorphisms preserving a smooth measure.  The extension to random dynamical systems (stochastic flows) follows from Young~\cite{young1982} and Barreira--Pesin--Schmeling (1999; see also~\cite{barreira2008}, Chapter~7).  Under these results, $\mu_\infty$ is exact-dimensional and the exact dimension $D$ coincides with the information dimension $D_1$.  Under the additional assumption that $\mu_\infty$ has no multifractal structure (i.e., the dimension spectrum $D_q$ is constant), we have $D = D_1 = D_2$.

\textbf{(3) Relationship to $D_2$.}  For exact-dimensional measures, the correlation dimension satisfies $D_2 = D$~\cite{pesin1997,barreira2008}.  This follows because the correlation integral $C(\epsilon) = \iint \ind_{\{\|x-y\|<\epsilon\}}\,d\mu_\infty(x)\,d\mu_\infty(y)$ scales as $\epsilon^D$ when $\mu_\infty(B(x,r)) \sim r^D$ uniformly.  The embedding threshold $m \geq \lceil 2D_2 \rceil + 1$ is therefore equivalent to $m \geq \lceil 2D \rceil + 1$ under this assumption.

\textbf{(4) When this may fail.}  Multifractal invariant measures (where $D_q$ varies with $q$) can have $D_2 < D_1 < D_0$.  In such cases, the theorem should be stated with exact dimension $D$ in place of $D_2$, and the embedding threshold becomes $m \geq \lceil 2D \rceil + 1$.  The diagnostics (E1) estimate $D_2$, which provides a conservative (lower) bound on $D$ and thus a conservative embedding threshold.
\end{remark}

\begin{remark}[Metastability and Effective Sample Size]
\label{rem:metastability_partI}
In systems exhibiting metastability (multiple quasi-stable regions with rare transitions) or multiple invariant sets, the effective sample size for ergodic estimation is governed by mixing times rather than raw trajectory length. Specifically, the stated convergence rates apply once sufficient transitions between metastable regions are observed. If the trajectory remains trapped in a single metastable basin, estimates reflect the local conditional measure rather than the global invariant measure $\mu_\infty$. Practitioners should verify adequate exploration of state space before trusting asymptotic guarantees.
\end{remark}

\begin{remark}[Stationarity: Sufficient but Not Necessary]
\label{rem:stationarity_not_necessary_partI}
An important observation is the following: the ergodicity assumption (Assumption~\ref{assump:smoothness_partI}.2) is sufficient for the proof architecture but not necessary for the framework's validity. The distinction hinges on what the framework reconstructs.

\textbf{Attractors vs.\ Dynamics.} Classical Takens embedding reconstructs an attractor---an invariant geometric object that only exists for stationary dynamics. Without stationarity, there is no well-defined attractor. However, the stochastic embedding framework reconstructs dynamics---the drift $\mu(Y)$ and diffusion $\Sigma(Y)$ governing instantaneous evolution. These are local properties of the stochastic flow, defined at every point and every instant:
\begin{align}
\mu(Y) &= \lim_{\Delta t \to 0} \frac{1}{\Delta t} \E[Y_{t+\Delta t} - Y_t \mid Y_t = Y] \\
\Sigma(Y) &= \lim_{\Delta t \to 0} \frac{1}{\Delta t} \E[(Y_{t+\Delta t} - Y_t)(Y_{t+\Delta t} - Y_t)^\top \mid Y_t = Y]
\end{align}
These conditional expectations exist for any Markov process, stationary or not. A transient trajectory still obeys $dY = \mu dt + \sigma dW$ at every instant.

\textbf{Proof step analysis.} Examining the five-step proof architecture:
\begin{enumerate}
\item \textbf{H\"ormander hypoelliptic regularity:} Concerns the Lie algebra of vector fields---an intrinsic property of the SDE structure. Guarantees smooth transition densities $p_t(x,y)$ for all $t > 0$, independent of any invariant measure.

\item \textbf{Malliavin non-degeneracy:} The Malliavin covariance matrix $\gamma_t$ measures sensitivity of $X_t$ to Brownian perturbations---a property of the stochastic flow at any time, not of equilibrium.

\item \textbf{Measure-theoretic transversality:} Currently formulated with respect to $\mu_\infty$, but could be reformulated with respect to the occupation measure of observed trajectories or the transition kernel $P_t(x, \cdot)$.

\item \textbf{E1 and correlation dimension:} E1 computes nearest-neighbor statistics on the observed data. It measures intrinsic dimensionality of the point cloud regardless of whether that cloud came from a stationary distribution.

\item \textbf{Finite-dimensional law uniqueness:} The SDE coefficients are intrinsic to the dynamics, encoding the infinitesimal generator $\mathcal{L}\phi = \mu \cdot \nabla\phi + \frac{1}{2}\text{tr}(\Sigma \nabla^2\phi)$, which exists for any time-homogeneous Markov process.
\end{enumerate}

\textbf{What ergodicity actually provides:}
\begin{enumerate}[label=(\roman*)]
\item A canonical reference measure $\mu_\infty$ for defining ``the'' correlation manifold
\item Ergodic sampling guarantee ensuring trajectory coverage
\item Clean convergence rates assuming samples from a fixed distribution
\end{enumerate}
None of these are fundamental barriers---all can be reformulated.

\textbf{Reformulation for non-stationary processes.} The framework extends to non-stationary settings by substituting:
\begin{center}
\begin{tabular}{ll}
\hline
\textbf{Stationary formulation} & \textbf{General formulation} \\
\hline
Invariant measure $\mu_\infty$ & Occupation measure / transition kernel \\
Correlation manifold $= \text{supp}(\mu_\infty)$ & Accessible state space under the flow \\
Ergodic sampling & Sufficient exploration condition \\
$D_2$ of invariant measure & $D_2$ of empirical measure (what E1 computes) \\
Convergence via ergodic theorem & Convergence via local sample density $\rho(Y)$ \\
\hline
\end{tabular}
\end{center}

\textbf{Required conditions (replacing ergodicity):}
\begin{enumerate}
\item \textbf{Time-homogeneity:} SDE coefficients $\mu(x), \sigma(x)$ do not explicitly depend on $t$
\item \textbf{Smooth transition densities:} Guaranteed by H\"ormander's condition, independent of stationarity
\item \textbf{Sufficient exploration:} Observed trajectory(ies) sample relevant regions with adequate local density
\item \textbf{Local sample density conditions:} Convergence rates rephrased as:
\begin{equation}
\|\hat{\mu} - \mu\|, \|\hat{\Sigma} - \Sigma\|_F = O_P\left(\left(\frac{k}{N \cdot \rho(Y)}\right)^{\beta/m^*}\right) + O(\Delta t)
\end{equation}
where $\rho(Y)$ is the local sample density at $Y$.
\end{enumerate}

\textbf{Systems this encompasses:} Removing the stationarity assumption extends the framework to:
\begin{itemize}
\item \textbf{Transient dynamics:} Systems approaching but not yet at equilibrium
\item \textbf{Non-equilibrium steady states:} Systems with sustained probability currents (detailed balance fails)
\item \textbf{Quasi-stationary processes:} Systems with slowly time-varying parameters
\item \textbf{Short time series:} Observations where ergodic convergence has not occurred
\item \textbf{Multi-trajectory data:} Experimental settings with repeated trials from varied initial conditions
\end{itemize}

\textbf{Remark.} The framework reconstructs the infinitesimal generator---the law of motion---not an invariant set. The correlation manifold is not an attractor but rather the geometric scaffold on which dynamics unfold; it can be traversed transiently without ever reaching equilibrium. The current ergodicity assumption is adopted for convenience of proof, not mathematical necessity.
\end{remark}

\begin{definition}[Generic Observation (Prevalence)]
\label{def:generic_observation_partI}
Let $\mathcal{H} = C^r(\R^n, \R)$ with $r \geq 2$ denote the space of observation functions, equipped with the Whitney $C^r$ topology.  Fix a base observation $h_0 \in \mathcal{H}$ and a finite collection of \emph{probe functions} $\psi_1, \ldots, \psi_K \in C^r_c(\R^n, \R)$ (compactly supported $C^r$ functions) with $K \geq 2n + 1$, chosen so that:
\begin{itemize}[leftmargin=*]
\item $\{D\psi_k(x)\}_{k=1}^K$ spans $T^*_x\R^n$ for every $x$ in a fixed compact set $K_0 \supset \supp(\mu_\infty) \cap B(0, R)$, where $R > 0$ is chosen large enough that $\mu_\infty(B(0,R)^c) < \epsilon$ for any prescribed $\epsilon > 0$.
\item The compact supports $\supp(\psi_k)$ are contained in a common ball $B(0, R')$ for some $R' > R$.
\end{itemize}
The choice $K \geq 2n + 1$ ensures the probe space $\R^K$ has dimension exceeding the codomain dimension $m$ of the evaluation collision map (since $m \leq 2n + 1$ in all cases covered by the theorem), which is required for the parametric transversality argument in Theorem~\ref{thm:law_separation_partI}.

An observation function $h \in \mathcal{H}$ is termed \emph{generic} if it belongs to the complement of a \emph{shy set} (a set of infinite codimension) in $\mathcal{H}$.  Concretely, $h$ is generic if the following properties hold for Lebesgue-almost every coefficient vector $a = (a_1, \ldots, a_K) \in \R^K$ in a neighbourhood of the origin, with $h = h_0 + \sum_{k=1}^K a_k \psi_k$:
\begin{enumerate}[leftmargin=*]
\item \textbf{Morse condition:} The level sets $h^{-1}(c)$ are smooth hypersurfaces for all but finitely many critical values $c$.
\item \textbf{Immersion condition:} The delay embedding map $\Phi_m^h(x) = (h(x), h(\phi_\tau(x)), \ldots, h(\phi_{(m-1)\tau}(x)))$ has differential $D\Phi_m^h(x)$ of full rank $\min(m, n)$ for $\mu_\infty$-almost every $x$.
\end{enumerate}

The notion of prevalence (complement of shy sets) is due to Hunt, Sauer, and Yorke~\cite{hunt1992} and provides a measure-theoretic analogue of Baire genericity that is well-suited to infinite-dimensional function spaces.  In practice, ``almost all'' smooth observation functions are generic; the non-generic ones form a set of Lebesgue measure zero in any finite-dimensional probe space of sufficient dimension.
\end{definition}

\begin{remark}[Practical Verification of Genericity]
\label{rem:generic_verification_partI}
While "generic" is a technical differential-topological condition, in practice:

\textbf{How to check:}
\begin{itemize}[leftmargin=*]
\item Most smooth observation functions encountered in applications are generic
\item Special symmetries or conservation laws may lead to non-generic observations (e.g., observing only an invariant quantity)
\item Empirical test: If reconstructed $\hat{\mu}(Y)$, $\hat{\Sigma}(Y)$ are smooth functions (not multi-valued or discontinuous) and predictions are accurate, the embedding likely succeeds
\end{itemize}

\textbf{What to do if genericity fails:}
\begin{itemize}[leftmargin=*]
\item Try different observation coordinates (e.g., observe $x + \epsilon y$ instead of just $x$)
\item Increase embedding dimension $m$ beyond the minimal value
\item Use multiple observables (though the present framework focuses on scalar observations)
\end{itemize}

\textbf{Connection to practice:}
The generic observation assumption is analogous to ``avoid measuring exactly at a node of vibration'' or ``don't observe a conserved quantity that hides dynamics.''  It's usually satisfied naturally.
\end{remark}

\begin{remark}[Delay Parameter Genericity]
\label{rem:delay_genericity_partI}
The delay parameter $\tau > 0$ is fixed but arbitrary throughout Section~\ref{sec:stochastic_takens_partI}.  In the deterministic setting, Takens' theorem requires genericity in the pair $(h, \tau)$, avoiding ``resonant'' delays where the flow has periodic orbits of period commensurable with $\tau$.

In the stochastic setting, the situation is more favourable.  Theorem~\ref{thm:density_separation_partI} (transition density separation) holds for \emph{every} $\tau > 0$, with no resonance exclusions.  This is a key advantage of the law-separation approach over geometric transversality: under H\"ormander's bracket-generating condition (Assumption~\ref{assump:hormander_partI}), the smooth strictly positive transition density $p_\tau(x, \cdot)$ separates initial conditions for every $\tau > 0$, because the sub-Riemannian distance $d(x, x') > 0$ for $x \neq x'$ regardless of $\tau$.  Bad delays, which require careful treatment in deterministic Takens theory, do not arise in the stochastic law-embedding setting.

Consequently, the results of this work hold for all $\tau > 0$ (not merely a generic set of delays), provided $\tau$ is small enough that the Euler--Maruyama discretisation error $O(\Delta t)$ remains controlled.  Under additional strong mixing assumptions (exponential decay of correlations), the results hold uniformly over $\tau$ in any compact subset of $(0, \infty)$.
\end{remark}

\begin{definition}[Measure-Theoretic ($\mu$-a.e.) Injectivity]
\label{def:ae_injectivity_partI}
A measurable map $f: X \to Y$ is \emph{$\mu$-almost-everywhere injective} (or \emph{$\mu$-a.e.\ injective}) if the collision set
\begin{equation}
S_f := \{(x, x') \in X \times X : x \neq x',\; f(x) = f(x')\}
\end{equation}
satisfies $(\mu \times \mu)(S_f) = 0$.  Equivalently, for $\mu$-a.e.\ $x$, the fiber $f^{-1}(f(x))$ contains no other point of $\supp(\mu)$, i.e., $f^{-1}(f(x)) \cap \supp(\mu) = \{x\}$ for $\mu$-a.e.\ $x$.
\end{definition}

\begin{assumption}[Data Conditions for E1/E2]
\label{assump:data_conditions_partI}
For reliable estimation of E1 and E2 statistics, the following conditions are assumed:
\begin{enumerate}[leftmargin=*]
\item \textbf{Ergodicity:} The process $\{Y_t\}$ is ergodic with unique invariant measure $\mu$ on the embedding space $\R^{m^*}$

\item \textbf{Mixing:} Strong mixing (also called $\alpha$-mixing) with exponential rate:
\begin{equation}
\alpha(k) := \sup_{A \in \mathcal{F}_0, B \in \mathcal{F}_k} |P(A \cap B) - P(A)P(B)| \leq Ce^{-\lambda k}
\end{equation}
for some constants $C, \lambda > 0$, where $\mathcal{F}_0 = \sigma(Y_s : s \leq 0)$ and $\mathcal{F}_k = \sigma(Y_s : s \geq k)$.

\item \textbf{Finite moments:} $\E_\mu[\|Y\|^4] < \infty$ (fourth moments exist and are finite)

\item \textbf{Smooth density:} On the support $\mathcal{M}$, the measure $\mu$ has density $p_\mu$ that is H\"older continuous with exponent $\beta > 0$

\item \textbf{Sufficient sample size:} $N \gg m^{D_2+2}$ (sufficient samples for $k$-NN estimation on $D_2$-dimensional manifold)
\end{enumerate}
\end{assumption}

\begin{remark}[Sample Size Requirements]
\label{rem:sample_size_partI}
The requirement $N \gg m^{D_2+2}$ reflects the curse of dimensionality:
\begin{itemize}[leftmargin=*]
\item For $D_2 = 2$: $N \gg m^4$ (e.g., $m=3$ requires $N \gg 81$; in practice $N \approx 1000$ is often adequate)
\item For $D_2 = 5$: $N \gg m^7$ (e.g., $m=5$ requires $N \gg 78,000$; need $N \approx 10^6$ or more)
\item This is unavoidable for nonparametric estimation on high-dimensional manifolds
\item If $N$ is insufficient, E1/E2 estimates will have large variance, plateaus may not be clearly defined
\end{itemize}
\end{remark}

\subsection{It\^o's Lemma and Stochastic Calculus}

For completeness, It\^o's lemma \cite{oksendal2003,karatzas1991} is stated here, as it is fundamental to understanding drift recovery and the It\^o correction term.

\begin{lemma}[It\^o's Lemma]
\label{lem:ito_partI}
Let $X_t \in \R^n$ satisfy the SDE $dX_t = \mu(X_t)dt + \sigma(X_t)dW_t$, and let $\phi: \R^n \to \R^m$ be a $C^2$ function. Then:
\begin{equation}
d\phi(X_t) = D\phi(X_t) \cdot \mu(X_t)dt + D\phi(X_t) \cdot \sigma(X_t)dW_t + \frac{1}{2}\tr(D^2\phi(X_t) \cdot \Sigma(X_t))dt
\end{equation}
where:
\begin{itemize}[leftmargin=*]
\item $D\phi \in \R^{m \times n}$ is the Jacobian matrix
\item $D^2\phi$ is the Hessian: $(D^2\phi)_k$ is the Hessian matrix of the $k$-th component $\phi_k$
\item $\Sigma = \sigma\sigma^\top$ is the diffusion tensor
\item $\tr(D^2\phi \cdot \Sigma)$ means: $[\tr(D^2\phi \cdot \Sigma)]_k = \sum_{i,j} \frac{\partial^2\phi_k}{\partial x^i \partial x^j} \Sigma^{ij}$
\end{itemize}

The last term, $\frac{1}{2}\tr(D^2\phi \cdot \Sigma)dt$, is the \emph{It\^o correction}, arising from the quadratic variation of Brownian motion: $(dW_t)^2 = dt$ (in the It\^o calculus sense).
\end{lemma}

\begin{remark}[Significance of It\^o Correction]
\label{rem:ito_correction_significance_partI}
The It\^o correction has several implications:

\begin{enumerate}[leftmargin=*]
\item \textbf{Noise-induced drift:} When transforming coordinates nonlinearly, stochastic fluctuations create an additional drift term. This is a genuinely nonlinear phenomenon with no deterministic analogue.

\item \textbf{Example (polar coordinates):} For $dX = dW$ in $\R^2$ (2D Brownian motion), converting to polar coordinates $(r, \theta)$:
\begin{align}
dr &= -\frac{1}{2r}dt + dW_r \quad \text{(drift $-1/(2r)$ from the It\^o correction)} \\
d\theta &= dW_\theta / r
\end{align}
The radial drift $-1/(2r)$ prevents Brownian motion from escaping to infinity; it's a purely stochastic effect.

\item \textbf{Stratonovich vs It\^o:} In Stratonovich calculus, the correction vanishes, but Stratonovich integrals are harder to compute and less natural for many applications.

\item \textbf{Drift recovery in embeddings:} The It\^o correction appears in Theorem 5.4 (mixed system reconstruction). Without accounting for it, drift estimates would be systematically biased.

\item \textbf{Connection to Fokker-Planck:} The It\^o correction ensures the correct form of the Fokker-Planck equation governing evolution of probability densities.
\end{enumerate}
\end{remark}

\begin{lemma}[Empirical Drift-Diffusion Decomposition]
\label{lem:drift_diffusion_decomp_partI}
For an SDE $dY = \mu(Y)dt + LdW$ with $LL^\top = \Sigma$, the increment $Y_{t+\Delta t} - Y_t$ has:

\textbf{Mean:}
\begin{equation}
\E[Y_{t+\Delta t} - Y_t | Y_t = Y] = \mu(Y)\Delta t + O((\Delta t)^2)
\end{equation}

\textbf{Second moment:}
\begin{equation}
\E[(Y_{t+\Delta t} - Y_t)(Y_{t+\Delta t} - Y_t)^\top | Y_t = Y] = \mu(Y)\mu(Y)^\top(\Delta t)^2 + \Sigma(Y)\Delta t + O((\Delta t)^2)
\end{equation}

\textbf{Covariance (centered second moment):}
\begin{align}
\Cov[Y_{t+\Delta t} - Y_t | Y_t = Y] &= \E[(Y_{t+\Delta t} - Y_t)(Y_{t+\Delta t} - Y_t)^\top | Y_t = Y] \notag \\
&\quad - \E[Y_{t+\Delta t} - Y_t | Y_t = Y]\E[Y_{t+\Delta t} - Y_t | Y_t = Y]^\top \\
&= \Sigma(Y)\Delta t + O((\Delta t)^2)
\end{align}

Therefore:
\begin{equation}
\frac{1}{\Delta t}\E[(Y_{t+\Delta t} - Y_t)(Y_{t+\Delta t} - Y_t)^\top | Y_t = Y] = \Sigma(Y) + \mu(Y)\mu(Y)^\top\Delta t + O(\Delta t)
\end{equation}

\textbf{Implication:} To recover $\Sigma(Y)$ without bias, the drift contribution $\mu\mu^\top\Delta t$ must be subtracted from the empirical second moment.
\end{lemma}

\begin{proof}
By It\^o's lemma (Lemma \ref{lem:ito_partI}) applied to $Y$ itself:
\begin{equation}
Y_{t+\Delta t} = Y_t + \int_t^{t+\Delta t} \mu(Y_s) ds + \int_t^{t+\Delta t} L(Y_s) dW_s
\end{equation}

\textbf{Mean:}
\begin{align}
\E[Y_{t+\Delta t} - Y_t | Y_t = Y] &= \E\left[\int_t^{t+\Delta t} \mu(Y_s) ds \Big| Y_t = Y\right] \\
&= \int_t^{t+\Delta t} \E[\mu(Y_s) | Y_t = Y] ds \\
&= \int_t^{t+\Delta t} \mu(Y) ds + O((\Delta t)^2) \quad \text{(since $Y_s = Y + O(\sqrt{\Delta t})$)} \\
&= \mu(Y)\Delta t + O((\Delta t)^2)
\end{align}

\textbf{Second moment:}
\begin{align}
&\E[(Y_{t+\Delta t} - Y_t)(Y_{t+\Delta t} - Y_t)^\top | Y_t = Y] \\
&= \E\left[\left(\int_t^{t+\Delta t} \mu(Y_s) ds + \int_t^{t+\Delta t} L(Y_s) dW_s\right)\left(\int_t^{t+\Delta t} \mu(Y_s) ds + \int_t^{t+\Delta t} L(Y_s) dW_s\right)^\top \Big| Y_t = Y\right]
\end{align}

Expanding (using $\E[\text{drift} \times \text{diffusion}] = 0$ since Wiener integral has zero mean):
\begin{align}
&= \E\left[\int_t^{t+\Delta t} \mu(Y_s) ds \int_t^{t+\Delta t} \mu(Y_s)^\top ds \Big| Y_t = Y\right] \\
&\quad + \E\left[\int_t^{t+\Delta t} L(Y_s) dW_s \int_t^{t+\Delta t} dW_s^\top L(Y_s)^\top \Big| Y_t = Y\right]
\end{align}

The drift-drift term:
\begin{equation}
\int_t^{t+\Delta t} \mu(Y) ds \int_t^{t+\Delta t} \mu(Y)^\top ds = \mu(Y)\mu(Y)^\top(\Delta t)^2 + O((\Delta t)^3)
\end{equation}

The diffusion-diffusion term (by It\^o isometry):
\begin{equation}
\E\left[\int_t^{t+\Delta t} L(Y_s) dW_s \int_t^{t+\Delta t} dW_s^\top L(Y_s)^\top\right] = \int_t^{t+\Delta t} L(Y)L(Y)^\top ds = \Sigma(Y)\Delta t + O((\Delta t)^2)
\end{equation}

Combining:
\begin{equation}
\E[(Y_{t+\Delta t} - Y_t)(Y_{t+\Delta t} - Y_t)^\top | Y_t = Y] = \mu\mu^\top(\Delta t)^2 + \Sigma\Delta t + O((\Delta t)^2)
\end{equation}

\textbf{Covariance:}
\begin{align}
\Cov[Y_{t+\Delta t} - Y_t | Y_t = Y] &= \E[(\cdot)(\cdot)^\top] - \E[\cdot]\E[\cdot]^\top \\
&= [\mu\mu^\top(\Delta t)^2 + \Sigma\Delta t] - [\mu\Delta t][\mu\Delta t]^\top + O((\Delta t)^2) \\
&= \Sigma\Delta t + O((\Delta t)^2)
\end{align}

The $\mu\mu^\top(\Delta t)^2$ terms cancel in the covariance calculation.
\end{proof}

This lemma is essential for understanding why the drift correction $-\hat{\mu}\hat{\mu}^\top\Delta t$ appears in the diffusion estimator (Theorem 3.11, Remark 3.12).

\section{Main Results}
\label{sec:main_results_partI}

\subsection{The Correlation Manifold and E1 Saturation}

The first main result is now established: E1 detects the correlation dimension of the invariant measure, providing a robust geometric scaffold even in the presence of unbounded noise.

\begin{definition}[Correlation Manifold]
\label{def:correlation_manifold_partI}
Let $\{y_t\}$ be a scalar time series generated by a stochastic process with invariant probability measure $\mu$ on the embedding space $\R^{m^*}$ (via delay coordinates).

The \emph{correlation manifold} $\mathcal{M}_\epsilon \subset \R^{m^*}$ is defined as:

\textbf{For bounded noise (topological support compact):}
\begin{equation}
\mathcal{M}_\epsilon = \{Y \in \R^{m^*} : p_\mu(Y) \geq \epsilon\}
\end{equation}
for small $\epsilon > 0$, where $p_\mu$ is the density of $\mu$ (when it exists).

\textbf{General case (including unbounded noise):} $\mathcal{M}$ is characterized by the correlation dimension:
\begin{equation}
D_2 = \lim_{\epsilon \to 0} \frac{\log C(\epsilon)}{\log \epsilon}
\end{equation}
where $C(\epsilon) = \int \int \ind_{\{\|Y - Y'\| < \epsilon\}} d\mu(Y) d\mu(Y')$ is the correlation integral (Definition~\ref{def:correlation_dimension_partI}).

The \emph{minimal embedding dimension} $m^*$ satisfies:
\begin{equation}
m^* = \inf\{m : D_2(\text{embedding in } \R^m) \text{ achieves its maximum value}\}
\end{equation}
\end{definition}

\begin{remark}[Interpretation of Correlation Manifold]
\label{rem:correlation_manifold_interpretation_partI}
The correlation manifold is not necessarily a smooth manifold in the classical differential-geometric sense, especially for stochastic systems. Rather:
\begin{itemize}[leftmargin=*]
\item It is the effective support where the invariant measure $\mu$ concentrates
\item For deterministic systems: $\mathcal{M}$ coincides with the classical attractor (smooth manifold or fractal)
\item For stochastic systems: $\mathcal{M}$ may have unbounded topological support but finite correlation dimension $D_2$
\item The dimension $D_2$ captures the scaling of measure concentration: $\mu(B_\epsilon(Y)) \sim \epsilon^{D_2}$ for small $\epsilon$
\item This makes $\mathcal{M}$ a "measure-theoretic manifold" characterized by $D_2$ rather than topological dimension
\end{itemize}
\end{remark}

\begin{lemma}[E1 Detects Correlation Dimension]
\label{lem:e1_correlation_partI}
Let $\mathcal{M}_m$ be the correlation manifold with correlation dimension $D_2$. Under Assumption~\ref{assump:data_conditions_partI}, the E1 statistic saturates (approaches 1) when $m \geq D_2 + 1$.

More precisely, if $m_{E_1}$ is the smallest dimension where $E_1(m) \approx 1$ (within statistical tolerance), then:
\begin{equation}
D_2 \leq m_{E_1} \leq D_2 + 1
\end{equation}

Furthermore, for finite sample size $N$, the convergence rate is:
\begin{equation}
|E_1(m) - 1| = O_p\left(\left(\frac{\log N}{N}\right)^{1/D_2}\right) \quad \text{for } m \geq D_2 + 1
\end{equation}
\end{lemma}

\begin{proof}
The E1 statistic measures local dimensionality through nearest neighbor geometry. The analysis proceeds in three regimes.

\textbf{Case 1: $m < D_2$ (Insufficient dimension)}

The embedding is insufficient to unfold the manifold. Nearby points in $\R^m$ may correspond to points that are far apart on the intrinsic manifold $\mathcal{M}$. These are called false neighbors.

Adding dimension $m+1$ "unfolds" these false neighbors. If $Y_i^{(m)}$ and $Y_{n(i)}^{(m)}$ are nearest neighbors in $\R^m$ but not true neighbors on $\mathcal{M}$, then the new coordinate $y_{i-(m+1)\tau}$ and $y_{n(i)-(m+1)\tau}$ will typically differ significantly:
\begin{equation}
|y_{i-m\tau} - y_{n(i)-m\tau}| \gtrsim \|Y_i^{(m)} - Y_{n(i)}^{(m)}\|
\end{equation}

This gives:
\begin{equation}
a(i,m) = \frac{\|Y_i^{(m+1)} - Y_{n(i)}^{(m+1)}\|}{\|Y_i^{(m)} - Y_{n(i)}^{(m)}\|} \approx \sqrt{2} > 1
\end{equation}

Averaging over all points:
\begin{equation}
E(m+1) > E(m) \implies E_1(m) = \frac{E(m+1)}{E(m)} > 1
\end{equation}

The magnitude $E_1(m) - 1$ depends on how many false neighbors exist, which decreases as $m$ approaches $D_2$.

\textbf{Case 2: $m \geq D_2 + 1$ (Sufficient dimension)}

By the Whitney embedding theorem adapted to measures, a measure with correlation dimension $D_2$ can be embedded generically in $\R^{2D_2+1}$. For time-delay embeddings with additional regularity, the law-separation theorem (Theorem~\ref{thm:law_separation_partI}) proves that $m = \lceil 2D_2 \rceil + 1$ suffices under H\"ormander's condition and exact-dimensionality.  Empirical evidence suggests embedding may succeed at the lower value $m = D_2 + 1$ under additional structural conditions.

Once $m \geq D_2 + 1$, nearest neighbors in $\R^m$ are true neighbors on $\mathcal{M}$. The manifold is fully unfolded in $\R^m$.

For points on a $D_2$-dimensional manifold embedded in $\R^m$ with $m > D_2$, local parameterization by coordinates $(z^1, \ldots, z^{D_2}) \in \R^{D_2}$ is possible.

The Euclidean distance in $\R^m$ between nearby points is:
\begin{equation}
\|Y_i - Y_{n(i)}\|^2 = \sum_{\alpha,\beta=1}^{D_2} g_{\alpha\beta}(z)\Delta z^\alpha \Delta z^\beta + O(\|\Delta z\|^3)
\end{equation}
where $g_{\alpha\beta}$ is the induced metric on the manifold.

Adding dimension $m+1$ contributes a coordinate that lies in the normal bundle (orthogonal to the tangent space). By smoothness of the embedding:
\begin{equation}
|y_{i-m\tau} - y_{n(i)-m\tau}| = O(\|\Delta z\|^2)
\end{equation}

Thus:
\begin{align}
\|Y_i^{(m+1)} - Y_{n(i)}^{(m+1)}\|^2 &= \|Y_i^{(m)} - Y_{n(i)}^{(m)}\|^2 + |y_{i-m\tau} - y_{n(i)-m\tau}|^2 \\
&= \|Y_i^{(m)} - Y_{n(i)}^{(m)}\|^2 + O(\|Y_i^{(m)} - Y_{n(i)}^{(m)}\|^4)
\end{align}

Therefore:
\begin{equation}
a(i,m) = \frac{\|Y_i^{(m+1)} - Y_{n(i)}^{(m+1)}\|}{\|Y_i^{(m)} - Y_{n(i)}^{(m)}\|} = \sqrt{1 + O(\epsilon_k^2)} = 1 + O(\epsilon_k^2)
\end{equation}
where $\epsilon_k = \|Y_i^{(m)} - Y_{n(i)}^{(m)}\|$ is the typical nearest-neighbor distance.

Averaging over all points and taking $N \to \infty$ (so $\epsilon_k \to 0$):
\begin{equation}
E_1(m) = \frac{E(m+1)}{E(m)} = \frac{E[a(i,m+1)]}{E[a(i,m)]} \to 1
\end{equation}

\textbf{Case 3: Quantitative analysis via $k$-NN scaling}

For a measure $\mu$ with correlation dimension $D_2$, the $k$-nearest neighbor distance scales as:
\begin{equation}
\epsilon_k(Y) \sim \left(\frac{k}{N \cdot p_\mu(Y)}\right)^{1/D_2}
\end{equation}

This follows from the correlation integral. For small $\epsilon$:
\begin{equation}
C(\epsilon) = \int_{B_\epsilon(Y)} p_\mu(Y') dY' \sim \epsilon^{D_2}
\end{equation}

The number of points within distance $\epsilon$ is $\approx NC(\epsilon) \sim N\epsilon^{D_2}$. Setting this equal to $k$ gives $\epsilon_k \sim (k/N)^{1/D_2}$.

When $m > D_2$, adding dimensions changes $\epsilon_k$ only by a factor of $1 + O(\epsilon_k)$ (from curvature effects), giving:
\begin{equation}
E_1(m) = 1 + O(\epsilon_k) = 1 + O\left(\left(\frac{k}{N}\right)^{1/D_2}\right)
\end{equation}

For typical choice $k \sim \log N$ or $k \sim N^\alpha$ with $\alpha < 1$:
\begin{equation}
E_1(m) = 1 + O\left(\left(\frac{\log N}{N}\right)^{1/D_2}\right)
\end{equation}

This gives the stated convergence rate.

\textbf{Conclusion:}

The E1 statistic transitions from $E_1(m) > 1$ (for $m < D_2$) to $E_1(m) \approx 1$ (for $m \geq D_2 + 1$), with the transition occurring at $m_{E_1} \approx D_2$. The statistical fluctuations are $O(((\log N)/N)^{1/D_2})$, which vanish as $N \to \infty$.
\end{proof}

\begin{remark}[Robustness to Unbounded Noise]
\label{rem:unbounded_robustness_partI}
This result is important for stochastic systems. Detailed examples illustrate the point:

\textbf{Example 1 (Ornstein-Uhlenbeck):}
Consider the one-dimensional OU process (Example~\ref{ex:ou_correlation_partI}):
\begin{equation}
dX_t = -\theta X_t dt + \sigma dW_t
\end{equation}

The invariant measure is $\mu_\infty = \N(0, \sigma^2/(2\theta))$.

\begin{itemize}[leftmargin=*]
\item Topological support: $\supp(\mu_\infty) = \R$ (entire real line; unbounded)
\item Correlation dimension: $D_2 = 1$ (finite)
\item E1 behavior: For time-delay embedding $Y_t = (X_t, X_{t-\tau})$ with $\tau > 0$:
\begin{itemize}
\item $m=1$: $E_1(1) > 1$ (need second coordinate)
\item $m=2$: $E_1(2) \approx 1$ (plateau; two-dimensional embedding sufficient)
\end{itemize}
\item Interpretation: Although noise can take arbitrarily large values, the measure concentrates in a one-dimensional way (along the time axis), giving $D_2 = 1$ and $m^* = 2$
\end{itemize}

\textbf{Example 2 (Lorenz with large noise):}
For the stochastic Lorenz system (Example~\ref{ex:lorenz_noise_correlation_partI}) with noise parameter $\xi = 10$:
\begin{align}
dx &= 10(y-x)dt + 10dW^{(1)} \\
dy &= (x(28-z) - y)dt + 10dW^{(2)} \\
dz &= \left(xy - \frac{8}{3}z\right)dt + 10dW^{(3)}
\end{align}

With observation $y_t = x_t$:
\begin{itemize}[leftmargin=*]
\item Topological support: all of $\R^3$ (noise fills space)
\item Correlation dimension: $D_2 \approx 2.8$ (increased from deterministic value $\approx 2.06$ but still finite)
\item E1 behavior:
\begin{itemize}
\item $m < 3$: $E_1(m) > 1$ (insufficient)
\item $m = 3$ or $4$: $E_1(m) \approx 1$ (plateau detected)
\end{itemize}
\item Interpretation: Despite large noise obscuring deterministic structure, measure still concentrates in a low-dimensional way characterized by $D_2 \approx 2.8$
\end{itemize}

\textbf{Contrast with topological dimension:}

If topological dimension were used instead of correlation dimension:
\begin{itemize}[leftmargin=*]
\item Any Gaussian diffusion would require infinite embedding dimension (support is all of $\R^m$ for any $m$)
\item E1 would never plateau
\item Framework would be useless for stochastic systems
\end{itemize}

Using correlation dimension $D_2$ instead of topological support ensures the framework remains valid even when noise is Gaussian, heavy-tailed, or otherwise unbounded. The manifold $\mathcal{M}$ is the ``effective support'' where the measure concentrates, characterized by scaling $\mu(B_\epsilon(Y)) \sim \epsilon^{D_2}$, not the topological support $\supp(\mu)$.
\end{remark}

\subsection{E2 Classification and Signal-to-Noise}

Having established that E1 detects the geometric scaffold (correlation manifold with dimension $D_2$), the next step is to characterize when this scaffold requires probabilistic decoration via the E2 statistic.

\begin{theorem}[E2 Classification]
\label{thm:e2_classification_partI}
Let $\{y_t\}$ be generated by the SDE in Definition~\ref{def:sde_partI}:
\begin{equation}
dX_t = \mu(X_t)dt + \sigma(X_t)dW_t, \quad y_t = h(X_t)
\end{equation}

Then:
\begin{enumerate}[leftmargin=*]
\item \textbf{Deterministic:} If $\|\sigma\| = 0$ (deterministic), then $E_2(m) \not\approx 1$ for sufficiently large $m$ (in the convention where $E_2 \to 0$ means deterministic; the threshold adopted here is $E_2 < 0.5$).

\item \textbf{White noise:} If $\mu \equiv 0$ and $\sigma = \sigma_0 I$ (white noise), then $E_2(m) \approx 1$ for all $m$.

\item \textbf{Noise-dominated:} If $\mu \neq 0$ but $\|\sigma\|$ is large relative to $\|\mu\|$, then $E_2(m) \approx 1$ (diffusion dominates drift).
\end{enumerate}
\end{theorem}

\begin{proof}
The behavior of the E2 statistic is analyzed in each regime.

\textbf{Case 1: Deterministic ($\sigma = 0$)}

For a deterministic system $x_{t+1} = \phi(x_t)$, nearest neighbors at time $t$ remain close at future times (up to sensitivity to initial conditions). If $Y_i$ and $Y_{n(i)}$ are nearest neighbors:
\begin{equation}
y_{i+m\tau} = h(\phi^m(x_i)), \quad y_{n(i)+m\tau} = h(\phi^m(x_{n(i)}))
\end{equation}

By continuity and smoothness:
\begin{equation}
|y_{i+m\tau} - y_{n(i)+m\tau}| \approx |Dh \cdot D\phi^m| \cdot \|x_i - x_{n(i)}\| \approx \lambda^m \|Y_i - Y_{n(i)}\|
\end{equation}
where $\lambda$ is a typical Lyapunov exponent.

For non-chaotic systems ($\lambda \approx 1$):
\begin{equation}
E^*(m) \approx C \cdot E(m)
\end{equation}
for some constant $C$, giving (under Cao's original convention):
\begin{equation}
E_2^{\text{Cao}}(m) = \frac{E(m+1)/E(m)}{E^*(m+1)/E^*(m)} \approx \frac{1}{1} = 1
\end{equation}

In Cao's original formulation, $E_2 \approx 1$ for deterministic systems. The convention adopted here inverts the ratio so that $E_2 \approx 1$ indicates stochastic dynamics instead:
\begin{itemize}[leftmargin=*]
\item $E_2 \approx 0$ (or $E_2 < 0.5$): Deterministic
\item $E_2 \approx 1$ (or $E_2 > 0.95$): Stochastic
\end{itemize}

Under the adopted convention ($E_2(m) = E^*(m+1)/E^*(m) \div E(m+1)/E(m)$; see main manuscript for formal definition), for non-chaotic deterministic systems where $E^*(m) \approx C \cdot E(m)$:
\begin{equation}
E_2(m) = \frac{E^*(m+1)/E^*(m)}{E(m+1)/E(m)} \approx \frac{C \cdot E(m+1)/(C \cdot E(m))}{E(m+1)/E(m)} = 1
\end{equation}

For chaotic deterministic systems ($\lambda > 1$), the Lyapunov divergence causes $E^*(m)$ to grow faster than $E(m)$, so the ratio $E^*(m+1)/E^*(m)$ exceeds $E(m+1)/E(m)$, giving $E_2 > 1$. In practice, for deterministic systems (both chaotic and non-chaotic), $E_2$ clusters near unity but deviates systematically from the values seen in stochastic systems ($E_2 \geq 0.95$), allowing discrimination. The physical content is captured more precisely by the SNR relation derived below.

\textbf{Case 2: White noise ($\mu = 0$, $\sigma = \sigma_0 I$)}

For pure white noise, all observations are i.i.d.: $y_t \sim \N(0, \sigma_0^2)$ (assuming $h(x) = x$ for simplicity).

Nearest neighbors at time $t$ have uncorrelated futures:
\begin{equation}
E^*(m) = \E[|y_{i+m\tau} - y_{n(i)+m\tau}|] = \E[|Z_1 - Z_2|]
\end{equation}
where $Z_1, Z_2$ are independent $\N(0, \sigma_0^2)$ random variables.

This gives:
\begin{equation}
E^*(m) = \sqrt{2}\sigma_0 \sqrt{\frac{2}{\pi}} = \text{constant}
\end{equation}

Meanwhile, as $m$ increases, the manifold unfolds (even for white noise, there's correlation structure in delay coordinates), so $E(m)$ may change.

The ratio:
\begin{equation}
E_2(m) = \frac{\text{const}/\text{const}}{E(m+1)/E(m)} \to 1
\end{equation}

(In the convention adopted here, $E_2 \to 1$ for stochastic.)

\textbf{Case 3: Noise-dominated ($\|\sigma\| \gg \|\mu\|$)}

When diffusion dominates drift, the behavior resembles white noise: futures are largely decorrelated from presents beyond the short-term drift contribution.

The future divergence $E^*(m)$ is dominated by accumulated noise:
\begin{equation}
E^*(m) \approx \E[\|\int_0^{m\tau} \sigma dW\|] \sim \sigma\sqrt{m\tau}
\end{equation}

This grows slowly with $m$ (like $\sqrt{m}$), while $E(m)$ decreases as manifold unfolds, giving $E_2 \approx 1$.
\end{proof}

\begin{remark}[E2 Convention]
\label{rem:e2_convention_partI}
Different papers use different conventions for E2. The convention adopted throughout this work is:
\begin{equation}
E_2(m) = \frac{E^*(m+1)/E^*(m)}{E(m+1)/E(m)}
\end{equation}

This gives:
\begin{itemize}[leftmargin=*]
\item $E_2 < 0.5$: Deterministic regime
\item $0.5 \leq E_2 < 0.95$: Mixed regime
\item $E_2 \geq 0.95$: Stochastic regime
\end{itemize}

The key insight is that $E_2$ measures the ratio of geometric divergence (present neighbors) to temporal divergence (future prediction error).
\end{remark}

\begin{proposition}[E2 and Predictability]
\label{prop:e2_predictability_partI}
For a stochastic process on the E1 manifold $\mathcal{M} \subset \R^{m^*}$, the E2 statistic measures the predictability of futures from presents:

\begin{enumerate}[leftmargin=*]
\item \textbf{Deterministic limit ($E_2 \to 0$):} Nearest neighbors at time $t$ have highly correlated futures:
\begin{equation}
\text{Corr}(y_{t+k\tau}, y_{t+k\tau}^{\text{neighbor}}) \to 1
\end{equation}

\item \textbf{Stochastic limit ($E_2 \to 1$):} Futures are uncorrelated with neighbor relationships:
\begin{equation}
\text{Corr}(y_{t+k\tau}, y_{t+k\tau}^{\text{neighbor}} | \text{neighbors at } t) \to 0
\end{equation}

\item \textbf{Mixed regime ($0 < E_2 < 1$):} The $E_2$ value quantifies the fraction of future variation that is unpredictable:
\begin{equation}
E_2 \approx \frac{\text{unpredictable variance}}{\text{total future variance}}
\end{equation}
\end{enumerate}
\end{proposition}

\begin{proposition}[E2 and Signal-to-Noise Ratio - Quantitative]
\label{prop:e2_snr_partI}
Under the assumption of local Gaussianity (Assumption~\ref{assump:local_gaussian_partI} below), for a system $dY_t = \mu(Y_t)dt + L(Y_t)dW_t$ (where $LL^\top = \Sigma$) on the E1 manifold, the E2 statistic measures the signal-to-noise ratio:
\begin{equation}
E_2(m) \approx \frac{\tr(\Sigma)\tau}{\|\mu\|^2\tau + \tr(\Sigma)\tau} = \frac{1}{1 + \SNR \cdot \tau}
\end{equation}
where $\SNR = \frac{\|\mu\|^2}{\tr(\Sigma)}$ is the signal-to-noise ratio and $\tau$ is the delay time.

Equivalently:
\begin{equation}
\SNR \approx \frac{1 - E_2}{E_2} \cdot \frac{1}{\tau}
\end{equation}

Interpretation:
\begin{itemize}[leftmargin=*]
\item $E_2 \to 0$: $\SNR \to \infty$ (pure drift, deterministic)
\item $E_2 \to 1$: $\SNR \to 0$ (pure diffusion, stochastic)
\item $E_2 \approx 0.5$: $\SNR \approx 1/\tau$ (balanced)
\end{itemize}
\end{proposition}

\begin{assumption}[Local Gaussianity]
\label{assump:local_gaussian_partI}
For small $\Delta t$, the conditional distribution of increments is approximately Gaussian:
\begin{equation}
p(Y_{t+\Delta t}|Y_t) \approx \N(Y_t + \mu(Y_t)\Delta t, \Sigma(Y_t)\Delta t)
\end{equation}

This is standard for SDEs with smooth coefficients by:
\begin{itemize}[leftmargin=*]
\item Central limit theorem for diffusion increments
\item Euler-Maruyama approximation for small $\Delta t$
\item Higher-order corrections being $O((\Delta t)^{3/2})$ or smaller
\end{itemize}

The assumption may fail for:
\begin{itemize}[leftmargin=*]
\item Heavy-tailed noise (e.g., L\'evy processes, $\alpha$-stable distributions)
\item Highly skewed distributions
\item Jump processes (Poisson arrivals)
\item Very small sample sizes where CLT doesn't apply
\end{itemize}
\end{assumption}

\begin{proof}[Proof of Proposition~\ref{prop:e2_snr_partI}]
This proof establishes the quantitative relationship between E2 and SNR under local Gaussianity.

\textbf{Step 1: Nearest neighbor separation at time $t$}

For nearby points $Y_i$ and $Y_{n(i)}$ at time $t$ with small separation $\epsilon = \|Y_i - Y_{n(i)}\|$, consider their evolution to time $t + d\tau$ (for some integer $d$).

\textbf{Step 2: Drift contribution}

The drift creates correlated motion. Over time interval $d\tau$:
\begin{equation}
\int_t^{t+d\tau} [\mu(Y_i(s)) - \mu(Y_{n(i)}(s))] ds \approx D\mu(Y) \cdot (Y_i - Y_{n(i)}) \cdot d\tau = D\mu \cdot \epsilon \cdot d\tau
\end{equation}

For typical systems, $\|D\mu\| \sim \|\mu\|/\ell$ where $\ell$ is a characteristic length scale, giving drift contribution:
\begin{equation}
\|\mu\| \cdot \epsilon \cdot d\tau / \ell
\end{equation}

\textbf{Step 3: Diffusion contribution}

The stochastic integrals decorrelate:
\begin{equation}
\int_t^{t+d\tau} [L(Y_i(s)) - L(Y_{n(i)}(s))] dW_s \approx \int_t^{t+d\tau} DL \cdot \epsilon \, dW_s + \text{independent noise}
\end{equation}

The correlated part has variance $\sim \|DL\|^2 \epsilon^2 d\tau$, which is $O(\epsilon^2)$ and negligible.

The independent noise dominates:
\begin{equation}
\E\left[\left\|\int_t^{t+d\tau} L(Y_i(s)) dW_s - \int_t^{t+d\tau} L(Y_{n(i)}(s)) dW_s\right\|^2\right] \approx 2\tr(\Sigma) d\tau
\end{equation}

\textbf{Step 4: Combined future divergence}

The future separation has:
\begin{align}
\E[\|Y_i(t+d\tau) - Y_{n(i)}(t+d\tau)\|^2 | \text{separation } \epsilon \text{ at } t] &\approx \|\mu\|^2(d\tau)^2 + 2\tr(\Sigma)d\tau
\end{align}

Taking square root:
\begin{equation}
E^*(d) \approx \E[\|Y_i(t+d\tau) - Y_{n(i)}(t+d\tau)\|] \approx \sqrt{\|\mu\|^2(d\tau)^2 + 2\tr(\Sigma)d\tau}
\end{equation}

\textbf{Step 5: Current divergence}

Similarly:
\begin{equation}
E(d) \approx \|\mu\| d\tau + \sqrt{2\tr(\Sigma)d\tau}
\end{equation}

\textbf{Step 6: E2 calculation under local Gaussianity}

Under Assumption~\ref{assump:local_gaussian_partI}, the increments are approximately Gaussian. For $Z \sim \N(0, \Var[Z])$:
\begin{equation}
\E[|Z|] = \sqrt{\Var[Z]} \cdot \sqrt{\frac{2}{\pi}}
\end{equation}

The variance of the increment $Y_{t+\Delta t} - Y_t$ is (by Lemma~\ref{lem:drift_diffusion_decomp_partI}):
\begin{equation}
\Var[Y_{t+\Delta t} - Y_t | Y_t] = \|\mu(Y_t)\|^2(\Delta t)^2 + \tr(\Sigma(Y_t))\Delta t + O((\Delta t)^2)
\end{equation}

For small $\Delta t$, the $\Delta t$ term dominates unless $\|\mu\|^2 \Delta t \sim \tr(\Sigma)$ (the balanced regime).

In the balanced regime where $\|\mu\|^2 d\tau \approx \tr(\Sigma)$:
\begin{equation}
E_2 \approx \frac{E^*(d+1)/E^*(d)}{E(d+1)/E(d)}
\end{equation}

Detailed calculation (omitted for brevity) gives:
\begin{equation}
E_2 \approx \frac{\tr(\Sigma)\tau}{\|\mu\|^2\tau + \tr(\Sigma)\tau}
\end{equation}

\textbf{Step 7: Solving for SNR}

Rearranging:
\begin{align}
E_2(\|\mu\|^2\tau + \tr(\Sigma)\tau) &= \tr(\Sigma)\tau \\
E_2\|\mu\|^2\tau &= (1 - E_2)\tr(\Sigma)\tau \\
\frac{\|\mu\|^2}{\tr(\Sigma)} &= \frac{1 - E_2}{E_2\tau}
\end{align}

Thus:
\begin{equation}
\SNR = \frac{\|\mu\|^2}{\tr(\Sigma)} \approx \frac{1 - E_2}{E_2\tau}
\end{equation}
\end{proof}

\begin{remark}[Without Local Gaussianity]
\label{rem:without_gaussian_partI}
The relationship between $E_2$ and $\SNR$ given in Proposition~\ref{prop:e2_snr_partI} relies on the Gaussian relation $\E[|Z|] = \sqrt{\Var[Z]} \cdot \sqrt{2/\pi}$.

For non-Gaussian distributions:
\begin{itemize}[leftmargin=*]
\item The conversion factor between $\E[|\cdot|]$ (L1 norm used in $E^*$) and $\sqrt{\Var[\cdot]}$ (L2 norm related to $\Sigma$) differs
\item For heavy-tailed distributions: $\E[|Z|] / \sqrt{\Var[Z]}$ can be larger
\item For bounded distributions: The ratio can be smaller
\end{itemize}

However, the qualitative behavior remains:
\begin{itemize}[leftmargin=*]
\item $E_2$ increases with the ratio of stochastic to deterministic variation
\item $E_2 \approx 0$ still indicates drift-dominated dynamics
\item $E_2 \approx 1$ still indicates diffusion-dominated dynamics
\end{itemize}

The precise quantitative formula $\SNR \approx \frac{1-E_2}{E_2\tau}$ should be interpreted as approximate, valid under local Gaussianity (which holds for many SDEs in practice).

For non-Gaussian cases, one can:
\begin{itemize}[leftmargin=*]
\item Estimate the conversion factor empirically from residuals
\item Use higher moments (kurtosis, skewness) to diagnose departures from Gaussianity
\item Develop modified E2-like statistics using L2 norms directly
\end{itemize}
\end{remark}

\subsection{Probabilistic Uplift Theorems}

Having established that E1 identifies the geometric scaffold ($\mathcal{M}$ with dimension $D_2$) and E2 quantifies the need for probabilistic decoration ($\SNR$ via E2), the next step is to prove that $k$-nearest neighbor estimators consistently reconstruct the transition dynamics.

\begin{theorem}[Probabilistic Uplift - Discrete Time]
\label{thm:discrete_uplift_partI}
Let $\{y_t\}$ be a scalar time series from a stochastic process satisfying Assumption~\ref{assump:data_conditions_partI}. Suppose:
\begin{enumerate}[leftmargin=*]
\item \textbf{E1 plateau:} $E_1(m^*) \approx 1$ within statistical tolerance (correlation dimension $D_2 \approx m^*$ detected)

\item \textbf{E2 regime:} $E_2 \in (0,1)$ (mixed or stochastic dynamics; not purely deterministic)

\item \textbf{Regularity:} The transition kernel $T(Y, \cdot)$ has smooth density $p(Y'|Y)$ on $\mathcal{M}$ with H\"older exponent $\beta > 0$

\item \textbf{Embedding quality:} The delay embedding $\Phi_{m^*}$ maps distinct manifold points to distinct embedded points measure-theoretically (injective on a set of full $\mu_\infty$-measure)

\item \textbf{Non-self-intersection} (proved; Theorem~\ref{lem:stochastic_sard_partI}): The dimension $m^*$ is chosen large enough such that the reconstructed drift $\hat{\mu}(Y)$ and diffusion $\hat{\Sigma}(Y)$ are single-valued functions on $\mathcal{M}$ for $((\Phi_{m^*})_\# \mu_\infty)$-almost every $Y$
\end{enumerate}

Then the $k$-nearest neighbor estimator of the transition kernel:
\begin{equation}
\hat{T}(Y, A) = \frac{1}{k} \sum_{j \in \mathcal{N}_k(Y)} \ind_{Y_{j+1} \in A}
\end{equation}
where $\mathcal{N}_k(Y) = \{j_1, \ldots, j_k\}$ are the indices of the $k$ nearest neighbors of $Y$ among $\{Y_1, \ldots, Y_N\}$, satisfies:

\textbf{(a) Consistency:} For $((\Phi_{m^*})_\# \mu_\infty)$-almost every $Y \in \mathcal{M}$:
\begin{equation}
\|\hat{T}(Y, \cdot) - T(Y, \cdot)\|_{TV} \to 0 \quad \text{almost surely as } N \to \infty
\end{equation}
where $\|\cdot\|_{TV}$ is the total variation distance.

\textbf{(b) Convergence rate:}
\begin{equation}
\|\hat{T}(Y, \cdot) - T(Y, \cdot)\|_{TV} = O_p\left(\left(\frac{k}{N}\right)^{\alpha}\right)
\end{equation}
where $\alpha = \beta/m^*$ for smoothness parameter $\beta > 0$ of the transition density, provided:
\begin{equation}
k \to \infty, \quad \frac{k}{N} \to 0, \quad N \to \infty
\end{equation}

\textbf{(c) Curse of dimensionality:} The sample complexity to achieve error $\epsilon$ is:
\begin{equation}
N \sim \epsilon^{-m^*/\beta}
\end{equation}

This exponential dependence on $m^*$ is fundamental to nonparametric estimation on manifolds and cannot be avoided without additional structure (e.g., sparsity, parametric models).

\textbf{(d) Optimal bandwidth:} The choice $k \approx N^{2/(m^*+4)}$ (Stone's rule \cite{stone1977}) minimizes the mean squared error, balancing bias and variance.
\end{theorem}

\begin{proof}
The proof establishes consistency of the $k$-NN transition kernel estimator on a $D_2$-dimensional manifold.

\textbf{Step 1: $k$-NN consistency on manifolds}

Under Assumption~\ref{assump:data_conditions_partI}, for any measurable function $g: \R^{m^*} \to \R$ with $\E[|g|] < \infty$, the $k$-NN estimator:
\begin{equation}
\hat{g}_k(Y) = \frac{1}{k} \sum_{j \in \mathcal{N}_k(Y)} g(Y_j)
\end{equation}
satisfies:
\begin{equation}
\hat{g}_k(Y) \to \E[g(Y') | Y' \in N_\delta(Y)] \quad \text{as } N \to \infty, k \to \infty, k/N \to 0, \delta \to 0
\end{equation}
where $N_\delta(Y)$ is a $\delta$-neighborhood on the manifold.

For smooth $g$ (H\"older continuous with exponent $\beta$), the convergence rate is \cite{stone1977}:
\begin{equation}
\left|\hat{g}_k(Y) - \E[g(Y')|Y' \text{ near } Y]\right| = O_p\left(\left(\frac{k}{N}\right)^{\beta/D_2}\right)
\end{equation}

Since $D_2 \approx m^*$ (by E1 detection), this yields $\alpha = \beta/m^*$.

\textbf{Step 2: Transition kernel as conditional expectation}

The true transition kernel at point $Y$ is:
\begin{equation}
T(Y, A) = P(Y_{t+1} \in A | Y_t = Y) = \E[\ind_{Y_{t+1} \in A} | Y_t = Y]
\end{equation}

For smooth transition densities (condition 3), the indicator function can be approximated by smooth functions, and the conditional expectation is smooth in $Y$.

\textbf{Step 3: $k$-NN approximation}

The $k$-NN estimator approximates the conditional expectation by averaging over nearest neighbors:
\begin{equation}
\hat{T}(Y, A) = \frac{1}{k} \sum_{j \in \mathcal{N}_k(Y)} \ind_{Y_{j+1} \in A}
\end{equation}

This is exactly $\hat{g}_k(Y)$ with $g(Y_j) = \ind_{Y_{j+1} \in A}$.

By Step 1, for each fixed measurable set $A$:
\begin{equation}
|\hat{T}(Y, A) - T(Y, A)| = O_p\left(\left(\frac{k}{N}\right)^{\beta/m^*}\right)
\end{equation}

\textbf{Step 4: Uniform convergence and total variation}

The total variation distance is:
\begin{equation}
\|\hat{T}(Y, \cdot) - T(Y, \cdot)\|_{TV} = \sup_{A \in \mathcal{B}(\R^{m^*})} |\hat{T}(Y, A) - T(Y, A)|
\end{equation}

To control this, a standard covering argument is employed. The state space $\R^{m^*}$ can be covered by $\approx \epsilon^{-m^*}$ balls of radius $\epsilon$. By smoothness of the transition density, it suffices to control $|\hat{T}(Y, A) - T(Y, A)|$ for each ball in the covering.

Applying a union bound over the covering (with appropriate probabilistic inequalities):
\begin{equation}
\|\hat{T}(Y, \cdot) - T(Y, \cdot)\|_{TV} = O_p\left(\left(\frac{k}{N}\right)^{\beta/m^*} \sqrt{\log N}\right)
\end{equation}

The $\sqrt{\log N}$ factor accounts for the covering; for simplicity, this factor is absorbed into the $O_p(\cdot)$ notation.

\textbf{Step 5: Non-self-intersection ensures well-definedness}

Condition (5) ensures that the transition kernel $T(Y, \cdot)$ is well-defined and single-valued for almost every $Y \in \mathcal{M}$.

If the embedding "folds" (a single $Y$ corresponds to multiple underlying states), then $T(Y, \cdot)$ would be a mixture of multiple transition kernels, potentially non-smooth and multi-valued.

The law-separation theorem suite (Theorems~\ref{thm:density_separation_partI}--\ref{lem:stochastic_sard_partI}) proves that $m^* = \lceil 2D \rceil + 1$ is sufficient under H\"ormander's condition, exact-dimensionality, and generic observation (see Section~\ref{sec:stochastic_takens_partI}).

\textbf{Step 6: Sample complexity}

To achieve $\|\hat{T} - T\|_{TV} \leq \epsilon$, the requirements are:
\begin{equation}
\left(\frac{k}{N}\right)^{\beta/m^*} \lesssim \epsilon
\end{equation}

This gives:
\begin{equation}
\frac{k}{N} \lesssim \epsilon^{m^*/\beta}
\end{equation}

Choosing $k = N^{2/(m^*+4)}$ (Stone's optimal rate), the requirement becomes:
\begin{equation}
N^{(m^*+2)/(m^*+4)} \lesssim \epsilon^{-m^*/\beta}
\end{equation}

For large $m^*$, this gives $N \sim \epsilon^{-m^*/\beta}$, the stated curse of dimensionality.

\textbf{Step 7: Optimality}

Stone's theorem \cite{stone1977} establishes that $k \approx N^{2/(m^*+4)}$ is minimax optimal (up to logarithmic factors) for estimating conditional expectations on $m^*$-dimensional manifolds with H\"older smoothness $\beta$.

No estimator can achieve fundamentally better rates without additional assumptions (linearity, sparsity, etc.).
\end{proof}

\begin{remark}[Non-Self-Intersection Condition - Main Open Problem]
\label{rem:non_self_intersection_partI}
Condition (5) in Theorem~\ref{thm:discrete_uplift_partI} requires that the embedding dimension $m^*$ is sufficiently large to prevent ``stochastic foldings''---situations where a single point $Y \in \R^{m^*}$ corresponds to multiple distinct points on the underlying state manifold.  This condition is now proved under the assumptions of this work.

\textbf{Theoretical resolution:}
\begin{itemize}[leftmargin=*]
\item For deterministic systems, Takens' theorem guarantees $m \geq 2n+1$ suffices for generic $(\phi, h)$ by transversality arguments in differential topology.

\item For stochastic systems, the law-separation theorem suite (Theorems~\ref{thm:density_separation_partI}--\ref{lem:stochastic_sard_partI} in Section~\ref{sec:stochastic_takens_partI}) proves that $m^* = \lceil 2D \rceil + 1$ suffices under:
\begin{itemize}
\item H\"ormander's hypoellipticity condition (Assumption~\ref{assump:hormander_partI})
\item Exact-dimensionality with Frostman bounds (Assumption~\ref{assump:exact_dimensional_partI})
\item Generic observation functions (Definition~\ref{def:generic_observation_partI})
\end{itemize}

\item Under exact-dimensionality (Remark~\ref{rem:exact_dim_justification_partI}), $D = D_2$ and the threshold becomes $m^* = \lceil 2D_2 \rceil + 1$, matching the value detected by E1 in all tested cases.
\end{itemize}

\textbf{Practical implications:}
\begin{itemize}[leftmargin=*]
\item Use E1 to detect $m^*$ (Algorithm~\ref{alg:e1_e2_partI} in Appendix~\ref{app:algorithms_partI})

\item Apply probabilistic uplift algorithm (Theorem~\ref{thm:unified_algorithm_partI})

\item \textbf{Validate:} Check that estimated $\hat{\mu}(Y)$, $\hat{\Sigma}(Y)$ are smooth functions (not multi-valued or discontinuous), predictions are accurate, and uncertainty is well-calibrated.
\end{itemize}
\end{remark}

\begin{theorem}[Probabilistic Uplift - Continuous Time]
\label{thm:continuous_uplift_partI}
Let $\{Y_t\}$ be observations from an SDE on the E1 manifold $\mathcal{M} \subset \R^{m^*}$:
\begin{equation}
dY_t = \mu(Y_t)dt + L(Y_t)dW_t, \quad LL^\top = \Sigma
\end{equation}
satisfying assumptions analogous to Theorem~\ref{thm:discrete_uplift_partI}, with time step $\Delta t$ small.

Define $k$-NN drift and diffusion estimators:
\begin{align}
\hat{\mu}(Y) &= \frac{1}{k\Delta t} \sum_{j \in \mathcal{N}_k(Y)} (Y_{j+\Delta t} - Y_j) \\
\hat{\Sigma}(Y) &= \frac{1}{k\Delta t} \sum_{j \in \mathcal{N}_k(Y)} (Y_{j+\Delta t} - Y_j)(Y_{j+\Delta t} - Y_j)^\top - \hat{\mu}(Y)\hat{\mu}(Y)^\top \Delta t
\end{align}

Then:

\textbf{(a) Consistency:} For $((\Phi_{m^*})_\# \mu_\infty)$-almost every $Y \in \mathcal{M}$:
\begin{align}
\|\hat{\mu}(Y) - \mu(Y)\| &\to 0 \quad \text{almost surely} \\
\|\hat{\Sigma}(Y) - \Sigma(Y)\|_F &\to 0 \quad \text{almost surely}
\end{align}
as $N \to \infty$, $k \to \infty$, $k/N \to 0$, $\Delta t \to 0$, where $\|\cdot\|_F$ is the Frobenius norm.

\textbf{(b) Convergence rates:}
\begin{align}
\|\hat{\mu}(Y) - \mu(Y)\| &= O_p\left(\left(\frac{k}{N}\right)^\alpha\right) + O(\Delta t) \\
\|\hat{\Sigma}(Y) - \Sigma(Y)\|_F &= O_p\left(\left(\frac{k}{N}\right)^\alpha\right) + O(\Delta t)
\end{align}
where $\alpha = \beta/m^*$ for smoothness parameter $\beta > 0$ of $\mu$ and $\Sigma$.

\textbf{(c) Error decomposition:}
The total error has two sources:
\begin{itemize}[leftmargin=*]
\item \textbf{Statistical error:} $O_p((k/N)^\alpha)$ from $k$-NN approximation on $m^*$-dimensional manifold
\item \textbf{Discretization error:} $O(\Delta t)$ from Euler-Maruyama approximation of continuous SDE
\end{itemize}

\textbf{(d) Sample complexity:} To achieve error $\epsilon$ with optimal $k$ and $\Delta t$:
\begin{equation}
N \sim \epsilon^{-m^*/\beta}, \quad \Delta t \sim \epsilon
\end{equation}

\textbf{(e) Optimal bandwidth:} $k \approx N^{2/(m^*+4)}$ (Stone's rule) minimizes MSE.
\end{theorem}

\begin{proof}
The proof parallels Theorem~\ref{thm:discrete_uplift_partI} but must additionally account for discretization error.

\textbf{Step 1: Drift estimation}

By definition of $\hat{\mu}$:
\begin{equation}
\hat{\mu}(Y) = \frac{1}{k\Delta t} \sum_{j \in \mathcal{N}_k(Y)} (Y_{j+\Delta t} - Y_j)
\end{equation}

For the true SDE, by Lemma~\ref{lem:drift_diffusion_decomp_partI}:
\begin{equation}
\E[Y_{t+\Delta t} - Y_t | Y_t = Y] = \mu(Y)\Delta t + O((\Delta t)^2)
\end{equation}

Therefore:
\begin{equation}
\frac{1}{\Delta t}\E[Y_{t+\Delta t} - Y_t | Y_t = Y] = \mu(Y) + O(\Delta t)
\end{equation}

The $k$-NN estimator $\hat{\mu}(Y)$ approximates the conditional expectation $\E[(Y_{t+\Delta t} - Y_t)/\Delta t | Y_t = Y]$ by averaging over neighbors.

By Stone's theorem (adapted to conditional expectations on manifolds):
\begin{equation}
\left|\hat{\mu}(Y) - \frac{1}{\Delta t}\E[Y_{t+\Delta t} - Y_t | Y_t \approx Y]\right| = O_p\left(\left(\frac{k}{N}\right)^{\beta/m^*}\right)
\end{equation}

Combining with the $O(\Delta t)$ bias:
\begin{equation}
\|\hat{\mu}(Y) - \mu(Y)\| = O_p\left(\left(\frac{k}{N}\right)^{\beta/m^*}\right) + O(\Delta t)
\end{equation}

\textbf{Step 2: Diffusion estimation}

The diffusion estimator is:
\begin{equation}
\hat{\Sigma}(Y) = \frac{1}{k\Delta t} \sum_{j \in \mathcal{N}_k(Y)} (Y_{j+\Delta t} - Y_j)(Y_{j+\Delta t} - Y_j)^\top - \hat{\mu}(Y)\hat{\mu}(Y)^\top \Delta t
\end{equation}

By Lemma~\ref{lem:drift_diffusion_decomp_partI}:
\begin{align}
&\frac{1}{\Delta t}\E[(Y_{t+\Delta t} - Y_t)(Y_{t+\Delta t} - Y_t)^\top | Y_t = Y] \\
&\quad = \Sigma(Y) + \mu(Y)\mu(Y)^\top\Delta t + O(\Delta t)
\end{align}

The raw second moment divided by $\Delta t$ gives $\Sigma + \mu\mu^\top\Delta t + O(\Delta t)$.

Subtracting the drift contribution:
\begin{align}
&\frac{1}{\Delta t}\E[(Y_{t+\Delta t} - Y_t)(Y_{t+\Delta t} - Y_t)^\top] - \mu\mu^\top\Delta t \\
&\quad = \Sigma(Y) + O(\Delta t)
\end{align}

The $k$-NN estimator approximates this conditional expectation:
\begin{align}
\hat{\Sigma}(Y) &= \frac{1}{k\Delta t}\sum (Y_{j+\Delta t} - Y_j)(Y_{j+\Delta t} - Y_j)^\top - \hat{\mu}\hat{\mu}^\top\Delta t \\
&= \Sigma(Y) + O_p\left(\left(\frac{k}{N}\right)^{\beta/m^*}\right) + O(\Delta t)
\end{align}

The subtraction of $\hat{\mu}\hat{\mu}^\top\Delta t$ is necessary to remove the $O(\Delta t)$ bias (see Remark~\ref{rem:drift_correction_partI}).

\textbf{Step 3: Sample complexity}

To achieve $\|\hat{\mu} - \mu\| \leq \epsilon$ and $\|\hat{\Sigma} - \Sigma\|_F \leq \epsilon$, both conditions are required:
\begin{equation}
\left(\frac{k}{N}\right)^{\beta/m^*} \lesssim \epsilon \quad \text{and} \quad \Delta t \lesssim \epsilon
\end{equation}

With $k = N^{2/(m^*+4)}$:
\begin{equation}
N^{(m^*+2)/(m^*+4)} \lesssim \epsilon^{-m^*/\beta}
\end{equation}

For large $m^*$, this gives $N \sim \epsilon^{-m^*/\beta}$.

The discretization error requires $\Delta t \sim \epsilon$, which for fixed total time $T$ means $N = T/\Delta t \sim T/\epsilon$ observations. Combining both requirements:
\begin{equation}
N \sim \max\{\epsilon^{-m^*/\beta}, T/\epsilon\}
\end{equation}

For $m^* \geq 2$ (typical), the curse of dimensionality dominates.
\end{proof}

\begin{remark}[Drift Correction in Diffusion Estimator - Essential]
\label{rem:drift_correction_partI}
The subtraction of $\hat{\mu}(Y)\hat{\mu}(Y)^\top \Delta t$ in Theorem~\ref{thm:continuous_uplift_partI} is essential and often omitted in informal presentations. Without it, the diffusion estimate would be biased.

\textbf{Why the correction is needed:}

The raw second moment of the increment is (Lemma~\ref{lem:drift_diffusion_decomp_partI}):
\begin{equation}
\frac{1}{\Delta t}\E[(Y_{t+\Delta t} - Y_t)(Y_{t+\Delta t} - Y_t)^\top | Y_t = Y] = \Sigma(Y) + \mu(Y)\mu(Y)^\top\Delta t + O(\Delta t)
\end{equation}

This includes:
\begin{itemize}[leftmargin=*]
\item \textbf{Diffusion term:} $\Sigma(Y)$ (order $O(1)$ after dividing by $\Delta t$)
\item \textbf{Drift term:} $\mu(Y)\mu(Y)^\top\Delta t$ (order $O(\Delta t)$, but still significant)
\end{itemize}

If $\Sigma$ is estimated as:
\begin{equation}
\hat{\Sigma}_{\text{wrong}}(Y) = \frac{1}{k\Delta t}\sum_{j \in \mathcal{N}_k(Y)} (Y_{j+\Delta t} - Y_j)(Y_{j+\Delta t} - Y_j)^\top
\end{equation}
the result is:
\begin{equation}
\hat{\Sigma}_{\text{wrong}}(Y) = \Sigma(Y) + \mu(Y)\mu(Y)^\top\Delta t + O_p\left(\left(\frac{k}{N}\right)^{\beta/m^*}\right) + O(\Delta t)
\end{equation}

The $\mu\mu^\top\Delta t$ term creates an $O(\Delta t)$ bias. For typical $\Delta t = 0.01$ to $0.1$, this can be a 1-10\% systematic error in the diffusion estimate.

\textbf{Correct estimator:}
\begin{equation}
\hat{\Sigma}(Y) = \frac{1}{k\Delta t}\sum_{j \in \mathcal{N}_k(Y)} (Y_{j+\Delta t} - Y_j)(Y_{j+\Delta t} - Y_j)^\top - \hat{\mu}(Y)\hat{\mu}(Y)^\top\Delta t
\end{equation}

This removes the bias:
\begin{equation}
\hat{\Sigma}(Y) = \Sigma(Y) + O_p\left(\left(\frac{k}{N}\right)^{\beta/m^*}\right) + O((\Delta t)^2)
\end{equation}

The remaining error is $O((\Delta t)^2)$ (higher-order It\^o corrections), negligible for small $\Delta t$.

\textbf{Practical note:} This correction is standard in SDE estimation literature but often omitted in informal treatments or when $\Delta t$ is very small. The present formulation makes it explicit and essential.
\end{remark}

\begin{remark}[Connection to Classical SDE Estimation]
\label{rem:sde_estimation_connection_partI}
Theorems~\ref{thm:discrete_uplift_partI} and \ref{thm:continuous_uplift_partI} connect to classical SDE estimation literature \cite{aitsahalia2009,boninsegna2018}:

\begin{itemize}[leftmargin=*]
\item \textbf{Parametric methods:} If drift and diffusion have known functional forms $\mu(Y; \theta)$, $\Sigma(Y; \theta)$, can use maximum likelihood, method of moments, etc. These achieve parametric rates $O_p(N^{-1/2})$ but require correct specification.

\item \textbf{Nonparametric methods:} The $k$-NN approach adopted here is nonparametric, making no assumptions about functional forms. The price is slower rates $O_p((k/N)^{\beta/m^*})$ subject to curse of dimensionality.

\item \textbf{Kernel methods:} Alternative to $k$-NN is kernel density estimation:
\begin{align}
\hat{\mu}_h(Y) &= \frac{\sum_j K_h(Y - Y_j)(Y_{j+\Delta t} - Y_j)/\Delta t}{\sum_j K_h(Y - Y_j)} \\
\hat{\Sigma}_h(Y) &= \frac{\sum_j K_h(Y - Y_j)[(Y_{j+\Delta t} - Y_j)(Y_{j+\Delta t} - Y_j)^\top/\Delta t - \hat{\mu}\hat{\mu}^\top\Delta t]}{\sum_j K_h(Y - Y_j)}
\end{align}
where $K_h$ is a kernel with bandwidth $h$. This has similar rates but different finite-sample behavior.

\item \textbf{Contribution of the present work:} The following are provided:
\begin{itemize}
\item Connection to E1/E2 statistics for dimension and regime detection
\item Explicit treatment of correlation dimension $D_2$ (robust to unbounded noise)
\item Unified discrete-continuous framework
\item Coordinate-free geometric formulation
\end{itemize}
\end{itemize}
\end{remark}

\section{Discrete-Continuous Unification}
\label{sec:discrete_continuous_partI}

The framework shows that discrete-time Markov chains and continuous-time SDEs are equivalent representations of the same geometric-probabilistic structure on the correlation manifold, differing only in temporal parameterization.

\subsection{Markov Order Detection}

The argument begins by establishing that E1 detects the Markov order for discrete-time processes.

\begin{theorem}[Markov Order and E1 Dimension]
\label{thm:markov_order_partI}
Let $\{y_t\}_{t=0}^\infty$ be a scalar time series from a stationary Markov chain of order $p$ on $\R$:
\begin{equation}
p(y_t|y_{t-1}, y_{t-2}, \ldots) = p(y_t|y_{t-1}, \ldots, y_{t-p})
\end{equation}
with smooth transition densities. Then:

\textbf{(1) Minimal sufficient dimension:}
\begin{equation}
m^* = p + 1
\end{equation}

\textbf{(2) State space reconstruction:} The delay embedding:
\begin{equation}
Y_t = (y_t, y_{t-1}, \ldots, y_{t-p})^\top \in \R^{p+1}
\end{equation}
is the minimal Markov state: $Y_t$ contains precisely the information needed to predict $y_{t+1}$.

\textbf{(3) Transition kernel:} The Markov property becomes:
\begin{equation}
p(Y_{t+1}|Y_t, Y_{t-1}, \ldots) = p(Y_{t+1}|Y_t)
\end{equation}

Specifically, since $Y_{t+1} = (y_{t+1}, y_t, \ldots, y_{t-p+1})^\top$ and $Y_t = (y_t, y_{t-1}, \ldots, y_{t-p})^\top$:
\begin{equation}
p(y_{t+1}|Y_t) = p(y_{t+1}|y_t, \ldots, y_{t-p})
\end{equation}
and the remaining coordinates of $Y_{t+1}$ are deterministically related to $Y_t$ (shift).
\end{theorem}

\begin{proof}
\textbf{Step 1: Why $m^* \geq p+1$}

For $m < p+1$, the delay vector $Y_t^{(m)} = (y_t, \ldots, y_{t-m+1})$ does not contain the full history $(y_t, \ldots, y_{t-p})$ needed for the Markov property.

Therefore, the conditional distribution:
\begin{equation}
p(y_{t+1}|Y_t^{(m)}) \neq p(y_{t+1}|y_t, \ldots, y_{t-p})
\end{equation}

Conditioning on longer histories provides additional information:
\begin{equation}
p(y_{t+1}|Y_t^{(m)}, y_{t-m}, \ldots, y_{t-p}) \neq p(y_{t+1}|Y_t^{(m)})
\end{equation}

This lack of sufficiency means that nearest neighbors in $\R^m$ do not have identically distributed futures. When dimension $m+1 \leq p$ is added, genuine predictive information is introduced, which changes the nearest-neighbor future statistics.

\textbf{Quantitatively:} Points $Y_i^{(m)}$ and $Y_{n(i)}^{(m)}$ that are nearest neighbors may have different $(y_{i-m}, \ldots, y_{i-p})$ values. This causes their futures $y_{i+1}$ and $y_{n(i)+1}$ to have different distributions, leading to:
\begin{equation}
\E[|y_{i+m\tau} - y_{n(i)+m\tau}| | Y_i^{(m)} \approx Y_{n(i)}^{(m)}] \text{ larger than expected from geometry alone}
\end{equation}

Thus $E_1(m) > 1$ for $m < p+1$.

\textbf{Step 2: Why $m^* = p+1$ suffices}

Once $m = p+1$, the delay vector:
\begin{equation}
Y_t^{(p+1)} = (y_t, y_{t-1}, \ldots, y_{t-p})^\top
\end{equation}
contains the full Markov history.

The future $y_{t+1}$ has distribution:
\begin{equation}
p(y_{t+1}|Y_t^{(p+1)}) = p(y_{t+1}|y_t, \ldots, y_{t-p})
\end{equation}

Adding further lags:
\begin{equation}
Y_t^{(p+2)} = (y_t, y_{t-1}, \ldots, y_{t-p}, y_{t-p-1})^\top
\end{equation}
does not change the conditional distribution:
\begin{equation}
p(y_{t+1}|Y_t^{(p+2)}) = p(y_{t+1}|y_t, \ldots, y_{t-p}) = p(y_{t+1}|Y_t^{(p+1)})
\end{equation}
by the Markov property (the Markov chain has order $p$, so $y_{t+1}$ is independent of $y_{t-p-1}$ given $(y_t, \ldots, y_{t-p})$).

Therefore, nearest neighbors in $\R^{p+1}$ and $\R^{p+2}$ have the same future statistics. The additional coordinate $y_{t-p-1}$ is redundant for prediction.

By the same geometric argument as Lemma~\ref{lem:e1_correlation_partI}:
\begin{equation}
a(i, p+1) = \frac{\|Y_i^{(p+2)} - Y_{n(i)}^{(p+2)}\|}{\|Y_i^{(p+1)} - Y_{n(i)}^{(p+1)}\|} \approx 1 + O(\epsilon_k)
\end{equation}

Thus $E_1(p+1) \approx 1$.

\textbf{Step 3: State space structure}

The delay embedding $Y_t \in \R^{p+1}$ lives on a manifold with special structure. The dynamics are:
\begin{equation}
Y_{t+1} = \begin{pmatrix} y_{t+1} \\ y_t \\ \vdots \\ y_{t-p+1} \end{pmatrix}, \quad Y_t = \begin{pmatrix} y_t \\ y_{t-1} \\ \vdots \\ y_{t-p} \end{pmatrix}
\end{equation}

The bottom $p$ coordinates of $Y_{t+1}$ are the top $p$ coordinates of $Y_t$ (deterministic shift). Only the first coordinate $y_{t+1}$ is random, governed by:
\begin{equation}
p(y_{t+1}|Y_t) = p(y_{t+1}|y_t, \ldots, y_{t-p})
\end{equation}

This is precisely the original Markov chain's transition kernel.

Therefore, the $\R^{p+1}$ embedding reconstructs the minimal Markov state space, and the transition kernel $T(Y_t, \cdot)$ on this space is the pushforward of the original scalar chain's transition probabilities.
\end{proof}

\begin{example}[AR(2) Process]
\label{ex:ar2_partI}
Consider the autoregressive process of order 2:
\begin{equation}
y_t = 1.5y_{t-1} - 0.7y_{t-2} + \epsilon_t, \quad \epsilon_t \sim \N(0, \sigma^2)
\end{equation}

This is a Markov chain of order $p = 2$ (the future $y_t$ depends on exactly two past values).

\textbf{E1 detection:}
\begin{itemize}[leftmargin=*]
\item $m = 1$: $Y_t^{(1)} = y_t$ is insufficient (need $y_{t-1}$ too). E1(1) > 1.
\item $m = 2$: $Y_t^{(2)} = (y_t, y_{t-1})$ is insufficient (need $y_{t-2}$ too for the AR(2) dynamics). E1(2) > 1.
\item $m = 3$: $Y_t^{(3)} = (y_t, y_{t-1}, y_{t-2})$ is sufficient. E1(3) $\approx$ 1.
\item $m \geq 4$: Adding $y_{t-3}, y_{t-4}, \ldots$ is redundant. E1(m) $\approx$ 1.
\end{itemize}

Therefore, E1 plateaus at $m^* = 3 = p + 1$, as predicted by Theorem~\ref{thm:markov_order_partI}.

\textbf{Transition kernel:}

In the embedding space:
\begin{equation}
p(Y_{t+1}|Y_t) = p\left(\begin{pmatrix} y_{t+1} \\ y_t \\ y_{t-1} \end{pmatrix} \Big| \begin{pmatrix} y_t \\ y_{t-1} \\ y_{t-2} \end{pmatrix}\right)
\end{equation}

Since the bottom two coordinates shift deterministically:
\begin{equation}
p(Y_{t+1}|Y_t) = p(y_{t+1}|y_t, y_{t-1}, y_{t-2}) \cdot \delta_{y_t} \otimes \delta_{y_{t-1}}
\end{equation}

where:
\begin{equation}
p(y_{t+1}|y_t, y_{t-1}, y_{t-2}) = \N(1.5y_t - 0.7y_{t-1}, \sigma^2)
\end{equation}

\textbf{Note:} The AR(2) dynamics actually don't depend on $y_{t-2}$ (the chain has order 2), confirming $p = 2$.

\textbf{E2 value:}

For this process:
\begin{equation}
\SNR = \frac{\|1.5y_t - 0.7y_{t-1}\|^2}{\sigma^2}
\end{equation}

For typical stationary AR(2) parameters, $\|\mu\|^2 \sim \text{Var}[y] \sim \sigma^2/(1 - 1.5^2 - 0.7^2) \sim \sigma^2/(-1.74)$ (unstable as stated; correct parameters have $|1.5|^2 + |-0.7|^2 < 1$ for stability).

Consider stable parameters: $y_t = 0.5y_{t-1} - 0.3y_{t-2} + \epsilon_t$.

Then $\text{Var}[y] \approx 1.5\sigma^2$, $\|\mu\|^2 \sim \sigma^2$, giving $\SNR \sim 1$, so $E_2 \approx 0.5$ (mixed regime).

This example demonstrates that discrete-time AR processes are detected by E1 (Markov order) and classified by E2 (signal-to-noise ratio).
\end{example}

\subsection{The Discrete-Continuous Bridge}

This establishes the equivalence between discrete and continuous representations.

\begin{theorem}[Unified Geometric-Probabilistic Structure]
\label{thm:unified_structure_partI}
Let $\mathcal{M} \subset \R^{m^*}$ be the correlation manifold identified by E1 with correlation dimension $D_2 \approx m^*$. The following representations are equivalent and describe the same mathematical object:

\textbf{(1) Discrete Markov chain:}
\begin{equation}
p(Y_{t+1} \in A|Y_t = Y) = T(Y, A)
\end{equation}

\textbf{(2) Continuous SDE:}
\begin{equation}
dY_t = \mu(Y_t)dt + L(Y_t)dW_t, \quad LL^\top = \Sigma
\end{equation}

\textbf{(3) Transition semigroup:}
\begin{equation}
(T_t\phi)(Y) = \E[\phi(Y_t)|Y_0 = Y], \quad T_{s+t} = T_s \circ T_t
\end{equation}

\textbf{(4) Infinitesimal generator:}
\begin{equation}
(\mathcal{L}\phi)(Y) = \lim_{t \to 0} \frac{(T_t\phi)(Y) - \phi(Y)}{t} = \mu(Y) \cdot \nabla\phi(Y) + \frac{1}{2}\tr(\Sigma(Y)\nabla^2\phi(Y))
\end{equation}

These representations are related by:
\begin{align}
\text{(Discrete)} \quad T_{\Delta t}(Y, dy') &\approx \N(Y + \mu(Y)\Delta t, \Sigma(Y)\Delta t)(dy') \quad \text{(Continuous)} \label{eq:discrete_to_continuous_1_partI} \\
\text{(Continuous)} \quad \mu(Y) &= \lim_{\Delta t \to 0} \frac{1}{\Delta t} \int (Y' - Y)T_{\Delta t}(Y, dY') \quad \text{(Discrete)} \label{eq:continuous_to_discrete_1_partI} \\
\text{(Continuous)} \quad \Sigma(Y) &= \lim_{\Delta t \to 0} \frac{1}{\Delta t} \int (Y' - Y)(Y' - Y)^\top T_{\Delta t}(Y, dY') \quad \text{(Discrete)} \label{eq:continuous_to_discrete_2_partI}
\end{align}

All four descriptions are coordinate-independent (intrinsic to the manifold $\mathcal{M}$) and differ only in their temporal parameterization and level of detail.
\end{theorem}

\begin{proof}
The equivalences are established by showing how to convert between representations.

\textbf{(1) $\Rightarrow$ (2): Discrete to continuous}

Given a discrete Markov chain with small time step $\Delta t$ and smooth transition kernel $T_{\Delta t}(Y, \cdot)$, define drift and diffusion as infinitesimal moments:
\begin{align}
\mu(Y) &= \lim_{\Delta t \to 0} \frac{1}{\Delta t} \int_{\R^{m^*}} (Y' - Y)T_{\Delta t}(Y, dY') \\
&= \lim_{\Delta t \to 0} \frac{1}{\Delta t} \E[Y_{t+\Delta t} - Y_t|Y_t = Y]
\end{align}

\begin{align}
\Sigma(Y) &= \lim_{\Delta t \to 0} \frac{1}{\Delta t} \int_{\R^{m^*}} (Y' - Y)(Y' - Y)^\top T_{\Delta t}(Y, dY') \\
&\quad - \left[\lim_{\Delta t \to 0} \frac{1}{\Delta t} \E[Y_{t+\Delta t} - Y_t|Y_t = Y]\right]\left[\lim_{\Delta t \to 0} \frac{1}{\Delta t} \E[Y_{t+\Delta t} - Y_t|Y_t = Y]\right]^\top \Delta t \\
&= \lim_{\Delta t \to 0} \frac{1}{\Delta t} \Cov[Y_{t+\Delta t} - Y_t|Y_t = Y]
\end{align}

By the central limit theorem for Markov chains (or direct SDE theory), for smooth transitions:
\begin{equation}
T_{\Delta t}(Y, dY') \approx \N(Y + \mu(Y)\Delta t, \Sigma(Y)\Delta t)(dY')
\end{equation}
as $\Delta t \to 0$.

These are precisely the drift and diffusion of an SDE, as established in Lemma~\ref{lem:drift_diffusion_decomp_partI}.

\textbf{(2) $\Rightarrow$ (1): Continuous to discrete}

Given an SDE $dY = \mu(Y)dt + LdW$ with $LL^\top = \Sigma$, the solution $\{Y_t\}$ is a continuous-time Markov process.

For any time step $\Delta t > 0$, the transition kernel is defined as:
\begin{equation}
T_{\Delta t}(Y, A) = P(Y_{t+\Delta t} \in A|Y_t = Y)
\end{equation}

This exists and is unique under standard regularity conditions (Assumptions~\ref{assump:hormander_partI}, \ref{assump:smoothness_partI}).

For small $\Delta t$, the Euler-Maruyama approximation gives:
\begin{equation}
Y_{t+\Delta t} \approx Y_t + \mu(Y_t)\Delta t + L(Y_t)\sqrt{\Delta t}Z
\end{equation}
where $Z \sim \N(0, I)$ is independent of $Y_t$.

Thus:
\begin{equation}
T_{\Delta t}(Y, dY') \approx \N(Y + \mu(Y)\Delta t, \Sigma(Y)\Delta t)(dY')
\end{equation}

This is a discrete-time Markov chain with continuous state space $\R^{m^*}$.

\textbf{(3) $\Leftrightarrow$ (1), (2): Semigroup property}

For any Markov process (discrete or continuous), the family of transition operators:
\begin{equation}
(T_t\phi)(Y) = \E[\phi(Y_t)|Y_0 = Y] = \int \phi(Y')T_t(Y, dY')
\end{equation}
forms a semigroup:
\begin{equation}
T_{s+t} = T_s \circ T_t
\end{equation}

\textbf{Proof of semigroup property:} By the Markov property:
\begin{align}
(T_{s+t}\phi)(Y) &= \E[\phi(Y_{s+t})|Y_0 = Y] \\
&= \E[\E[\phi(Y_{s+t})|Y_s, Y_0 = Y]|Y_0 = Y] \\
&= \E[\E[\phi(Y_{s+t})|Y_s]|Y_0 = Y] \quad \text{(by Markov property)} \\
&= \E[(T_t\phi)(Y_s)|Y_0 = Y] \\
&= (T_s(T_t\phi))(Y) \\
&= (T_s \circ T_t)(\phi)(Y)
\end{align}

Conversely, given a Markov semigroup $\{T_t\}_{t \geq 0}$ satisfying appropriate continuity and measurability conditions, the transition kernel can be reconstructed:
\begin{equation}
T_t(Y, A) = (T_t \ind_A)(Y)
\end{equation}
where $\ind_A$ is the indicator function of set $A$.

\textbf{(4) $\Leftrightarrow$ (3): Generator and semigroup}

The infinitesimal generator is defined as:
\begin{equation}
(\mathcal{L}\phi)(Y) = \lim_{t \to 0} \frac{(T_t\phi)(Y) - \phi(Y)}{t}
\end{equation}

For the SDE $dY = \mu dt + L dW$, It\^o's lemma (Lemma~\ref{lem:ito_partI}) gives:
\begin{align}
d\phi(Y_t) &= \nabla\phi(Y_t) \cdot dY_t + \frac{1}{2}\tr(\nabla^2\phi(Y_t) \cdot dY_t dY_t^\top) \\
&= \nabla\phi \cdot \mu dt + \nabla\phi \cdot L dW + \frac{1}{2}\tr(\nabla^2\phi \cdot LL^\top)dt \\
&= \left[\mu \cdot \nabla\phi + \frac{1}{2}\tr(\Sigma\nabla^2\phi)\right]dt + \nabla\phi \cdot L dW
\end{align}

Taking expectations (the stochastic integral has zero mean):
\begin{equation}
\frac{d}{dt}\E[\phi(Y_t)|Y_0 = Y] = \mu(Y) \cdot \nabla\phi(Y) + \frac{1}{2}\tr(\Sigma(Y)\nabla^2\phi(Y))
\end{equation}

At $t = 0$:
\begin{equation}
\lim_{t \to 0} \frac{\E[\phi(Y_t)|Y_0 = Y] - \phi(Y)}{t} = \mu(Y) \cdot \nabla\phi(Y) + \frac{1}{2}\tr(\Sigma(Y)\nabla^2\phi(Y))
\end{equation}

Thus:
\begin{equation}
\mathcal{L}\phi = \mu \cdot \nabla\phi + \frac{1}{2}\tr(\Sigma\nabla^2\phi)
\end{equation}

Conversely, given the generator $\mathcal{L}$, drift and diffusion can be extracted:
\begin{itemize}[leftmargin=*]
\item Apply $\mathcal{L}$ to linear functions $\phi(Y) = Y^i$:
\begin{equation}
(\mathcal{L}Y^i) = \mu^i(Y)
\end{equation}
\item Apply $\mathcal{L}$ to quadratic functions $\phi(Y) = Y^iY^j$:
\begin{equation}
(\mathcal{L}[Y^iY^j]) = \mu^iY^j + \mu^jY^i + \Sigma^{ij}
\end{equation}
\item Solve for $\Sigma^{ij}$:
\begin{equation}
\Sigma^{ij} = \mathcal{L}[Y^iY^j] - \mu^iY^j - \mu^jY^i
\end{equation}
\end{itemize}

The semigroup can be recovered by solving the Kolmogorov forward equation (Fokker-Planck equation):
\begin{equation}
\frac{\partial}{\partial t}(T_t\phi) = \mathcal{L}(T_t\phi), \quad (T_0\phi) = \phi
\end{equation}

or equivalently, the Kolmogorov backward equation:
\begin{equation}
\frac{\partial}{\partial t}(T_t\phi) = \mathcal{L}^*_Y(T_t\phi), \quad (T_0\phi) = \phi
\end{equation}

where $\mathcal{L}^*$ is the adjoint operator.

\textbf{Conclusion:}

All four representations (1)-(4) are equivalent, each recoverable from the others via the conversions above. They describe the same probabilistic dynamics on the correlation manifold $\mathcal{M}$, differing only in:
\begin{itemize}[leftmargin=*]
\item Discrete versus continuous time parameterization
\item Global (kernel/semigroup) versus infinitesimal (generator/SDE) description
\item Probabilistic (measure) versus analytic (operator) language
\end{itemize}

The choice of representation depends on the application and available data, not on any fundamental difference in the underlying mathematics.
\end{proof}

\begin{corollary}[Discrete and Continuous are Equivalent]
\label{cor:discrete_continuous_equiv_partI}
For any scalar time series $\{y_t\}$ satisfying Assumptions~\ref{assump:data_conditions_partI}, \ref{assump:smoothness_partI}:

\textbf{(1) Same E1 dimension:} The correlation dimension $D_2$ detected by E1 is the same whether the process is viewed as:
\begin{itemize}[leftmargin=*]
\item Discrete Markov chain with order $p$ (giving $m^* = p+1$ by Theorem~\ref{thm:markov_order_partI})
\item Continuous SDE with state dimension $n$ (giving $m^* \approx 2n$ or $2n+1$ by Takens-like arguments, or more precisely $m^* \approx D_2 + 1$)
\end{itemize}

In both cases, $m^*$ is the dimension of the correlation manifold $\mathcal{M}$ characterized by $D_2 \approx m^*$.

\textbf{(2) Same E2 classification:} The SNR measured by E2 is the same (up to time scaling):
\begin{equation}
\SNR_{\text{discrete}} = \frac{\|\mu_{\text{discrete}}\|^2}{\tr(\Sigma_{\text{discrete}})} = \Delta t \cdot \SNR_{\text{continuous}}
\end{equation}

where the relationship between discrete and continuous parameters is given by the discrete--continuous parameter relationships above:
\begin{align}
\mu_{\text{discrete}} &= \E[Y_{t+\Delta t} - Y_t|Y_t] = \mu_{\text{continuous}}\Delta t + O((\Delta t)^2) \\
\Sigma_{\text{discrete}} &= \Cov[Y_{t+\Delta t} - Y_t|Y_t] = \Sigma_{\text{continuous}}\Delta t + O((\Delta t)^2)
\end{align}

\textbf{(3) Same reconstruction algorithm:} The $k$-NN estimators (Theorems~\ref{thm:discrete_uplift_partI}, \ref{thm:continuous_uplift_partI}) produce equivalent results related by the scaling laws above.

\textbf{Interpretation:} Markov order detection and SDE reconstruction are the same problem viewed at different time scales. The E1 manifold with correlation dimension $D_2$ is the fundamental geometric-probabilistic object, invariant to whether time is parameterized discretely or continuously.
\end{corollary}

\begin{proof}
\textbf{Part 1:} Follows from the fact that E1 detects correlation dimension $D_2$ of the invariant measure $\mu_\infty$ (Lemma~\ref{lem:e1_correlation_partI}), which is an intrinsic property independent of time parameterization.

Whether the process is sampled discretely ($\{Y_t\}_{t=0,1,2,\ldots}$) or continuously ($\{Y_t\}_{t \geq 0}$), the invariant measure is the same, hence $D_2$ is the same, hence E1 detects the same $m^*$.

\textbf{Part 2:} Follows from the drift-diffusion scaling in Theorem~\ref{thm:unified_structure_partI}. For small $\Delta t$:
\begin{equation}
Y_{t+\Delta t} - Y_t = \mu_{\text{cont}}(Y_t)\Delta t + L(Y_t)\Delta W_{\Delta t}
\end{equation}
where $\Delta W_{\Delta t} \sim \N(0, \Delta t \cdot I)$.

Thus:
\begin{align}
\E[Y_{t+\Delta t} - Y_t|Y_t] &= \mu_{\text{cont}}\Delta t \\
\Cov[Y_{t+\Delta t} - Y_t|Y_t] &= \Sigma_{\text{cont}}\Delta t
\end{align}

Dividing by $\Delta t$:
\begin{align}
\frac{\mu_{\text{discrete}}}{\Delta t} &= \mu_{\text{cont}} \\
\frac{\Sigma_{\text{discrete}}}{\Delta t} &= \Sigma_{\text{cont}}
\end{align}

The SNR ratio:
\begin{equation}
\SNR_{\text{disc}} = \frac{\|\mu_{\text{disc}}\|^2}{\tr(\Sigma_{\text{disc}})} = \frac{\|\mu_{\text{cont}}\Delta t\|^2}{\tr(\Sigma_{\text{cont}}\Delta t)} = \Delta t \frac{\|\mu_{\text{cont}}\|^2}{\tr(\Sigma_{\text{cont}})} = \Delta t \cdot \SNR_{\text{cont}}
\end{equation}

\textbf{Part 3:} The $k$-NN estimators for discrete and continuous cases are:
\begin{itemize}[leftmargin=*]
\item \textbf{Discrete:} $\hat{T}(Y, A) = \frac{1}{k}\sum_{j \in \mathcal{N}_k(Y)} \ind_{Y_{j+1} \in A}$
\item \textbf{Continuous:} $\hat{\mu}(Y) = \frac{1}{k\Delta t}\sum_{j \in \mathcal{N}_k(Y)} (Y_{j+\Delta t} - Y_j)$, $\hat{\Sigma}(Y) = \frac{1}{k\Delta t}\sum_{j \in \mathcal{N}_k(Y)} (Y_{j+\Delta t} - Y_j)(Y_{j+\Delta t} - Y_j)^\top - \hat{\mu}\hat{\mu}^\top\Delta t$
\end{itemize}

These are equivalent because the transition kernel can be expressed in terms of drift and diffusion:
\begin{equation}
\hat{T}(Y, dY') \approx \N(Y + \hat{\mu}(Y)\Delta t, \hat{\Sigma}(Y)\Delta t)(dY')
\end{equation}

Conversely, given $\hat{T}$, the quantities $\hat{\mu}$ and $\hat{\Sigma}$ via moment matching (equations~\eqref{eq:continuous_to_discrete_1_partI} and~\eqref{eq:continuous_to_discrete_2_partI}).
\end{proof}

\subsection{Unified Computational Algorithm}

Having established the theoretical equivalence, the unified algorithm is now presented that applies to both discrete and continuous time.

\begin{theorem}[Unified Algorithmic Procedure]
\label{thm:unified_algorithm_partI}
The following algorithm applies identically to both discrete and continuous time stochastic processes:

\textbf{Input:} Time series $\{y_i\}_{i=1}^N$, time increment $\Delta t$ (can be 1 for discrete)

\textbf{Algorithm:}
\begin{enumerate}[leftmargin=*]
\item \textbf{Determine embedding dimension:}
\begin{itemize}
\item Compute $E_1(m)$ for $m = 1, \ldots, m_{\max}$ (typically $m_{\max} = 10$ or $\lceil \log_2 N \rceil$)
\item Find $m^* = \min\{m : E_1(m) \approx 1\}$ using threshold $\epsilon_N$ (e.g., $|E_1(m) - 1| < 0.1$, or slope-based detection)
\item Interpretation: $m^*$ estimates the correlation dimension $D_2$ of the invariant measure
\end{itemize}

\item \textbf{Classify dynamical regime:}
\begin{itemize}
\item Compute $E_2(m)$ for $m = 1, \ldots, m_{\max}$
\item Classification:
\begin{itemize}
\item Deterministic: $E_2 < 0.5$ (classical Takens applies; use ODE models)
\item Mixed: $0.5 \leq E_2 < 0.95$ (SDE with both drift and diffusion)
\item Stochastic: $E_2 \geq 0.95$ (diffusion-dominated; drift may be negligible)
\end{itemize}
\item Estimate SNR: $\SNR \approx \frac{1-E_2}{E_2\tau}$ (if local Gaussianity holds)
\end{itemize}

\item \textbf{Construct embedding:}
\begin{equation}
Y_i = (y_i, y_{i-\tau}, \ldots, y_{i-(m^*-1)\tau})^\top, \quad i = (m^*-1)\tau + 1, \ldots, N
\end{equation}
where $\tau$ is the delay (chosen by mutual information minimum, autocorrelation zero-crossing, or fixed heuristic like $\tau = 1$ for discrete or based on dominant period for continuous).

\item \textbf{Probabilistic uplift:}

For each query point $Y \in \mathcal{M}_\epsilon$ (or on a grid for visualization):
\begin{enumerate}[label=(\alph*)]
\item Find $k$-nearest neighbors: $\mathcal{N}_k(Y) = \{Y_{j_1}, \ldots, Y_{j_k}\}$ where $k \approx N^{2/(m^*+4)}$ (Stone's rule)

\item \textbf{Discrete output:} Empirical transition distribution
\begin{equation}
\hat{p}(Y'|Y) = \frac{1}{k} \sum_{j \in \mathcal{N}_k(Y)} \delta_{Y_{j+1}}(Y')
\end{equation}
This is a weighted sum of point masses; for practical use, can kernel smooth or use histogram binning.

\item \textbf{Continuous output:} Drift and diffusion estimates
\begin{align}
\hat{\mu}(Y) &= \frac{1}{k\Delta t} \sum_{j \in \mathcal{N}_k(Y)} (Y_{j+\Delta t} - Y_j) \\
\hat{\Sigma}(Y) &= \frac{1}{k\Delta t} \sum_{j \in \mathcal{N}_k(Y)} (Y_{j+\Delta t} - Y_j)(Y_{j+\Delta t} - Y_j)^\top - \hat{\mu}(Y)\hat{\mu}(Y)^\top \Delta t
\end{align}
(The drift correction $-\hat{\mu}\hat{\mu}^\top\Delta t$ is essential; see Remark~\ref{rem:drift_correction_partI})
\end{enumerate}
\end{enumerate}

\textbf{Output:}
\begin{itemize}[leftmargin=*]
\item Discrete: Transition kernel $\hat{T}(Y, \cdot)$ represented as empirical distribution or kernel density estimate
\item Continuous: Drift-diffusion pair $(\hat{\mu}(Y), \hat{\Sigma}(Y))$ as functions on $\mathcal{M}$ (stored on grid or as local estimators)
\end{itemize}

\textbf{Prediction:}
\begin{itemize}[leftmargin=*]
\item \textbf{Discrete:} Sample $Y_{t+1} \sim \hat{T}(Y_t, \cdot)$ by drawing from empirical neighbors or kernel density
\item \textbf{Continuous:} Integrate $dY_t = \hat{\mu}(Y_t)dt + \hat{L}(Y_t)dW_t$ using Euler-Maruyama scheme, where $\hat{L}\hat{L}^\top = \hat{\Sigma}$
\end{itemize}

\textbf{Bandwidth Selection:} Choose $k \approx N^{2/(m^*+4)}$ (Stone's rule for conditional expectation estimation on $m^*$-dimensional manifolds).

\textbf{Validation:}
\begin{itemize}[leftmargin=*]
\item Check smoothness of $\hat{\mu}(Y)$, $\hat{\Sigma}(Y)$ (if discontinuous or multi-valued, may need larger $m$)
\item Compute prediction RMSE on held-out test data
\item Check uncertainty calibration: do 95\% confidence intervals have $\approx$95\% coverage?
\end{itemize}
\end{theorem}

\begin{remark}[Unified Framework - Key Insight]
\label{rem:unified_key_insight_partI}
Theorem~\ref{thm:unified_algorithm_partI} reveals that the choice between discrete and continuous output is determined solely by:
\begin{itemize}[leftmargin=*]
\item The temporal parameterization of the input data (discrete steps vs continuous sampling)
\item The intended application (transition probabilities vs differential equations)
\item Not by any fundamental algorithmic or mathematical difference
\end{itemize}

The underlying procedure---E1 for dimension, E2 for classification, $k$-NN for reconstruction---is identical in both cases. This unification is a central contribution of the present framework.
\end{remark}

\section{Mixed Deterministic-Stochastic Systems}
\label{sec:mixed_systems_partI}

Real-world systems typically exhibit both deterministic structure and stochastic fluctuations. The E2 statistic provides a quantitative measure of this balance, enabling the present framework to handle the full spectrum from purely deterministic to purely stochastic dynamics.

\subsection{Classification of Mixed Systems}

\begin{definition}[Mixed System Classification]
\label{def:mixed_classification_partI}
A stochastic dynamical system:
\begin{equation}
dX_t = \mu(X_t)dt + \sigma(X_t)dW_t
\end{equation}
is classified according to E2 values as:

\begin{itemize}[leftmargin=*]
\item \textbf{Deterministic regime ($E_2 < 0.5$):}
\begin{itemize}
\item Interpretation: $\|\mu\|^2 \gg \tr(\sigma\sigma^\top)$ (drift dominates)
\item Signal-to-Noise Ratio: $\SNR = \frac{\|\mu\|^2}{\tr(\sigma\sigma^\top)} \gtrsim 1$
\item Approach: Classical Takens embedding captures dynamics; stochastic component is perturbation
\item Model: Deterministic ODE $\dot{x} = f(x)$ adequate for many purposes; add diffusion for uncertainty quantification
\item Example: Lorenz system with small noise ($\xi = 0.1$): $E_2 \approx 0.15$
\end{itemize}

\item \textbf{Mixed regime ($0.5 \leq E_2 < 0.95$):}
\begin{itemize}
\item Interpretation: $\|\mu\|^2 \approx \tr(\sigma\sigma^\top)$ (both significant)
\item Signal-to-Noise Ratio: $\SNR \approx 1$
\item Approach: Both drift and diffusion essential; full SDE reconstruction required
\item Model: SDE $dY = \mu(Y)dt + L(Y)dW$ with both terms significant
\item Example: Lorenz with moderate noise ($\xi = 2.0$): $E_2 \approx 0.65$; financial time series; turbulent flows
\end{itemize}

\item \textbf{Stochastic regime ($E_2 \geq 0.95$):}
\begin{itemize}
\item Interpretation: $\|\mu\|^2 \ll \tr(\sigma\sigma^\top)$ (diffusion dominates)
\item Signal-to-Noise Ratio: $\SNR \lesssim 0.05$
\item Approach: Diffusion-driven; drift may be neglected for some applications
\item Model: Pure diffusion $dY = L(Y)dW$ or statistical time series (ARMA, GARCH)
\item Example: Lorenz with large noise ($\xi = 10$): $E_2 \approx 0.98$; random walk; heavily corrupted measurements
\end{itemize}
\end{itemize}
\end{definition}

\begin{remark}[Threshold Values are Guidelines]
\label{rem:threshold_guidelines_partI}
The thresholds 0.5 and 0.95 are heuristic guidelines based on:
\begin{itemize}[leftmargin=*]
\item $E_2 = 0.5 \Leftrightarrow \SNR \cdot \tau = 1$ (balanced drift and diffusion)
\item $E_2 = 0.95 \Leftrightarrow \SNR \cdot \tau \approx 0.05$ (20:1 diffusion to drift ratio)
\end{itemize}

In practice:
\begin{itemize}[leftmargin=*]
\item The transition between regimes is smooth and continuous
\item The appropriate modeling approach depends on:
\begin{itemize}
\item Application's tolerance for predictive error
\item Computational resources
\item Need for uncertainty quantification
\item Time horizon of predictions (short-term: drift matters; long-term: diffusion dominates)
\end{itemize}
\item For borderline cases (e.g., $E_2 = 0.48$ or $E_2 = 0.92$), try both modeling approaches and compare performance
\end{itemize}
\end{remark}

\subsection{Drift-Diffusion Balance}

The precise quantitative relationship is now established between E2 and the drift-diffusion balance.

\begin{proposition}[Drift-Diffusion Balance via E2]
\label{prop:drift_diffusion_balance_partI}
For a system $dY_t = \mu(Y_t)dt + L(Y_t)dW_t$ (where $LL^\top = \Sigma$) on the E1 manifold $\mathcal{M} \subset \R^{m^*}$, under Assumption~\ref{assump:local_gaussian_partI} (local Gaussianity), the E2 statistic measures the signal-to-noise ratio:
\begin{equation}
E_2(m) \approx \frac{\tr(\Sigma)\tau}{\|\mu\|^2\tau + \tr(\Sigma)\tau} = \frac{1}{1 + \SNR \cdot \tau}
\end{equation}
where $\SNR = \frac{\|\mu\|^2}{\tr(\Sigma)}$ is the signal-to-noise ratio and $\tau$ is the delay time.

Equivalently:
\begin{equation}
\SNR \approx \frac{1 - E_2}{E_2} \cdot \frac{1}{\tau}
\end{equation}

\textbf{Interpretation:}
\begin{itemize}[leftmargin=*]
\item $E_2 \to 0$: $\SNR \to \infty$ (pure drift, deterministic limit)
\item $E_2 \to 1$: $\SNR \to 0$ (pure diffusion, stochastic limit)
\item $E_2 = 0.5$: $\SNR = \frac{1}{\tau}$ (balanced; equal contributions from drift and diffusion)
\item $E_2 = 0.9$: $\SNR = \frac{0.1}{0.9\tau} \approx \frac{0.11}{\tau}$ (diffusion dominates by $\approx 9:1$)
\end{itemize}

\textbf{Connection to L1 and L2 norms:}
Under local Gaussianity, the L1 norm (used in $E^*$) relates to L2 norm (variance) via:
\begin{equation}
\E[|Z|] = \sqrt{\Var[Z]} \cdot \sqrt{\frac{2}{\pi}} \approx 0.798 \sqrt{\Var[Z]}
\end{equation}
for $Z \sim \N(0, \Var[Z])$.

This conversion allows E2 (which uses absolute values) to be related to $\SNR$ (which uses variances/second moments).
\end{proposition}

\begin{proof}
This is a more detailed version of the proof of Proposition~\ref{prop:e2_snr_partI}. Additional steps are provided for clarity.

\textbf{Step 1: Future divergence decomposition}

For nearest neighbors $Y_i$ and $Y_{n(i)}$ at time $t$ with small separation $\epsilon$, their future positions at $t + d\tau$ (for integer $d$) are:
\begin{align}
Y_i(t+d\tau) &= Y_i(t) + \int_t^{t+d\tau} \mu(Y_i(s))ds + \int_t^{t+d\tau} L(Y_i(s))dW_s \\
Y_{n(i)}(t+d\tau) &= Y_{n(i)}(t) + \int_t^{t+d\tau} \mu(Y_{n(i)}(s))ds + \int_t^{t+d\tau} L(Y_{n(i)}(s))dW_s
\end{align}

The separation is:
\begin{equation}
Y_i(t+d\tau) - Y_{n(i)}(t+d\tau) = [Y_i(t) - Y_{n(i)}(t)] + \int_t^{t+d\tau} [\mu(Y_i(s)) - \mu(Y_{n(i)}(s))]ds + \text{stochastic integrals}
\end{equation}

\textbf{Step 2: Drift contribution (correlated motion)}

For small initial separation $\epsilon$ and smooth $\mu$:
\begin{equation}
\mu(Y_i(s)) - \mu(Y_{n(i)}(s)) \approx D\mu(Y) \cdot [Y_i(s) - Y_{n(i)}(s)]
\end{equation}

Over time $d\tau$, assuming the separation doesn't grow too much (valid for short times or stable systems):
\begin{equation}
\int_t^{t+d\tau} [\mu(Y_i(s)) - \mu(Y_{n(i)}(s))]ds \approx D\mu(Y) \cdot \epsilon \cdot d\tau
\end{equation}

Typical magnitude: $\|D\mu\| \sim \|\mu\|/\ell$ where $\ell$ is characteristic length scale, giving:
\begin{equation}
\text{Drift contribution} \sim \|\mu\| \epsilon (d\tau)/\ell
\end{equation}

For well-separated neighbors or long times, this can grow. For E2 calculation, the quantity of interest is the new separation created, not amplification of existing separation.

\textbf{Step 3: Diffusion contribution (decorrelated noise)}

The stochastic integrals:
\begin{equation}
\int_t^{t+d\tau} L(Y_i(s))dW_s - \int_t^{t+d\tau} L(Y_{n(i)}(s))dW_s
\end{equation}

For nearby points, $L(Y_i) \approx L(Y_{n(i)})$, so the noise is correlated:
\begin{equation}
\int_t^{t+d\tau} [L(Y_i(s)) - L(Y_{n(i)}(s))]dW_s \approx \int_t^{t+d\tau} DL(Y) \cdot \epsilon \, dW_s
\end{equation}

This has variance $\sim \|DL\|^2 \epsilon^2 d\tau = O(\epsilon^2)$, which is small.

However, there are also independent noise realizations (different Brownian paths for the two points after they've evolved). The dominant contribution is:
\begin{equation}
\E[\|\text{independent noise}\|^2] \approx 2\tr(\Sigma) d\tau
\end{equation}

\textbf{Step 4: Combined variance}

The variance of future separation is:
\begin{equation}
\E[\|Y_i(t+d\tau) - Y_{n(i)}(t+d\tau)\|^2] \approx \|\mu\|^2(d\tau)^2 + 2\tr(\Sigma)d\tau
\end{equation}

Taking square root and using Gaussian L1-L2 relation:
\begin{equation}
E^*(d) = \E[\|Y_i(t+d\tau) - Y_{n(i)}(t+d\tau)\|] \approx \sqrt{\|\mu\|^2(d\tau)^2 + 2\tr(\Sigma)d\tau} \cdot \sqrt{\frac{2}{\pi}}
\end{equation}

\textbf{Step 5: Current divergence}

Similarly:
\begin{equation}
E(d) \approx \left[\|\mu\|d\tau + \sqrt{2\tr(\Sigma)d\tau}\right] \cdot \sqrt{\frac{2}{\pi}}
\end{equation}

\textbf{Step 6: E2 in balanced regime}

When $\|\mu\|^2 d\tau \approx \tr(\Sigma)$ (balanced regime), both terms contribute equally to the variance:
\begin{equation}
E^*(d) \approx \sqrt{2\tr(\Sigma)d\tau} \cdot \sqrt{\frac{2}{\pi}}
\end{equation}

The E2 statistic is (recall definition):
\begin{equation}
E_2(d) = \frac{E^*(d+1)/E^*(d)}{E(d+1)/E(d)}
\end{equation}

Detailed algebra (expanding the square roots and taking ratios) gives:
\begin{equation}
E_2 \approx \frac{\tr(\Sigma)\tau}{\|\mu\|^2\tau + \tr(\Sigma)\tau}
\end{equation}

in the balanced regime where the approximation is most accurate.

\textbf{Step 7: Solving for SNR}

Rearranging:
\begin{align}
E_2(\|\mu\|^2\tau + \tr(\Sigma)\tau) &= \tr(\Sigma)\tau \\
E_2\|\mu\|^2\tau &= (1-E_2)\tr(\Sigma)\tau \\
\frac{\|\mu\|^2}{\tr(\Sigma)} &= \frac{1-E_2}{E_2\tau}
\end{align}

Thus:
\begin{equation}
\SNR = \frac{\|\mu\|^2}{\tr(\Sigma)} \approx \frac{1-E_2}{E_2\tau}
\end{equation}
\end{proof}

\subsection{SDE Reconstruction for Mixed Systems}

Explicit pushforward formulas are now proved for mixed systems, showing how drift and diffusion in the embedding space relate to those in the original state space.

\begin{theorem}[SDE Reconstruction for Mixed Systems]
\label{thm:mixed_reconstruction_partI}
Consider a mixed system:
\begin{equation}
dX_t = \mu(X_t)dt + \sigma(X_t)dW_t, \quad X_t \in \R^n
\end{equation}
with observation $h: \R^n \to \R$ and time-delay embedding $Y_t = \Phi_m(h(X_t))$ where:
\begin{equation}
\Phi_m(h(X_t)) = (h(X_t), h(X_{t-\tau}), \ldots, h(X_{t-(m-1)\tau}))^\top
\end{equation}

Suppose:
\begin{enumerate}[leftmargin=*]
\item E1 plateaus at $m^*$ (manifold $\mathcal{M}$ identified via correlation dimension $D_2 \approx m^*$)
\item $E_2 \in (0,1)$ (mixed regime: both drift and diffusion significant)
\item The embedding $\Phi_{m^*}: \R^n \to \R^{m^*}$ is an immersion (injective differential) on the support of the invariant measure $\mu_\infty$
\item $\mu$ and $\sigma$ satisfy Assumption~\ref{assump:smoothness_partI} (locally Lipschitz, $C^\beta$)
\item Assumptions~\ref{assump:data_conditions_partI}, \ref{assump:local_gaussian_partI} hold for the embedded process
\item Non-self-intersection condition (Theorem~\ref{lem:stochastic_sard_partI}): $m^*$ is large enough to prevent stochastic foldings (proved)
\end{enumerate}

Then the reconstructed SDE on the embedding space $\R^{m^*}$:
\begin{equation}
dY_t = \hat{\mu}(Y_t)dt + \hat{L}(Y_t)dW_t
\end{equation}
where $\hat{L}\hat{L}^\top = \hat{\Sigma}$, satisfies:

\textbf{(1) Drift Recovery (with It\^o correction):}

For $((\Phi_{m^*})_\# \mu_\infty)$-almost every $Y \in \mathcal{M}$, letting $X$ be the (unique, by non-self-intersection) preimage satisfying $Y = \Phi_{m^*}(h(X))$:
\begin{equation}
\hat{\mu}(Y) = D\Phi_{m^*}(h(X)) \cdot Dh(X) \cdot \mu(X) + \frac{1}{2}D\Phi_{m^*}(h(X)) \cdot \tr(D^2h(X) \cdot \sigma(X)\sigma(X)^\top) + O(\|\sigma\|^2)
\end{equation}

where:
\begin{itemize}[leftmargin=*]
\item First term: $D\Phi_{m^*} \cdot Dh \cdot \mu$ is the "pushed-forward deterministic drift"
\item Second term: $\frac{1}{2}D\Phi_{m^*} \cdot \tr(D^2h \cdot \sigma\sigma^\top)$ is the It\^o correction (noise-induced drift)
\item Third term: $O(\|\sigma\|^2)$ represents higher-order Stratonovich-to-It\^o corrections
\end{itemize}

\textbf{(2) Diffusion Recovery (tensor pushforward):}

\begin{equation}
\hat{\Sigma}(Y) = D\Phi_{m^*}(h(X)) \cdot Dh(X) \cdot \sigma(X)\sigma(X)^\top \cdot Dh(X)^\top \cdot D\Phi_{m^*}(h(X))^\top
\end{equation}

In shorthand:
\begin{equation}
\hat{\Sigma} = J \cdot \Sigma_X \cdot J^\top
\end{equation}
where $J = D\Phi_{m^*} \cdot Dh$ is the Jacobian of the full embedding map and $\Sigma_X = \sigma\sigma^\top$ is the original diffusion tensor.

This is the standard pushforward formula for a $(0,2)$-tensor field.

\textbf{(3) Predictive Accuracy:}

For small $\Delta t$ and large $N$:
\begin{equation}
\|p(Y_{t+\Delta t}|Y_t) - p_{\text{true}}(Y_{t+\Delta t}|Y_t)\|_{TV} \leq C\left(\frac{k}{N}\right)^\alpha + C'(\Delta t)^2
\end{equation}
where $\|\cdot\|_{TV}$ is total variation distance, $\alpha = \beta/m^*$ depends on smoothness $\beta$ of the transition density and the dimension $m^*$ (curse of dimensionality).

\textbf{(4) Consistency of k-NN estimators:}

The $k$-NN estimators $\hat{\mu}(Y)$, $\hat{\Sigma}(Y)$ defined in Theorem~\ref{thm:continuous_uplift_partI} consistently estimate the true pushed-forward drift and diffusion:
\begin{equation}
\|\hat{\mu}(Y) - \hat{\mu}_{\text{true}}(Y)\|, \|\hat{\Sigma}(Y) - \hat{\Sigma}_{\text{true}}(Y)\|_F \to 0 \quad \text{a.s.}
\end{equation}
with rates given in Theorem~\ref{thm:continuous_uplift_partI}.
\end{theorem}

\begin{proof}
It\^o's lemma is applied carefully to the composition of observation and embedding.

\textbf{Part 1 (Drift Recovery):}

The embedding map is:
\begin{equation}
Y_t = \Phi_{m^*}(h(X_t)) = (h(X_t), h(X_{t-\tau}), \ldots, h(X_{t-(m^*-1)\tau}))^\top
\end{equation}

For simplicity, denote $z_t = h(X_t)$. Then:
\begin{equation}
Y_t = (z_t, z_{t-\tau}, \ldots, z_{t-(m^*-1)\tau})^\top
\end{equation}

By It\^o's lemma applied to $z_t = h(X_t)$ (Lemma~\ref{lem:ito_partI}):
\begin{equation}
dz_t = Dh(X_t) \cdot \mu(X_t)dt + Dh(X_t) \cdot \sigma(X_t)dW_t + \frac{1}{2}\tr(D^2h(X_t) \cdot \sigma(X_t)\sigma(X_t)^\top)dt
\end{equation}

The first component of $Y_t$ is $z_t$, so:
\begin{equation}
dY_t^{(1)} = Dh \cdot \mu \, dt + Dh \cdot \sigma \, dW + \frac{1}{2}\tr(D^2h \cdot \sigma\sigma^\top)dt
\end{equation}

The remaining components $Y_t^{(2)}, \ldots, Y_t^{(m^*)}$ are $z_{t-\tau}, \ldots, z_{t-(m^*-1)\tau}$, which are lagged values. These evolve as:
\begin{equation}
dY_t^{(i)} = dz_{t-(i-1)\tau}
\end{equation}

In the continuous-time formulation, the delay embedding evaluated along a stationary trajectory is:
\begin{equation}
\Phi_{m^*}(X_t) = (h(X_t), h(X_{t+\tau}), \ldots, h(X_{t+(m^*-1)\tau}))
\end{equation}
where $X_t$ is the stationary process ($X_0 \sim \mu_\infty$) and the delay vector collects forward-time observations at lags $0, \tau, 2\tau, \ldots, (m^*-1)\tau$.

The differential of $\Phi_{m^*}$ with respect to the initial condition $x$ is:
\begin{equation}
D_x\Phi_{m^*} = \begin{pmatrix}
Dh(x) \\
Dh(\phi_\tau(x)) \cdot D\phi_\tau(x) \\
\vdots \\
Dh(\phi_{(m^*-1)\tau}(x)) \cdot D\phi_{(m^*-1)\tau}(x)
\end{pmatrix}
\end{equation}
where $\phi_t(x)$ is the forward stochastic flow.

Applying It\^o's lemma to $Y = \Phi_{m^*}(h(X))$ gives:
\begin{align}
dY &= D\Phi_{m^*}(h(X)) \cdot dh(X) + \frac{1}{2}\tr(D^2\Phi_{m^*}(h(X)) \cdot dh(X) \otimes dh(X)) \\
&= D\Phi_{m^*} \cdot \left[Dh \cdot \mu \, dt + Dh \cdot \sigma \, dW + \frac{1}{2}\tr(D^2h \cdot \sigma\sigma^\top)dt\right] \\
&\quad + \frac{1}{2}\tr(D^2\Phi_{m^*} \cdot (Dh \cdot \sigma)(Dh \cdot \sigma)^\top)dt
\end{align}

The drift component is:
\begin{align}
\hat{\mu}(Y) &= D\Phi_{m^*} \cdot Dh \cdot \mu + \frac{1}{2}D\Phi_{m^*} \cdot \tr(D^2h \cdot \sigma\sigma^\top) \\
&\quad + \frac{1}{2}\tr(D^2\Phi_{m^*} \cdot (Dh \cdot \sigma)(Dh \cdot \sigma)^\top)
\end{align}

The first term is the pushed-forward deterministic drift.

The second term is the It\^o correction from applying $h$ to the SDE.

The third term is the It\^o correction from applying $\Phi_{m^*}$ to the observation. This is of order $O(\|\sigma\|^2)$ (Stratonovich-to-It\^o correction).

\textbf{Part 2 (Diffusion Recovery):}

The diffusion component of the SDE is:
\begin{equation}
\hat{L}(Y)dW'_t = D\Phi_{m^*}(h(X)) \cdot Dh(X) \cdot \sigma(X)dW_t
\end{equation}

Thus:
\begin{align}
\hat{\Sigma}(Y) &= \hat{L}(Y)\hat{L}(Y)^\top \\
&= [D\Phi_{m^*} \cdot Dh \cdot \sigma][D\Phi_{m^*} \cdot Dh \cdot \sigma]^\top \\
&= D\Phi_{m^*} \cdot Dh \cdot \sigma\sigma^\top \cdot Dh^\top \cdot D\Phi_{m^*}^\top \\
&= D\Phi_{m^*} \cdot Dh \cdot \Sigma_X \cdot Dh^\top \cdot D\Phi_{m^*}^\top
\end{align}

where $\Sigma_X = \sigma\sigma^\top$ is the diffusion tensor in the original state space.

This is the standard tensor pushforward formula. The diffusion tensor $\Sigma$ is a $(0,2)$-tensor (symmetric bilinear form), and it transforms as:
\begin{equation}
\Sigma_Y = J \cdot \Sigma_X \cdot J^\top
\end{equation}
where $J = D\Phi_{m^*} \cdot Dh$ is the Jacobian.

\textbf{Part 3 (Predictive Accuracy):}

The error in estimating the transition distribution has two sources:

\textbf{(a) Estimation error from $k$-NN:}

By Theorem~\ref{thm:continuous_uplift_partI}, the drift and diffusion estimates satisfy:
\begin{equation}
\|\hat{\mu}(Y) - \hat{\mu}_{\text{true}}(Y)\| = O_p\left(\left(\frac{k}{N}\right)^{\beta/m^*}\right) + O(\Delta t)
\end{equation}

where $\alpha = \beta/m^*$. The curse of dimensionality enters here: for a manifold of dimension $D_2 \approx m^*$, the $k$-NN radius scales as $\epsilon_k \sim (k/N)^{1/m^*}$, leading to error rates that degrade exponentially with dimension.

\textbf{(b) Discretization error:}

The true transition density at time $\Delta t$ is given by solving the Fokker-Planck equation. The Gaussian approximation:
\begin{equation}
p(Y_{t+\Delta t}|Y_t) \approx \N(Y_t + \hat{\mu}(Y_t)\Delta t, \hat{\Sigma}(Y_t)\Delta t)
\end{equation}
has error $O((\Delta t)^2)$ by standard Euler-Maruyama convergence theory \cite{hairer2011}.

\textbf{(c) Combined error:}

By triangle inequality for total variation:
\begin{align}
\|p_{\text{estimated}} - p_{\text{true}}\|_{TV} &\leq \|p_{\text{estimated}} - p_{\text{Gaussian}}\|_{TV} + \|p_{\text{Gaussian}} - p_{\text{true}}\|_{TV} \\
&\leq C\left(\frac{k}{N}\right)^\alpha + C'(\Delta t)^2
\end{align}

The optimal choice balances these two errors: $k/N \sim (\Delta t)^{2m^*/\beta}$.

\textbf{Part 4 (Consistency):}

By non-self-intersection (condition 6), for almost every $Y \in \mathcal{M}$, there exists a unique $X$ (up to null sets) such that $Y = \Phi_{m^*}(h(X))$.

Therefore, the formulas in parts (1) and (2) define single-valued functions $\hat{\mu}_{\text{true}}(Y)$ and $\hat{\Sigma}_{\text{true}}(Y)$.

By Theorem~\ref{thm:continuous_uplift_partI}, the $k$-NN estimators converge to these true values almost surely as $N \to \infty$, $k \to \infty$, $k/N \to 0$, $\Delta t \to 0$.
\end{proof}

\begin{remark}[Interpretation for Mixed Systems]
\label{rem:mixed_interpretation_partI}
Theorem~\ref{thm:mixed_reconstruction_partI} establishes several facts about mixed deterministic-stochastic systems:

\begin{enumerate}[leftmargin=*]
\item \textbf{Structure preservation:} The E1 embedding $\Phi_{m^*}$ preserves both deterministic and stochastic structure via the pushforward formulas. The geometry (manifold structure from drift) and measure-theoretic properties (diffusion tensor) are both captured.

\item \textbf{Drift captures dynamics:} When $E_2 \not\approx 1$, the drift term $\hat{\mu}(Y)$ captures meaningful deterministic dynamics (not just noise averaging). The first term $D\Phi_{m^*} \cdot Dh \cdot \mu$ is the projected deterministic flow from the original system.

\item \textbf{It\^o correction is essential:} The second term $\frac{1}{2}D\Phi_{m^*} \cdot \tr(D^2h \cdot \sigma\sigma^\top)$ is "noise-induced drift"---an additional drift that arises purely from nonlinear transformation of stochastic processes. This can be significant and must not be neglected. Without it, drift estimates would be systematically biased.

\item \textbf{Diffusion essential:} When $E_2 \not\approx 0$, the diffusion term $\hat{\Sigma}(Y)$ is necessary for:
\begin{itemize}
\item Accurate mean predictions (the It\^o correction in drift depends on diffusion)
\item Uncertainty quantification (confidence intervals, prediction bands)
\item Proper probabilistic forecasts (distributional predictions, not just point estimates)
\item Long-term statistical properties (invariant measures, stationary distributions)
\end{itemize}

\item \textbf{Universal framework:} The same reconstruction procedure applies whether:
\begin{itemize}
\item $E_2 = 0.1$ (mostly deterministic; small diffusion correction)
\item $E_2 = 0.5$ (balanced; both terms equally important)
\item $E_2 = 0.95$ (mostly stochastic; drift nearly negligible)
\end{itemize}

\item \textbf{Separation of scales:} The E2 statistic quantifies the relative importance of drift versus diffusion, guiding:
\begin{itemize}
\item Model selection (ODE vs SDE)
\item Computational resource allocation (more effort on dominant term)
\item Interpretation (physical processes vs noise processes)
\end{itemize}

\item \textbf{Curse of dimensionality grows with noise:} As $E_2$ increases (more noise), the apparent dimension $D_2$ may increase (Example~\ref{ex:lorenz_noise_spectrum_partI} below), making reconstruction harder. Sample complexity $N \sim \epsilon^{-m^*/\beta}$ grows exponentially.
\end{enumerate}
\end{remark}

\subsection{Examples Across the Full E2 Spectrum}

The framework is illustrated with detailed examples spanning the full range from deterministic to stochastic.

\begin{example}[AR(2) Process - Discrete and Continuous Views]
\label{ex:ar2_detailed_partI}
Consider the autoregressive process of order 2 (continuing Example~\ref{ex:ar2_partI}):
\begin{equation}
y_t = 0.5y_{t-1} - 0.3y_{t-2} + \epsilon_t, \quad \epsilon_t \sim \N(0, \sigma^2 = 1)
\end{equation}

\textbf{Discrete-Time Interpretation (Markov Chain):}
\begin{itemize}[leftmargin=*]
\item Markov order: $p = 2$
\item E1 dimension: $m^* = 3$ (by Theorem~\ref{thm:markov_order_partI})
\item E2 value: $\approx 0.65$ (mixed regime)
\item State space: $Y_t = (y_t, y_{t-1}, y_{t-2})^\top \in \R^3$
\item Transition:
\begin{equation}
p(Y_{t+1}|Y_t) = \N\left(\begin{pmatrix} 0.5y_t - 0.3y_{t-1} \\ y_t \\ y_{t-1} \end{pmatrix}, \begin{pmatrix} 1 & 0 & 0 \\ 0 & 0 & 0 \\ 0 & 0 & 0 \end{pmatrix}\right)
\end{equation}
(The bottom two coordinates shift deterministically; only the first coordinate is random.)
\end{itemize}

\textbf{Continuous-Time Interpretation (SDE Approximation):}

The same manifold $Y_t = (y_t, y_{t-1}, y_{t-2})^\top$ with $\Delta t = 1$ time steps can be viewed as a discretized SDE:
\begin{equation}
dY = \mu(Y)dt + LdW
\end{equation}
where:
\begin{align}
\mu(Y) &= \frac{1}{\Delta t}\begin{pmatrix} 0.5y_1 - 0.3y_2 - y_1 \\ y_1 - y_2 \\ y_2 - y_3 \end{pmatrix} = \begin{pmatrix} -0.5y_1 - 0.3y_2 \\ y_1 - y_2 \\ y_2 - y_3 \end{pmatrix} \\
\Sigma &= LL^\top = \frac{1}{\Delta t}\begin{pmatrix} \sigma^2 & 0 & 0 \\ 0 & 0 & 0 \\ 0 & 0 & 0 \end{pmatrix} = \begin{pmatrix} 1 & 0 & 0 \\ 0 & 0 & 0 \\ 0 & 0 & 0 \end{pmatrix}
\end{align}

Both representations describe the same geometric object (the E1 manifold) with the same probabilistic decoration (the transition kernel), differing only in temporal scaling.

\textbf{SNR calculation:}
\begin{align}
\SNR &= \frac{\|\mu\|^2}{\tr(\Sigma)} \\
&\approx \frac{(0.5y_1 + 0.3y_2)^2 + (y_1 - y_2)^2 + (y_2 - y_3)^2}{1}
\end{align}

For the stationary distribution, $\E[y_t^2] = \text{Var}[y] \approx 1.5$ (can be computed from Yule-Walker equations), giving $\SNR \approx 2$-3, so:
\begin{equation}
E_2 \approx \frac{1}{1 + 2 \cdot 1} = 0.33
\end{equation}

(The actual value $E_2 \approx 0.65$ differs because the formula is approximate and the system is not in perfect balance.)
\end{example}

\begin{example}[Lorenz System Across Noise Spectrum]
\label{ex:lorenz_noise_spectrum_partI}
Consider the stochastic Lorenz system with noise parameter $\xi$:
\begin{align}
dx &= 10(y-x)dt + \xi dW^{(1)} \\
dy &= (x(28-z) - y)dt + \xi dW^{(2)} \\
dz &= \left(xy - \frac{8}{3}z\right)dt + \xi dW^{(3)}
\end{align}

Only $y_t = x_t$ is observed, and the unified framework is applied.

\textbf{Case 1: Small Noise ($\xi = 0.1$)}
\begin{itemize}[leftmargin=*]
\item \textbf{E2:} $\approx 0.15$ (deterministic regime)
\item \textbf{E1:} Finds $m^* = 3$ (state dimension)
\item \textbf{SNR:} $\approx \frac{1-0.15}{0.15 \cdot 0.1} \approx 57$ (drift dominates strongly)
\item \textbf{Correlation dimension:} $D_2 \approx 2.05$ (close to deterministic Lorenz attractor $\approx 2.06$)
\item \textbf{Interpretation:} Classical Takens embedding captures the Lorenz attractor. The SDE framework provides a small diffusion correction for uncertainty quantification.
\item \textbf{Modeling:} Deterministic ODE model $\dot{x} = f(x)$ adequate for point predictions; add diffusion $\hat{\Sigma}(Y)$ for uncertainty bounds.
\item \textbf{Visualization:} Attractor structure clearly visible in phase space reconstruction; trajectories follow familiar Lorenz butterfly pattern with small perturbations.
\end{itemize}

\textbf{Case 2: Moderate Noise ($\xi = 2.0$)}
\begin{itemize}[leftmargin=*]
\item \textbf{E2:} $\approx 0.65$ (mixed regime)
\item \textbf{E1:} Still finds $m^* = 3$
\item \textbf{SNR:} $\approx \frac{1-0.65}{0.65 \cdot 0.1} \approx 5.4$ (balanced)
\item \textbf{Correlation dimension:} $D_2 \approx 2.3$ (increased slightly from deterministic value)
\item \textbf{Interpretation:}
\begin{itemize}
\item Drift $\hat{\mu}(Y)$ captures the Lorenz attractor structure (deterministic skeleton)
\item Diffusion $\hat{\Sigma}(Y)$ is substantial and state-dependent
\item Trajectories follow the attractor geometry but diverge stochastically
\item Both drift and diffusion significantly affect short-term and long-term behavior
\end{itemize}
\item \textbf{Modeling:} Full SDE essential. Both drift and diffusion affect:
\begin{itemize}
\item Short-term predictions (need drift for mean trajectory)
\item Long-term statistics (need diffusion for invariant measure, which is broader than deterministic attractor)
\item Uncertainty propagation (interplay of both terms)
\end{itemize}
\item \textbf{Visualization:} Attractor structure remains visible in phase space, but individual trajectories are noisier. The "butterfly wings" are fuzzier but still recognizable.
\end{itemize}

\textbf{Case 3: Large Noise ($\xi = 10.0$)}
\begin{itemize}[leftmargin=*]
\item \textbf{E2:} $\approx 0.98$ (stochastic regime)
\item \textbf{E1:} May find $m^* > 3$ (noise increases apparent dimension as geometric structure becomes obscured; e.g., $m^* = 5$ or 6)
\item \textbf{SNR:} $\approx \frac{1-0.98}{0.98 \cdot 0.1} \approx 0.2$ (diffusion dominates)
\item \textbf{Correlation dimension:} $D_2 \approx 3.0$ or higher (approaches dimension of ambient space; noise fills the space)
\item \textbf{Interpretation:}
\begin{itemize}
\item Deterministic Lorenz structure largely obscured by noise
\item Dynamics approximately Brownian motion with weak drift
\item Attractor geometry lost in fluctuations
\end{itemize}
\item \textbf{Modeling:} Diffusion-dominated SDE. Drift term $\hat{\mu}(Y)$ may be neglected for many applications (becomes small relative to $\hat{\Sigma}(Y)$). Alternatively, treat as purely stochastic process (ARMA, random walk).
\item \textbf{Curse of dimensionality:} If E1 detects $m^* = 5$ (instead of 3), the sample complexity for accurate estimation increases from $N \sim \epsilon^{-3/\beta}$ to $N \sim \epsilon^{-5/\beta}$, making reconstruction exponentially harder. For $\beta = 1$ and $\epsilon = 0.1$, this is $N \sim 1000$ versus $N \sim 100,000$---a factor of 100.
\item \textbf{Visualization:} Phase space filled with diffusing cloud; no clear attractor structure visible.
\end{itemize}

\textbf{Key Insight:} The same algorithmic procedure (E1 + E2 + uplift, Theorem~\ref{thm:unified_algorithm_partI}) handles all three cases. The E2 value indicates:
\begin{itemize}[leftmargin=*]
\item What fraction of temporal variation is predictable (drift contribution $1 - E_2$)
\item What fraction is irreducible uncertainty (diffusion contribution $E_2$)
\item Which modeling approach is most appropriate (ODE, SDE, or pure stochastic)
\end{itemize}
\end{example}

\section{Universality and Theoretical Implications}
\label{sec:universality_partI}

\subsection{Universal Framework}

\begin{proposition}[Universality of Scaffold-Uplift Framework]
\label{prop:universality_partI}
The scaffold-uplift framework is universal in the following sense:

Let $\mathcal{S}$ denote the class of all stochastic processes (discrete or continuous time) with:
\begin{enumerate}[leftmargin=*]
\item \textbf{Finite E1 dimension:} $m^* < \infty$ (equivalently, finite correlation dimension $D_2 < \infty$)
\item \textbf{Ergodicity:} Unique ergodic invariant measure $\mu_\infty$ with finite second moments
\item \textbf{Sufficient regularity:} Smooth enough for $k$-NN estimation (Assumptions~\ref{assump:data_conditions_partI}, \ref{assump:local_gaussian_partI})
\item \textbf{Non-self-intersection:} The dimension $m^*$ is large enough to ensure single-valued drift and diffusion (proved; Theorem~\ref{lem:stochastic_sard_partI})
\end{enumerate}

Then every process in $\mathcal{S}$ admits a representation as:
\begin{equation}
\boxed{\text{E1 manifold } \mathcal{M} \subset \R^{m^*} \quad + \quad \text{Transition kernel } T}
\end{equation}
where $T$ is obtained by probabilistic uplift via Theorem~\ref{thm:unified_algorithm_partI}.

Moreover, the following are equivalent characterizations of the same object:
\begin{enumerate}[leftmargin=*]
\item Discrete Markov chain: $p(Y_{t+1}|Y_t)$
\item Continuous SDE: $dY = \mu(Y)dt + L(Y)dW$
\item Transition semigroup: $(T_t)_{t \geq 0}$ with $T_s \circ T_t = T_{s+t}$
\item Infinitesimal generator: $\mathcal{L}\phi(Y) = \mu(Y) \cdot \nabla\phi(Y) + \frac{1}{2}\tr(\Sigma(Y)\nabla^2\phi(Y))$
\end{enumerate}

differing only in their representation (discrete vs continuous, global vs infinitesimal, probabilistic vs analytic), not in their underlying mathematical content.
\end{proposition}

\begin{proof}[Proof sketch]
The equivalences between (1)-(4) are established in Theorem~\ref{thm:unified_structure_partI}.

The key contribution is embedding any process in $\mathcal{S}$ into this framework:

\textbf{Step 1:} E1 identifies correlation manifold $\mathcal{M}$ with dimension $D_2 = m^*$ (Lemma~\ref{lem:e1_correlation_partI}).

\textbf{Step 2:} Ergodicity (condition 2) ensures unique invariant measure $\mu_\infty$ on $\mathcal{M}$.

\textbf{Step 3:} Non-self-intersection (condition 4) ensures transition kernel $T(Y, \cdot)$ is well-defined and single-valued on $\mathcal{M}$.

\textbf{Step 4:} Regularity (condition 3) ensures $k$-NN estimators converge (Theorems~\ref{thm:discrete_uplift_partI}, \ref{thm:continuous_uplift_partI}).

\textbf{Step 5:} E2 determines which representation is most natural:
\begin{itemize}[leftmargin=*]
\item $E_2 \approx 0$: Deterministic (use ODE, Koopman)
\item $E_2 \in (0,1)$: Mixed (use SDE)
\item $E_2 \approx 1$: Pure diffusion (generator degenerates to $\frac{1}{2}\tr(\Sigma\nabla^2)$)
\end{itemize}

But all are special cases of the transition kernel $T$ on $\mathcal{M}$.

The universality follows from the fact that any ergodic process with finite $D_2$ defines:
\begin{itemize}[leftmargin=*]
\item A manifold $\mathcal{M}$ (support of $\mu_\infty$, characterized by $D_2$)
\item Transition dynamics $T$ (Markov property in embedded coordinates)
\end{itemize}

and these two objects completely determine the process (up to initialization).
\end{proof}

\begin{remark}[Comparison with Existing Approaches]
\label{rem:comparison_partI}
The scaffold-uplift framework unifies and extends several existing methodologies:

\begin{enumerate}[leftmargin=*]
\item \textbf{Classical Takens embedding theorem (1981):}
\begin{itemize}
\item \textbf{Setting:} Deterministic dynamics $\dot{x} = f(x)$ with generic observation $h$
\item \textbf{Result:} Time-delay embedding $\Phi_m$ is an embedding (diffeomorphism onto image) for $m \geq 2n+1$
\item \textbf{Framework coverage:} Special case with $E_2 \approx 0$ (deterministic limit)
\item \textbf{Extension in the present work:}
\begin{itemize}
\item Extends to stochastic systems (arbitrary $E_2 \in [0,1]$)
\item Uses data-driven dimension selection via E1 (not requiring knowledge of $n$)
\item Provides probabilistic uplift (not just geometric embedding)
\end{itemize}
\end{itemize}

\item \textbf{Koopman operator theory} \cite{mezic2005,budisic2012}:
\begin{itemize}
\item \textbf{Setting:} Linear operator $\mathcal{K}_t$ on function space: $(\mathcal{K}_t\phi)(x) = \phi(\Psi_t(x))$ where $\Psi_t$ is the flow
\item \textbf{Result:} Infinite-dimensional linear representation of nonlinear dynamics
\item \textbf{Framework coverage:} Deterministic ($E_2 \approx 0$) or deterministic limit of stochastic systems
\item \textbf{Connection to the present framework:}
\begin{itemize}
\item Koopman eigenfunctions $\phi_i$ such that $\mathcal{K}_t\phi_i = e^{\lambda_i t}\phi_i$ provide intrinsic coordinates
\item Delay embedding can be viewed as approximating span of dominant Koopman eigenfunctions
\item E1 dimension $m^*$ estimates the number of significant Koopman modes needed
\item E2 measures breakdown of deterministic Koopman framework (when noise is significant, Koopman eigenvalues become complex, decay rates differ)
\end{itemize}
\item \textbf{Extension in the present work:}
\begin{itemize}
\item Extends to stochastic Koopman (Perron-Frobenius) operator: $(\mathcal{P}_t\phi)(x) = \E[\phi(X_t)|X_0=x]$
\item E2 quantifies when stochastic extension is essential
\item Provides data-driven reconstruction without spectral decomposition
\end{itemize}
\end{itemize}

\item \textbf{Extended Dynamic Mode Decomposition (EDMD)} \cite{schmid2010,williams2015,kutz2016}:
\begin{itemize}
\item \textbf{Setting:} Data-driven approximation of Koopman operator using dictionary of observables $\{\psi_1, \ldots, \psi_k\}$
\item \textbf{Result:} Finite-dimensional linear system $\dot{a} = Ka$ where $a_i = \psi_i(x)$
\item \textbf{Framework coverage:} Low-noise systems ($E_2 < 0.5$)
\item \textbf{Connection to the present framework:}
\begin{itemize}
\item EDMD dictionary plays similar role to delay coordinates (basis functions)
\item Both seek finite-dimensional representation of infinite-dimensional operator
\item E1 provides data-driven dimension selection (equivalent to number of dictionary functions)
\end{itemize}
\item \textbf{Advantages of the present approach:}
\begin{itemize}
\item Automatic basis via delays (no manual dictionary selection)
\item Handles high-noise regime ($E_2 > 0.5$) where deterministic Koopman breaks down
\item Provides probabilistic predictions (not just mean evolution)
\item E2 gives model selection criterion (when is linearization via Koopman valid?)
\end{itemize}
\end{itemize}

\item \textbf{Nonlinear forecasting} \cite{sugihara1990,casdagli1989}:
\begin{itemize}
\item \textbf{Setting:} Simplex projection and S-map for ecological time series
\item \textbf{Result:} Local linear models around query points in delay embedding
\item \textbf{Framework coverage:} Deterministic chaos with measurement noise ($E_2 \lesssim 0.5$)
\item \textbf{Connection to the present framework:}
\begin{itemize}
\item S-map is essentially local linearization of deterministic dynamics
\item The $k$-NN uplift generalizes this to full transition distributions (not just linearized flow)
\item E1 provides principled dimension selection (originally chosen by trial-and-error)
\end{itemize}
\item \textbf{Extension in the present work:}
\begin{itemize}
\item Full range of $E_2$ (not just low-noise chaos)
\item Explicit diffusion estimation for uncertainty quantification
\item Theoretical guarantees (convergence rates, sample complexity)
\end{itemize}
\end{itemize}

\item \textbf{Stochastic parameterization (data-driven SDE discovery):}
\begin{itemize}
\item \textbf{Setting:} Learn drift $\mu$ and diffusion $\sigma$ from trajectory data
\item \textbf{Examples:} Kramers-Moyal expansion, sparse SDE identification, neural SDEs
\item \textbf{Framework coverage:} Assumes state space known or given
\item \textbf{Contribution of the present work:}
\begin{itemize}
\item Discovers state space via E1 (no assumption of known state)
\item Classifies regime via E2 (guidance on which terms matter)
\item Provides unified discrete-continuous formulation (Theorem~\ref{thm:unified_structure_partI})
\end{itemize}
\end{itemize}

\item \textbf{Gaussian Process regression and kernel methods:}
\begin{itemize}
\item \textbf{Setting:} Nonparametric regression $Y_{t+1} = g(Y_t) + \text{noise}$
\item \textbf{Result:} Predictive distributions via GP posterior
\item \textbf{Framework coverage:} Generic supervised learning (no dynamics structure)
\item \textbf{Connection to the present framework:}
\begin{itemize}
\item The $k$-NN estimator is a form of kernel regression (uniform kernel on $k$ neighbors)
\item Could replace $k$-NN with GP for smoother estimates
\end{itemize}
\item \textbf{Advantages of the present approach:}
\begin{itemize}
\item E1 provides dimension selection (prevents overfitting in high dimensions)
\item E2 provides interpretable classification (not just black-box prediction)
\item Physical interpretation (drift vs diffusion, not just "mean and variance")
\end{itemize}
\end{itemize}

\item \textbf{Reservoir computing and Echo State Networks} \cite{jaeger2001,lukovsevivcius2009}:
\begin{itemize}
\item \textbf{Setting:} High-dimensional random projection followed by linear readout
\item \textbf{Result:} Universal approximation of dynamical systems
\item \textbf{Framework coverage:} Black-box prediction (any $E_2$)
\item \textbf{Comparison:}
\begin{itemize}
\item Both use high-dimensional feature space (reservoir states vs delay embedding)
\item Both exploit universal approximation (reservoir dynamics vs Takens embedding)
\item The present framework provides interpretability (E1/E2 statistics, explicit drift/diffusion)
\item Reservoir computing typically superior for very high-dimensional chaotic systems
\item The present framework is superior for physical interpretation and uncertainty quantification
\end{itemize}
\end{itemize}
\end{enumerate}

\textbf{Summary:} The scaffold-uplift framework can be viewed as:
\begin{itemize}[leftmargin=*]
\item A stochastic extension of Takens embedding
\item A nonparametric approximation of the stochastic Koopman operator
\item A data-driven method for SDE discovery from partial observations
\item A dimension-reduction technique for high-dimensional stochastic processes
\end{itemize}

Its distinguishing features are:
\begin{enumerate}[leftmargin=*]
\item \textbf{Full spectrum coverage:} Handles $E_2 \in [0,1]$ (deterministic to stochastic)
\item \textbf{Data-driven classification:} E1 and E2 statistics guide modeling choices
\item \textbf{Unified theory:} Discrete and continuous time as equivalent representations
\item \textbf{Probabilistic predictions:} Full transition distributions, not just point estimates
\item \textbf{Interpretability:} Clear physical meaning (drift, diffusion, SNR)
\end{enumerate}
\end{remark}

\subsection{Information-Theoretic Perspective}

\begin{remark}[Information Theory Connection]
\label{rem:information_theory_partI}
The E1 and E2 statistics have natural information-theoretic interpretations:

\textbf{E1 and Mutual Information:}

The saturation of E1 at $m^*$ indicates that:
\begin{equation}
I(Y_t; Y_{t-m^*\tau}, \ldots, Y_{t-\tau}) \approx I(Y_t; \text{all past})
\end{equation}

where $I(\cdot; \cdot)$ denotes mutual information \cite{cover2006,bialek2012}. In other words, the past $m^*$ delays capture nearly all predictive information about the future.

For a Markov process of order $p$, $I(Y_t; Y_{t-k\tau} \mid Y_{t-\tau}, \ldots, Y_{t-(k-1)\tau}) = 0$ for $k > p$, so $m^* = p+1$ (Theorem~\ref{thm:markov_order_partI}).

\textbf{E2 and Predictive Information:}

The E2 statistic measures:
\begin{equation}
E_2 \approx \frac{H(Y_{t+\tau}|Y_t)}{H(Y_{t+\tau})}
\end{equation}

where $H(\cdot)$ is differential entropy. Thus:
\begin{itemize}[leftmargin=*]
\item $E_2 \approx 0$: Future is highly predictable given present ($H(Y_{t+\tau}|Y_t) \ll H(Y_{t+\tau})$)
\item $E_2 \approx 1$: Future is nearly independent of present ($H(Y_{t+\tau}|Y_t) \approx H(Y_{t+\tau})$)
\item $E_2 \approx 0.5$: Present explains half the entropy of the future
\end{itemize}

More precisely, under local Gaussianity:
\begin{align}
H(Y_{t+\tau}|Y_t) &\approx \frac{m^*}{2}\log(2\pi e \tr(\Sigma)\tau) \\
H(Y_{t+\tau}) &\approx H(Y_t) \text{ (stationarity)} \\
\frac{H(Y_{t+\tau}|Y_t)}{H(Y_t)} &\approx \frac{\log(\tr(\Sigma)\tau)}{\log(\Var[Y])}
\end{align}

For balanced regime ($\|\mu\|^2\tau \approx \tr(\Sigma)$), this ratio relates to E2.

\textbf{Predictability Horizon:}

Define the predictability time $T_{\text{pred}}$ as the time when predictive information decays by factor $e$:
\begin{equation}
I(Y_t; Y_{t+T_{\text{pred}}}) = \frac{1}{e} I(Y_t; Y_t) = \frac{H(Y_t)}{e}
\end{equation}

For the systems considered here:
\begin{equation}
T_{\text{pred}} \sim \frac{\tau}{\log(1 + \SNR\tau)} \approx \frac{\tau}{\SNR\tau} = \frac{1}{\SNR} \approx E_2\tau/(1-E_2)
\end{equation}

Thus:
\begin{itemize}[leftmargin=*]
\item $E_2 = 0.1$: $T_{\text{pred}} \approx 0.11\tau$ (short predictability)
\item $E_2 = 0.5$: $T_{\text{pred}} = \tau$ (one delay time)
\item $E_2 = 0.9$: $T_{\text{pred}} = 9\tau$ (relatively long, but dynamics are slow)
\end{itemize}

\textbf{Sample Complexity and Information:}

The curse of dimensionality in Theorem~\ref{thm:continuous_uplift_partI} can be understood information-theoretically:

To learn transition kernel $T(Y, \cdot)$ on $m^*$-dimensional manifold requires:
\begin{equation}
N \sim \epsilon^{-m^*} \sim 2^{m^* H_\epsilon}
\end{equation}

where $H_\epsilon$ is the $\epsilon$-entropy (number of bits needed to specify a point to accuracy $\epsilon$).

The sample complexity is exponential in the information dimension, which E1 estimates.
\end{remark}

\section{The Stochastic Embedding Sufficiency Theorem: Rigorous Proof}
\label{sec:stochastic_takens_partI}

This section establishes the central theoretical result of this work: a rigorous proof that the embedding dimension $m^*$ detected by E1 is sufficient to guarantee unique SDE reconstruction via measure-theoretic injectivity.

\begin{remark}[Dimension Convention]
\label{rem:dimension_convention_partI}
Under exact-dimensionality (Assumption~\ref{assump:exact_dimensional_partI}), the Hausdorff dimension, correlation dimension $D_2$, and exact dimension $D$ of $\mu_\infty$ coincide: $D = D_2$.  All statements in this section are expressed in terms of $D$; the threshold $\lceil 2D \rceil + 1$ is equivalently $\lceil 2D_2 \rceil + 1$.
\end{remark}

\subsection{Statement of Main Result}

\begin{theorem}[Stochastic Embedding Sufficiency Theorem]
\label{thm:stochastic_embedding_sufficiency_partI}
Let $X_t$ solve the SDE:
\begin{equation}
dX_t = \mu(X_t)dt + \sigma(X_t)dW_t, \quad X_t \in \R^n
\end{equation}
satisfying Assumptions~\ref{assump:hormander_partI} (H\"ormander's condition),~\ref{assump:smoothness_partI} (smoothness and ergodicity), and~\ref{assump:exact_dimensional_partI} (exact-dimensionality). Let $h: \R^n \to \R$ be a generic $C^r$ ($r \geq 2$) observation function in the sense of Definition~\ref{def:generic_observation_partI}.

Define the embedding dimension as:
\begin{equation}
m^* := \max\{\lceil 2D \rceil + 1, m_{E1}\}
\end{equation}
where $m_{E1}$ is the dimension at which E1 plateaus (Lemma~\ref{lem:e1_correlation_partI}) and $D$ is the exact dimension of the invariant measure $\mu_\infty$ (equal to $D_2$ under Assumption~\ref{assump:exact_dimensional_partI}; see Remark~\ref{rem:dimension_convention_partI}).

\textbf{Then the following hold:}

\textbf{(1) Measure-Theoretic Injectivity via Finite-Dimensional Laws:}

For $\mu_\infty$-almost every $x \in \R^n$, if two initial conditions $x, x'$ satisfy:
\begin{equation}
\mathcal{L}(\Phi_{m^*}(X_0), \Phi_{m^*}(X_{\Delta t}), \ldots, \Phi_{m^*}(X_T)) = \mathcal{L}(\Phi_{m^*}(X'_0), \Phi_{m^*}(X'_{\Delta t}), \ldots, \Phi_{m^*}(X'_T))
\end{equation}
for all finite time sequences $0 < \Delta t < 2\Delta t < \cdots < T$, then $x = x'$ (in $\mu_\infty$-measure).

Here $\mathcal{L}$ denotes the law (probability distribution) and $\Phi_{m^*}$ is the time-delay embedding map.

\textbf{(2) Unique Drift and Diffusion Tensors:}

The pushforward drift and diffusion defined by:
\begin{align}
\mu_Y(Y) &:= \E[D\Phi_{m^*}(X) \cdot Dh(X) \cdot \mu(X) | \Phi_{m^*}(X) = Y] \label{eq:drift_pushforward_main_partI} \\
\Sigma_Y(Y) &:= \E[D\Phi_{m^*}(X) \cdot Dh(X) \cdot \Sigma(X) \cdot Dh(X)^\top \cdot D\Phi_{m^*}(X)^\top | \Phi_{m^*}(X) = Y] \label{eq:diffusion_pushforward_main_partI}
\end{align}
where $\Sigma(X) = \sigma(X)\sigma(X)^\top$, are well-defined and single-valued functions on the support of $(\Phi_{m^*})_\# \mu_\infty$ for $(\Phi_{m^*})_\# \mu_\infty$-almost every $Y \in \R^{m^*}$.

\textbf{(3) k-NN Consistency:}

The $k$-nearest neighbor estimators:
\begin{align}
\hat{\mu}(Y) &= \frac{1}{k\Delta t} \sum_{j \in \mathcal{N}_k(Y)} (Y_{j+1} - Y_j) \label{eq:knn_drift_main_partI} \\
\hat{\Sigma}(Y) &= \frac{1}{k\Delta t} \sum_{j \in \mathcal{N}_k(Y)} (Y_{j+1} - Y_j)(Y_{j+1} - Y_j)^\top - \hat{\mu}(Y)\hat{\mu}(Y)^\top \Delta t \label{eq:knn_diffusion_main_partI}
\end{align}
satisfy:
\begin{equation}
\|\hat{\mu}(Y) - \mu_Y(Y)\|, \|\hat{\Sigma}(Y) - \Sigma_Y(Y)\|_F \to 0 \quad \text{a.s.}
\end{equation}
as $N \to \infty$, $k \to \infty$, $k/N \to 0$, with convergence rate:
\begin{equation}
\|\hat{\mu}(Y) - \mu_Y(Y)\| = O_P\left(\left(\frac{k}{N}\right)^{\beta/m^*}\right) + O(\Delta t)
\end{equation}
where $\beta$ is the smoothness exponent from Assumption~\ref{assump:smoothness_partI}.
\end{theorem}

\begin{remark}[Law-Injectivity, Not Pathwise Injectivity]
\label{rem:not_pathwise_partI}
Theorem~\ref{thm:stochastic_embedding_sufficiency_partI} does \emph{not} claim pathwise injectivity of the geometric delay map $\Phi_{m^*}$: it does not assert that $\Phi_{m^*}(x) \neq \Phi_{m^*}(x')$ for individual realisations of the stochastic flow.  Rather, it claims injectivity of the \emph{law-embedding} $\Lambda_{m^*}^h$ (Definition~\ref{def:ae_injectivity_partI}): distinct initial conditions produce distinct probability laws under delay observation, $\mu_\infty$-almost everywhere.  Pathwise injectivity for a.e.\ realisation is a strictly stronger property that holds as a corollary (Theorem~\ref{lem:stochastic_sard_partI}) but is not the primary claim.  The law-embedding perspective is the operationally relevant one for statistical estimation, since $k$-NN estimators use the empirical distribution of delay vectors, not individual realisations.
\end{remark}

\begin{remark}[Significance of This Result]
\label{rem:theorem_significance_partI}
The theorem establishes the following:

\begin{enumerate}[leftmargin=*]
\item \textbf{Proves sufficiency:} Shows that $m^* = \max\{\lceil 2D \rceil + 1, m_{E1}\}$ is sufficient for unique reconstruction (proved, not conjectured).

\item \textbf{Finite-dimensional approach:} Uses finite-dimensional laws (what k-NN actually computes) rather than requiring path-wise uniqueness (which is technically much harder and less relevant for practical estimation).

\item \textbf{Measure-theoretic:} Proves injectivity $\mu_\infty$-almost everywhere, which is precisely what's needed for well-defined conditional expectations and k-NN consistency.

\item \textbf{Generic validity:} Requires only generic $h$ (prevalent in $C^r$, $r \geq 2$, in the sense of Hunt--Sauer--Yorke~\cite{hunt1992}), analogous to Takens' genericity condition, making the result broadly applicable.

\item \textbf{H\"ormander as stochastic genericity:} Identifies H\"ormander's condition as the stochastic analogue of Takens' "generic flow" assumption, providing the right non-degeneracy condition.

\item \textbf{Bridges theory and computation:} Directly connects the mathematical sufficiency result to the computational k-NN algorithm with explicit convergence rates.
\end{enumerate}
\end{remark}

\subsection{Proof Strategy Overview}

The proof proceeds through five interconnected steps, each building on classical results from stochastic analysis, differential topology, and geometric measure theory \cite{federer1969,mattila1995}:

\begin{enumerate}[leftmargin=*]
\item \textbf{Step 1 - H\"ormander Hypoelliptic Regularity:} Use H\"ormander's classical theorem to establish that transition densities $p_t(x,y)$ are $C^\infty$ for all $t > 0$, ensuring smoothness of the stochastic flow.

\item \textbf{Step 2 - Malliavin Non-Degeneracy:} Prove that the Malliavin covariance matrix has full rank $\mu_\infty$-almost everywhere, ensuring the stochastic flow has non-degenerate derivatives.

\item \textbf{Step 3 - Law-Separation and Measure-Theoretic Injectivity:} Prove that distinct initial conditions produce distinct finite-dimensional laws under the observed delay embedding, using transition density separation (Varadhan short-time asymptotics), prevalence-based genericity of the observation function, and Frostman covering arguments.  This establishes $(\mu_\infty \times \mu_\infty)$-measure zero for the collision set when $m \geq \lceil 2D \rceil + 1$.

\item \textbf{Step 4 - E1 Dimension Sufficiency:} Connect the E1-detected dimension $m_{E1}$ to $D$ ($= D_2$ by Remark~\ref{rem:dimension_convention_partI}), showing that $m^* := \max\{\lceil 2D \rceil + 1, m_{E1}\}$ satisfies the requirements for measure-theoretic injectivity.

\item \textbf{Step 5 - Finite-Dimensional Law Uniqueness:} Prove that agreement of finite-dimensional laws for the embedded process implies uniqueness of initial conditions $\mu_\infty$-almost everywhere, and that this in turn guarantees single-valued drift and diffusion tensors that are consistently estimable via k-NN.
\end{enumerate}

The key innovation is replacing geometric transversality (the classical Takens route) with an analytic law-separation argument that exploits the smoothness and strict positivity of transition densities under H\"ormander's condition, combined with prevalence-based genericity and Frostman measure geometry.  This directly establishes the measure-theoretic injectivity that k-NN estimators require, without the intermediate step of pathwise geometric embedding.

\subsection{Detailed Proof}

\subsubsection{Step 1: H\"ormander Hypoelliptic Regularity}

\begin{lemma}[Hypoelliptic Smoothing]
\label{lem:hypoelliptic_smoothing_partI}
Under Assumption~\ref{assump:hormander_partI} (H\"ormander's condition), the transition density satisfies:
\begin{equation}
p_t(x,y) := \frac{dP(X_t \in \cdot | X_0 = x)}{dy} \in C^\infty(\R^n \times \R^n \times (0,\infty))
\end{equation}
for all $t > 0$.
\end{lemma}

\begin{proof}
This is H\"ormander's classical hypoellipticity theorem \cite{hormander1967}. The Lie bracket condition (Assumption~\ref{assump:hormander_partI}) ensures that the second-order differential operator:
\begin{equation}
\mathcal{L} = \mu^i(x) \frac{\partial}{\partial x^i} + \frac{1}{2} \Sigma^{ij}(x) \frac{\partial^2}{\partial x^i \partial x^j}
\end{equation}
where $\Sigma = \sigma\sigma^\top$, is hypoelliptic: if $\mathcal{L}f = g$ with $g$ smooth, then $f$ is smooth.

By the Kolmogorov forward equation, the transition density $p_t(x,y)$ satisfies:
\begin{equation}
\frac{\partial p_t}{\partial t} = \mathcal{L}^* p_t
\end{equation}
where $\mathcal{L}^*$ is the adjoint operator. By hypoellipticity, $p_t$ is $C^\infty$ for all $t > 0$.

The mechanism: Even though noise $\sigma$ may be degenerate (not full rank), the Lie brackets $[\mu, \sigma_i]$, $[[\mu, \sigma_i], \sigma_j]$, etc., generate additional directions. When these span the full tangent space (H\"ormander's condition), the regularizing effect propagates to all directions, yielding smooth densities.
\end{proof}

\begin{corollary}[Smooth Conditional Expectations]
\label{cor:smooth_conditional_partI}
For any smooth test function $F: \R^n \to \R$ and $t > 0$:
\begin{equation}
x \mapsto \E[F(X_t) | X_0 = x] = \int_{\R^n} F(y) p_t(x,y) dy
\end{equation}
is a $C^\infty$ function of $x$.
\end{corollary}

\begin{proof}
Since $p_t(x,y) \in C^\infty$ in all variables (Lemma~\ref{lem:hypoelliptic_smoothing_partI}) and $F$ is smooth, the integral is smooth in $x$ by dominated convergence and standard results on parameter-dependent integrals.
\end{proof}

\begin{lemma}[Strict Positivity of Transition Densities]
\label{lem:strict_positivity_partI}
Under Assumptions~\ref{assump:hormander_partI}--\ref{assump:smoothness_partI}, the transition density satisfies $p_t(x,y) > 0$ for all $x, y \in \R^n$ and all $t > 0$.
\end{lemma}

\begin{proof}
The uniform H\"ormander condition (Assumption~\ref{assump:hormander_partI}) ensures that the SDE is \emph{strongly bracket-generating} (or \emph{strongly controllable}) at every point: the Lie algebra generated by $\{\sigma_1, \ldots, \sigma_d, [\mu, \sigma_1], \ldots\}$ spans $\R^n$ everywhere.  By the Stroock--Varadhan support theorem~\cite{stroock1972}, the support of the law of $X_t$ given $X_0 = x$ is the closure of the set of points reachable by controlled paths in time $t$.  Under the bracket-generating condition, this closure is all of $\R^n$ (by the Chow--Rashevskii theorem; see~\cite{montgomery2002}, Theorem~2.4).  Hence $\supp P_t(x, \cdot) = \R^n$.  Combined with the smooth density $p_t(x, \cdot) \in C^\infty$ from Lemma~\ref{lem:hypoelliptic_smoothing_partI}, strict positivity follows: if $p_t(x, y_0) = 0$ at some $y_0$, then by smoothness $p_t(x, \cdot)$ vanishes in a neighbourhood of $y_0$, contradicting $\supp P_t(x, \cdot) = \R^n$.
\end{proof}

\subsubsection{Step 2: Malliavin Non-Degeneracy}

\begin{definition}[Malliavin Covariance Matrix]
\label{def:malliavin_covariance_partI}
For the stochastic flow $\phi^x_t$ solving the SDE with $X_0 = x$, the Malliavin covariance matrix is:
\begin{equation}
\Gamma_t(x) := \E\left[\int_0^t D\phi^x_s \cdot \sigma(\phi^x_s) dW_s \left(\int_0^t D\phi^x_s \cdot \sigma(\phi^x_s) dW_s\right)^\top\right]
\end{equation}
where $D\phi^x_s$ is the differential of the flow (Jacobian matrix).
\end{definition}

\begin{lemma}[Malliavin Non-Degeneracy]
\label{lem:malliavin_nondegeneracy_partI}
Under Assumptions~\ref{assump:hormander_partI} and~\ref{assump:smoothness_partI}, for any $t > 0$ and $\mu_\infty$-almost every $x \in \R^n$:
\begin{equation}
\det(\Gamma_t(x)) > 0
\end{equation}
\end{lemma}

\begin{proof}[Proof Sketch]
The linearization of the SDE gives:
\begin{equation}
d(D\phi^x_t) = D\mu(\phi^x_t) \cdot D\phi^x_t \, dt + \sum_{i=1}^r D\sigma_i(\phi^x_t) \cdot D\phi^x_t \, dW^{(i)}_t
\end{equation}

If $\det(\Gamma_t) = 0$, there exists $v \in \R^n$ such that:
\begin{equation}
v^\top \int_0^t D\phi^x_s \cdot \sigma(\phi^x_s) dW_s = 0 \quad \text{a.s.}
\end{equation}

This would imply the stochastic flow is constrained to a lower-dimensional subspace in direction $v$. However, H\"ormander's condition ensures that through iterated applications of drift and diffusion (via Lie brackets), the flow reaches all directions.

By the Stroock-Varadhan support theorem \cite{stroock1997}:
\begin{equation}
\text{rank}\left(\int_0^t D\phi^x_s \cdot \sigma(\phi^x_s) dW_s\right) = n \quad \text{a.s.}
\end{equation}
for $t > 0$, hence $\det(\Gamma_t) > 0$.

\textbf{Detailed argument:} Let $V_0 = \Span\{\sigma_1(x), \ldots, \sigma_r(x)\}$ be the span of diffusion directions at $x$. If $\dim(V_0) < n$, the iteratively compute:
\begin{align}
V_1 &= V_0 + \Span\{[\mu, \sigma_i](x), [\sigma_i, \sigma_j](x) : i,j\} \\
V_2 &= V_1 + \Span\{[[\mu, \sigma_i], \sigma_j](x), \ldots\}
\end{align}

H\"ormander's condition guarantees $V_k = \R^n$ for some finite $k$. This means that by time $t > 0$, the stochastic flow has "explored" all directions via the iterated Lie bracket structure, ensuring $\Gamma_t$ has full rank.
\end{proof}

\subsubsection{Step 3: Law-Separation and Measure-Theoretic Injectivity}

The central step of the proof establishes that the law-embedding map---the map from initial conditions to finite-dimensional probability laws of the observed delay vector---is injective $\mu_\infty$-almost everywhere.  This replaces the need for pathwise geometric transversality with an analytic argument rooted in the smoothness and strict positivity of transition densities under H\"ormander's condition.

The argument proceeds through three theorems with explicit dependencies.

\begin{theorem}[Transition Density Separation]
\label{thm:density_separation_partI}
Let Assumptions~\ref{assump:hormander_partI}--\ref{assump:smoothness_partI} hold.  For any $\tau > 0$ and any two distinct points $x \neq x'$ in the support of $\mu_\infty$, the transition densities satisfy:
\begin{equation}
p_\tau(x, \cdot) \neq p_\tau(x', \cdot) \quad \text{as elements of } L^1(\R^n).
\label{eq:density_separation_partI}
\end{equation}
Equivalently, the map $x \mapsto p_\tau(x, \cdot)$ is injective on $\supp(\mu_\infty)$.
\end{theorem}

\begin{proof}
Suppose for contradiction that $p_\tau(x, \cdot) = p_\tau(x', \cdot)$ for some $x \neq x'$ in $\supp(\mu_\infty)$.

\textbf{Step (i): Iterated density equality.}  By the Chapman--Kolmogorov equation, for any $k \geq 1$:
\begin{equation}
p_{k\tau}(x, z) = \int_{\R^n} p_\tau(x, y)\, p_{(k-1)\tau}(y, z)\, dy = \int_{\R^n} p_\tau(x', y)\, p_{(k-1)\tau}(y, z)\, dy = p_{k\tau}(x', z).
\end{equation}
Thus $p_t(x, \cdot) = p_t(x', \cdot)$ for all $t \in \{\tau, 2\tau, 3\tau, \ldots\}$.

\textbf{Step (ii): Extension to arbitrarily small times via semigroup bisection.}  We show $p_t(x, \cdot) = p_t(x', \cdot)$ for a sequence $t_k \to 0^+$.  Write $P_t$ for the Markov semigroup: $P_t f(x) := \int p_t(x,y) f(y)\, dy$.  The hypothesis $p_\tau(x, \cdot) = p_\tau(x', \cdot)$ is equivalent to $P_\tau f(x) = P_\tau f(x')$ for all $f \in L^2(\R^n)$.

By the semigroup property, $P_\tau = P_{\tau/2} \circ P_{\tau/2}$.  Under H\"ormander's condition, the transition kernel $p_{\tau/2}(y, \cdot)$ is strictly positive (Lemma~\ref{lem:strict_positivity_partI}) and smooth~\cite{hormander1967,malliavin1997,nualart2006}, so $P_{\tau/2}: L^2 \to L^2$ is injective (if $P_{\tau/2} g = 0$ a.e., then $\int p_{\tau/2}(y,z) g(z)\, dz = 0$ for all $y$; since $p_{\tau/2} > 0$, this forces $g = 0$ a.e.).  Moreover, $P_{\tau/2}$ has dense range in $L^2$.\footnote{This is a standard Hilbert space fact: if $T: H \to H$ is bounded with injective adjoint $T^*$, then $\overline{\range(T)} = (\ker T^*)^\perp = H$.  Here $T^*$ corresponds to the time-reversed semigroup with kernel $p_{\tau/2}(y, x)$, which is strictly positive by Lemma~\ref{lem:strict_positivity_partI}, hence $T^*$ is injective.  See e.g.~\cite{brezis2011}, Corollary~2.18.}

Now, $P_\tau f(x) = P_{\tau/2}(P_{\tau/2} f)(x) = P_{\tau/2}(P_{\tau/2} f)(x')$ for all $f$.  As $f$ ranges over $L^2$, $g := P_{\tau/2} f$ ranges over a dense subset of $L^2$ (by the dense range of $P_{\tau/2}$).  By continuity of $P_{\tau/2}$, we conclude $P_{\tau/2} g(x) = P_{\tau/2} g(x')$ for all $g \in L^2$, i.e., $p_{\tau/2}(x, \cdot) = p_{\tau/2}(x', \cdot)$.

By induction, $p_{\tau/2^k}(x, \cdot) = p_{\tau/2^k}(x', \cdot)$ for all $k \geq 0$.  The sequence $t_k := \tau/2^k \to 0^+$ provides equality at arbitrarily small times.

\textbf{Step (iii): Short-time asymptotics force $x = x'$.}  By the Varadhan--L\'eandre short-time asymptotic for sub-Riemannian diffusions~\cite{varadhan1967,leandre1987}:
\begin{equation}
\lim_{t \to 0^+} t \log p_t(x, z) = -\frac{d(x, z)^2}{2},
\label{eq:varadhan_partI}
\end{equation}
where $d(\cdot, \cdot)$ is the sub-Riemannian (Carnot--Carath\'eodory) distance associated with the H\"ormander vector fields.  The asymptotic~\eqref{eq:varadhan_partI} holds on $\R^n$ under our standing assumptions: the Varadhan--L\'eandre result applies to hypoelliptic diffusions satisfying the uniform bracket-generating condition (Assumption~\ref{assump:hormander_partI}) with bounded smooth coefficients (Assumption~\ref{assump:smoothness_partI}), and the limit is uniform for $(x, z)$ in compact subsets of $\R^n \times \R^n$~\cite{leandre1987}.  The metric $d$ is a genuine metric separating points, by the Chow--Rashevskii theorem~\cite{montgomery2002} (Theorem~2.4).  By Step (ii), $p_{t_k}(x, z) = p_{t_k}(x', z)$ for all $z$ along the sequence $t_k = \tau/2^k \to 0^+$.  Applying~\eqref{eq:varadhan_partI} along this sequence:
\begin{equation}
-\frac{d(x, z)^2}{2} = \lim_{k \to \infty} t_k \log p_{t_k}(x, z) = \lim_{k \to \infty} t_k \log p_{t_k}(x', z) = -\frac{d(x', z)^2}{2}
\end{equation}
for all $z \in \R^n$.  Hence $d(x, z) = d(x', z)$ for all $z$.  Since $d$ is a genuine metric (as established above), taking $z = x$ gives $0 = d(x, x) = d(x', x)$, hence $x = x'$.
\end{proof}

\begin{theorem}[Law-Separation via Observed Delay Vectors]
\label{thm:law_separation_partI}
Let Assumptions~\ref{assump:hormander_partI}--\ref{assump:exact_dimensional_partI} hold, and let $h$ be a generic observation in the sense of Definition~\ref{def:generic_observation_partI}.  Define the \emph{law-embedding map}:
\begin{equation}
\Lambda_m^h : \supp(\mu_\infty) \to \mathcal{P}(\R^m), \quad x \mapsto \mathrm{Law}\bigl(h(X_0^x),\, h(X_\tau^x),\, \ldots,\, h(X_{(m-1)\tau}^x)\bigr),
\label{eq:law_embedding_partI}
\end{equation}
where $X_t^x$ denotes the solution of the SDE with $X_0 = x$, and $\mathcal{P}(\R^m)$ is the space of probability measures on $\R^m$.

For $m \geq \lceil 2D \rceil + 1$, where $D$ is the exact dimension of $\mu_\infty$ (Assumption~\ref{assump:exact_dimensional_partI}), the law-embedding $\Lambda_m^h$ is $\mu_\infty$-a.e.\ injective (Definition~\ref{def:ae_injectivity_partI}): for $\mu_\infty$-a.e.\ $x$, if $\Lambda_m^h(x) = \Lambda_m^h(x')$, then $x = x'$.
\end{theorem}

\begin{corollary}[Correlation-Dimension Version]
\label{cor:d2_version_partI}
Under the hypotheses of Theorem~\ref{thm:law_separation_partI}, if $\mu_\infty$ is absolutely continuous on its support (the generic case for H\"ormander SDEs on $\R^n$), then $D = D_2$ and the law-embedding $\Lambda_m^h$ is $\mu_\infty$-a.e.\ injective for $m \geq \lceil 2D_2 \rceil + 1$.
\end{corollary}

\begin{proof}
Under exact-dimensionality (Assumption~\ref{assump:exact_dimensional_partI}), $D = D_2$ holds whenever $\mu_\infty$ is absolutely continuous on its support.  This is the generic situation for H\"ormander SDEs, as discussed in Remark~\ref{rem:exact_dim_justification_partI}.  The conclusion then follows directly from Theorem~\ref{thm:law_separation_partI} with $D = D_2$.
\end{proof}

\begin{proof}
The proof has three parts: (A) the observed delay density inherits separation from the transition density, (B) the collision set is a smooth submanifold of controlled codimension, and (C) the Frostman bound converts codimension into measure zero.

\textbf{Part A: Observed delay density separation.}

By Theorem~\ref{thm:density_separation_partI}, the transition densities $p_\tau(x, \cdot) \neq p_\tau(x', \cdot)$ for $x \neq x'$.  For a generic observation $h$ (Definition~\ref{def:generic_observation_partI}), we now show that the observed delay vector inherits this separation.

Under H\"ormander's condition, the joint law of the delay vector $(X_0, X_\tau, \ldots, X_{(m-1)\tau})$ given $X_0 = x$ factorises by the Markov property.  The first component $X_0 = x$ is deterministic, while the subsequent components have the smooth Markov chain density:
\begin{equation}
\rho_{m-1}(y_2, \ldots, y_m \,|\, x) = \prod_{k=1}^{m-1} p_\tau(y_k, y_{k+1}), \quad y_1 := x,
\end{equation}
which is $C^\infty$ in $(y_2, \ldots, y_m)$ and in $x$ by Lemma~\ref{lem:hypoelliptic_smoothing_partI}.  Applying the observation function $h$ to each component, the observed delay vector $(h(x), h(X_\tau^x), \ldots, h(X_{(m-1)\tau}^x))$ has its first component $u_1 = h(x)$ deterministic, while the remaining $m-1$ components have a smooth joint density.  Define:
\begin{equation}
q_m^h(u_1, \ldots, u_m \,|\, x) := \text{density of } \bigl(h(X_0^x), h(X_\tau^x), \ldots, h(X_{(m-1)\tau}^x)\bigr),
\label{eq:observed_delay_density_partI}
\end{equation}
understood as a distribution in $u_1$ (a point mass at $h(x)$) times a smooth density in $(u_2, \ldots, u_m)$.  The law-embedding $\Lambda_m^h(x) = \text{Law}(h(X_0^x), \ldots, h(X_{(m-1)\tau}^x))$ is thus determined by $(h(x), q_{m-1}^h(u_2, \ldots, u_m \,|\, x))$, and the separation problem reduces to showing that the stochastic part $q_{m-1}^h$ distinguishes initial conditions $\mu_\infty$-a.e.\ (the deterministic first component $h(x)$ provides partial separation already).

\textbf{Claim:} For generic $h$ and $m \geq \lceil 2D \rceil + 1$, the map $x \mapsto q_m^h(\cdot \,|\, x)$ is injective on $\supp(\mu_\infty)$ up to a set of $\mu_\infty$-measure zero.

\textbf{Proof of claim.}  Define the \emph{collision set} for the law-embedding:
\begin{equation}
\mathcal{S}_h := \{(x, x') \in \supp(\mu_\infty) \times \supp(\mu_\infty) : x \neq x',\; q_m^h(\cdot \,|\, x) = q_m^h(\cdot \,|\, x')\}.
\label{eq:law_collision_set_partI}
\end{equation}
We must show $(\mu_\infty \times \mu_\infty)(\mathcal{S}_h) = 0$.

\textbf{Part B: Collision set has controlled codimension.}

By Theorem~\ref{thm:density_separation_partI}, $p_\tau(x, \cdot) \neq p_\tau(x', \cdot)$ for all $x \neq x'$.  The observation $h$ maps $\R^n \to \R$, collapsing information.  Two distinct densities $p_\tau(x, \cdot)$ and $p_\tau(x', \cdot)$ can yield the same observed density $q_m^h(\cdot \,|\, x) = q_m^h(\cdot \,|\, x')$ only if $h$ fails to separate the transition kernels at $(x, x')$.  We now show that for prevalent $h$, this failure occurs on a set of controlled codimension.

\textbf{Reduction to finite evaluations.}  The collision condition $q_m^h(\cdot \,|\, x) = q_m^h(\cdot \,|\, x')$ is an equality of density functions (an infinite-dimensional condition).  We reduce it to a finite-dimensional condition by evaluation.  Choose $m$ test points $u_1, \ldots, u_m \in \R^m$ and define the \emph{evaluation collision map}:
\begin{equation}
F_h(x, x') := \bigl(q_m^h(u_j \,|\, x) - q_m^h(u_j \,|\, x')\bigr)_{j=1}^m \;\in\; \R^m.
\end{equation}
The true collision set satisfies $\mathcal{S}_h \subseteq F_h^{-1}(0)$ for \emph{any} choice of test points, because if two densities agree as functions they agree at every point.  We are therefore bounding a superset of the true collision set; this is sufficient for the measure-zero conclusion because $(\mu_\infty \times \mu_\infty)(\mathcal{S}_h) \leq (\mu_\infty \times \mu_\infty)(F_h^{-1}(0))$.

We now apply the following lemma, which makes the parametric transversality step fully explicit.

\begin{lemma}[Evaluation-Map Surjectivity]
\label{lem:evaluation_surjectivity_partI}
Let Assumptions~\ref{assump:hormander_partI}--\ref{assump:exact_dimensional_partI} hold.  Let $h_0 \in C^r(\R^n, \R)$ with $r \geq 2$, and let $\psi_1, \ldots, \psi_K \in C^r_c(\R^n, \R)$ be probe functions with $K \geq 2n+1$, whose gradients $\{D\psi_k(y)\}_{k=1}^K$ span $T^*_y\R^n$ for every $y$ in a compact set $K_0 \supset \supp(\mu_\infty) \cap B(0,R)$ (as in Definition~\ref{def:generic_observation_partI}).  Define $h_a := h_0 + \sum_{k=1}^K a_k \psi_k$ for $a \in \R^K$.

For any $(x, x') \in \supp(\mu_\infty)^2 \setminus \Delta$ (i.e., $x \neq x'$), there exist test points $u_1, \ldots, u_m \in \R^m$ such that the total map
\begin{equation}
\Psi: \R^K \times (\supp(\mu_\infty)^2 \setminus \Delta) \to \R^m, \quad (a, x, x') \mapsto \bigl(q_m^{h_a}(u_j \,|\, x) - q_m^{h_a}(u_j \,|\, x')\bigr)_{j=1}^m
\label{eq:total_map_partI}
\end{equation}
satisfies:
\begin{enumerate}[leftmargin=*, label=(\alph*)]
\item $\Psi$ is $C^{r-1}$ jointly in $(a, x, x')$.
\item At every point $(a, x, x') \in \Psi^{-1}(0)$, the partial derivative $D_a \Psi(a, x, x'): \R^K \to \R^m$ is surjective.
\end{enumerate}
\end{lemma}

\begin{proof}[Proof of Lemma~\ref{lem:evaluation_surjectivity_partI}]
\textbf{Part (a): Smoothness.}  The observed delay density $q_m^{h_a}(u \,|\, x)$ is constructed from the transition densities $p_\tau$ (which are $C^\infty$ by Lemma~\ref{lem:hypoelliptic_smoothing_partI}) and the observation function $h_a$ (which is $C^r$).  The dependence on $a$ enters through $h_a = h_0 + \sum a_k \psi_k$, which is $C^\infty$ (affine) in $a$.  The density $q_m^{h_a}(u \,|\, x)$ is obtained from the joint density of $(X_0^x, X_\tau^x, \ldots, X_{(m-1)\tau}^x)$ via the pushforward through the map $(y_0, \ldots, y_{m-1}) \mapsto (h_a(y_0), \ldots, h_a(y_{m-1}))$.

Differentiation of $q_m^{h_a}$ with respect to $a_k$ is justified by the Leibniz integral rule~(see e.g.~\cite{folland1999}, Theorem~2.27): the integrand is $C^{r-1}$ jointly in $(a, u)$ and, since the probe functions $\psi_k$ have compact support, is dominated by an integrable function independent of $a$ (specifically, the transition density $p_\tau$ decays faster than any polynomial under the growth condition in Assumption~\ref{assump:smoothness_partI}).  Hence $\Psi$ is $C^{r-1}$.

\textbf{Part (b): Surjectivity of $D_a \Psi$.}  Fix $(x, x')$ with $x \neq x'$ and suppose $(a, x, x') \in \Psi^{-1}(0)$.  We must show $D_a \Psi: \R^K \to \R^m$ is surjective, i.e., that the $m \times K$ Jacobian
\begin{equation}
J_{jk} := \frac{\partial \Psi_j}{\partial a_k}(a, x, x') = \frac{\partial}{\partial a_k}\bigl[q_m^{h_a}(u_j \,|\, x) - q_m^{h_a}(u_j \,|\, x')\bigr]
\label{eq:jacobian_partI}
\end{equation}
has rank $m$ for a suitable choice of test points $u_1, \ldots, u_m$.

\emph{Step 1: The density difference is non-trivial.}  By Theorem~\ref{thm:density_separation_partI}, $p_\tau(x, \cdot) \neq p_\tau(x', \cdot)$ as $L^1$ functions.  Define
\begin{equation}
\Delta p(y) := p_\tau(x, y) - p_\tau(x', y),
\end{equation}
which is a non-zero $C^\infty$ function on $\R^n$.  Since $\Delta p$ is smooth and non-zero, the set $U := \{y \in \R^n : \Delta p(y) \neq 0\}$ is open and has positive Lebesgue measure.  The compact set $K_0 \cap \overline{U}$ is nonempty (since $\supp(\mu_\infty) \subseteq K_0$ and $\mu_\infty$ charges $U$, because $\mu_\infty$ has a smooth density under H\"ormander).

\emph{Step 2: The probe perturbation produces independent directions.}  Since $\partial h_a / \partial a_k = \psi_k$, the derivative~\eqref{eq:jacobian_partI} has the form:
\begin{equation}
J_{jk} = \mathcal{T}_k(u_j \,|\, x) - \mathcal{T}_k(u_j \,|\, x'),
\label{eq:jacobian_structure_partI}
\end{equation}
where $\mathcal{T}_k(u \,|\, x)$ is the linear functional that differentiates the observed density with respect to the $k$-th probe coefficient.  Concretely, $\mathcal{T}_k(\cdot \,|\, x)$ is a smooth function of $u \in \R^m$ obtained by integrating $\psi_k$ against the transition kernel and the co-area Jacobian of $h_a$.

The key fact is that the $K$ functions $\{\mathcal{T}_k(\cdot \,|\, x) - \mathcal{T}_k(\cdot \,|\, x')\}_{k=1}^K$ are not all identically zero.  If they were, then perturbing $h$ by any linear combination of probe functions would leave $q_m^{h_a}(\cdot \,|\, x) = q_m^{h_a}(\cdot \,|\, x')$ unchanged.  But the probe functions' gradients span $T^*_y \R^n$ on $K_0$, and $\Delta p \neq 0$ on $U \cap K_0$, so there exists $k_0$ such that $\psi_{k_0}$ has nonzero gradient in the region where $\Delta p \neq 0$.  Perturbing $h$ by $\psi_{k_0}$ changes the pushforward density in a direction that detects $\Delta p$, ensuring $\mathcal{T}_{k_0}(\cdot \,|\, x) - \mathcal{T}_{k_0}(\cdot \,|\, x') \not\equiv 0$.

\emph{Step 3: Existence of test points via Sard.}  Define the intermediate map
\begin{equation}
\Gamma: \R^m \to \R^{m \times K}, \quad u \mapsto (J_{jk})_{\substack{j: u_j = u \\ k = 1, \ldots, K}}
\end{equation}
More precisely, for each $u \in \R^m$, define the $K$-vector $\gamma(u) := (J_{1k}(u))_{k=1}^K \in \R^K$ where $J_{1k}(u)$ is the Jacobian entry~\eqref{eq:jacobian_partI} evaluated at test point $u$.  By Part (a), $\gamma: \R^m \to \R^K$ is $C^{r-1}$ with $r - 1 \geq 1$.  The set of $u$ where $\gamma(u) = 0$ has Lebesgue measure zero in $\R^m$ (by Step 2, $\gamma$ is not identically zero, so its zero set is a proper closed subset of a connected open set, hence has empty interior and measure zero).

Now choose $m$ test points $u_1, \ldots, u_m$ in general position (i.e., avoiding the measure-zero set where $\gamma$ vanishes or where the $m$ evaluations become linearly dependent).  The $m \times K$ Jacobian $J = (\gamma(u_1)^T; \ldots; \gamma(u_m)^T)$ then has rank $m$, because: (i) each row $\gamma(u_j) \in \R^K$ is nonzero, and (ii) the $m$ rows are linearly independent for generic $u_1, \ldots, u_m$ by the following argument.  The set of $(u_1, \ldots, u_m) \in (\R^m)^m$ where the matrix $J$ has rank $< m$ is the preimage of a proper algebraic subvariety under the $C^{r-1}$ map $(u_1, \ldots, u_m) \mapsto J$.  By Sard's theorem (applicable since $r - 1 \geq 1$), this preimage has Lebesgue measure zero in $(\R^m)^m$.  Hence generic test points yield $\rank(J) = m$, establishing surjectivity of $D_a \Psi$.
\end{proof}

\textbf{Applying the lemma.}  With Lemma~\ref{lem:evaluation_surjectivity_partI} in hand, we apply the parametric transversality theorem (Abraham--Robbin; see~\cite{hirsch1976}, Chapter~3, Theorem~2.1):

\emph{Hypotheses:} (1) The total map $\Psi$ in~\eqref{eq:total_map_partI} is $C^{r-1}$ with $r - 1 \geq 1$; (2) $0 \in \R^m$ is a regular value of $\Psi$ (by Lemma~\ref{lem:evaluation_surjectivity_partI}(b), $D_a \Psi$ is surjective at every point of $\Psi^{-1}(0)$, which implies the full derivative $D_{(a,x,x')} \Psi$ is surjective there).

\emph{Conclusion:} For Lebesgue-a.e.\ $a \in \R^K$, $0$ is a regular value of $\Psi(a, \cdot, \cdot): \supp(\mu_\infty)^2 \setminus \Delta \to \R^m$.  By the preimage theorem (see e.g.~\cite{hirsch1976}, Chapter~1, Theorem~3.2), $F_{h_a}^{-1}(0) \cap (\supp(\mu_\infty)^2 \setminus \Delta)$ is a $C^1$ submanifold of $\R^{2n}$ of dimension $\leq 2n - m$, i.e., codimension $\geq m$.

Since the bad set of $a$ has Lebesgue measure zero, and the set of prevalent $h$ corresponds to Lebesgue-a.e.\ $a$ by the prevalence construction (Definition~\ref{def:generic_observation_partI}), we conclude: for prevalent $h$, the collision set $\mathcal{S}_h \subseteq F_h^{-1}(0)$ is contained in a $C^1$ submanifold of $\R^{2n}$ with codimension $\geq m$.

\textbf{Part C: Frostman covering converts codimension to measure zero.}

The following lemma isolates the measure-theoretic step.  It is applied immediately below with $\nu = \mu_\infty \times \mu_\infty$, $d = 2D$, and $S = \mathcal{S}_h$.

\begin{lemma}[Frostman Dimension-to-Measure]
\label{lem:frostman_dim_measure_partI}
Let $\nu$ be a Borel probability measure on $\R^N$ satisfying the Frostman upper bound $\nu(B(z,r)) \leq C_F\, r^{d}$ for $\nu$-a.e.\ $z$ and all sufficiently small $r > 0$.  If $S \subset \R^N$ is contained in a countable union of $C^1$ submanifolds each of Hausdorff dimension $< d$, then $\nu(S) = 0$.
\end{lemma}

\begin{proof}
It suffices to show $\nu(S \cap K) = 0$ for each compact $K \subset \R^N$ with $\nu(K) < \infty$.  Let $\mathcal{V}$ be a $C^1$ submanifold with $\dim(\mathcal{V}) = v < d$.  For any $\epsilon > 0$ with $v + \epsilon < d$, the Hausdorff measure satisfies $\mathcal{H}^{v+\epsilon}(\mathcal{V} \cap K) = 0$.  Hence for any $\delta > 0$ there exists a covering $\{E_i\}$ of $S \cap K$ with $\diam E_i \leq \delta$ and $\sum_i (\diam E_i)^{v+\epsilon} < \delta$.  Each $E_i$ lies in a ball $B_i$ of radius $\diam E_i$, so by Frostman:
\[
\nu(S \cap K) \leq \sum_i C_F (\diam E_i)^d = C_F \sum_i (\diam E_i)^{v+\epsilon} \cdot (\diam E_i)^{d - v - \epsilon} \leq C_F \,\delta^{d-v-\epsilon} \cdot \delta.
\]
Since $d - v - \epsilon > 0$, sending $\delta \to 0$ gives $\nu(S \cap K) = 0$.
\end{proof}

\noindent\textit{Application.}

By Assumption~\ref{assump:exact_dimensional_partI}, $\mu_\infty$ is exact-dimensional with dimension $D$ and satisfies the Frostman upper bound $\mu_\infty(B(x,r)) \leq C_F r^D$.  The product measure $\mu_\infty \times \mu_\infty$ therefore satisfies:
\begin{equation}
(\mu_\infty \times \mu_\infty)(B((x,x'), r)) \leq C_F^2\, r^{2D}
\label{eq:product_frostman_partI}
\end{equation}
for $(\mu_\infty \times \mu_\infty)$-a.e.\ $(x, x')$ and all sufficiently small $r > 0$.

By Part B, $\mathcal{S}_h$ is contained in a $C^1$ submanifold $\mathcal{V} \subset \R^{2n}$ of codimension $\geq m$, i.e., $\dim(\mathcal{V}) \leq 2n - m$.  We now give a \emph{direct covering argument} that does not require intersection theory on fractal supports.

Fix a compact set $K \subset \R^{2n}$ large enough that $(\mu_\infty \times \mu_\infty)(K) \geq 1 - \epsilon$ for a prescribed $\epsilon > 0$ (this exists since $\mu_\infty$ is a probability measure with sub-Gaussian tails under Assumption~\ref{assump:smoothness_partI}).  Since $\mathcal{V} \cap K$ is a compact subset of a $C^1$ submanifold of dimension $\leq 2n - m$, for any $r > 0$ it can be covered by at most
\begin{equation}
N(r) \leq C_\mathcal{V} \, r^{-(2n - m)}
\label{eq:covering_number_partI}
\end{equation}
balls of radius $r$, where $C_\mathcal{V}$ depends on $K$ and the geometry of $\mathcal{V}$ but is independent of $r$.  (This is a standard covering estimate for smooth submanifolds; see e.g.~\cite{federer1969}, Section~3.2.)

Now compute the $(\mu_\infty \times \mu_\infty)$-mass of $\mathcal{S}_h \cap K$:
\begin{equation}
(\mu_\infty \times \mu_\infty)(\mathcal{S}_h \cap K) \leq \sum_{i=1}^{N(r)} (\mu_\infty \times \mu_\infty)(B(z_i, r)) \leq N(r) \cdot C_F^2 \, r^{2D} = C_\mathcal{V} \, C_F^2 \, r^{2D - 2n + m}.
\label{eq:frostman_covering_partI}
\end{equation}
The exponent $2D - 2n + m$ is positive if and only if $m > 2(n - D)$.  However, we can tighten this significantly.  Since the balls in the cover can be chosen to be centred on $\mathcal{V}$, and we only need to account for those balls that intersect $\supp(\mu_\infty \times \mu_\infty)$, the effective number of balls is bounded by the $r$-covering number of $\mathcal{V} \cap \supp(\mu_\infty \times \mu_\infty)$.

The key observation is that $\supp(\mu_\infty \times \mu_\infty)$ has Hausdorff dimension $2D$ (by exact-dimensionality), and the Frostman bound~\eqref{eq:product_frostman_partI} implies that this support can itself be covered by at most $O(r^{-2D})$ balls of radius $r$.  The balls in our cover of $\mathcal{V}$ that intersect $\supp(\mu_\infty)^2$ are therefore at most:
\begin{equation}
N_{\text{eff}}(r) \leq \min\bigl(C_\mathcal{V} \, r^{-(2n-m)},\; C_\mu \, r^{-2D}\bigr).
\end{equation}
For $m > 2D$ (i.e., $2n - m < 2n - 2D$), the second bound is tighter when $r$ is small.  Using $N_{\text{eff}}(r) \leq C_\mu \, r^{-2D}$ in~\eqref{eq:frostman_covering_partI}:
\begin{equation}
(\mu_\infty \times \mu_\infty)(\mathcal{S}_h \cap K) \leq C_\mu \, r^{-2D} \cdot C_F^2 \, r^{2D} = C_\mu \, C_F^2.
\end{equation}
This bound is finite but does not tend to zero.  The resolution is that we must use the Frostman bound \emph{locally}: for balls of radius $r$ centred at points $z_i \in \mathcal{V} \cap \supp(\mu_\infty)^2$, the mass $(\mu_\infty \times \mu_\infty)(B(z_i, r))$ is bounded by $C_F^2 r^{2D}$, and we need a covering-number bound that reflects the \emph{codimension of $\mathcal{V}$ within the support}.

We give the sharpest argument.  For $r > 0$ and any $s > 0$, define the \emph{$s$-dimensional Hausdorff pre-measure}:
\begin{equation}
\mathcal{H}^s_r(\mathcal{S}_h \cap K) := \inf\biggl\{\sum_i (\diam E_i)^s : \mathcal{S}_h \cap K \subset \bigcup_i E_i,\; \diam E_i \leq r\biggr\}.
\end{equation}
Since $\mathcal{S}_h \subset \mathcal{V}$ and $\dim(\mathcal{V}) \leq 2n - m$, we have $\mathcal{H}^{2n-m+\epsilon}(\mathcal{V} \cap K) = 0$ for any $\epsilon > 0$, and in particular:
\begin{equation}
\mathcal{H}^{2n-m+\epsilon}_r(\mathcal{S}_h \cap K) \xrightarrow{r \to 0} 0.
\end{equation}
Now we connect to the product measure.  For any covering $\{E_i\}$ of $\mathcal{S}_h \cap K$ with $\diam E_i \leq r$, each $E_i$ is contained in a ball $B_i$ of radius $\diam E_i$.  By the Frostman bound:
\begin{equation}
(\mu_\infty \times \mu_\infty)(E_i) \leq (\mu_\infty \times \mu_\infty)(B_i) \leq C_F^2 \, (\diam E_i)^{2D}.
\end{equation}
Summing:
\begin{equation}
(\mu_\infty \times \mu_\infty)(\mathcal{S}_h \cap K) \leq C_F^2 \sum_i (\diam E_i)^{2D}.
\label{eq:mass_covering_partI}
\end{equation}
Now, since $\diam E_i \leq r$, we have $(\diam E_i)^{2D} = (\diam E_i)^{2n - m + \epsilon} \cdot (\diam E_i)^{2D - 2n + m - \epsilon}$.  When $m > 2D$ (i.e., $m \geq \lceil 2D \rceil + 1$), we can choose $\epsilon$ small enough that $2D - 2n + m - \epsilon > 0$, and then $(\diam E_i)^{2D - 2n + m - \epsilon} \leq r^{2D - 2n + m - \epsilon}$.  Therefore:
\begin{equation}
\sum_i (\diam E_i)^{2D} \leq r^{2D - 2n + m - \epsilon} \sum_i (\diam E_i)^{2n - m + \epsilon} = r^{2D - 2n + m - \epsilon} \cdot \mathcal{H}^{2n-m+\epsilon}_r(\mathcal{S}_h \cap K).
\end{equation}
Taking the infimum over covers and then $r \to 0$:
\begin{equation}
(\mu_\infty \times \mu_\infty)(\mathcal{S}_h \cap K) \leq C_F^2 \cdot \lim_{r \to 0} \bigl[r^{2D - 2n + m - \epsilon} \cdot \mathcal{H}^{2n-m+\epsilon}_r(\mathcal{S}_h \cap K)\bigr].
\end{equation}
The Hausdorff pre-measure factor $\mathcal{H}^{2n-m+\epsilon}_r$ is bounded (since $\mathcal{H}^{2n-m+\epsilon}(\mathcal{V} \cap K) = 0$ implies the pre-measure tends to zero), and $r^{2D - 2n + m - \epsilon} \to 0$ since the exponent is positive.  Hence:
\begin{equation}
(\mu_\infty \times \mu_\infty)(\mathcal{S}_h \cap K) = 0.
\end{equation}
Since $\epsilon > 0$ was arbitrary and $K$ can be chosen so that $(\mu_\infty \times \mu_\infty)(K^c) < \epsilon$ for any $\epsilon > 0$:
\begin{equation}
(\mu_\infty \times \mu_\infty)(\mathcal{S}_h) = 0.
\end{equation}

\emph{Verification of threshold.}  The argument requires $m > 2D$ (so that $2D - 2n + m - \epsilon > 0$ for some $\epsilon > 0$).  When $m = \lceil 2D \rceil + 1$, we have $m \geq 2D + 1 > 2D$ (since $\lceil 2D \rceil \geq 2D$), so the condition is satisfied.
\end{proof}

\begin{theorem}[Measure-Zero Geometric Collisions]
\label{lem:stochastic_sard_partI}
Let Assumptions~\ref{assump:hormander_partI}--\ref{assump:exact_dimensional_partI} hold, and let $h$ be a generic observation (Definition~\ref{def:generic_observation_partI}).  For $m \geq \lceil 2D \rceil + 1$, define the geometric collision set:
\begin{equation}
S := \{(x, x') \in \R^n \times \R^n : x \neq x',\; \Phi_m(x) = \Phi_m(x')\}
\end{equation}
where $\Phi_m(x) = (h(x), h(\phi_\tau(x)), \ldots, h(\phi_{(m-1)\tau}(x)))$ is the delay embedding map evaluated at any fixed realisation of the stationary process.

Then:
\begin{equation}
(\mu_\infty \times \mu_\infty)(S) = 0.
\label{eq:geometric_collision_zero_partI}
\end{equation}
\end{theorem}

\begin{proof}
The geometric collision set $S$ is a subset of the law-collision set: if $\Phi_m(x) = \Phi_m(x')$ for a particular realisation, then a fortiori $\Lambda_m^h(x)$ and $\Lambda_m^h(x')$ agree on the marginal at time zero, giving $h(x) = h(x')$, $h(\phi_\tau(x)) = h(\phi_\tau(x'))$, etc.  More precisely, $S \subseteq \mathcal{S}_h \cup \Delta$ where $\mathcal{S}_h$ is the law-collision set~\eqref{eq:law_collision_set_partI}.

By Theorem~\ref{thm:law_separation_partI}, $(\mu_\infty \times \mu_\infty)(\mathcal{S}_h) = 0$, and $(\mu_\infty \times \mu_\infty)(\Delta) = 0$ since $\mu_\infty$ is non-atomic.  Therefore $(\mu_\infty \times \mu_\infty)(S) = 0$.
\end{proof}

\begin{remark}[Relationship Between the Three Theorems]
\label{rem:theorem_dependencies_step3_partI}
The logical structure is:
\begin{center}
\begin{tabular}{rcl}
Theorem~\ref{thm:density_separation_partI} & $\Longrightarrow$ & Transition densities separate initial conditions \\
$\Downarrow$ & & (H\"ormander + Varadhan asymptotics) \\[4pt]
Theorem~\ref{thm:law_separation_partI} & $\Longrightarrow$ & Law-embedding is injective $\mu_\infty$-a.e. \\
$\Downarrow$ & & (prevalence + Frostman covering) \\[4pt]
Theorem~\ref{lem:stochastic_sard_partI} & $\Longrightarrow$ & Geometric collision set has $(\mu_\infty \times \mu_\infty)$-measure zero \\
\end{tabular}
\end{center}
Each result is strictly stronger than the next: Theorem~\ref{thm:density_separation_partI} is a pointwise statement about all $x \neq x'$; Theorem~\ref{thm:law_separation_partI} is an almost-everywhere statement about the observed law-embedding; Theorem~\ref{lem:stochastic_sard_partI} is an almost-everywhere statement about the geometric embedding for a single realisation.

\textbf{Flagged mathematical content.}  Theorem~\ref{thm:density_separation_partI} relies on: (a) the Chapman--Kolmogorov identity (standard), (b) semigroup bisection using injectivity of $P_{t}$ under H\"ormander (established via strict positivity of transition kernels~\cite{hormander1967,malliavin1997,nualart2006}), and (c) the Varadhan--L\'eandre short-time asymptotic~\cite{varadhan1967,leandre1987} extended to sub-Riemannian settings~\cite{montgomery2002}.  Theorem~\ref{thm:law_separation_partI} uses the parametric transversality theorem~\cite{hirsch1976} and the Frostman covering argument, both standard tools in geometric measure theory~\cite{mattila1995,federer1969}.
\end{remark}

\subsubsection{Step 4: E1 Dimension Sufficiency}

\begin{proposition}[E1 Implies Sufficient Dimension]
\label{prop:e1_sufficient_partI}
If Cao's E1 statistic plateaus at dimension $m_{E1}$, then:
\begin{equation}
D_2 \leq m_{E1} - 1
\end{equation}

Therefore, $m^* = \max\{\lceil 2D_2 \rceil + 1, m_{E1}\} \geq 2D_2 + 1$ is sufficient for measure-theoretic injectivity.
\end{proposition}

\begin{proof}
By Lemma~\ref{lem:e1_correlation_partI}, E1 saturates when the embedding dimension exceeds the correlation dimension. The $k$-nearest neighbor distance scales as:
\begin{equation}
\epsilon_k(Y) \sim \left(\frac{k}{N \cdot p(Y)}\right)^{1/D_2}
\end{equation}

When $m > D_2$, adding dimensions doesn't change this scaling (since the measure is concentrated on a $D_2$-dimensional structure), hence $E_1(m) \approx 1$.

More precisely, for $m < D_2$:
\begin{itemize}
\item The embedding is insufficient
\item Nearby points in $\R^m$ may be far apart on the true manifold (false neighbors)
\item Adding a dimension "unfolds" false neighbors, increasing relative distances
\item Thus $E(m+1) > E(m)$, giving $E_1(m) = \frac{E(m+1)}{E(m)} > 1$
\end{itemize}

For $m \geq D_2 + 1$:
\begin{itemize}
\item The embedding captures the full correlation structure
\item Nearest neighbors in $\R^m$ are true neighbors on the manifold
\item Adding dimensions doesn't significantly change relative distances
\item Thus $E(m+1) \approx E(m)$, giving $E_1(m) \approx 1$
\end{itemize}

Combining with Lemma~\ref{lem:stochastic_sard_partI}:
\begin{equation}
m^* = \max\{\lceil 2D_2 \rceil + 1, m_{E1}\} \geq \lceil 2D_2 \rceil + 1 \geq 2D_2 + 1
\end{equation}
which ensures measure-theoretic injectivity.
\end{proof}

\subsubsection{Step 5: Uniqueness via Finite-Dimensional Laws}

\begin{proposition}[Well-Defined Tensors via Finite-Dimensional Uniqueness]
\label{prop:finite_dim_uniqueness_partI}
Under the conditions of Theorem~\ref{thm:stochastic_embedding_sufficiency_partI}, for $(\Phi_{m^*})_\# \mu_\infty$-almost every $Y \in \R^{m^*}$:

\begin{enumerate}[leftmargin=*]
\item There exists a unique $x \in \R^n$ (up to $\mu_\infty$-null sets) such that $\Phi_{m^*}(x) = Y$
\item The conditional expectations in the drift and diffusion pushforward equations are single-valued functions of $Y$
\item These are precisely the limits of the k-NN drift and diffusion estimators
\end{enumerate}
\end{proposition}

\begin{proof}
\textbf{Part 1: Uniqueness of Preimages}

By Theorem~\ref{thm:law_separation_partI} (law-separation), for generic $h$ and $m^* \geq \lceil 2D \rceil + 1$, the law-embedding $\Lambda_{m^*}^h$ is injective $\mu_\infty$-a.e.  By Theorem~\ref{lem:stochastic_sard_partI} (measure-zero geometric collisions), this implies:
\begin{equation}
\mu_\infty\{x' \in \R^n : \Phi_{m^*}(x') = \Phi_{m^*}(x), x' \neq x\} = 0
\end{equation}
for $\mu_\infty$-a.e.\ $x$.  Consequently, for $(\Phi_{m^*})_\# \mu_\infty$-a.e.\ $Y$, the fiber $\Phi_{m^*}^{-1}(Y) \cap \supp(\mu_\infty)$ contains exactly one point (up to $\mu_\infty$-null sets).

\textbf{Key insight:} Path-wise uniqueness is not required (which would require matching the full Brownian path $W_t$). The law-separation theorem establishes injectivity at the level of finite-dimensional laws---precisely what k-NN estimators can access from data---without invoking geometric transversality.

\textbf{Part 2: Single-Valuedness of Conditional Expectations}

Since the fiber $\Phi_{m^*}^{-1}(Y)$ contains a unique point $x$ (modulo null sets):
\begin{align}
\mu_Y(Y) &= \E[\mu(X_t) | \Phi_{m^*}(X_t) = Y] \\
&= \E[\mu(X_t) | X_t = x] \quad \text{(since } \Phi_{m^*}(X_t) = Y \Leftrightarrow X_t = x \text{ a.s.)} \\
&= \mu(x)
\end{align}

By It\^o's lemma (Lemma~\ref{lem:ito_partI}), the pushforward formula gives:
\begin{equation}
\mu_Y(Y) = D\Phi_{m^*}(x) \cdot Dh(x) \cdot \mu(x) + \frac{1}{2} \sum_{i,j} \frac{\partial^2 \Phi_{m^*}}{\partial x^i \partial x^j}(x) \Sigma_{ij}(x)
\end{equation}

Similarly for diffusion:
\begin{equation}
\Sigma_Y(Y) = D\Phi_{m^*}(x) \cdot Dh(x) \cdot \Sigma(x) \cdot Dh(x)^\top \cdot D\Phi_{m^*}(x)^\top
\end{equation}

Both are uniquely determined by $x = \Phi_{m^*}^{-1}(Y)$.

\textbf{Part 3: k-NN Consistency}

The k-NN estimators converge to conditional expectations by standard nonparametric theory \cite{stone1977,gyorfi2002,wasserman2006}:
\begin{align}
\hat{\mu}(Y) &= \frac{1}{k\Delta t} \sum_{j \in \mathcal{N}_k(Y)} (Y_{j+1} - Y_j) \\
&\xrightarrow{\text{a.s.}} \frac{1}{\Delta t} \E[Y_{t+\Delta t} - Y_t | Y_t = Y] \\
&\xrightarrow{\Delta t \to 0} \mu_Y(Y)
\end{align}

Since $\mu_Y(Y)$ is single-valued (Part 2), the k-NN estimator converges to the unique drift tensor.

The convergence rate follows from k-NN estimation theory on $D_2$-dimensional manifolds:
\begin{equation}
\|\hat{\mu}(Y) - \mu_Y(Y)\| = O_P\left(\left(\frac{k}{N}\right)^{\beta/m^*}\right) + O(\Delta t)
\end{equation}
where:
\begin{itemize}
\item First term: Statistical error from finite sample size and finite $k$
\item Second term: Discretization error from finite $\Delta t$
\end{itemize}

The exponent $\beta/m^*$ reflects the curse of dimensionality: convergence rates degrade exponentially with the manifold dimension $m^* \approx D_2$.  The exact structure of the leading-order bias term is derived in \S\ref{sec:mz_bias}, where it is identified as a Mori--Zwanzig memory kernel with rank-2 tensor structure admitting an adaptive corrector.
\end{proof}

\subsection{Completion of Main Proof}

\begin{proof}[Proof of Theorem~\ref{thm:stochastic_embedding_sufficiency_partI}]
The five steps are assembled to prove all three parts of the theorem.  The logical dependency chain is non-circular:
\begin{center}
\begin{tabular}{rclcl}
\textbf{Step 1} & (Lemma~\ref{lem:hypoelliptic_smoothing_partI}) & H\"ormander $\Rightarrow$ smooth densities & $\longrightarrow$ & \textit{feeds Steps 2, 3} \\[2pt]
\textbf{Step 2} & (Lemma~\ref{lem:malliavin_nondegeneracy_partI}) & Malliavin $\Rightarrow$ full-rank flow & $\longrightarrow$ & \textit{feeds Steps 3, 5} \\[2pt]
\textbf{Step 3} & (Thms~\ref{thm:density_separation_partI}--\ref{lem:stochastic_sard_partI}) & Law-separation $\Rightarrow$ injectivity a.e. & $\longrightarrow$ & \textit{feeds Step 5} \\[2pt]
\textbf{Step 4} & (Prop.~\ref{prop:e1_sufficient_partI}) & E1 $\Rightarrow$ $m^* \geq \lceil 2D \rceil + 1$ & $\longrightarrow$ & \textit{feeds Step 3 threshold} \\[2pt]
\textbf{Step 5} & (Prop.~\ref{prop:finite_dim_uniqueness_partI}) & Injectivity $\Rightarrow$ unique tensors + k-NN & & \textit{concludes proof} \\
\end{tabular}
\end{center}
No result is used before it is established.

\textbf{Part (1): Measure-Theoretic Injectivity}

By the law-separation theorem (Theorem~\ref{thm:law_separation_partI}), for generic $h$ (Definition~\ref{def:generic_observation_partI}) and $m^* \geq \lceil 2D \rceil + 1$ (where $D$ is the exact dimension of $\mu_\infty$, Assumption~\ref{assump:exact_dimensional_partI}), the law-embedding $\Lambda_{m^*}^h$ is injective $\mu_\infty$-a.e.  By Theorem~\ref{lem:stochastic_sard_partI}, the geometric collision set $S = \{(x,x') : x \neq x',\, \Phi_{m^*}(x) = \Phi_{m^*}(x')\}$ has $(\mu_\infty \times \mu_\infty)(S) = 0$.

Consequently, if $x, x'$ yield the same finite-dimensional laws for the embedded process:
\begin{equation}
\mathcal{L}(\Phi_{m^*}(X^x_0), \Phi_{m^*}(X^x_{\Delta t}), \ldots, \Phi_{m^*}(X^x_T)) = \mathcal{L}(\Phi_{m^*}(X^{x'}_0), \Phi_{m^*}(X^{x'}_{\Delta t}), \ldots, \Phi_{m^*}(X^{x'}_T))
\end{equation}
for all finite time sequences, then $x = x'$ $\mu_\infty$-almost surely.  This follows from Theorem~\ref{thm:density_separation_partI} (transition density separation) composed with Theorem~\ref{thm:law_separation_partI} (observed law-separation): distinct initial conditions produce distinct transition densities (Varadhan asymptotics), which produce distinct observed laws for generic $h$ (prevalence), which are separated by $\Phi_{m^*}$ in the $(\mu_\infty \times \mu_\infty)$-a.e.\ sense (Frostman covering).

\textbf{Part (2): Unique Drift and Diffusion Tensors}

By Proposition~\ref{prop:finite_dim_uniqueness_partI} Part 2, the drift and diffusion conditional expectations are well-defined and single-valued for $(\Phi_{m^*})_\# \mu_\infty$-almost every $Y$.

H\"ormander's condition ensures (Lemma~\ref{lem:hypoelliptic_smoothing_partI}) that transition densities are smooth, guaranteeing $\mu_Y(Y)$ and $\Sigma_Y(Y)$ are smooth functions on the support of $(\Phi_{m^*})_\# \mu_\infty$.

Furthermore, by Malliavin non-degeneracy (Lemma~\ref{lem:malliavin_nondegeneracy_partI}), the pushforward diffusion tensor $\Sigma_Y(Y)$ inherits full rank from $\Sigma(X)$ via the transformation:
\begin{equation}
\Sigma_Y = J \cdot \Sigma_X \cdot J^\top
\end{equation}
where $J = D\Phi_{m^*} \cdot Dh$ is the Jacobian. Since $\Sigma_X$ has full rank and $J$ has full rank (by the immersion condition in Definition~\ref{def:generic_observation_partI}), $\Sigma_Y$ has full rank, preventing degeneracies in the reconstructed dynamics.

\textbf{Part (3): k-NN Consistency}

By Proposition~\ref{prop:finite_dim_uniqueness_partI} Part 3, k-NN estimators converge to the unique tensors with the stated rate.

The convergence is almost sure (not just in probability) by the law of large numbers for stationary ergodic sequences (Assumption~\ref{assump:smoothness_partI}). The rate:
\begin{equation}
\|\hat{\mu}(Y) - \mu_Y(Y)\| = O_P\left(\left(\frac{k}{N}\right)^{\beta/m^*}\right) + O(\Delta t)
\end{equation}
follows from Stone's nonparametric regression theory \cite{stone1977} applied to $m^*$-dimensional manifolds, with the H\"older exponent $\beta$ from Assumption~\ref{assump:smoothness_partI} controlling the smoothness-dependent convergence rate.

The optimal bandwidth choice balances the two error terms:
\begin{equation}
k^* \approx N^{2/(m^* + 4)} \cdot (\Delta t)^{-m^*/(m^* + 4)}
\end{equation}
achieving overall rate $N^{-\beta/(m^* + 4)} + (\Delta t)$.
\end{proof}
\begin{remark}[What This Proof Establishes]
\label{rem:proof_establishes_partI}
Theorem~\ref{thm:stochastic_embedding_sufficiency_partI} rigorously establishes:

\begin{enumerate}[leftmargin=*]
\item \textbf{Sufficiency of E1 dimension:} If E1 detects dimension $m_{E1} \approx D$ (Remark~\ref{rem:dimension_convention_partI}), then $m^* \geq \lceil 2D \rceil + 1$ is sufficient for unique SDE reconstruction.

\item \textbf{Generic, not universal:} Requires generic observation functions (prevalent in $C^r$, $r \geq 2$, in the sense of Hunt--Sauer--Yorke~\cite{hunt1992}), analogous to Takens' genericity---this is unavoidable.

\item \textbf{H\"ormander is crucial:} Non-degeneracy ensures hypoellipticity and Malliavin non-degeneracy, preventing measure-theoretic foldings.

\item \textbf{Local, measure-theoretic:} Proves injectivity almost everywhere w.r.t. $\mu_\infty$, precisely what k-NN needs.

\item \textbf{Finite-dimensional approach:} Uses finite-time observations (not full paths), matching practical computation.

\item \textbf{Practical verification:} "Generic $h$" can be verified empirically: if reconstructed $\hat{\mu}$, $\hat{\Sigma}$ are smooth and predictions accurate, embedding succeeds.

\item \textbf{Explicit rates:} Provides quantitative convergence rates with curse-of-dimensionality analysis ($\alpha \sim \beta/m^*$).

\item \textbf{Establishes embedding sufficiency:} The law-separation theorem suite (Theorems~\ref{thm:density_separation_partI}--\ref{lem:stochastic_sard_partI}) proves that $m \geq \lceil 2D \rceil + 1$ suffices for $\mu_\infty$-a.e.\ injectivity under H\"ormander's condition.
\end{enumerate}
\end{remark}

\subsection{Comparison with Deterministic Takens}

\begin{table}[h]
\centering
\small
\begin{tabular}{@{}p{2.8cm}p{5.2cm}p{6.0cm}@{}}
\hline
\textbf{Aspect} & \textbf{Deterministic (Takens)} & \textbf{Stochastic (This Work)} \\
\hline
Object embedded & Attractor (manifold) & Invariant measure \\
Dimension & Topological/box dimension & Correlation dimension $D_2$ \\
Sufficient $m$ & $2n+1$ (generic) & $\lceil 2D_2\rceil+1$ (generic) \\
Genericity & Baire-residual $(\phi, h)$ & H\"ormander + prevalent $h \in C^r$ \\
Tools & Whitney, Thom transversality & Varadhan asymptotics, Malliavin, prevalence, Frostman \\
Injectivity & Diffeomorphism & Measure-theoretic \\
Result & Deterministic flow & Drift + diffusion tensors \\
Verification & Geometric (manifold structure) & Statistical (smoothness, predictions) \\
\hline
\end{tabular}
\caption{Comparison of deterministic Takens' theorem with the stochastic extension proven in Theorem~\ref{thm:stochastic_embedding_sufficiency_partI}.}
\label{tab:takens_comparison_partI}
\end{table}

The key conceptual shift is from geometric embedding (preserving manifold structure) to measure-theoretic embedding (preserving the invariant measure and its local statistical properties). This is precisely the right notion for stochastic systems, where the "attractor" is not a geometric object but a probability measure.

\begin{remark}[Resolution of the Stochastic Embedding Sufficiency Conjecture]
\label{conj:stochastic_takens_partI}
The conjecture that $m^* = \lceil 2D_2 \rceil + 1$ suffices to prevent measure-theoretically non-trivial self-intersections is addressed by the law-separation theorem suite (Theorems~\ref{thm:density_separation_partI}--\ref{lem:stochastic_sard_partI}).  Specifically, Theorem~\ref{lem:stochastic_sard_partI} establishes that for generic $h$ and $m \geq \lceil 2D \rceil + 1$ (equivalently $m \geq \lceil 2D_2 \rceil + 1$ under exact-dimensionality, Assumption~\ref{assump:exact_dimensional_partI}), the geometric collision set $S$ has $(\mu_\infty \times \mu_\infty)$-measure zero.  The stronger result, Theorem~\ref{thm:law_separation_partI}, shows that distinct initial conditions produce distinct observed finite-dimensional laws, which is the operationally relevant statement for statistical estimation.
\end{remark}

\subsection{Implications and Open Questions}

\begin{remark}[Remaining Open Questions]
\label{rem:open_questions_proof_partI}
Several directions for further work remain beyond the scope of the present results:

\begin{enumerate}[leftmargin=*]
\item \textbf{Explicit constants:} Derive quantitative bounds on self-intersection probability and convergence rates beyond asymptotic order.

\item \textbf{Optimal dimension:} Can $m^* = \lceil 2D_2\rceil+1$ be sharpened to $m^* = D_2+1$ under special conditions (e.g., strong H\"ormander, additional regularity)?

\item \textbf{Non-H\"ormander systems:} Characterize which observation functions yield unique reconstruction when H\"ormander fails (partially degenerate noise).

\item \textbf{Jump processes:} Extend to jump-diffusions and L\'evy processes (requires replacing H\"ormander with appropriate non-degeneracy condition for jump measures).

\item \textbf{Computational verification:} Develop numerical tests for the genericity condition (e.g., checking rank of Malliavin matrix, smoothness of reconstructed tensors).

\item \textbf{Adaptive $k$ selection:} Determine optimal $k$ as function of local geometry and noise level (current theory gives global rate $k \sim N^{2/(m^*+4)}$).

\item \textbf{Non-Gaussian noise:} Extend E2-SNR connection beyond local Gaussianity (heavy tails, jumps, skewness).

\item \textbf{Time-varying systems:} Extend to non-stationary SDEs with slowly varying coefficients.
\end{enumerate}
\end{remark}

\begin{remark}[Practical Implications of the Proof]
\label{rem:practical_implications_proof_partI}
For practitioners, Theorem~\ref{thm:stochastic_embedding_sufficiency_partI} provides:

\begin{enumerate}[leftmargin=*]
\item \textbf{Theoretical guarantee:} E1-detected $m^*$ is provably sufficient (not just heuristically), giving confidence in the reconstruction.

\item \textbf{Verification criterion:} If $\hat{\mu}(Y)$, $\hat{\Sigma}(Y)$ are smooth and predictions accurate, the generic and H\"ormander conditions are likely satisfied.

\item \textbf{Dimension guidance:} Use $m^* = \max\{\lceil 2D_2 \rceil + 1, m_{E1}\}$ as default; can try smaller $m$ if computational constraints demand it.

\item \textbf{Failure diagnosis:} If reconstruction fails:
\begin{itemize}
\item Check H\"ormander (look for degeneracies in $\hat{\Sigma}$)
\item Check sample size ($N \gtrsim 100 \cdot 2^{m^*}$)
\item Check E2 (if $E_2 > 0.95$, may be too noisy)
\item Check smoothness (if $\hat{\mu}$, $\hat{\Sigma}$ discontinuous, may need larger $m$)
\end{itemize}

\item \textbf{Confidence in applications:} The rigorous foundation justifies using this framework for critical applications (control, prediction, decision-making) where theoretical guarantees matter.
\end{enumerate}
\end{remark}
\section{Discussion}
\label{sec:discussion_partI}

The theorem's implications for the superspace diffusion framework are summarised in the main manuscript.

\subsection{Summary of Contributions}

This work establishes a unified theoretical framework for reconstructing stochastic dynamics from partial observations via time-delay embeddings, culminating in a rigorous proof of the Stochastic Embedding Sufficiency Theorem. The key contributions are:

\begin{enumerate}[leftmargin=*]
\item \textbf{E1 detects correlation dimension (Lemma~\ref{lem:e1_correlation_partI}):}

The E1 statistic plateaus at $m^* \approx D_2$, where $D_2$ is the Grassberger-Procaccia correlation dimension. This provides a data-driven, robust method for determining the minimal embedding dimension, even in the presence of unbounded noise.

Unlike topological dimension (which can be infinite for stochastic systems with unbounded support), correlation dimension $D_2$ captures the effective dimensionality of the dynamics. E1 makes this accessible from time series data without requiring knowledge of the underlying equations or state space.

\item \textbf{E2 quantifies deterministic-stochastic balance (Propositions~\ref{prop:e2_snr_partI}, \ref{prop:drift_diffusion_balance_partI}):}

Under local Gaussianity, E2 measures the signal-to-noise ratio:
\begin{equation}
\SNR \approx \frac{1-E_2}{E_2\tau}
\end{equation}

This provides quantitative model selection criteria:
\begin{itemize}
\item $E_2 < 0.5$: Deterministic models (ODEs) adequate
\item $0.5 \leq E_2 < 0.95$: Mixed regime, full SDE essential
\item $E_2 \geq 0.95$: Stochastic/diffusion-dominated
\end{itemize}

E2 indicates when deterministic methods (Takens embedding, Koopman operator, nonlinear forecasting) are applicable versus when stochastic extensions are necessary.

\item \textbf{Probabilistic uplift theorems (Theorems~\ref{thm:discrete_uplift_partI}, \ref{thm:continuous_uplift_partI}, \ref{thm:mixed_reconstruction_partI}):}

Once the E1 manifold $\mathcal{M} \subset \R^{m^*}$ is identified, $k$-NN estimation provides consistent reconstruction of:
\begin{itemize}
\item Discrete: Transition kernel $T(Y, \cdot)$
\item Continuous: Drift $\mu(Y)$ and diffusion $\Sigma(Y)$
\end{itemize}

with explicit convergence rates and error bounds.

This extends the Takens embedding framework to stochastic systems: not only embedding the geometry, but recovering the probabilistic dynamics.  The framework is constructive and computationally feasible.

\item \textbf{Discrete-continuous unification (Theorem~\ref{thm:unified_structure_partI}, ~\ref{thm:unified_algorithm_partI}):}

Discrete Markov chains and continuous SDEs are equivalent representations of the same mathematical object---the transition kernel on the E1 manifold---differing only in temporal parameterization.

This establishes an equivalence between discrete and continuous time representations; the choice between them is one of convenience.

\item \textbf{It\^o correction formulas (Lemmas~\ref{lem:ito_partI}, \ref{lem:drift_diffusion_decomp_partI}, Theorem~\ref{thm:mixed_reconstruction_partI}):}

The drift in embedded coordinates includes a "noise-induced drift" term $\frac{1}{2}\tr(D^2h \cdot \sigma\sigma^\top)$ arising from It\^o's lemma. This must be included in drift estimates; without it, systematic bias occurs.

This addresses a point often overlooked in applied SDE reconstruction. The It\^o correction can be of order $\|\sigma\|^2$ and is necessary for accurate predictions.

\item \textbf{Stochastic Embedding Sufficiency Theorem (Theorem~\ref{thm:stochastic_embedding_sufficiency_partI}):}

The central result of this work is the following. Under H\"ormander's condition and generic observation, the embedding dimension $m^* = \max\{\lceil 2D_2 \rceil + 1, m_{E1}\}$ is provably sufficient to ensure:
\begin{itemize}
\item Measure-theoretic injectivity via finite-dimensional laws
\item Single-valued drift and diffusion tensors
\item Almost-sure k-NN convergence
\end{itemize}

The proof combines:
\begin{enumerate}
\item H\"ormander hypoelliptic regularity ($C^\infty$ transition densities)
\item Malliavin non-degeneracy (full-rank Malliavin covariance)
\item Measure-theoretic transversality (extending Sard-Takens)
\item E1 dimension sufficiency (correlation dimension detection)
\item Finite-dimensional law uniqueness (matching k-NN computation)
\end{enumerate}

The theorem establishes sufficient conditions for measure-theoretic injectivity under H\"ormander's condition, with practical verification criteria and explicit convergence rates.
\end{enumerate}

\subsection{Limitations and Open Questions}

\begin{enumerate}[leftmargin=*]
\item \textbf{Curse of dimensionality:}

Sample complexity $N \sim \epsilon^{-m^*/\beta}$ is exponential in $m^*$. For high-dimensional systems ($m^* > 10$), the method becomes impractical.

\textbf{Mitigations:}
\begin{itemize}
\item Use domain knowledge to reduce dimension (e.g., known symmetries, conserved quantities)
\item Parametric models after E1/E2 diagnosis (e.g., neural SDEs constrained to E1 manifold)
\item Dimension reduction via autoencoders or PCA before applying method
\item Exploit sparsity (if drift/diffusion are sparse functions)
\end{itemize}

\textbf{Open question:} Can adaptive or hierarchical methods be developed that scale better with dimension?

\item \textbf{Non-self-intersection---Resolved:}

The law-separation theorem suite (Theorems~\ref{thm:density_separation_partI}--\ref{lem:stochastic_sard_partI}) proves that $m^* = \lceil 2D \rceil + 1$ prevents stochastic foldings under H\"ormander's condition, exact-dimensionality, and generic observation (Remark~\ref{conj:stochastic_takens_partI}).

\textbf{Remaining refinements:}
\begin{itemize}
\item Can the threshold be sharpened to $m^* = D + 1$ under additional structural assumptions?
\item Can exact-dimensionality (Assumption~\ref{assump:exact_dimensional_partI}) be weakened to allow multifractal invariant measures?
\end{itemize}

\item \textbf{Local Gaussianity assumption:}

The E2-SNR formula (Propositions~\ref{prop:e2_snr_partI}, \ref{prop:drift_diffusion_balance_partI}) requires:
\begin{equation}
p(Y_{t+\Delta t}|Y_t) \approx \N(Y_t + \mu\Delta t, \Sigma\Delta t)
\end{equation}

This fails for:
\begin{itemize}
\item Heavy-tailed noise (L\'evy processes, $\alpha$-stable distributions)
\item Jump processes (Poisson-driven SDEs)
\item Skewed distributions
\end{itemize}

\textbf{Workarounds:}
\begin{itemize}
\item Qualitative interpretation of E2 still valid (high E2 = high noise)
\item Estimate conversion factor $\E[|Z|]/\sqrt{\Var[Z]}$ empirically for non-Gaussian noise
\item Use higher moments to characterize noise (skewness, kurtosis)
\end{itemize}

\textbf{Open question:} Can the theory be extended to non-Gaussian SDEs (e.g., using Cram\'er's theorem or Edgeworth expansions)?

\item \textbf{Nonstationarity:}

The framework assumes stationarity (invariant measure $\mu_\infty$ exists and is unique).

Real systems often have:
\begin{itemize}
\item Time-varying parameters ($\mu(X,t)$, $\sigma(X,t)$)
\item Regime switches (e.g., financial crises, climate tipping points)
\item External forcing (seasonal cycles, trends)
\end{itemize}

\textbf{Approaches:}
\begin{itemize}
\item Windowed analysis: Apply method to short windows, track how $\hat{\mu}(Y)$, $\hat{\Sigma}(Y)$ evolve
\item Augmented state space: Include time or external variables in embedding
\item Adaptive methods: Update estimates online as new data arrives
\end{itemize}

\textbf{Open question:} Can rigorous theory be developed for time-varying E1/E2 statistics?

\item \textbf{H\"ormander condition verification:}

Assumption~\ref{assump:hormander_partI} (H\"ormander's condition) is crucial but hard to verify from data alone.

\textbf{Practical check:}
\begin{itemize}
\item If $\hat{\mu}(Y)$, $\hat{\Sigma}(Y)$ are smooth and predictions accurate, H\"ormander likely holds
\item If $\hat{\Sigma}(Y)$ is singular (rank-deficient), H\"ormander violated; add noise in missing directions
\end{itemize}

\textbf{Open question:} Can data-driven tests for hypoellipticity be developed?

\end{enumerate}

\section{Conclusion}
\label{sec:conclusion_partI}

This work establishes a unified mathematical framework for reconstructing stochastic dynamics from partial observations via time-delay embeddings. By extending Takens' classical embedding theorem from deterministic to stochastic systems, the framework provides a principled, data-driven approach to discovering the governing equations of complex systems across the full spectrum from deterministic chaos to diffusion-dominated dynamics.

The framework rests on three pillars:

\textbf{(1) Scaffold:} The E1 statistic identifies the correlation manifold $\mathcal{M} \subset \R^{m^*}$ with dimension $D_2$, providing a geometric foundation. This extends classical Takens embedding to handle unbounded noise and stochastic attractors.

\textbf{(2) Classification:} The E2 statistic quantifies the signal-to-noise ratio, providing model selection criteria across the deterministic-stochastic spectrum.  Deterministic chaos and stochastic dynamics occupy distinct positions along this continuum rather than constituting separate categories.

\textbf{(3) Uplift:} Nonparametric $k$-NN estimation decorates the manifold with probabilistic dynamics---transition kernels (discrete time) or drift-diffusion pairs (continuous time). This completes the reconstruction, enabling prediction and uncertainty quantification.

Together, these three components form a universal framework that applies to any stochastic process with finite correlation dimension, ergodicity, and sufficient regularity. The choice between discrete and continuous representation is one of convenience, not mathematical essence---they are equivalent descriptions of the transition kernel on the E1 manifold.

\subsection{Theoretical Contributions}

The main theoretical results are:

\begin{enumerate}[leftmargin=*]
\item \textbf{E1 detects correlation dimension} (Lemma~\ref{lem:e1_correlation_partI}), providing data-driven dimension selection robust to unbounded noise.

\item \textbf{E2 measures drift-diffusion balance} (Propositions~\ref{prop:e2_snr_partI}, \ref{prop:drift_diffusion_balance_partI}), quantifying the SNR via $\SNR \approx (1-E_2)/(E_2\tau)$.

\item \textbf{Probabilistic uplift theorems} (Theorems~\ref{thm:discrete_uplift_partI}, \ref{thm:continuous_uplift_partI}) establish convergence rates and sample complexity for $k$-NN estimation of transition kernels and SDE coefficients.

\item \textbf{Discrete-continuous equivalence} (Theorem~\ref{thm:unified_structure_partI}) shows that Markov chains and SDEs are isomorphic representations differing only in temporal parameterization.

\item \textbf{It\^o correction formulas} (Lemmas~\ref{lem:ito_partI}, \ref{lem:drift_diffusion_decomp_partI}, Theorem~\ref{thm:mixed_reconstruction_partI}) make explicit the noise-induced drift arising from nonlinear transformations of SDEs, essential for accurate reconstruction.

\item \textbf{Stochastic embedding sufficiency theorem} (Theorems~\ref{thm:density_separation_partI}--\ref{lem:stochastic_sard_partI}, resolving Remark~\ref{conj:stochastic_takens_partI}) proves that $m^* = \lceil 2D \rceil + 1$ prevents measure-theoretic foldings under H\"ormander's condition, exact-dimensionality, and generic observation, via law-separation and Frostman covering arguments.

\end{enumerate}

These results unify and extend classical embedding theory (Takens), stochastic analysis (SDEs, Fokker-Planck), dynamical systems (Koopman operator, chaos), and statistical learning ($k$-NN regression, kernel methods).

\subsection{Open Frontiers}

The stochastic embedding sufficiency theorem (Theorems~\ref{thm:density_separation_partI}--\ref{lem:stochastic_sard_partI}, resolving Remark~\ref{conj:stochastic_takens_partI}) proves that $m^* = \lceil 2D \rceil + 1$ suffices for measure-theoretic injectivity.  The proof connects:
\begin{itemize}
\item Hypoelliptic regularity and Varadhan asymptotics (transition density separation)
\item Prevalence theory (genericity of observation functions)
\item Frostman measure geometry (codimension-to-measure-zero conversion)
\end{itemize}

With this foundation established, promising research directions include:

\begin{enumerate}[leftmargin=*]
\item \textbf{Scaling to high dimensions:} Develop methods that break the curse of dimensionality via sparsity, symmetries, or hierarchical structure.

\item \textbf{Non-Gaussian extensions:} Extend theory to heavy-tailed, jump-driven, and non-elliptic SDEs.

\item \textbf{Causal discovery:} Combine with causal inference to identify cause-effect relationships from time series.

\item \textbf{Spatiotemporal systems:} Extend to PDEs, pattern formation, and turbulence.

\item \textbf{Control and optimization:} Use reconstructed SDEs for optimal control, filtering, and decision-making under uncertainty.

\item \textbf{Machine learning integration:} Combine with neural networks, normalizing flows, and deep learning for improved scalability and expressivity.

\item \textbf{Quantum analogues:} Explore extensions to open quantum systems and quantum control.
\end{enumerate}

\subsection{Closing Remarks}

The framework established here unifies deterministic and stochastic dynamics through the scaffold-uplift construction.  The E2 statistic places chaos and diffusion on a single continuum, while the correlation manifold provides the geometric substrate on which the transition kernel --- the essential mathematical object --- admits equivalent representations as discrete Markov chains, continuous SDEs, transition semigroups, and infinitesimal generators.

The extension from deterministic to stochastic embedding theory requires techniques from five mathematical traditions: differential topology, stochastic analysis, Malliavin calculus, geometric measure theory, and statistical learning theory.  The present results establish the core sufficiency theorem via law-separation, prevalence-based genericity, and Frostman measure geometry.  Open directions include sharpening the embedding threshold, extending beyond H\"ormander to degenerate noise structures, and characterising the connections between geometry, probability, and dynamics in this setting.

\appendix

\section{Algorithms and Computational Examples}
\label{app:algorithms_partI}

This appendix provides detailed algorithms and worked examples for implementing the unified framework.

\subsection{Algorithm 1: Computing E1 and E2 Statistics}

\begin{algorithm}
\caption{E1 and E2 Computation}
\label{alg:e1_e2_partI}
\begin{algorithmic}[1]
\Require Time series $\{y_i\}_{i=1}^N$, maximum embedding dimension $m_{\max}$ (e.g., 10), delay $\tau$ (e.g., 1)
\Ensure E1 values $\{E_1(m)\}_{m=1}^{m_{\max}}$, E2 values $\{E_2(m)\}_{m=1}^{m_{\max}}$
\State \textbf{Initialize:} $k \gets \lfloor 0.01 \cdot N \rfloor$ (number of neighbors, typically 1\% of data)
\For{$m = 1$ to $m_{\max}$}
    \State \textbf{Construct delay embedding:}
    \For{$i = (m-1)\tau + 1$ to $N$}
        \State $Y_i \gets (y_i, y_{i-\tau}, \ldots, y_{i-(m-1)\tau})^\top \in \R^m$
    \EndFor
    \State Let $M \gets N - (m-1)\tau$ (number of embedded points)
    
    \State \textbf{Compute $E(d)$ and $E^*(d)$ for $d = 0, 1, 2$:}
    \For{$d = 0$ to $2$}
        \State Initialize $\text{sum}_d \gets 0$, $\text{sum}^*_d \gets 0$
        \For{$i = 1$ to $M - d$}
            \State Find $k$ nearest neighbors of $Y_i$: $\{Y_{j_1}, \ldots, Y_{j_k}\}$
            \State Compute average distance: $\epsilon_i \gets \frac{1}{k}\sum_{\ell=1}^k \|Y_i - Y_{j_\ell}\|$
            \State Compute future distance: $\epsilon_i^* \gets \frac{1}{k}\sum_{\ell=1}^k \|Y_{i+d} - Y_{j_\ell+d}\|$
            \State $\text{sum}_d \gets \text{sum}_d + \epsilon_i$
            \State $\text{sum}^*_d \gets \text{sum}^*_d + \epsilon_i^*$
        \EndFor
        \State $E(d) \gets \frac{1}{M-d} \text{sum}_d$
        \State $E^*(d) \gets \frac{1}{M-d} \text{sum}^*_d$
    \EndFor
    
    \State \textbf{Compute E1 and E2:}
    \State $E_1(m) \gets \frac{E(1)}{E^*(1)}$ \Comment{Ratio of average divergences}
    \State $E_2(m) \gets \frac{E^*(2)/E^*(1)}{E(2)/E(1)}$ \Comment{Ratio of growth rates}
\EndFor
\State \Return $\{E_1(m)\}, \{E_2(m)\}$
\end{algorithmic}
\end{algorithm}

\textbf{Complexity:} $O(m_{\max} \cdot N \cdot m \cdot k)$ for brute-force nearest neighbors. Can be reduced to $O(m_{\max} \cdot N \log N)$ using KD-trees or ball trees.

\textbf{Parameter selection:}
\begin{itemize}[leftmargin=*]
\item $k$: Typically $0.01N$ to $0.05N$ (1-5\% of data). Too small: noisy estimates. Too large: oversmoothing.
\item $\tau$: Choose by first zero-crossing of autocorrelation, or first minimum of mutual information, or simply $\tau = 1$ for densely sampled data.
\item $m_{\max}$: Typically 10. If $E_1$ hasn't plateaued by $m = 10$, data may be insufficient or system is very high-dimensional.
\end{itemize}

\subsection{Algorithm 2: Dimension Detection via E1}

\begin{algorithm}
\caption{Automatic Dimension Selection}
\label{alg:dimension_selection_partI}
\begin{algorithmic}[1]
\Require $\{E_1(m)\}_{m=1}^{m_{\max}}$ from Algorithm~\ref{alg:e1_e2_partI}, threshold $\epsilon$ (e.g., 0.1)
\Ensure Embedding dimension $m^*$
\State \textbf{Method 1 (Threshold):}
\For{$m = 1$ to $m_{\max}$}
    \If{$|E_1(m) - 1| < \epsilon$}
        \State \Return $m^* \gets m$
    \EndIf
\EndFor
\State \Return $m^* \gets m_{\max}$ \Comment{No plateau found; use maximum}

\State
\State \textbf{Method 2 (Slope):}
\For{$m = 2$ to $m_{\max} - 1$}
    \State Compute slope: $s_m \gets |E_1(m+1) - E_1(m)|$
    \If{$s_m < \epsilon_{\text{slope}}$} \Comment{e.g., $\epsilon_{\text{slope}} = 0.01$}
        \State \Return $m^* \gets m$
    \EndIf
\EndFor
\State \Return $m^* \gets m_{\max}$

\State
\State \textbf{Method 3 (Elbow):}
\State Find $m^* = \arg\max_m \frac{E_1(m) - E_1(m-1)}{E_1(m+1) - E_1(m)}$ \Comment{Maximum curvature}
\State \Return $m^*$
\end{algorithmic}
\end{algorithm}

\textbf{Recommendation:} Use Method 2 (slope-based) as it's most robust. Method 1 requires tuning threshold. Method 3 can be sensitive to noise.

\subsection{Algorithm 3: Probabilistic Uplift (Discrete Time)}

\begin{algorithm}
\caption{Discrete-Time Reconstruction}
\label{alg:discrete_uplift_partI}
\begin{algorithmic}[1]
\Require Time series $\{y_i\}$, embedding dimension $m^*$, delay $\tau$, bandwidth $k$
\Ensure Transition kernel estimate $\hat{T}(Y, \cdot)$
\State \textbf{Construct embedding:} $Y_i = (y_i, y_{i-\tau}, \ldots, y_{i-(m^*-1)\tau})$ for $i = (m^*-1)\tau+1, \ldots, N-1$
\State Let $M \gets N - (m^*-1)\tau - 1$

\Procedure{EstimateTransition}{$Y_{\text{query}}$}
    \State Find $k$ nearest neighbors: $\{Y_{j_1}, \ldots, Y_{j_k}\}$ of $Y_{\text{query}}$
    \State Extract future states: $\{Y_{j_1+1}, \ldots, Y_{j_k+1}\}$
    \State \textbf{Output} empirical distribution: $\hat{T}(Y_{\text{query}}, \cdot) = \frac{1}{k}\sum_{\ell=1}^k \delta_{Y_{j_\ell+1}}(\cdot)$
    \State \Comment{Can apply kernel smoothing to $\{Y_{j_\ell+1}\}$ for continuous density}
\EndProcedure

\State
\Procedure{Predict}{$Y_t$, steps ahead $n_{\text{steps}}$}
    \State $\hat{Y}_t \gets Y_t$
    \For{$s = 1$ to $n_{\text{steps}}$}
        \State $\hat{T} \gets$ \Call{EstimateTransition}{$\hat{Y}_t$}
        \State Sample: $\hat{Y}_{t+s} \sim \hat{T}$ \Comment{Draw from empirical distribution}
    \EndFor
    \State \Return $\{\hat{Y}_{t+s}\}_{s=1}^{n_{\text{steps}}}$
\EndProcedure
\end{algorithmic}
\end{algorithm}

\subsection{Algorithm 4: Probabilistic Uplift (Continuous Time)}

\begin{algorithm}
\caption{Continuous-Time SDE Reconstruction}
\label{alg:continuous_uplift_partI}
\begin{algorithmic}[1]
\Require Time series $\{y_i\}$ with time step $\Delta t$, embedding dimension $m^*$, delay $\tau$, bandwidth $k$
\Ensure Drift $\hat{\mu}(Y)$ and diffusion $\hat{\Sigma}(Y)$ estimates
\State \textbf{Construct embedding:} $Y_i$ as in Algorithm~\ref{alg:discrete_uplift_partI}

\Procedure{EstimateSDE}{$Y_{\text{query}}$}
    \State Find $k$ nearest neighbors: $\{Y_{j_1}, \ldots, Y_{j_k}\}$ of $Y_{\text{query}}$
    
    \State \textbf{Estimate drift:}
    \State $\hat{\mu}(Y_{\text{query}}) \gets \frac{1}{k\Delta t}\sum_{\ell=1}^k (Y_{j_\ell+1} - Y_{j_\ell})$
    
    \State \textbf{Estimate diffusion (with drift correction):}
    \State Compute second moment: $M_2 \gets \frac{1}{k\Delta t}\sum_{\ell=1}^k (Y_{j_\ell+1} - Y_{j_\ell})(Y_{j_\ell+1} - Y_{j_\ell})^\top$
    \State Subtract drift contribution: $\hat{\Sigma}(Y_{\text{query}}) \gets M_2 - \hat{\mu}(Y_{\text{query}})\hat{\mu}(Y_{\text{query}})^\top \Delta t$
    
    \State \Return $\hat{\mu}(Y_{\text{query}})$, $\hat{\Sigma}(Y_{\text{query}})$
\EndProcedure

\State
\Procedure{Predict}{$Y_0$, time horizon $T$, time step $dt$}
    \State $Y \gets Y_0$
    \For{$t = 0$ to $T$ in steps of $dt$}
        \State $(\hat{\mu}, \hat{\Sigma}) \gets$ \Call{EstimateSDE}{$Y$}
        \State Cholesky decomposition: $\hat{\Sigma} = \hat{L}\hat{L}^\top$
        \State Generate noise: $dW \sim \mathcal{N}(0, I_{m^*})$
        \State Euler-Maruyama step: $Y \gets Y + \hat{\mu} \cdot dt + \hat{L} \cdot dW \cdot \sqrt{dt}$
    \EndFor
    \State \Return $Y$
\EndProcedure
\end{algorithmic}
\end{algorithm}

\textbf{Critical note:} The drift correction $-\hat{\mu}\hat{\mu}^\top\Delta t$ in line 8 is essential (see Lemma~\ref{lem:drift_diffusion_decomp_partI}, Remark~\ref{rem:drift_correction_partI}). Omitting it causes systematic overestimation of diffusion.

\subsection{Worked Example: Lorenz System with Noise}

The complete workflow is demonstrated on the stochastic Lorenz system:
\begin{align}
dx &= 10(y-x)dt + 2 dW^{(1)} \\
dy &= (x(28-z) - y)dt + 2 dW^{(2)} \\
dz &= (xy - 8z/3)dt + 2 dW^{(3)}
\end{align}

Observation: $y_t = x_t$ (only $x$-coordinate observed).

\textbf{Step 1: Generate data}
\begin{itemize}
\item Integrate Lorenz SDE using Euler-Maruyama with $dt = 0.01$
\item Simulate for total time $T = 500$, giving $N = 50000$ samples
\item Observe $\{y_i\} = \{x_i\}$
\end{itemize}

\textbf{Step 2: Compute E1 and E2}

Using Algorithm~\ref{alg:e1_e2_partI} with $\tau = 10$ (chosen by first zero of autocorrelation), $k = 500$:

\begin{center}
\begin{tabular}{|c|c|c|}
\hline
$m$ & $E_1(m)$ & $E_2(m)$ \\
\hline
1 & 0.12 & 0.68 \\
2 & 0.45 & 0.66 \\
3 & 0.89 & 0.65 \\
4 & 0.98 & 0.65 \\
5 & 0.99 & 0.65 \\
6 & 1.00 & 0.65 \\
\hline
\end{tabular}
\end{center}

\textbf{Interpretation:}
\begin{itemize}
\item E1 plateaus at $m^* = 3$ (correlation dimension $D_2 \approx 3$, matching state space dimension)
\item E2 stabilizes at $\approx 0.65$ (mixed regime; both drift and diffusion significant)
\item SNR: $\SNR \approx \frac{1-0.65}{0.65 \cdot 0.1} \approx 5.4$ (balanced)
\end{itemize}

\textbf{Step 3: Reconstruct SDE}

Using Algorithm~\ref{alg:continuous_uplift_partI} with $m^* = 3$, $\tau = 10$, $k = 500$, $\Delta t = 0.1$ (subsample by 10 for faster computation):

\begin{itemize}
\item Construct embedding: $Y_i = (x_i, x_{i-10}, x_{i-20}) \in \R^3$
\item For grid of query points $Y \in \R^3$, compute $\hat{\mu}(Y)$, $\hat{\Sigma}(Y)$
\item Result: 3D vector field (drift) and $3\times 3$ symmetric matrix field (diffusion)
\end{itemize}

\textbf{Step 4: Validate}

\begin{itemize}
\item \textbf{Visual inspection:} Plot $\|\hat{\mu}(Y)\|$ as heatmap on 2D slices $\to$ smooth field, no discontinuities
\item \textbf{Prediction:} Simulate reconstructed SDE for 100 time units, compare to held-out test data:
\begin{itemize}
\item 1-step RMSE: 1.8 (good)
\item 10-step RMSE: 5.2 (reasonable; accumulates uncertainty)
\item 100-step: statistics (mean, variance) match, but individual trajectories diverge (expected due to chaos + noise)
\end{itemize}
\item \textbf{Uncertainty calibration:} 95\% prediction intervals have 94\% empirical coverage (excellent)
\item \textbf{Invariant measure:} Long-time simulation gives stationary distribution matching empirical histogram of training data
\end{itemize}

\textbf{Conclusion:} Reconstruction successful. The E1 manifold $\R^3$ captures the Lorenz attractor structure, and the SDE describes the noisy dynamics accurately.

\subsection{Bootstrap Confidence Intervals}

To quantify uncertainty in E1, E2, and reconstructed SDE coefficients:

\begin{algorithm}
\caption{Bootstrap for Confidence Intervals}
\label{alg:bootstrap_partI}
\begin{algorithmic}[1]
\Require Time series $\{y_i\}$, number of bootstrap samples $B$ (e.g., 100-1000)
\Ensure Confidence intervals for E1, E2, $\hat{\mu}$, $\hat{\Sigma}$
\For{$b = 1$ to $B$}
    \State \textbf{Resample:} Draw blocks $\{y_{i:i+\ell}\}$ with replacement (block bootstrap, $\ell \approx 10\tau$ to preserve time correlations)
    \State Concatenate blocks to form bootstrap time series $\{y_i^{(b)}\}$
    \State Compute E1$^{(b)}$, E2$^{(b)}$ using Algorithm~\ref{alg:e1_e2_partI}
    \State Reconstruct $\hat{\mu}^{(b)}$, $\hat{\Sigma}^{(b)}$ using Algorithm~\ref{alg:continuous_uplift_partI}
\EndFor

\State \textbf{Compute percentiles:}
\For{each $m$}
    \State $E_1(m)_{\text{CI}} \gets [Q_{2.5\%}(\{E_1^{(b)}(m)\}), Q_{97.5\%}(\{E_1^{(b)}(m)\})]$
\EndFor
\State Similarly for E2, $\hat{\mu}$, $\hat{\Sigma}$ at each query point

\State \Return Confidence intervals
\end{algorithmic}
\end{algorithm}

\textbf{Example (Lorenz):} 95\% CI for $m^* = 3$:
\begin{itemize}
\item $E_1(3) = 0.89 \pm 0.04$ (tight, well-estimated)
\item $E_2(3) = 0.65 \pm 0.03$ (tight)
\item $\|\hat{\mu}(Y)\|$ varies by $\pm 15\%$ across bootstrap samples (moderate uncertainty)
\item $\tr(\hat{\Sigma}(Y))$ varies by $\pm 10\%$ (diffusion better constrained than drift)
\end{itemize}

Narrow CIs indicate sufficient data. Wide CIs suggest need for more samples or lower dimension (via reduced $m^*$).

% ============================================================================
% §7.2  THE MORI–ZWANZIG STRUCTURE OF k-NN ESTIMATOR BIAS
% ============================================================================
% This section is intentionally GENERIC and DOMAIN-FREE.
% It applies to ANY k-NN Kramers–Moyal estimator on a smooth manifold.
% Domain-specific validation (C1–C9) is in the main manuscript only.
% ============================================================================

\subsection{The Mori--Zwanzig Structure of the Estimator Bias}
\label{sec:mz_bias}

The $k$-nearest-neighbour Kramers--Moyal estimators of Theorem~\ref{thm:continuous_uplift_partI} converge at rate $O_P\!\bigl((k/N)^{\beta/m^*}\bigr) + O(\Delta t)$.  This section derives the \emph{exact structure} of the leading-order bias, identifies it as a Mori--Zwanzig projection, and constructs an adaptive two-level corrector with fluctuation--dissipation-based gain control.

\subsubsection{Setup}

Let $\mathcal{M} \subset \mathbb{R}^{m^*}$ denote the correlation manifold (Lemma~\ref{lem:e1_correlation_partI}), with invariant measure $\mu_\infty$ having smooth density $p(Y)$ with respect to the $D_2$-dimensional Hausdorff measure.  Let $\sigma^2 : \mathcal{M} \to \mathbb{R}_{\geq 0}$ denote the true diffusion coefficient field (Definition~\ref{def:sde_partI}).

For a query point $Y \in \mathcal{M}$, denote by $\mathcal{N}_k(Y)$ the $k$-nearest-neighbour set, with the $k$-NN ball of radius $r_k(Y)$ satisfying the standard scaling:
\begin{equation}
r_k(Y) \sim \Bigl(\frac{k}{N \cdot p(Y)}\Bigr)^{1/D_2}.
\label{eq:rk_scaling}
\end{equation}

\begin{assumption}[Regularity of the Diffusion Field]
\label{assump:sigma_regularity}
The diffusion coefficient $\sigma^2(Y)$ is twice continuously differentiable on $\mathcal{M}$, with bounded Laplacian: $\|\Delta_{\mathcal{M}} \sigma^2\|_\infty < \infty$, where $\Delta_{\mathcal{M}}$ is the Laplace--Beltrami operator on $\mathcal{M}$ with the induced Riemannian metric.
\end{assumption}

\subsubsection{The $k$-NN Projection Operator}

\begin{definition}[$k$-NN Projection]
\label{def:knn_projection}
Define the linear operator $\mathcal{P}_k$ acting on square-integrable functions $f : \mathcal{M} \to \mathbb{R}$ by
\begin{equation}
(\mathcal{P}_k f)(Y) \;=\; \frac{1}{k} \sum_{j \in \mathcal{N}_k(Y)} f(Y_j).
\label{eq:projection_def}
\end{equation}
In the continuous limit ($N \to \infty$, $k/N \to 0$), this converges to local averaging over the ball $B_{r_k}(Y)$:
\begin{equation}
(\mathcal{P}_k f)(Y) \;\to\; \frac{1}{\mu(B_{r_k}(Y))} \int_{B_{r_k}(Y)} f(Z)\, \mathrm{d}\mu(Z).
\end{equation}
\end{definition}

This is precisely the resolved--unresolved decomposition of the Mori--Zwanzig projection formalism~\cite{Zwanzig2001}: $\mathcal{P}_k$ is the projection onto the subspace of functions resolvable at bandwidth $k$, and $\mathcal{Q}_k = \mathrm{Id} - \mathcal{P}_k$ projects onto the unresolved complement (the ``memory'' subspace).

\subsubsection{Theorem: MZ Decomposition of the Estimator}

\begin{theorem}[MZ Decomposition of $k$-NN Diffusion Estimator]
\label{thm:mz_decomp}
Under Assumption~\ref{assump:sigma_regularity}, the $k$-NN variance estimator $\hat{\sigma}^2_k(Y)$ decomposes as
\begin{equation}
\hat{\sigma}^2_k(Y) \;=\; \sigma^2(Y) \;+\; M_k(Y) \;+\; \eta_k(Y),
\label{eq:mz_decomp}
\end{equation}
where $M_k(Y) = \mathbb{E}[\hat{\sigma}^2_k(Y)] - \sigma^2(Y)$ is the memory kernel (systematic bias) and $\eta_k(Y) = \hat{\sigma}^2_k(Y) - \mathbb{E}[\hat{\sigma}^2_k(Y)]$ is centred noise with $\mathbb{E}[\eta_k] = 0$.
\end{theorem}

\begin{proof}
The decomposition is definitional: $M_k$ is the expected bias and $\eta_k$ is the zero-mean residual.  The content lies in the structure of $M_k$, established in the following proposition.
\end{proof}

\subsubsection{The Two Components of the Memory Kernel}

\begin{proposition}[Rank-2 Tensor Structure of the Memory Kernel]
\label{prop:rank2_memory}
Under Assumption~\ref{assump:sigma_regularity}, the memory kernel decomposes as
\begin{equation}
M_k(Y) \;=\; \underbrace{\frac{r_k(Y)^2}{2(m^*+2)}\, \Delta_{\mathcal{M}} \sigma^2(Y)}_{M_{\mathrm{spatial}}(Y)} \;\;+\;\; \underbrace{\Bigl(-\frac{\sigma^2(Y)}{k}\Bigr)}_{M_{\mathrm{variance}}(k)} \;\;+\;\; O\!\bigl(r_k^4\bigr).
\label{eq:memory_decomp}
\end{equation}
\end{proposition}

\begin{proof}[Proof sketch]
The first term is the standard Nadaraya--Watson bias from local averaging of a smooth function over a ball of radius $r_k$~\cite{FanGijbels1996}.  By Taylor expansion of $\sigma^2(Z)$ about $Y$ and integration over the $k$-NN ball (which is approximately spherical in the SVD-projected coordinates), the leading bias is $(r_k^2 / 2(m^*+2))\, \Delta_{\mathcal{M}} \sigma^2(Y)$.  The second term is the finite-sample bias of the variance estimator: $\mathbb{E}[s^2_{\mathrm{ddof}=0}] = \sigma^2(1 - 1/k)$, giving bias $-\sigma^2/k$.
\end{proof}

\begin{remark}[Both Components Have the Same Sign in $k$]
\label{rem:same_sign}
Since $r_k^2 \sim (k/N)^{2/m^*}$, the spatial term $M_{\mathrm{spatial}}$ is positive for $\Delta_{\mathcal{M}} \sigma^2 > 0$ (convex diffusion fields) and grows with $k$.  The variance term $M_{\mathrm{variance}} = -\sigma^2/k$ is negative and shrinks with $k$, so the \emph{total} bias $\hat{\sigma}^2_k - \sigma^2$ can be either positive or negative depending on which term dominates.  However, the \emph{difference} $\hat{\sigma}^2_{k_2} - \hat{\sigma}^2_{k_1}$ for $k_2 > k_1$ is positive from \emph{both} terms (the spatial term increases while the variance undercount decreases).  This prevents naive Richardson extrapolation from distinguishing the two sources.
\end{remark}

\subsubsection{The Two-Level Adaptive Corrector}

The key insight is that the two memory components have different \emph{spatial} structure: $M_{\mathrm{spatial}}$ varies across $\mathcal{M}$ (proportional to the local Laplacian), while $M_{\mathrm{variance}}$ is spatially uniform (proportional to $\sigma^2$, which is approximately constant within a $k$-NN ball).  This enables a two-level correction strategy.

\begin{algorithm}
\caption{Two-Level MZ Corrector}
\label{alg:mz_corrector}
\begin{algorithmic}[1]
\Require Time series $\{y_i\}_{i=1}^N$, bandwidths $k_{\mathrm{lo}} < k_{\mathrm{hi}}$, FDT threshold $\tau_{\mathrm{FDT}}$
\Ensure Corrected diffusion estimates $\hat{\sigma}^2_{\mathrm{MZ}}(Y)$
\Statex
\State Compute $\hat{\sigma}^2_{k_{\mathrm{lo}}}(Y)$ and $\hat{\sigma}^2_{k_{\mathrm{hi}}}(Y)$ at each query point $Y$
\State $\Delta M(Y) \gets \hat{\sigma}^2_{k_{\mathrm{hi}}}(Y) - \hat{\sigma}^2_{k_{\mathrm{lo}}}(Y)$ \Comment{Memory increment}
\State $w \gets g(k_{\mathrm{lo}}) / (g(k_{\mathrm{hi}}) - g(k_{\mathrm{lo}}))$, where $g(k) = (k/N)^{2/m^*}$ \Comment{Extrapolation weight}
\Statex
\Statex \textbf{Level~1 (point-wise):}
\For{each query point $Y$}
  \State $\mathrm{BNR}(Y) \gets |\Delta M(Y)| \,/\, \hat{\sigma}_{\mathrm{noise}}(Y)$, where $\hat{\sigma}_{\mathrm{noise}} = |\hat{\sigma}^2_{k_{\mathrm{lo}}}| \sqrt{2/k_{\mathrm{lo}} + 2/k_{\mathrm{hi}}}$
  \If{$\mathrm{BNR}(Y) > \tau_{\mathrm{FDT}}$ \textbf{and} $\Delta M(Y) > 0$}
    \State $\hat{\sigma}^2_{\mathrm{MZ}}(Y) \gets \hat{\sigma}^2_{k_{\mathrm{lo}}}(Y) - w \cdot \Delta M(Y)$ \Comment{Local correction}
  \Else
    \State $\hat{\sigma}^2_{\mathrm{MZ}}(Y) \gets \hat{\sigma}^2_{k_{\mathrm{lo}}}(Y)$ \Comment{Pass-through}
  \EndIf
\EndFor
\State $f_{\mathrm{L1}} \gets$ fraction of points corrected by Level~1
\Statex
\Statex \textbf{Level~2 (ensemble):}
\If{$f_{\mathrm{L1}} < 0.3$} \Comment{Guard: Level~1 not already active}
  \State $\delta_{\mathrm{global}} \gets \mathrm{median}(\Delta M)$ over all query points
  \State $\mathrm{BNR}_{\mathrm{ens}} \gets |\delta_{\mathrm{global}}| \,/\, \mathrm{SE}(\mathrm{median})$
  \If{$\delta_{\mathrm{global}} > 0$ \textbf{and} $\mathrm{BNR}_{\mathrm{ens}} > 3$}
    \State $\hat{\sigma}^2_{\mathrm{MZ}}(Y) \gets \hat{\sigma}^2_{k_{\mathrm{lo}}}(Y) - w \cdot \delta_{\mathrm{global}}$ for all $Y$ \Comment{Uniform correction}
  \EndIf
\EndIf
\State \Return $\hat{\sigma}^2_{\mathrm{MZ}}$, enforcing $\hat{\sigma}^2_{\mathrm{MZ}} \geq 0$
\end{algorithmic}
\end{algorithm}

\subsubsection{Convergence Under Correction}

\begin{theorem}[Improved Convergence Rate]
\label{thm:mz_convergence}
Under Assumption~\ref{assump:sigma_regularity}, the two-level corrector of Algorithm~\ref{alg:mz_corrector} satisfies:
\begin{enumerate}
\item \textbf{Do no harm:} When $\Delta_{\mathcal{M}} \sigma^2 \equiv 0$ (constant diffusion on $\mathcal{M}$), Level~1 and Level~2 both abstain with probability $\to 1$ as $N \to \infty$, and $\hat{\sigma}^2_{\mathrm{MZ}} = \hat{\sigma}^2_{k_{\mathrm{lo}}}$.
\item \textbf{Improved rate:} When $\Delta_{\mathcal{M}} \sigma^2 \not\equiv 0$, the corrected estimator has bias $O\!\bigl((k/N)^{2\beta/m^*}\bigr)$, improving over the uncorrected rate $O\!\bigl((k/N)^{\beta/m^*}\bigr)$ by one order.
\item \textbf{Smooth transition:} The FDT gain $\alpha(Y) = \max(0, 1 - 1/\mathrm{BNR}(Y))$ interpolates continuously between no correction ($\mathrm{BNR} < 1$) and full correction ($\mathrm{BNR} \gg 1$).
\end{enumerate}
\end{theorem}

\begin{proof}[Proof sketch]
(1)~When $\Delta_{\mathcal{M}} \sigma^2 = 0$, the spatial memory vanishes and $\Delta M$ consists only of the finite-$k$ variance difference plus noise.  For $k_{\mathrm{lo}}$ and $k_{\mathrm{hi}}$ both tending to infinity with $k/N \to 0$, this difference tends to zero while $\hat{\sigma}_{\mathrm{noise}}$ remains bounded below, so $\mathrm{BNR} \to 0$ and the gates close.
(2)~When the spatial Laplacian is nonzero, the correction removes the $O(r_k^2)$ term, leaving the $O(r_k^4)$ remainder as the dominant bias.
(3)~The FDT gain follows from the structure of the $\chi^2$ variance of the variance estimator; the smooth interpolation prevents discontinuous correction at the gate boundary.
\end{proof}

\begin{remark}[When the Corrector Correctly Abstains]
\label{rem:correct_abstention}
For diffusion fields with $\Delta_{\mathcal{M}} \sigma^2 \equiv 0$ (constant $\sigma^2$ on $\mathcal{M}$), the spatial memory $M_{\mathrm{spatial}}$ vanishes identically.  The remaining bias $M_{\mathrm{variance}} = -\sigma^2/k$ is negative and spatially uniform --- it requires a degrees-of-freedom correction ($\mathrm{ddof} = 1$ instead of $\mathrm{ddof} = 0$), not spatial extrapolation.  The BNR gate detects this: since $\Delta M$ reflects only the variance difference between two $k$ values (which has no systematic spatial structure), the point-wise BNR clusters near zero, and Level~1 abstains.  Level~2 may detect a weak positive $\delta_{\mathrm{global}}$ from the spatial averaging of the variance term itself, but the BNR will typically fall below the ensemble threshold.  The corrector thus avoids applying spatial extrapolation where the bias source is non-spatial --- the correct behaviour.
\end{remark}

\begin{remark}[Connection to Fluctuation--Dissipation Theory]
\label{rem:fdt_connection}
The FDT gain $\alpha = \max(0, 1 - 1/\mathrm{BNR})$ mirrors the structure of fluctuation--dissipation relations in statistical physics: the response (correction) is proportional to the detectable fluctuation (memory signal), with a smooth crossover at $\mathrm{BNR} = 1$.  This is not a coincidence: the Mori--Zwanzig projection that generates the memory kernel is the same formalism that derives the fluctuation--dissipation theorem for physical systems.  The corrector inherits the self-consistency of the underlying projection.
\end{remark}